%% file: draft_IslamHistory.tex
\documentclass[a4,12pt]{article}
\usepackage{titletoc}
\usepackage{titlesec}
\titleformat{\section}[block]{\Large\bfseries\filcenter}{\thesection}{1em}{}
\usepackage{multibib}
\newcites{App}{References for the Appendix}
\usepackage{amsmath}
\usepackage{amsthm}
\usepackage{amsfonts}
\usepackage{latexsym}
\usepackage{multirow}
\usepackage{caption}
\usepackage{lscape}
\usepackage[dvipdfmx]{graphicx}
\usepackage{url}
\usepackage{setspace}
\usepackage{color}
\usepackage{booktabs} % for much better looking tables
\usepackage{array} % for better arrays (eg matrices) in maths
\usepackage{paralist} % very flexible & customisable lists (eg. enumerate/itemize, etc.)
\usepackage{verbatim} % adds environment for commenting out blocks of text & for better verbatim
\usepackage{subfig} % make it possible to include more than one captioned figure/table in a single float
\usepackage{amsmath}
\usepackage{graphicx}
\usepackage[flushleft]{threeparttable} %added by MK
\usepackage{amssymb}
\usepackage{threeparttable}
\usepackage{booktabs,tabularx}
\usepackage{longtable}
\usepackage{adjustbox}

\newcommand{\quotes}[1]{``#1''}
\newcommand{\ST}[1]{\textcolor{red!81!black}{[\textbf{ST:} #1]}}
\newcommand{\MK}[1]{\textcolor{blue!81!black}{[\textbf{MK:} #1]}}
\usepackage{enumitem,amssymb}
\newlist{todolist}{itemize}{3}
\setlist[todolist]{label=$\square$}
\usepackage{todonotes}
\newcommand{\STmemo}[1]{\vspace{0.75cm} \todo[inline,color=red!20!white]{\textbf{ST:} #1}}

\newcommand{\blue}{\textcolor{blue}}
\newcommand{\red}{\textcolor{red}}

\usepackage[T1]{fontenc}
\usepackage{color}
\definecolor{myblue}{rgb}{0,0,0.7}
\definecolor{myred}{rgb}{0.7,0,0}
\definecolor{mygreen}{rgb}{0,0.42,0}
\usepackage[linktocpage=true]{hyperref}
\hypersetup{ colorlinks, linkcolor=myblue, urlcolor=myblue, citecolor=myblue }

\usepackage[authoryear]{natbib}
\usepackage{delarray}
\usepackage[top=31.75mm,bottom=31.75mm,left=31.75mm,right=31.75mm]{geometry} % 1.25 inch, adjusting to the ECTA format % Miyauchi-san's WP has 34 lines per page. Same as ours. So, onehalf space would be ok.

\usepackage{times}     % This also saves space
\title{\textbf{Jihad over Centuries}\footnote{
\scriptsize
We are grateful to Andrew Foster and Stelios Michalopoulos for continuous guidance and support, and Shuhei Kitamura and Jared Rubin for invaluable discussions.
For helpful comments, we thank Eren Arbatli, Alberto Bisin, Dan Bj{\"o}rkegren, Gabriel Brown, Thomas Cornelissen, Jon Denton-Schneider, Han Dorussen, Emma Duchini, Steven Durlauf, Èric Roca Fernández, John Friedman, Tomohiro Hara, Peter Hull, Taylor Jaworski, Donia Kamel, Erik Kimbrough, Murat Kirdar, Hisaki Kono, Motohiro Kumagai, Takashi Kurosaki, Etienne Le Rossignol, Sunghun Lim, Avital Livny, Jordan Loper, Sara Lowes, Masaru Nagashima, Mike Neubauer, Henrique Pita Barros, Steve Pfaff, Louis Putterman, Michele Rosenberg, Devesh Rustagi, Yu Sasaki, Sou Shinomoto, Bryce Steinberg, Yoshito Takasaki, Matt Turner, David Weil, Reed Wood, Junichi Yamasaki, Yanos Zylberberg, and audiences at ASREC 2023 (Harvard), BÆM Development Workshop (Bristol), Brown (Growth Lab, Development group, and Applied Micro Lunch), CAC 2025 (Lyon), CERDI, Chapman (IRES workshop), CSAE 2024 (Oxford), Durham DEW 2025, Essex, Hitotsubashi Summer Institute 2023, the 4th JADE conference, Manchester (Lewis Lab Workshop), SEA 94th Annual Meeting, and WE-SPICE.
We thank Stelios Michalopoulos for sharing the shape file of historical trade routes and Nick Drake for sharing the shape file of ancient water sources.
We thank Sourliish Ravindran for outstanding research assistance.
Kubo acknowledges the Bravo Center Research Funding at Brown University.
Tsuda acknowledges financial support from the Interdisciplinary Opportunity Dissertation Completion Fellowship at Brown University and funding from the Murata Science Foundation.
All remaining errors are ours.
First draft: November 9, 2022.
}}
\author{Masahiro Kubo\footnote{\scriptsize Université Clermont Auvergne, CNRS, IRD, CERDI, F63000, Clermont-Ferrand. E-mail: masahiro.kubo@uca.fr}
\\{\it CERDI}
\and Shunsuke Tsuda\footnote{\scriptsize University of Essex. E-mail: shunsuke.tsuda@essex.ac.uk}
\\{\it University of Essex}
}
\begin{document}
\maketitle

% landlocked = inland? ニュアンスの違いある？
% https://en.wikipedia.org/wiki/Marabout
% https://en.wikipedia.org/wiki/Taqiyya

\begin{abstract}
% Michele: Origins of heterogeneity in the strength of cultural attachment across places
% Culture is important for behavior and shaping economic outcomes.
% Conflict is important --> Jihadist violence share is rising
% To study, we focus on jihad, which is a type of cultural phenomenon
% Conflicts fought based on identity
% (Brown tea) more general introduction?: Conflicts fought based on identity (e.g., Christian in England?)
% More economics citation at the 2nd paragraph in Intro.
% First sentence in abstract is not good. Save more space and explain details more.
%% ECTA 150 words (2026APRIL)
This paper investigates the origins of Islamist insurgencies as a form of cultural revival in West Africa. Exploiting variation in access to ancient water sources, which have largely disappeared, as an instrument, we show that the decline of trans-Saharan cities---once-prosperous under pre-colonial Islamic states---led to contemporary hotspots of Islamist violence. Contemporary violence is concentrated not where colonial resistance by Islamic states was fiercest, but where overwhelming military asymmetries induced outward submission, a pattern supported by historical evidence on weapon access. This strategic adaptation allowed radical Islamism to survive defeat and persist as a latent legacy. Qualitative evidence suggests ideological transmission was sustained through a religious practice of internally preparing to reassert Islamic purity. This mechanism is further supported by a dynamic model of conflict and individual-level surveys examining extreme religious ideologies. Moreover, the concentration of Islamist violence in areas that experienced reversals of fortune mirrors a global pattern.
\\
\\
% https://www.aeaweb.org/jel/guide/jel.php
%N37: Labor and Consumers, Demography, Education, Health, Welfare, Income, Wealth, Religion, and Philanthropy --Africa  Oceania
%N47  Government, War, Law, International Relations, and Regulation--Africa  Oceania
%O13: Economic Development--Agriculture, Natural Resources, Energy, Environment, Other Primary Products
%P51:	Comparative Analysis of Economic Systems
%P52: Comparative Studies of Particular Economies
%Z12: Cultural Economics, Economic Sociology, Economic Anthropology--Religion
%\\
%\\ \blue{\textbf{Blue parts:} candidates for dropping or revising to save space}
%\\\\\\
%\end{footnotesize}
\end{abstract}

%\doublespacing

%\clearpage
%\startcontents[sections]
%\printcontents[sections]{l}{1}{\setcounter{tocdepth}{2}}
%\begin{small}
%\tableofcontents
%\end{small}
\clearpage
%\begin{spacing}{1.5}
\begin{onehalfspace}  % REStud
%\begin{doublespace}  % QJE
\section{Introduction}
%It has drawn global attention since the September 11 attacks perpetrated by Al Qaeda.
%The recent rise of the Islamic State of Iraq and Syria (ISIS) has exacerbated this threat.
%Although the ISIS has lost substantial territories in Syria and Iraq, jihadist violence remains prevalent in other regions, including Africa.
%Generally, there is substantial variation in jihadist activities {\it within} the Islamic world.
%This threat, however, is not uniform {\it within} the Islamic world, exhibiting substantial regional variation and expanding beyond the Middle East, particularly into Africa.
%However, despite the severity of jihadist violence worldwide,
%%%Role models of the motivating paragraph: Bazzi et al. (2020QJE); Ali et al. (2018EJ)
%As determinants of jihad, focus on the change in the spatial dist of econ activities over the course of history
%Legacy: colonization, decline, violence
%Shuhei's comments: (a) Change in ancient water sources; (b) Historical Islamic states; (c) European colonization; (d) Technological change (from camel to ship)
% GPT: Why does jihadist violence emerge in some places but not others within the Islamic world? Despite extensive research on the proximate determinants of insurgency—such as state capacity, economic shocks, and counterinsurgency strategies—far less is known about the deep historical forces shaping the spatial distribution of jihadist violence.
%%%
Islamic extremist conflict and violence, commonly referred to as {\it jihad,}\footnote{
The literal meaning of ``jihad'' in Arabic is ``striving or exerting oneself (with regard to one's religion)'' (\citealt{Cook2015}).
It also has other meanings, such as ``the effort to lead a good life, to make society more moral and just, and to spread Islam through preaching, teaching, or armed struggle'' (\citealt{Esposito1999}).
%See \citet{Cook2015} for detailed discussions.
Although there is controversy over whether the word's interpretation is exclusively spiritual or if it includes military action, this paper uses the terms ``jihadist violence'', ``jihad'', and ``Islamist violence'' interchangeably.
We broadly define a jihadist organization as a non-state group that aims to topple a government or to govern a particular region to establish an Islamic caliphate based on a strict interpretation of Shariah law. %See Appendix \ref{app_heteg_groups} for more detailed organization-specific ideologies and goals.
}
pose a global threat.
Beyond its global salience since the September 11 attacks in 2001 and the rise of the Islamic State of Iraq and Syria (ISIS) since the mid-2010s, jihadist violence displays
substantial spatial variation, even \textit{within} the Islamic world, and expands beyond the Middle East, particularly into Africa.\footnote{
%Jihadist organizations are major drivers of instability in Africa.
According to authors' calculation using the Armed Conflict Location and Event Data (ACLED), between 2001 and 2019, approximately 30\% of violent events involving non-state rebels and militias were linked to jihadist groups, resulting in more than 100,000 total fatalities.
}
%striking heterogeneity across space, even \textit{within} the Islamic world.
%This threat, however, is not uniform {\it within} the Islamic world, exhibiting substantial regional variation and expanding beyond the Middle East, particularly into Africa.
Moreover, while jihadist activity has risen sharply in recent decades, it is not a merely contemporary phenomenon but has been cyclic over the centuries (\citealt{Cook2015,Sanneh2016}).
Understanding its origins is important not only for the economics of conflict (\citealt{Blattman2023}; \citealt{RVV2025}), but also for broader questions about how cultural attachment, religious identity, and ideology shape behavior (\citealt{BV2011}; \citealt{LMM2025}).
%Moreover, while jihadist activity has risen sharply in recent decades, it is not a merely contemporary phenomenon, but part of a longer historical cycle.
%Moreover, despite a notable rise in jihadist violence in recent years, jihad itself is not merely a contemporary phenomenon but has been cyclic over the centuries.
Yet, the roles that geography and history
%\blue{---particularly prehistoric nature, pre-colonial economy and culture, and colonial legacies---}
play in explaining this variation and persistence remain poorly understood.
\par
We approach this question through the lens of past prosperity, decline, and cultural revival.
Cultural revival refers to the reassertion of group identity through the recovery of traditions perceived as lost due to colonization, displacement, oppression, or modernization (\citealt{Anderson1983,HR1983,Smith1991}).
Revival movements have also occurred with violence across diverse contexts, such as ``The Troubles'' in Northern Ireland, the Khmer Rouge in Cambodia, and the January 6 United States Capitol Attack.\footnote{Generally, various contexts illustrate political and cultural consequences following declines from past prosperity, including Europe (\citealt{CS2018}; \citealt{FV2023}; \citealt{HH2024}; \citealt{NS2023}; \citealt{OR2024}), the United States (\citealt{ADHM2020}; \citealt{BW2021}), and Islamic contexts in the Middle East and Asia (\citealt{BC2017}; \citealt{Chen2010}).
%Within the Islamic context, similar revival dynamics have been observed in the Middle East (\citealt{BC2017}; \citealt{Chaney2013}), Bangladesh (\red{Riaz 2008; Fair et al. 2017}), and Southeast Asia (\citealt{Chen2010}; \red{Pepinsky et al. 2018}).
}
Jihadist movements can likewise be framed as radical forms of cultural revival against Westernization and secularization (\citealt{Lewis1990,Sounaye2017,Yates2007}).
%While the phenomenon is widely studied in history, political science, and sociology, recent theoretical frameworks in economics also provide valuable insights (\citealt{CRS2024}; \citealt{IRS2021}).
%\textcolor{red}{However, the origins of cultural revival and their long-run pathways to resurgence have been empirically underexplored.}
While the phenomenon is widely studied in history, political science, and sociology, recent theoretical work in economics also offers valuable insights (\citealt{CRS2024}; \citealt{IRS2021}).
However, the historical roots and long-run pathways leading to cultural revival and its political or ideological resurgence remain empirically underexplored.
%%Yet, the empirical foundations of cultural revival and its long-run pathways to political or ideological resurgence remain underexplored.
%%Yet, the historical roots and long-run pathways leading to cultural revival and its political or ideological resurgence remain underexplored.
%%Within the Islamic context, similar revival dynamics have been observed in the Middle East (\citealt{BC2017}; \citealt{Chaney2013}), Bangladesh (\red{Riaz 2008; Fair et al. 2017}), and Southeast Asia (\citealt{Chen2010}; \red{Pepinsky et al. 2018}).
%}
%%, reflecting historical memories of Islamic civilization’s past prosperity and subsequent decline (\citealt{Lewis1990,Sounaye2017,Yates2007}).
%%\\\red{MEMO: Cite \citet{Carvalho2020} \citet{Grosjean2014} somewhere?}
\par
In this paper, we investigate the origins of Islamist insurgencies, focusing on the historical rise and decline of Islamic civilization.
We exploit the historical transformation of pre-colonial trans-Saharan trade hubs that once flourished under Islamic states but subsequently became peripheral, following environmental change and colonial conquest. %, and shifts in trade technologies.
%We document how once-prosperous inland trade cities along West Africa’s trans-Saharan routes—central to Islamic economic life prior to European colonization—later became peripheral following environmental change, colonial conquest, and shifts in trade technologies.
Using an instrumental variable (IV) strategy, we estimate the persistent effects of these declined centers on contemporary jihadist violence. %, which escalated markedly in the 2010s.
We then propose a mechanism linking this revival to power relations between Islamic states and European colonizers during the early colonial era, which shaped long-run ideological persistence. We further explore whether this spatial pattern of jihadist violence aligns with a broader, global phenomenon.
%\ST{GEMINI1}
%We exploit the trajectory of pre-colonial trans-Saharan trade hubs that once flourished under Islamic states but subsequently became peripheral, following the desiccation of water sources and the colonial reconfiguration of trade.
%Using an instrumental variable strategy, we estimate the persistent legacy of these 'declined cities' on contemporary jihadist violence, specifically explaining its marked escalation in the 2010s. We propose a mechanism rooted in the military asymmetries between Islamic states and colonizers during the conquest, and finally, we demonstrate that this spatial pattern of violence mirrors a broader global phenomenon.
%\ST{GEMINI2}
%We focus on the locations of historical trans-Saharan trade nodes—established prior to European colonization—that lost their comparative advantage following the shrinkage of water sources and technological shifts in transport.
%Employing an instrumental variable strategy, we identify the causal link between these now-marginalized sites and the recent surge in jihadist violence. We attribute this persistence to a specific colonial legacy shaped by military power relations during the era of conquest, and further document that this 'past-core-and-present-periphery' pattern extends globally.
\par
West Africa experienced a stark reversal of fortune and today suffers disproportionately from jihadist violence.
%In the pre-colonial era, economic activities was concentrated in landlocked regions under the influence of Islamic states, which controlled trade routes across the Sahara, facilitating extensive trade in goods and slaves.
In the pre-colonial era, economic activities were concentrated in inland areas under the influence of Islamic states, which controlled trans-Saharan trade routes---founded up to the 1800s.\footnote{
%The relationship between state centralization and trade has been extensively observed not only across the Sahara but also across pre-colonial Africa (\citealt{Fenske2014}).
%Pre-colonial states in Africa are closely associated with historical trade and contemporary development (\citealt{Fenske2014,MP2013}).
The link between pre-colonial state centralization and trade networks is well-documented not only in the trans-Saharan region but broadly across pre-colonial Africa (\citealt{Fenske2014}).
Pre-colonial state structures are also closely associated with contemporary economic development (\citealt{MP2013}).
}
%However, with the arrival of Europeans—missionaries, scholars, and merchants—along the West African coast, economic activity gradually shifted inland-to-coast, accompanied by changes in slave trading patterns, public investments, and the adoption of modern trading technologies.
%However, once Europeans---such as missionaries, scholars, and merchants---arrived at coastal West Africa, economic activity gradually shifted from inland to coastal areas, along with slave trading, and public investments, and the advent of modern trading technologies.
%\blue{However, the arrival of Europeans triggered a fundamental spatial shift. As trade reoriented toward the Atlantic coast—driven by maritime commerce and colonial infrastructure—the inland regions lost their comparative advantage.}
However, with the arrival of Europeans, economic activity gradually shifted from inland to coastal areas, as trade patterns and technologies changed.
Consequently, the inland regions lost their comparative advantage, and many pre-colonial core cities declined.
Today, economic activity is concentrated in a few large cities, leaving vast inland regions economically marginalized, and violent events involving jihadist groups have substantially increased, both in frequency and geographical scope.\footnote{
%The presence of jihadist groups is even more pronounced in West Africa.
In this region, jihadist groups were involved in more than 40\% of conflict and violent events during 2001--2019.
These incidents claimed approximately 40,000 lives, accounting for more than one-third of the total jihadist-related fatalities across the entire African continent.
}
%Today, West Africa has experienced a substantial increase in jihadist violence, both in frequency and geographic scope.
%Today, economic activity in West Africa is concentrated in a small number of large cities, leaving vast inland regions economically marginalized.
%These regions have also experienced a substantial increase in jihadist violence, both in frequency and geographic scope.
%Violent events involving jihadist groups in these regions have substantially increased, both in frequency and geographical scope since 2001.
%Today, economic activity in many developing countries, including those in West Africa, tends to be concentrated in a few large cities, leaving other regions economically stagnant.
%In West Africa specifically, this spatial inequality is exacerbated by insurgent activities.
%Over the past decade, violent events involving jihadist groups have substantially increased, both in frequency and geographical scope.
%During the past decade in West Africa, the number of violent Islamic groups, the number of violent events involving them, and their geographical coverage have all been increasing.
%West Africa is becoming a new hotspot for jihad in recent years.
\par
\begin{figure}[t]
\begin{center}
\includegraphics[width=7.5cm]{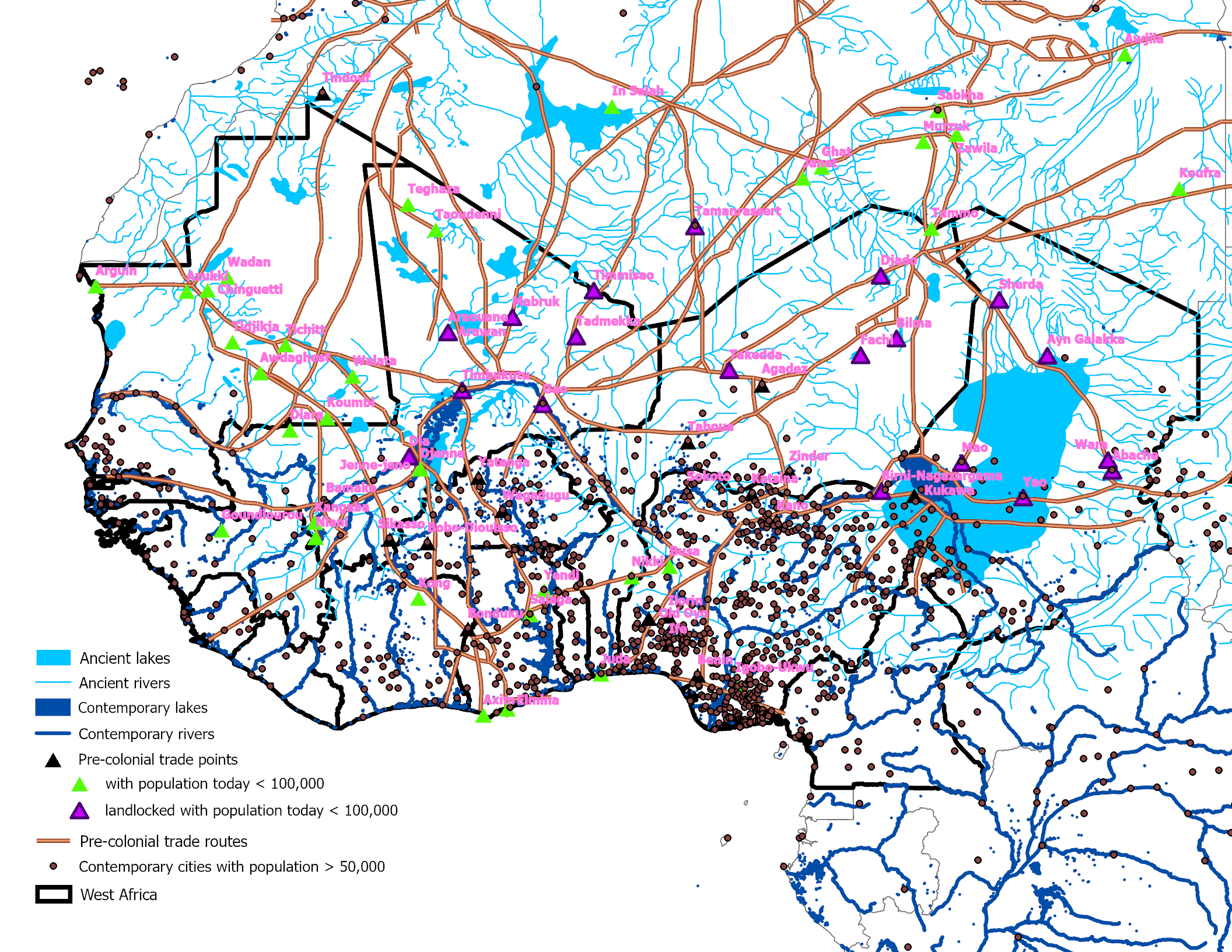}
\includegraphics[width=7.5cm]{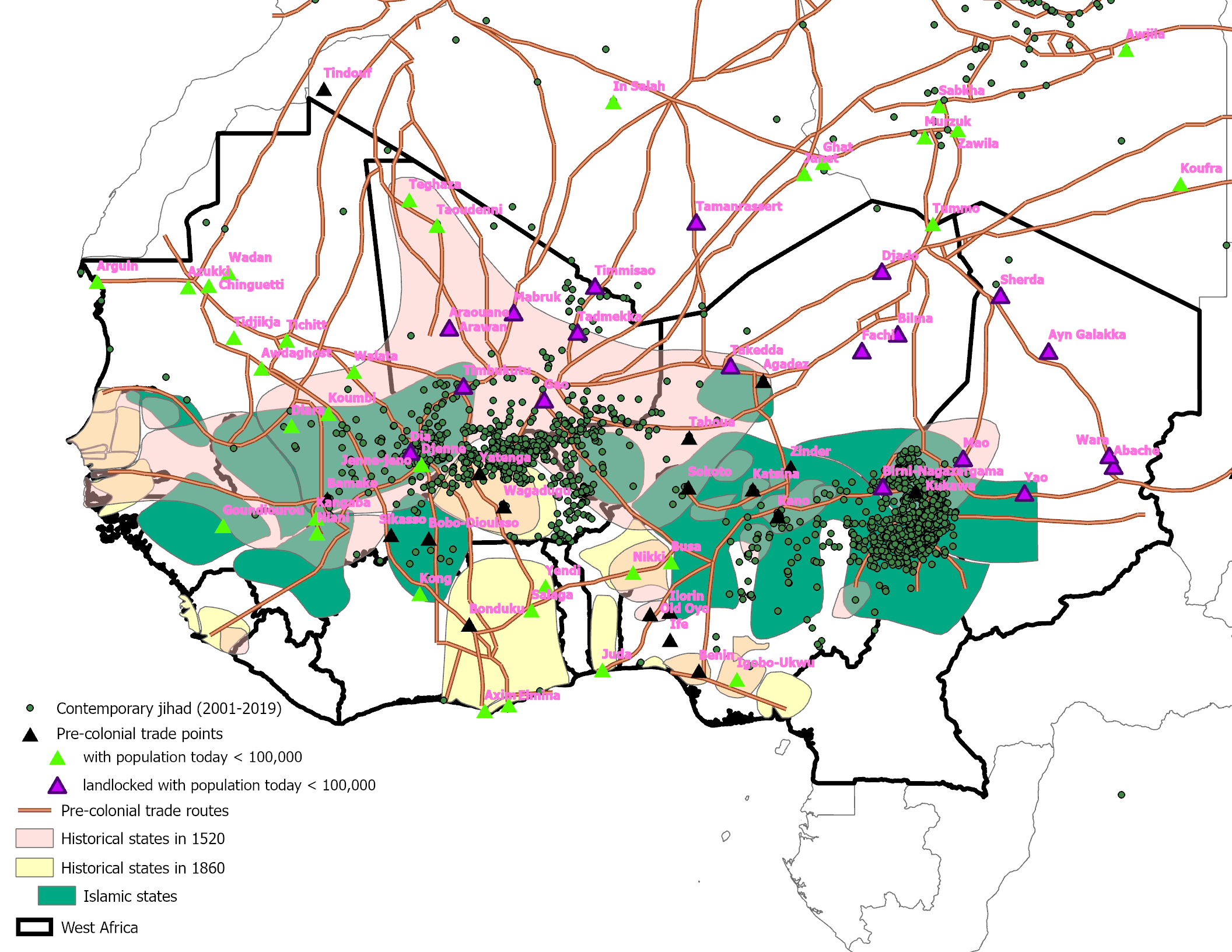}
%\includegraphics[width=14.5cm]{maps/map_water_sources_cities_past_present.png}\\
%\caption{Water Sources and Cities---Past and Present}
\caption{Water Sources, Historical Islamic Economy, and Contemporary Jihadist Violence}
%\caption{Geography, History, and Contemporary Jihadist Violence \MK{Better title?}}
%{\footnotesize(Source: ACLED)}
\label{map_geography_history_jihad}
%\label{map_water_cities}
{\parbox[t]{\textwidth}{
{\scriptsize\begin{singlespace}
\textit{Notes}:
This left map shows ancient and contemporary water sources, historical trade points, and contemporary cities. %\ST{\textbf{Which looks better, 50,000 or 100,000? Also, no need to show historical city name labels..?}}
Data sources include \citet{DBABW2011},  \href{https://www.naturalearthdata.com}{Natural Earth}, \href{https://www.hydrosheds.org/products/hydrolakes}{HydroLAKES}, \citet{OBrien1999}, \citet{Kennedy2002}, \citet{Bossard2014}, \href{https://ghsl.jrc.ec.europa.eu/ucdb2018Overview.php}{Urban Centre Database}, and \citet{MNP2018}. %, and section \ref{sec_data} provides details.
This right map shows historical states (circa 1520 and 1860), historical trade points, and contemporary jihadist violent events.
Data sources include \href{https://www.culturesofwestafrica.com/maps/}{Cultures of West Africa}, \citet{OBrien1999}, \citet{Kennedy2002}, \citet{Bossard2014}, \citet{MNP2018}, and ACLED (\citealt{RLHK2010}).
Section \ref{sec_data} provides details for these data sources.
\end{singlespace}}}}
\end{center}
\end{figure}

Mapping our data reveals two key facts.
First, as the left map of Figure \ref{map_geography_history_jihad} shows, many pre-colonial, landlocked cities that later declined were located near ancient water sources, yet today few populated cities remain around them.\footnote{
Timbuktu in Mali (Figure \ref{app_fig_Timbuktu}) provides a notable example. It was a major trade hub under the Songhai Empire in the 15th and 16th centuries and thus known as the ``golden city.''
However, this once-prosperous city later fell into the periphery, and its economy remains relatively underdeveloped in the modern era.
%Section \ref{sec_history_cities} provides more details about the development of pre-colonial states and cities.
}
%First, there is a notable relationship between water sources and city formation over time.
%Figure \ref{map_water_cities} shows ancient and contemporary water sources (lakes and rivers), core trading points along the trans-Saharan caravan routes, and contemporary cities.
%Many pre-colonial, landlocked cities that have since declined were situated near ancient water sources, such as lakes %(e.g., Timbuktu in Mali, Bilma in Niger, and several locations around Lake Chad)
%or rivers.
%Yet, today we rarely observe populated cities around these ancient water sources.
\if0
\begin{figure}[htbp]
\begin{center}
%\includegraphics[width=16cm]{maps/map_historical_trade_jihad_today.png}\\
%\vspace{0.3cm}
\includegraphics[width=12cm]{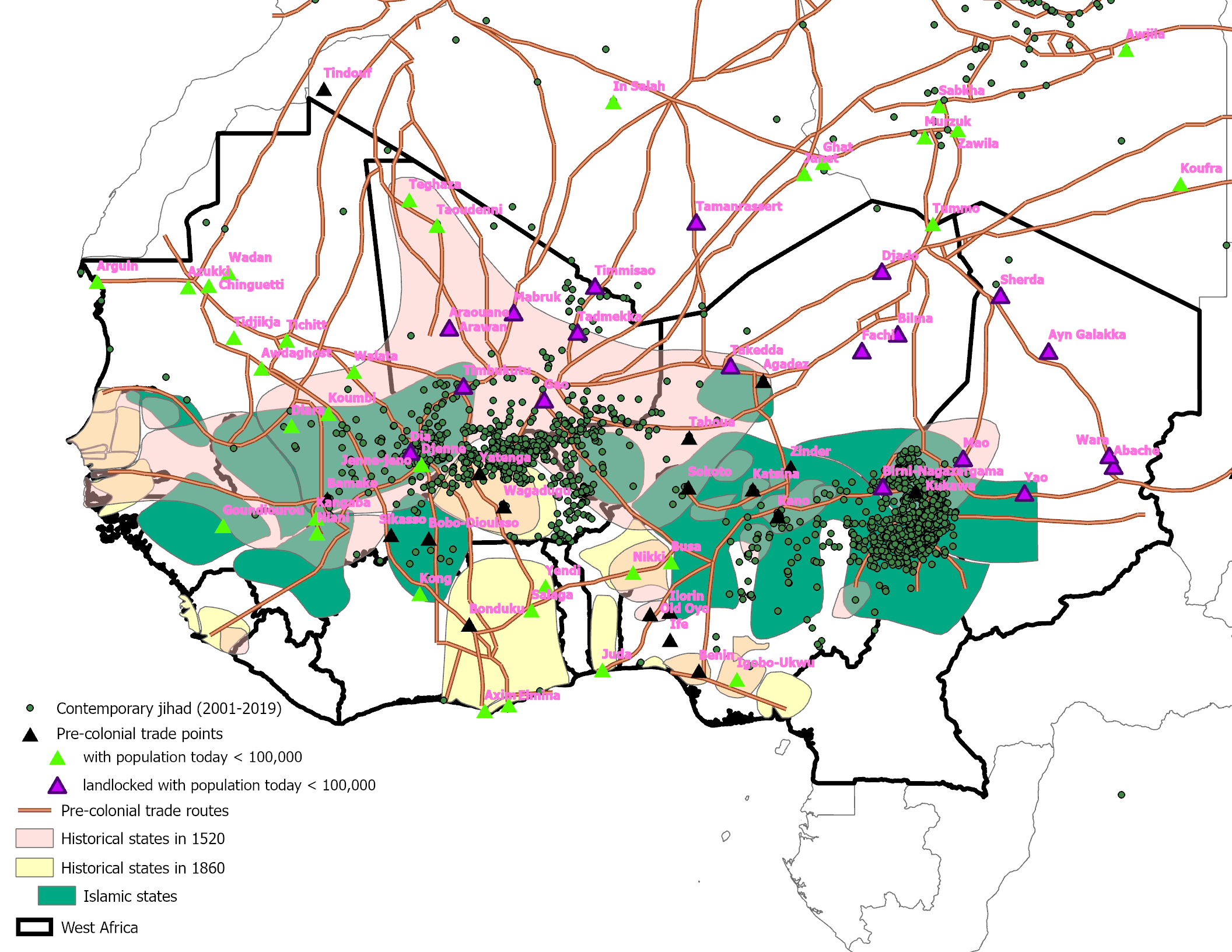}\\
%\includegraphics[width=14.5cm]{maps/map_historical_trade_jihad_today.png}\\
\caption{Pre-Colonial States, Trade Points, and Contemporary Jihad}
%{\footnotesize(Source: ACLED)}
\label{map_historical_trade_jihad_today}
{\parbox[t]{170mm}{{\footnotesize{\it Notes}:
This figure shows historical states (circa 1520 and 1860), historical trade points, and contemporary jihadist violent events.
Data sources include \href{https://www.culturesofwestafrica.com/maps/}{Cultures of West Africa}, \citet{OBrien1999}, \citet{Kennedy2002}, \citet{Bossard2014}, \citet{MNP2018}, and ACLED (\citealt{RLHK2010}), and section \ref{sec_data} provides details.
}}}
\end{center}
\end{figure}
\fi
%Second, as Figure \ref{map_historical_trade_jihad_today} illustrates,
Second, as the right map illustrates,
contemporary Islamist violence is concentrated around several historically significant but now-declined trade points and in regions once governed by specific Islamic states. %, particularly pronounced in locations tied to specific Islamic states.
%These patterns underscore the enduring influence of declined cities and the legacies of distinct Islamic states.
%%we observe concentrated violence around several historically significant but now-declined inland trade points.
%%Likewise, contemporary Islamic conflicts cluster in regions once governed by Islamic states, but not in areas historically controlled by non-Islamic states.
%%Moreover, conflicts are particularly pronounced in locations tied to specific Islamic states.
%there is a clear connection between historical Islamic civilizations and contemporary jihad.
%as Figure \ref{map_historical_trade_jihad_today} shows, we observe concentrated violence around several historically significant but now-declined inland trade points. %(e.g., Timbuktu, Dia, northeastern Mali, and the Lake Chad region),
%along with spillovers into neighboring areas. %(e.g., the Burkina Faso–Niger border).
%Similarly, contemporary Islamic conflicts tend to cluster in regions formerly occupied by Islamic states, but not in areas historically governed by non-Islamic states.
%Moreover, contemporary conflicts concentrate in specific locations linked to particular historical Islamic states.
%%These patterns highlight the importance of understanding the enduring influence of now-deserted historical cities and the legacies of different Islamic states.
%\red{These patterns highlight the importance of understanding the enduring influence of now-deserted historical cities and examining heterogeneity across different Islamic states legacies and contemporary factors.}
\par
%\ST{REVISE AND FOCUS ON THE PERSISTENT EFFECTS OF THE DECLINE} \blue{Motivated by the historical background and facts, our empirical analysis takes two steps.
%First, we aim to identify the ancient origins (or lack thereof) of pre-colonial and contemporary city formation, focusing on initial natural geography (specifically, ancient lakes) and its evolution over time (the shrinking of water sources).
%Second, using an instrumental variable strategy, we estimate the persistent influence of core pre-colonial trading cities that flourished during the height of Islamic state power and have since declined on contemporary Islamist insurgencies.
%\blue{These cities were the core along pre-colonial trade networks, while coastal cities were peripheral before the advent of modern trading technologies and the economic shift from inland to coastal areas following colonization.}  % ST: This is already explained above (Historical background paragraph), so omitted here.
Motivated by the historical background and facts, we estimate the persistent influence of core pre-colonial trading cities that flourished during the height of Islamic state power and have since declined on contemporary Islamist insurgencies, as violent religious revival.
%\blue{These cities were the core along pre-colonial trade networks, while coastal cities were peripheral before the advent of modern trading technologies and the economic shift from inland to coastal areas following colonization.}  % ST: This is already explained above (Historical background paragraph), so omitted here.
As a proxy for access to such declined cities, we focus on proximity to inland pre-colonial core cities with small contemporary populations.
For empirical analysis, we construct artificial 0.5 × 0.5 degree (about 55km × 55km) grid cells covering the entirety of West Africa.
\par
%As an instrument, we use access to ancient water sources that mostly shrank by today, which in theory predicts both the past prosperity and the subsequent decline of pre-colonial cities.
As an instrument for the decline of pre-colonial core cities, we use access to ancient water sources, which have mostly shrank.
%As an instrument, we use access to ancient water sources that mostly shrank by today.
%In the second stage, we employ ancient water accessibility as an instrument for the proximity to pre-colonial trading cities.
By controlling for contemporary water access, this IV essentially captures variation in access to ancient water sources that have disappeared.
Beyond the predetermined nature of water access in ancient periods, two identifying assumptions are required for causal inference.
First, ancient water access affects contemporary outcomes only through its influence on economic activities in historical states before colonization.
Second, conditional on observable geographical characteristics, unobservable factors driving contemporary Islamist insurgencies are uncorrelated with ancient water access.
The following arguments support these assumptions.
Most ancient lakes have vanished due to exogenous long-term climate and environmental changes, making it unlikely that they directly affect present-day jihadist activities.
Furthermore, we empirically demonstrate that ancient water access is largely uncorrelated with predetermined features, including geographical conditions and pre-colonial cultural and institutional variables.
\par
\begin{figure}[t]
\begin{center}
\includegraphics[width=13cm]{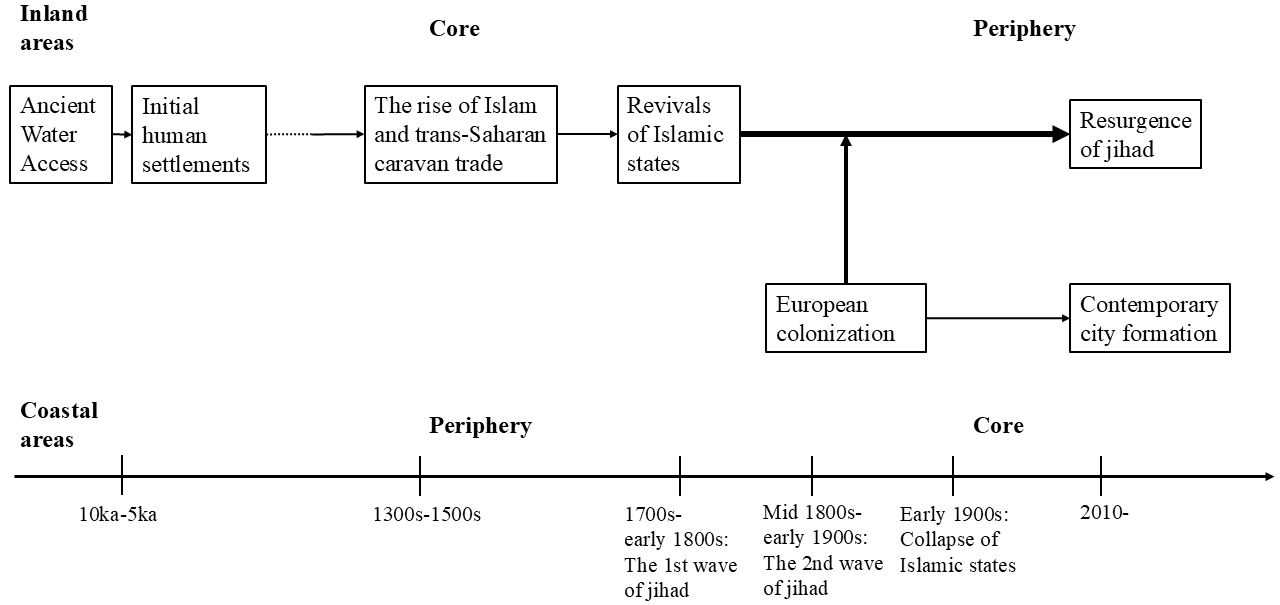}
\caption{Summary of History, Geography, and Empirical Results}
\label{fig_conceptual}
\end{center}
\end{figure}

Figure \ref{fig_conceptual} summarizes our empirical results. %, along with the history and geography in our context.
%\blue{Our main finding, represented by the bold arrow, is that grid cells located closer to declined cities, which thrived under the historical Islamic states, are significantly more likely to be proximate to jihadist violence and to experience its onset with greater intensity over the past decade.}
Our main finding, illustrated by the bold arrow, is that the historical decline of core trans-Saharan cities led to contemporary hotspots of Islamist violence. Specifically, grid cells closer to these formerly prosperous cities exhibit a higher probability of conflict onset, greater violence intensity, and closer spatial proximity to jihadist activity since 2010.
%, which thrived under the historical Islamic states, are significantly more likely to be proximate to jihadist violence and to experience its onset with greater intensity over the past decade.
%To establish this result using the aforementioned IV specification,
To support the relevance of our instrument underlying this result, we show that proximity to ancient lakes that disappeared strongly predicts the locations of these declined cities, with substantially greater explanatory power than alternative combinations of historical cities. %, supporting the relevance of our instrument.
Supplementing this, we also observe that the initial geography does not predict contemporary economic outcomes. %, which are instead shaped by colonial history and contemporary natural geography.
%\ST{One takeaway sentence focusing on the main result and the connection to next}
\par
\par
We argue that the primary mechanism behind the main result is the persistence of jihadist ideology as a colonial legacy, shaped by the power relations between Islamic states and European colonial forces. %, coupled with a contemporary shock. \ST{No need to have this phrase at the beginning of this para, while we explain this inside.}
Islamic state forces with better access to weapons adopted confrontation strategies and engaged in more intense fighting with colonizers.
As a result, the European forces militarily defeated such states, and the seeds of jihadist ideology within them were also diminished.
%Consequently, these states ceased to exist, and the seeds of jihadist ideology within them were also diminished.
%\red{ % GEMINI good but very different from original. Adopted GPT version for now.
%In contrast, facing stark military asymmetries, Islamic states in areas with limited access to weapons did not resist European forces as fiercely. Instead, they adopted strategies of alliance, acquiescence, or submission. Consequently, rather than being eradicated by defeat, radical Islamist ideology persisted in these regions, serving as a latent catalyst for contemporary insurgencies.}
In contrast, Islamic state forces in areas with limited weapon access did not engage as intensively with the colonial invasion, given the pronounced asymmetry in military capacity. Under such conditions, strategies of alliance, acquiescence, or outward submission to colonial authorities were often adopted. Consequently, the ideological foundations of radical Islamism in these areas
%survived defeat and were {\it not} extinguished,
survived defeat and were preserved underground, rather than being extinguished,
fueling future jihadist mobilization.
%\blue{In contrast, Islamic state forces in areas with limited access to weapons did not engage as fiercely with European forces due to the stark asymmetry in military capacity. Under such circumstances, these states resorted to strategies of alliance, acquiescence, or submission to colonizers. Therefore, even under defeat, the seeds of jihadist ideology around these areas were {\it not} diminished, potentially fueling future jihadist conflicts.}
%\blue{While these states also ceased to exist, the seeds of jihadist ideology within them were {\it not} diminished, potentially fueling future jihadist conflicts.}
However, while such ideological persistence may be a necessary condition for contemporary jihadist activities, it is not sufficient on its own.
This persistence, when coupled with a contemporary shock that strengthens insurgent forces (e.g., the inflow of fighters and weapons from a neighboring country), can trigger a sudden surge in contemporary jihad.
%This mechanism aligns with the religious practice called \textit{taqiyya}, prompting Muslims to outwardly adapt to situations they could not change while internally preparing to reassert Islamic purity.

We support this mechanism in four complementary ways.
%First, we examine the cyclical nature of jihad, marked by distinct spatial distributions across centuries, varying intensities of historical jihads across locations, and historical records of strategies adopted by Islamic states against colonization.   \ST{This is redundant, as it is explained in the next sentence: marked by distinct spatial distributions across centuries}
%\blue{First, we document the cyclical nature of jihad, using historical records on the spatial variation in the intensity of historical jihads and the strategies adopted by Islamic states in response to colonization. These observations reveal that contemporary jihads appear to arise in areas \textit{without} intense armed resistance against colonial invasions.}  %%% LATEST %%%
%%These observations reveal a reversal of the conventional view that conflict breeds conflict: contemporary jihads appear to arise in areas \textit{without} intense armed resistance against colonial invasions.
%These observations reveal a striking reversal of the conventional view that conflict breeds conflict: contemporary jihads disproportionately arise in areas \textit{without} intense armed resistance against colonial invasions.}
%These observations reveal a striking reverse pattern relative to the common view that conflicts tend to recur in the same locations: contemporary jihads often arise in areas where armed resistance against colonial invasions was \textit{not} present.
First,
%quantitative evidence on historical weapon access, combined with qualitative evidence on the religious practice of \textit{taqiyya}, supports the mechanism:
quantitative and qualitative evidence on historical weapon access, responses to colonial invasion, and the religious practice of \textit{taqiyya}, support the mechanism:
overwhelming military asymmetries limited intense armed resistance by Islamic states in ``past-core'' areas, instead inducing outward adaptation to circumstances that could not be immediately altered while internally preserving the intention to reassert Islamic purity when the time is ripe.
Consistently, neither ancient water access nor proximity to past-core cities significantly predicts the location of historical jihads. Together, these observations imply that contemporary Islamist violence is concentrated in areas \textit{without} intense armed resistance against colonial invasions.
%, representing the cyclical nature of jihad with distinct spatial distributions.}
%Consistently, the ancient water access and the past core areas do not significantly predict the spatial distribution of
%\ST{ADD A LINE about historical records on the spatial variation in the intensity of historical jihads? Like ``consistently, ancient water access and the past core areas do not predict the locations of historical jihads''?}
%Second, qualitative evidence on the religious practice, combined with quantitative evidence using historical weapon access, supports these patterns, indicating that overwhelming military asymmetries limited intense armed resistance by Islamic states and led instead to the persistence of jihadist ideology.

Second, we rationalize the observed patterns using a simple dynamic model of conflict between a colonizer and an Islamic state, characterized by asymmetric military capabilities and territorial endowments, building on \citet{BS2020}.

%Third, \ST{MOSQUE HERE.} \blue{we rationalize the observed patterns using a simple dynamic model of conflict between a colonizer and an Islamic state, characterized by asymmetric military capabilities and territorial endowments..} \ST{Now we can use 4 lines here. Would it suffice or do we want 1 more line..? Now we can use 5 lines here, which may finally suffice?}

Third, we provide quantitative evidence, using individual-level survey data from Muslims, that extreme religious ideologies---such as excluding other religions, governing by religious law, and restricting female education---are concentrated near declined cities.

Fourth, if place-based ideological persistence drives jihadist violence, a natural implication is that violent groups operating there rely on localized recruitment.
To test this, we examine heterogeneity across contemporary jihadist groups, finding that the persistent influence of declined cities on violence is stronger for Islamic State (IS) affiliates and Boko Haram than for Al Qaeda affiliates.
Additional supportive evidence suggests IS and Boko Haram’s greater dependence on localized recruitment, supporting the proposed mechanism.
%Together, these findings also support the proposed mechanism.
%%Moreover, to assess whether recruitment takes place around these areas, we study heterogeneity across contemporary jihadist organizations.%, focusing on their reliance on localized recruitment strategies.
%%We find a clear contrast: the persistent influence of declined cities on violent events is stronger for Islamic State (IS) factions and Boko Haram than for Al Qaeda factions.
%%Additional supportive evidence suggests this reflects IS and Boko Haram’s greater dependence on localized recruitment, reinforcing the mechanism by which persistent jihadist ideology fuels contemporary violence.
%%%We provide further supporting evidence that this pattern reflects greater reliance on localized recruitment, which reinforces the mechanism that the persistent jihadist ideology is driving contemporary jihadist violence.

We rule out prominent alternative mechanisms and conduct extensive robustness checks.
We demonstrate that %the relationship between declined pre-colonial inland cities and contemporary jihadist violence is
the main results are
not fully driven by contemporary environmental factors (such as current water availability or precipitation), the historical spread of Islam per se, or other colonial activities (such as national borders, infrastructure investments, European settlement, or missionaries) and state capacity measures.
The results also persist when restricting the sample to predominantly Muslim regions, ruling out explanations based on religious composition. %or Christian-Muslim differences.
%, and they do not extend to general, non-jihadist conflicts, serving as a placebo test.
%To rule out the influence of religious composition, we restrict our analysis to predominantly Muslim regions, confirming that our results are not merely capturing differences between Christian and Muslim populations.
We further validate %the specificity of our
the mechanism through a placebo test, showing that the persistent influence of declined cities does not extend to general, non-jihadist conflicts.
Our estimates are also robust to alternative measures of ancient water access and city decline, a different conflict dataset, and adjustments for spatial autocorrelation.

Finally, we argue that the spatial pattern revealed in West Africa---where jihadist activities concentrate in locations that experienced reversals of fortune over the centuries---extends globally.
%%\MK{Thanks, I totally agree (I think I've suggested before). In the next sentence beginning with "Regions that were...", since religious revivalism is already a broad concept, it seems sufficient to use ``religious revivalism'' on its own rather than``forms of religious revivalism'' or we can be more specific such as ``violent religious revivalism''}
%%Taken together, our results imply that jihadist activities concentrate in %\red{areas characterized as ``past-core-and-present-periphery,'' that is,}
%%locations that experienced reversals of fortune over the centuries.
%%In other words, contemporary jihads emerge primarily in areas that experienced reversals of fortune over the centuries.
%%\ST{TOPIC SENTENCE} \blue{Finally, we argue that this spatial pattern also extends globally, drawing on global-level information on historical populations, overland trade routes, and contemporary Muslim populations and jihadist activities.}
%%\MK{Revise a bit. Religious aspect + Colonial aspect}
Regions that were once prosperous and interconnected through religious and commercial networks, including prominent cases such as Afghanistan and Syria, tend to experience violent religious revivalism today, shaped in part by colonial legacies. %and unfolding beyond the boundaries of modern nation-states.
At the same time, %our findings underscore that
the detailed mechanisms translating long-run decline into contemporary violence might be inherently local, reflecting context-specific historical trajectories. %, institutional arrangements, and modes of ideological transmission.
Understanding jihadist violence would thus require combining a global perspective on structural reversals with close attention to the local pathways through which cultural revival emerges.
%%\ST{To preempt against the common critique that we received from referees about global vs. local, shall we add some words saying that even though the global pattern is common, local mechanisms driving this may differ and be underexplored?}
%The cases of Afghanistan, Pakistan, Iraq, and Syria, as well as historical trade routes connecting Asia and Europe, reinforce this interpretation.
%%\\{\color{red} Heterogeneity by groups}
%%\par
%%To understand the spillovers of jihadist events beyond the stylized locations, %i.e., deviations from historically expected pathways,
%%we examine two dimensions of heterogeneity.
%%First, the explanatory power of historical determinants varies considerably across different contemporary periods, notably, weakening after intensified competition between Al Qaeda and the Islamic State (IS), the two dominant factions in the global jihadist movement.
%%Second, the degree of persistence is highly heterogeneous across jihadist groups affiliated with the different factions.
%%Although identifying the exact mechanism behind this heterogeneity is beyond the scope of this paper, we discuss interpretations consistent with these findings, focusing on organizational structures and operational strategies of these factions.
%%, by bringing in an additional district-level dataset.
%%These findings stress the importance of examining the interaction between historical and contemporary factors to explain the Islamic conflicts.
%\\
%\par
%\textbf{Related literature.}

This paper contributes to three strands of literature.
%First, it relates to
The first is the literature on the economics of conflict, particularly research on the historical roots of contemporary violence
(\citealt{ACAGK2020,BR2014,DO2020,Jha2013,MP2016,MNR2020}).
Closest to our work are \citet{DepetrisChauvin2015} and \citet{Heldring2021}, which examine the legacy of historical states on modern conflict.\footnote{
\citet{DepetrisChauvin2015} documents the negative relationship between the political centralization of historical states in 1000--1850 CE and contemporary civil conflicts.
\citet{Heldring2021} examines the effect of exposure to historical state institutions under centralized rule one century earlier on contemporary violence in Rwanda and argues the culturally transmitted norm of obedience as the primary mechanism.%, supported by differential impacts interacted with contemporary government policies and by lab-in-the field experiments to measure rule-following behavior.
}
%These papers focus on the historical presences of state-like institutions and their long-term persistence.
While these papers focus on the long-term influence of state presence, we exploit variation within a specific institutional context, Islamic states, and uncover a long-run link between religious institutions and conflict, emphasizing how the decline of historical civilization shapes contemporary violence.\footnote{
Focusing on the decline of pre-colonial states also contributes to the literature on the long-term effects of pre-colonial centralized states on economic prosperity %(e.g., \citealt{GR2007, MP2013, AE2017, DLQ2018}).
(e.g., \citealt{DLQ2018, GR2007, MP2013}).
While it has primarily focused on the enduring effects of historically flourishing states, it has paid little attention to the process of their subsequent decline and its consequences.
%While this body of work has largely focused on periods of state flourishing, it has paid little attention to the consequences of their decline.
}
%how the collapse of these states shaped contemporary patterns of violence.}
%\red{Our study differs because we focus on a particular institution, Islam, among historical states.}
This paper also relates to theoretical and empirical works on conflict cycles and persistence (e.g., \citealt{AZ2014,DFO2019,RTZ2013b,VV2012}). %and empirical evidence on long-run cyclical patterns (e.g., \citealt{DFO2019,VV2012}).
%\red{Empirical investigations are scarce, with the notable exception \citet{BR2014}, which documents the positive correlation between historical and post-colonial conflict in Africa.}
%\red{The long-run cyclical nature of conflict and violence has also been documented empirically (e.g., \citealt{ACAGK2020,DFO2019,VV2012}).
%Most directly
In particular, \citet{BR2014} show a positive relationship between historical and post-colonial conflicts in Africa.
In contrast, we document that Islamist violence is more prevalent in areas that lacked intense colonial resistance. %intense armed resistance during the colonial conquest.
%In contrast, this paper highlights the reverse relationship: contemporary Islamist violence and conflict are more prevalent in areas where armed resistance was \textit{not} present during the colonial conquest.
%More broadly, we contribute by identifying the mechanism through which historical constraints shape the spatial distribution of contemporary violence, operating through the localized ideological persistence.
More broadly, we contribute by identifying the mechanism through which historical constraints shape the localized ideological persistence and consequently, the spatial distribution of contemporary violence.
\par
%Second, this paper relates to
The second is on the deep roots of development and the persistent effects of historical institutions in Africa (see \citealt{MP2020} for a review), particularly the legacy of colonization (e.g., \citealt{AJR2001}; \citealt{BPW2022}; \citealt{HR2018}; \citealt{Huillery2009}; \citealt{MP2013}; \citealt{NP2012}; \citealt{Okoye2019}).
We advance this literature on three fronts.
%This paper makes three key contributions to this literature.
First, we examine the persistence of extreme ideology and the eruption of related violence long after the colonial oppression.
%While prior work has documented backlash to physical or cultural oppression primarily in the short run (\citealt{GW2011,Fouka2020}),
While prior work primarily documents short-run backlash to oppression (\citealt{GW2011,Fouka2020}),
we uncover a more delayed and severe response: the surge of violence aimed at Islamic revival emerging in subsequent generations, roughly fifty years after independence.
We further emphasize how a religious practice facilitated this ideological transmission, mirroring the joint evolution of culture and institutions (e.g., \citealt{LNRW2017, BV2011}).
%These dynamics, culminating in violence aimed at Islamic revival, echo the joint evolution of culture and institutions (e.g., \citealt{LNRW2017, BV2011}). %from colonial rule.
%%This phenomenon reflects the long-term persistence of ideologies with deep historical roots through a mechanism different from the case of anti-Semitism in Nazi Germany (\citealt{VV2012}).
%%This phenomenon reflects the long-term persistence of ideologies with deep historical roots, comparable to anti-Semitism in Nazi Germany (\citealt{VV2012}), though emerging through a different mechanism.
%%These jihadist events are underpinned by the long-term persistence of ideologies with deep historical origins as in the case of anti-Semitism in Nazi Germany (\citealt{VV2012}).
%Second, this paper focuses on the period of colonial invasion rather than colonial policies and institutions.
%We shed light on the interaction between pre-colonial states and colonial forces, demonstrating how variation in the responses of pre-colonial Islamic states to invasion---before colonial rule was fully consolidated---shaped contemporary patterns of violence.
Second, we shift the focus from consolidated colonial policies to the initial period of colonial invasion.
We highlight how interactions between pre-colonial states and colonial forces, prior to the establishment of colonial institutions, shaped long-run patterns of violence through strategic responses to military asymmetry.
%\blue{We demonstrate that the strategic responses of pre-colonial Islamic states to European invasion, prior to the consolidation of colonial rule, have profound legacy effects on contemporary violence.}
%before the extensive consolidation of colonial rule
%Third, \red{this paper exploits variation within the Islamic world by examining pre-colonial trade circa 1800 and the power asymmetries between Islamic states and colonization forces.
%By analyzing different strategies of resistance and accommodation against colonizers and their consequences, we offer new insights into the colonial legacy for contemporary economic and political outcomes.}
%This contribution emphasizes the importance of incorporating political factors, such as incentives for coordination or engagement in conflicts, in a framework of economic geography.
%In this respect, this paper is most closely related to \citet{ABH2023} and \citet{KL2021}.
Third, we explore dynamic natural geography,
%addressing a limitation in prior work that typically relies on static geographic features.
departing from standard approaches that rely on static geographic features.
%Prior studies of pre-modern human settlements typically relied on contemporary geographic data, implicitly assuming that natural geography is fixed (e.g., \citealt{Michalopoulos2012}). In contrast,
%\red{This paper highlights the distinctive role of ancient lakes, rather than contemporary water sources, in shaping pre-modern settlements and economic activity} \ST{Not looks great. Revise a bit?}.
%By documenting the distinctive role of ancient, now-depleted water sources, we isolate a historical shock that shaped the rise and fall of pre-colonial cities but has no direct, persistent effect on modern agglomeration.%
We document the distinctive role of ancient water sources, relative to contemporary ones, in shaping both the rise and decline of pre-colonial cities, without generating persistent effects on modern city formation.%
\footnote{% Cite \citet{Grosjean2014} somewhere?
%This paper also connects to the long-standing literature about economic geography focusing on persistence and path dependence (\citealt{AD2021}). In particular, it is closely tied
In this respect, this paper also connects to research on the historical origins of cities and economic activities (e.g., \citealt{BMPR2021,BL2012,Ellingsen2021,HSSW2018,JKM2017,MV2016,MR2018}), and in particular to recent exceptions that leverage changing natural geography (\citealt{ABH2023,JHZR2022,Seror2020}).
%(e.g., \citealt{AS2020,BMPR2021,BL2012,BB2017,BC2020,Ellingsen2021,HSSW2018,JKM2017,MV2016,MR2018,Nagy2023,RSW2011}).
%Like these studies, we investigate both first- and second-nature forces (natural geography, especially water sources, and local historical shocks, especially colonization and technological change in trade) to understand the persistence and path dependence of economic activities.
%This paper exploits the effects of changing natural geography, while most previous research has only examined the changing effects of fixed natural geography over time, albeit with some recent exceptions (\citealt{ABH2023,JHZR2022,Seror2020}).
%Crucially, we document a striking dichotomy: while initial natural geography does not persist in shaping modern economic activity, its legacy remains deeply entrenched in contemporary Islamist violence.
Crucially, our findings point not only to a lack of persistence of initial geography in shaping modern economic activity, but also to its persistence on non-economic outcomes, such as jihadist movements.
%peculiar activities (jihadist movements).
}
\par
%Lastly, this paper contributes to the literature on the economics of religion and the Islamic economy (see \citealt{Iyer2016} and \citealt{Kuran2018} for reviews).
Finally, this paper contributes to the literature on the economics of religion, particularly studies of the Islamic world and the rise of violent extremism.
%\red{Several studies have examined the contemporary determinants of jihadist violence, focusing on counterinsurgency forces and strategies (e.g., \citealt{BSF2011,FSVW2021}) as well as various exogenous shocks, including climate change (\citealt{MN2025}), election timing (\citealt{CLSW2018}), marriage markets (\citealt{Rexer2021}), mineral resources (\citealt{Limodio2022}), and unemployment (\citealt{BDJBJ2022}). While these factors can in theory be applied to insurgent violence more broadly, we offer a more targeted explanation for jihadist violence.}
Much of the economics research on violent extremism examines proximate causes, such as counterinsurgency strategies (\citealt{BSF2011,FSVW2021}) and contemporary shocks (\citealt{BDJBJ2022,CLSW2018,MN2025,Limodio2022,Rexer2022}), or the organizational logic and structure of insurgent groups %as providers of club goods and social services
(\citealt{Berman2011}; \citealt{TW2019}).\footnote{See \citet{Iyer2016} and \citet{Kuran2018} for reviews of broader studies of the Islamic economy. Several studies investigated the historical, geographical, and institutional determinants of the spread (e.g., \citealt{BKM2020,MNP2018}), the politics (e.g., \citealt{Chaney2013}), and the economic performance (e.g., \citealt{BBV2013,CY2015,Rubin2017}) of Islam.}
%However, there is little research on the determinants of jihadist violence, without which it is challenging to fully characterize the state of the contemporary Islamic economy.
%A key unanswered question, however, is why jihadist violence takes root in some parts of the Islamic world but not others.
%We address this gap by offering a deep historical explanation for this spatial heterogeneity.
By offering a deep historical explanation, we address a key unanswered question: why jihadist violence takes root in some parts of the Islamic world but not others.
We trace the geography of contemporary jihad back to its ancient origins, long predating the arrival of Islam, and to the intersection of pre-colonial economic prosperity with the colonial encounter.
By framing jihadist movements as radical forms of cultural revival, %against Westernization,
our analysis also speaks to %connects the West African case to
a broader pattern of cultural backlash and revival that emerges in response to modernization and perceived oppression across diverse global contexts.
\section{Historical Background}\label{sec_history}
%{\color{red} [MK will complete this section. Maybe it's good to show the evolution of historical states (from culture of...) from the initial settlement, not only pre-colonial periods.]}
%Memo
%MUST READ! ''Trimingham, J. S. (1962): A History of Islam in West Africa. Oxford University Press, London.'' BUT, CANNOT GET IT BESIDES PURCHASE. NOT AVAILABLE AT BROWN.
%{\color{red} The structure will be revised.}
%\blue{We provide historical background of Islamic states and Muslim communities in West Africa from the pre-colonial period through independence.}
%First, we describe the spread of Islam and the development of pre-colonial Islamic states and cities, drawing on historical accounts.
%Second, we describe the emergence of 19th-century Islamic states and their access to weapons.
%Finally, we summarize how these states reacted to European colonization and the experience of Muslim communities during the colonial and post-independence periods.

%\subsection{Pre-Colonial Cities and Islam in West Africa}\label{sec_history_cities}
\subsubsection*{The spread of Islam and pre-colonial economic prosperity}\label{sec_history_cities}
%{\color{red} Relevant sources: \citet{Austen2010}; Chapter 14 of \citet{Battuta2004}; \citet{Connah1987}; \citet{Kane2017}}
%\\
%\par
%{\color{red} By the way, just a stylistic issue. This may be how to directly cite:
%\begin{itemize}
%\item[]``E.g.) Contents in Ibn Battuta's records...In 800, only four decades after its founding, Baghdad had become a metropolis of more than 300,000 inhabitants...it was the center of economic and political power in the Islam world(just brought from a later section to see how it looks)''
%\end{itemize}
%though there may be other options in LaTex, since we want the same right space as left (See, e.g., \citealt{BKM2020})?}

Thousands of years ago, with ample lakes and rivers, the Sahara attracted human settlements (\citealt{DBABW2011}).
After the African Humid Period (AHP), water sources in the Sahara became gradually depleted, which resulted in the Sahara Desert.
Nevertheless, the Sahara Desert, ``one of the world's greatest barriers to human movement'' was bridged by trade, which led to the births of the core trading cities (\citealt{Bovill1968}; \citealt{Connah1987}, p.98).
\par
%In the first millennium AD, a strategic geographic position and abundant natural resources, such as gold and salt exchanged with North African empires, underpinned the political power and economic prosperity of the Ghana Empire  (\citealt{Trimingham1962}; \citealt{CS1965}).
In the first millennium AD, a strategic geographic position and lucrative trans-Saharan trade in gold and salt with North African empires underpinned the prosperity of the Ghana Empire  (\citealt{Trimingham1962}).
Following its collapse in 1235, the Mali Empire was established under the Muslim ruler Sundiata.
Located in the savanna and endowed with rich natural resources, including control over gold-bearing areas, the empire expanded through trade links with North Africa and facilitated the spread of Islam across diverse subject populations (\citealt{Trimingham1962}, p.61).
%With the trade contact with North Africa, the Mali Empire spread Islam, embracing large numbers of subject states of diverse populations (\citealt{Trimingham1962}, p.61).
In the 15th and 16th centuries, as the Mali Empire declined, the Songhai Empire rose to prominence under Muslim kings, fostering a remarkable period of civilization (\citealt{Trimingham1962}; \citealt{KN1997}).
%Through the traders, Islam spread across the Sahara into the Sudan belt.
%In the eleventh century and the twelfth century, Islam was adopted by the rulers of Mali Empire and Kanem Empire, which %empowered Muslims to the governance of the core trading cities in the subsequent centuries (\citealt{Trimingham1968}, p. 34-36).
\par
Timbuktu, located in present-day Mali, exemplifies a core trading cities under Islamic rule.
It served as the second capital of the Songhai Empire and experienced the ``Golden Age'' during the 15th and 16th centuries (\citealt{Singleton2004}; \citealt{Austen2010}, p.57), before being invaded by Moroccans in 1590s.
The shaded areas in the right map of Figure \ref{map_geography_history_jihad} depict historical states in the 16th century, including the Songhai Empire at the center.
Writing about nearby Jenne in 1655, al-Sa'di observed that ``caravans flock to Timbuktu from all points of horizon'' (translated by \citealt{Connah1987}, p.97).
Beyond its role as a major trade hub, Timbuktu was also a prominent center of Islamic scholarship (\citealt{Kane2017}).
\par
Near Timbuktu, Gao was also a major center of Saharan trade even before the Islamic era and later became an early hub of trans-Saharan commerce (\citealt{Austen2010}, p.57).
In 1353, during the height of the Mali Empire (depicted in the top-left map of Figure \ref{map_historical states}), Ibn Battyuta traveled down the Niger River, referred to in his account as the ``Nile,'' from Timbuktu (Tumbuktá) to Gao (Gawgaw), describing the prosperity in \citet{Battuta2004}:\footnote{In addition, Bilma and Taoudenni were key trading centers known for salt production.Trans-Saharan caravans exchanged goods for salt at these locations and passed through them as important transit points toward the Sudan (\citealt{Austen2010}, p.38).}
%Sailing down the Niger river (he was calling it ``Nile'') from Timbuktu (Tumbuktá), Ibn Battuta reached Gao (Gawgaw) in 1353 at the age of Mali Empire depicted in the top left map in Figure \ref{map_historical states}. He describes the prosperity in \citet{Battuta2004}:
\vspace{-0.2cm}
\begin{center}
{\parbox[t]{145mm}{``I went on from there to Gawgaw [Gogo], which is a large city on the Nile, and one of the finest towns in the Negrolands. It is also one of their biggest and best-provisioned towns, with rice in plenty, milk, and fish, and there is a species of cucumber there called \textit{‘inán} which has no equal. The buying and selling of its inhabitants is done with cowry-shells, and the same is the case at Mállí.''}}
\end{center}
\vspace{0.1cm}
%\par
%In addition, Bilma and Taoudenni were key trading centers known for salt production.
%%Trans-Saharan caravans exchanged their Mediterranean goods for salt and passed through as a transit point to the Sudan (\citealt{Austen2010}, p.38).
%Trans-Saharan caravans exchanged goods for salt at these locations and passed through them as important transit points toward the Sudan (\citealt{Austen2010}, p.38).
\par
Pre-colonial core cities gradually declined as Europeans reached to coastal trading posts.
\citet{Austen2010} notes ``the beginning of the twentieth century clearly marks the end of trans-Saharan trade as a significant avenue of international commerce'' and that economic activity in West Africa has shifted away from the desert toward the Atlantic Ocean. %(\citealt{Austen2010}, p.119).
%\par
%Takedda is an another example.
%Ibn Battuta visited Takedda and describes that the inhabitants of Takedda have no occupation except trade %(\citealt{Battuta2004}).

\subsubsection*{Islamic states in the 19th century}\label{sec_history_19th}
From the 17th to the early 19th century, the ``first wave'' of jihad spread across Africa, primarily aimed for purification and extension of Islam and the enforcement of Islamic law %(\citealt{Curtin1971};
(\citealt{WM2017}; \citealt{RNF2004}, p.74-75).
\citet{Lovejoy2016} notes, ``the idea of jihad was rooted in the confrontation of established political authority through the purification of Islamic practice and the imposition of governments that were forcefully committed to governance on the basis of Islamic law and tradition.''
The green areas in Figure \ref{map_geography_history_jihad} represent pre-colonial Islamic states in the 19th century. %, some of which engaged in jihad against the colonization forces.
\par
The Futa Jallon jihad was led by Muslim settlers and Fulani pastoralists against the dominant Jalonke landlords, to whom they paid taxes on trade and cattle (\citealt{Lapidus2002}, p.418).
The Futa Toro jihad was conducted by religious leaders with itinerant beggars, rebelling against the local dynasty in protest of fiscal oppression and the lack of protection from Mauritanian raids (\citealt{Lapidus2002}, p.419).
The Sokoto Califate was founded by Islamic scholar, 'Uthman Don Fodio, who lead a jihad against the rulers of Gobir.
\par
In contrast, the Kong and Samori (Wassoulou) empires, contemporaneous Islamic states, were established by Muslim merchants and traders.
Rather than pursuing jihad, their aim was to control trade independently of existing states.
As a result, commercial considerations outweighed religious motivations in the state formation process (\citealt{Azarya1980}, p.428).
% How they got weapons: Europeans and Ottman Empire
%Islamic states
%(\citealt{Nunn2008}, p142-143).
\par
From the 15th century onward, prior to colonization, the Atlantic slave trade intensified and fostered the spread of weapons in West Africa.
Slave raiding created strong incentives for individuals and communities to acquire arms, such as iron knives, spears, swords, and firearms to defend themselves.
These weapons were often obtained from Europeans in exchange for slaves, reinforcing a self-perpetuating cycle in which increased access to arms fueled further slave raiding.
%As a result, slave raids intensified and to protect oneself, individuals and communities seek for weapons.
This cycle has been described by historians as the ``gun-slave cycle'' (e.g., \citealt{Lovejoy2011}).  %or the ``iron-slave cycle'' (e.g., \citealt{Hawthorne2003}).
Beyond local communities, states also engaged in slave raiding to finance military expansion, purchasing firearms and horses supplied in large quantities by European traders (\citealt{Law1976}, p.72).
For example, Samori, the leader of the Wassoulou Empire, financed arms acquisitions through the exchange of slaves for horses in the Sahel and Mossi regions (\citealt{Boahen1985}, p.123). Similarly, Bornu actively conducted slave raids to finance trade with the Ottoman Empire in exchange for weapons and luxury goods (e.g., \citealt{Lovejoy2011}, p.69; \citealt{Lapidus2002}, p.405).

\subsubsection*{Muslim communities under colonial rule and post independence}\label{sec_history_colonial}
Between 1880 and 1914, Europeans %, primarily the French and the British,
brought nearly all of West Africa, with the exception of Liberia, under colonial rule.
The French relied predominantly on direct military conquest, whereas the British more often established control through treaties of protectorate (\citealt{Boahen1985} p.117).\footnote{There are discussions about how aggressive the French imperialists were (\citealt{Mcgowan1981}, p.245).}
Africans resorted to three strategies against colonization: confrontation, alliance, and acquiescence or submission (\citealt{Boahen1985} p.117). %\footnote{Appendix \ref{app_strategies} provides historical evidence of strategies taken by Islamic states.}
The subjugation of Islamic territories by non-Muslim colonial powers provoked confrontations by some Islamic states, often described as the ``second wave'' of jihad (\citealt{WM2017}).

%%%REVISE?%%%
%While some Muslim responses to French and British invasion were militant, armed resistance largely subsided following colonial consolidation.
While initial responses to French and British invasions were sometimes militant, armed resistance largely subsided following colonial consolidation.
Nevertheless, opposition to colonial rule persisted in indirect forms, particularly through schools and reform movements (\citealt{Lapidus2002}, p.737; \citealt{Nyang1984}).
For example, in Ibadan (Nigeria), the Bamidele movement promoted the preservation of the Arabic language, Muslim attire, and reformed Islamic practices.
Furthermore, Muslim ethnic groups such as the Hausa and Yoruba organized to safeguard their religious identity.
These forms of cultural and ideological resistance endured across generations, as Muslims often interpreted colonial rule as a temporary setback (\textit{fitna}) rather than a justification for cultural surrender (\citealt{Nyang1984}).

%The continuity of
The Muslim identity was further sustained through \textit{taqiyya}, a practice that permitted outward compliance with colonial rules while maintaining internal commitment and waiting patiently for the tides to turn (\citealt{Hiskett1994}; \citealt{Umar2006}).
``\textit{Taqiyya} is based on the Qur’anic injunction about the obligation of remaining `faithful' and true to Islam in situations of danger and hostility, with dispensation for withholding the truth and hiding one's true intentions to ride out the challenge'' (Qur’an 3: 28, 19:18; 49:13; \citealt{Sanneh2016}, p.262).

%At the same time, colonial authorities often regarded Muslims as culturally and educationally more advanced than non-Muslim Africans, and appointed Muslim chiefs and clerks as administrators in non-Muslim areas (\citealt{Lapidus2002}, p.736).
Meanwhile, colonial authorities often regarded Muslims as relatively more advanced in administrative and educational terms and therefore appointed Muslim chiefs and clerks to govern non-Muslim populations (\citealt{Lapidus2002}, p.736).
%However, most Sudanic and West African peoples were and are ruled by narrow---often military---elites, in the name of interests and ideologies that do not, with some exceptions, reflect the values and identities of the masses.
However, political authority typically remained concentrated in narrow elites, often military, whose interests %and ideologies
did not necessarily reflect those of the broader population.
These new elites, commonly non-Muslims, prioritized political and economic modernization and treated Islam as a ``personal religion'' on a par with Christianity, rather than as a foundation for political order (\citealt{Lapidus2002}).\footnote{
The Muslim community gradually became larger with significant conversions to Islam among pagan peoples. The Muslim population of West Africa approximately doubled between 1900 and 1960, and continues to grow substantially in later periods (\citealt{Lapidus2002}, p.736).
}
\par
% Directly check Brigaglia2012 and Loimeier2016
% MK cite these since sakai (2015, 2018) mentions but not directly check due to unaveilability
Colonial arrangements generated frictions within Muslim communities.
By the 1950s, attacks against Muslim leaders (\textit{marabouts}) who cooperated with colonial administrations began to surface (\citealt{Loimeier2003}).
In Nigeria, though the reintroduction of Shariah criminal law extended to twelve northern states in 2001, it largely functioned as an instrument of political competition among elites (\citealt{Loimeier2016}).
Amid growing disillusionment with the secular state and increasing desire for Islamic revival, radical movements emerged among young Salafists, eventually leading to Boko Haram's insurgency (\citealt{Brigaglia2012}).
Not only in Nigeria but more broadly, resistance to secularization and Westernization---processes intensified under colonial rule---fueled the rise of reformist movements and Salafism across West Africa, which advocated for the purity and revival of Islam (\citealt{Sounaye2017}).
%Between 1900 and 1960, the Muslim population of West Africa approximately doubled, and continues to grow substantially in later periods (\citealt{Lapidus2002}, p.736).
%Changes in the economy broke down lineage bonds, brought individuals into large-scale market arrangements, promoted mobility, and generated a demand for individual, as opposed to collectively shared, wealth (\citealt{Lapidus2002}, p734).\\

%%%%%%%%%%%%%%%%%%%%%%%%%%%%%%%%%%%%%%%%%%%%%%%%%%%%%%%%%%%%%%%%%%%%%%%%%%%%%%%%%%%%%%%%%%%%%%
\section{Data}\label{sec_data}
The main data sources are as follows. Section \ref{sec_mechanism} and Appendix \ref{app_data} describe other data.
%\par
%\ST{Footnote when we first rely on the historical state info (5.1).} \blue{\textbf{Pre-colonial Islamic states.}
%\href{https://www.culturesofwestafrica.com/maps/}{Cultures of West Africa} creates the maps that show spatial locations of historical states before colonization as well as modern countries after independence by using multiple sources of references.\footnote{The references and maps are available in the \href{https://web.archive.org/web/20190603215326/https://www.culturesofwestafrica.com/wp-content/uploads/2018/09/HistoryWestAfrica.pdf}{website}.}
%We digitize maps of historical states over the centuries from pre-colonial periods to the colonial era (Figure \ref{map_historical states}). In Appendix \ref{app_empires}, we describe how to identify Islamic states in detail.}
\par
\textbf{Pre-colonial trade routes and points.}
We draw on three sources: \citet{OBrien1999}, \citet{Kennedy2002}, and \citet{Bossard2014}, which provide historical trade routes, trade points, and cities in the pre-colonial period.
For trade routes, we use mapped routes before 1800 from \citet{Kennedy2002}, supplemented by \citet{OBrien1999}.
These data were digitized by \citet{MNP2018}. %\footnote{We appreciate the authors for generously sharing the digitized data.}  % ALREADY IN acknowledgement in the 1st page
To identify cities that have declined, we rely on \citet{Bossard2014} (Map 1.15 p. 39), which overlays historical and present-day cities along pre-colonial routes based on multiple sources.
We further use \citet{OBrien1999} and \citet{Kennedy2002} to complement this information by incorporating contemporary population data.
\par
%\textbf{History of Sahara.}
\textbf{Ancient water sources.}
We use the map of ancient lakes (more than 5 thousand years ago) constructed by \citet{DBABW2011}.\footnote{
%\ST{DROP OR REDUCE?}
%During the ``African Humid Period'' from around 10,000 years ago, the Sahara enjoyed climatic and environmental conditions favourable for human habitation and cattle raising and it has been referred as the ``Green Sahara'' (e.g., \citealt{deMenocal_etal2000}; \citealt{Dunne_etal2012}).
We digitize the map of ancient water sources depicted by a cartographer, Carl Churchill (\href{https://www.flickr.com/photos/cchurchili/40921572803/}{link}).
}
%\citet{DB2016}, and \citet{LRR2013}.
They assess and map the paleohydrology of the entire Sahara.
In particular, they use a digital elevation model (DEM) and Landsat satellite imageries to identify ancient river channels and lake shorelines.
Ancient lake areas were then basically estimated from the shorelines identified by the DEM.
As a complement, remote sensing is also being used to map lake sediment outcrops, which are readily distinguished from other materials observed in the satellite imageries.
For more technical details, see \citet{DB2006} and the Supporting Information of \citet{DBABW2011}.
%{\color{red} It is better if we can track the shrinkage of lakes over time (or argue some geological/geographical between areas with and without ancient lakes/rivers), but it may not be realistic.}

%\textbf{Jihadist groups in the contemporary world.}
%\textbf{Contemporary conflict events involving jihadist groups.}
\textbf{Contemporary Islamist violence.}
The main data source for contemporary conflict events and actors involved in conflicts is Armed Conflict Location and Event Data (ACLED, \citealt{RLHK2010}).
Each actor appeared in ACLED is classified into an Islamist group or not by hand.\footnote{
Information about violent Islamist group are drawn from several sources, including ACLED reports, \href{https://africacenter.org/}{Africa Center for Strategic Studies (ACSS)}, \href{https://cisac.fsi.stanford.edu/mappingmilitants}{Mapping Militants Project (MMP)}, \href{https://www.jessicamaves.com/forge.html}{the Foundations of Rebel Group Emergence (FORGE) Dataset}, and \citet{WM2017}.}
Appendix \ref{app_heteg_groups} lists major jihadist groups and their stated ideologies and goals.
%\textbf{Contemporary conflict events involving jihadist groups.}
%We construct event-level measures of jihadist violence using ACLED data and restrict the sample as follows.
%First, we focus on violent events involving jihadist groups, classified as rebel groups or political militias between 2001 and 2019 (spanning the post-9/11 period up to the year preceding the COVID-19 pandemic).\footnote{
We focus on violent events involving jihadist groups between 2001 and 2019 (spanning the post-9/11 period up to the year preceding the COVID-19 pandemic).%\footnote{See the codebook \citet{ACLED2019} for detailed classifications of conflict actors. Rebel groups are defined as ``political organizations whose goal is to counter an established national governing regime by violent acts.''Political militias are defined as ``a more diverse set of violent actors, who are often created for a specific purpose or during a specific time period and for the furtherance of a political purpose by violence.''}
We retain two categories of violence involving these groups.
The first includes violence against state forces, encompassing both domestic government actors and external forces, such as international organizations, foreign state militaries, private security firms, and independent mercenaries.
%In this type, opponents include both government actors in the country where each event is observed and external/other forces (international organizations, state forces active outside their main country of operation, private security firms and their armed employees, and hired mercenaries acting independently).
%According to ACLED classifications,
This category includes both battles and explosions or remote violence.
The second category includes violence against civilians. %, covering both direct attacks and explosions or remote violence when civilians are targeted.
No minimum fatality threshold is imposed.\footnote{
\citet{ACLED2019} defines a battle as ``a violent interaction between two politically organized armed groups at a particular time and location,''
explosions/remote violence as ``one-sided violent events in which the tool for engaging in conflict creates asymmetry by taking away the ability of the target to respond,''
and violence against civilians as ``violent events where an organized armed group deliberately inflicts violence upon unarmed non-combatants.''
We exclude battles between jihadist groups and other non-state actors.
Indeed, the selected two categories account for over 90\% of all violent events involving Islamist groups.
}
Figure \ref{app_fig_groups_AQ_IS_WA} shows jihadist violence in West Africa from 2001 to 2019.

\textbf{Afrobarometer.}
%We leverage the Afrobarometer surveys to examine ideological persistence as the mechanism.
To examine ideological persistence as the mechanism, we use the Afrobarometer data, which comprise nationally representative, individual-level surveys conducted across several African countries, with geo-coded information available in each enumeration area (EA). We use two waves (rounds 6 and 7), implemented between 2014 and 2018, which include relevant variables for this study. Appendix \ref{app_Afrobarometer} provides details.

\section{Persistent Effects of Declined Cities}\label{sec_emp}
%\section{Empirical Analysis}\label{sec_emp}
%\subsection{Persistent Effects of the Declined Cities on Contemporary Jihad}\label{sec_emp_IV}
%\subsection{\textcolor{red}{Legacy of Declined Cities on Contemporary Islamist Insurgencies}}\label{sec_emp_IV}
%\subsection{Persistent Influence of Pre-Colonial Islamic States on Contemporary Islamist Insurgencies}\label{sec_emp_IV}
In this section, we empirically examine the persistent influence of core trans-Saharan cities that flourished under pre-colonial Islamic states but have since declined on contemporary Islamist insurgencies.
As the locations of these declined cities were unlikely to have been determined randomly, we employ an instrumental variable strategy.
This section proceeds as follows.
First, we define the empirical specification, with 0.5 × 0.5 degree (about 55km × 55km) grid cells covering the entire West Africa.
%Unless otherwise noted, each grid cell is the unit of analysis.
Second, we describe the logic and validity of the instrumental variable.
Finally, we present the empirical results.

\subsection{Empirical Specification}\label{sec_emp_IV_strategy}
%We construct 0.5 × 0.5 degree (about 55km × 55km) grid cells covering the entirety of West Africa. Unless otherwise noted, each grid cell is the unit of analysis.
%throughout the empirical analysis.

\subsubsection*{Defining city decline}
The primary variable of interest captures proximity to declined historical cities in the trans-Saharan caravan routes founded up to the 1800s when historical Islamic states played
significant economic roles before European colonization.
As systematic data on historical city populations is not available, it is infeasible to measure decline based on population changes directly.
Instead, we construct a credible proxy based on the shifting core-periphery structure of the region.
Specifically, we proxy a ``declined historical city'' by a landlocked pre-colonial city with a contemporary population of less than 100,000.
This definition rests on two premises.
First, during the era of historical Islamic states, core cities were predominantly landlocked, driven by overland trade.
Second, contemporary cities with populations below 100,000 can reasonably be classified as peripheral in today’s economy.
We then measure access to these locations by defining $CityDecline_o$ as the straight-line distance from the centroid of grid cell $o$ to the nearest declined historical city.
Furthermore, as Section \ref{sec_alt_mechanisms} illustrates, we also verify that our empirical results are robust to alternative definitions of $CityDecline_o$, including different contemporary population thresholds and alternative definitions of ``landlocked'' (benchmarked as >1,000 km from the coast).\footnote{
The set of declined historical cities includes prominent examples such as Gao, Tadmakka, Takedda, Timbuktu, and Wara, which were major urban and religious centers in the pre-colonial period but have experienced substantial decline over time (\citealt{Austen2010, Gomez2018, JD2008}). %, %Nixon2013}).
These locations also hosted important historical mosques, reflecting their role as centers of Islamic learning and religious life (\citealt{Pradines2022}).
To demonstrate that our results are not driven solely by pre-colonial religiosity, Section \ref{sec_alt_mechanisms} conducts robustness checks that explicitly control for proximity to historical mosques.
%\MK{Most important cities are included (city names; citation). Also write about historical mosques briefly, and then we refer to the robustness section for more details.}
%\ST{In the robustness section: defend against Christian-Muslim difference + Coastal-Inland difference; Write some notes}
}
%These variants include different contemporary population thresholds, alternative definitions of ``landlocked'' (benchmarked as >1,000 km from the coast),
%\ST{[WE ARE NOT DOING THIS NOW, RIGHT? DROP THIS?] and a weighted accessibility measure defined by $CityDecline_o  = \sum_s \frac{1}{(Distance_{os})^\delta}$, where $Distance_{os}$ is the distance from grid cell $o$ to declined historical city $s$}. %, instead of the distance to the nearest declined city.

\subsubsection*{The instrumental variable exploiting the \textit{shrinkage} of ancient water sources}
As an instrument for $CityDecline_o$, we exploit variation in proximity to water resources in ancient periods (more than five thousand years ago).
This subsection focuses on the definition and empirical specification, while the following subsection details the underlying identification logic.
The instrument, $AncientWaterAccess_o$, is defined as the straight-line distance from grid cell $o$ to the nearest ancient lake (specifically, the nearest boundary between land and lake).
Alternatively, we also construct a size-weighted measure of ancient water access, $\sum_w \frac{AncientLakeArea_{w}}{Distance_{ow}}$,
%\begin{eqnarray*}
%AncientWaterAccess_o = \sum_w \frac{AncientLakeArea_{w}}{(Distance_{ow})^\delta}
%\end{eqnarray*}
where $Distance_{ow}$ is the distance from cell $o$ to lake $w$.
This measure accounts for the surface area of each ancient lake, $AncientLakeArea_{w}$, reflecting that larger lakes offered superior water resource accessibility.
%Again, Section \ref{sec_alt_mechanisms} shows robustness of our empirical results.
%\blue{We first test whether proximity to water resources in ancient periods (more than five thousand years ago) predicts the formation of core trading cities in the trans-Saharan caravan routes founded up to the 1800s when historical Islamic states played significant economic roles before European colonization.}

Importantly, because we control for contemporary water access throughout the analysis, the instrument essentially captures \textit{variation in access to ancient lakes that have since shrunk.}
Indeed, Section \ref{sec_alt_mechanisms} also confirms that the empirical results are robust to alternative measures that directly capture the proximity to only ancient lake areas where water sources have disappeared today.
%Nonetheless, Section \ref{sec_alt_mechanisms} also confirms that the empirical results are robust to alternative measures that directly capture the proximity to only areas where ancient lakes once existed but have since disappeared.
However, we do not use these measures in the main specification.
This is because they are constructed using information on contemporary lakes and rivers, whose locations and extents may also be potentially shaped by human economic activity, including deforestation, dam construction, and water infrastructure.
%\MK{Briefly explain why we are not using the distance to the lost water areas for the main specification.}

Note also that we capture walking distance for the following reasons.
First, there were no modern roads in the ancient periods and the pre-colonization age.
Second, there were no modern cars in these periods.
Camels were the means of transportation for the trans-Saharan trade.
Third, walking distance matters even today in insurgent activities.
Rebel groups tend to move not only through roads but also through off-road (e.g., \citealt{TSFTW2016}).
Throughout, we primarily use the straight-line distance measures for both $CityDecline_o$ and $AncientWaterAccess_o$ because these measures are the simplest and most straightforward ones to capture walking distance without specific assumptions about travel costs.
Section \ref{sec_alt_mechanisms} shows that our empirical results are robust to alternative accessibility measures.
%We also check the robustness of the empirical results using alternative accessibility measures in Section \ref{sec_alt_mechanisms}.

%\ST{(Already mentioned these info) For the remainder of our analysis, We primarily use the straight-line distance measures for both $CityDecline_o$ and $AncientWaterAccess_o$ because these measures are the simplest and most straightforward ones to capture walking distance without specific assumptions about travel costs. We also check the robustness of the empirical results using alternative accessibility measures in Section \ref{sec_alt_mechanisms}.}

%\textbf{Cost distance based on the Human Mobility Index (days).} The third measure incorporates the Human Mobility Index proposed by \citet{Ozak2010,Ozak2018}.
%The Human Mobility Index computes travel time to cross any square kilometer on land, taking into account slope and terrain conditions, climate conditions, etc.
%Based on this index, we construct the travel time from each grid cell centroid to the nearest ancient lake or the nearest core trading city.

\subsubsection*{Estimating equations}
%\textbf{The first stage.} The first-stage regression specification is as follows:
We first test whether ancient water access can predict past prosperity and decline of the historical cities by taking logarithms and estimating the following first-stage regression:
\begin{eqnarray}\label{eq_city_origin}
\log(CityDecline_o) = \gamma_0 + \gamma_1\log(AncientWaterAccess_o) + \gamma_2X_o + \phi_c + u_o
\end{eqnarray}
where
$o$ represents a grid cell, $X_o$ is a vector of cell-level geographical controls,\footnote{
Grid cell-level geographical controls include a landlocked dummy, average malaria suitability, average caloric suitability in post 1500, average elevation, terrain ruggedness, proximity to contemporary water sources, and contemporary populations.
}
$\phi_c$ is a contemporary country fixed effect, and $u_o$ is an error term.
%The dependent variable, $CityDecline_o$, is the accessibility of a trading point in the trans-Saharan caravan routes.
%In our main specification, $CityDecline_o$ takes the accessibility of a declined landlocked city, proxied by distance to an inland pre-colonial city that has contemporary populations less than 100,000.
%$AncientWaterAccess_o$, is the accessibility of an ancient water source.

%\textbf{The second stage.}
%Given the endogeneity of the locations of pre-colonial cities that flourished under pre-colonial Islamic states but have since declined, we instrument $\mbox{CityDecline}_o$ (proxied by the log of one plus distance to an inland pre-colonial city that has contemporary populations less than 100,000) by proximity to an ancient lake.
%\blue{Notably, by controlling for contemporary water access, the instrument essentially captures variation in access to ancient lakes that have since shrunk.}
We then use the predicted proximity to a declined city from the first stage to estimate the following two-stage least squares:
%We thus use the predicted trading city access from the first stage to estimate the following two-stage least squares:
\begin{eqnarray}\label{eq_jihad_iv}
Y_o = \beta_0 + \beta_1 \log(CityDecline_o) + \beta_2X_o + \phi_c + \epsilon_o
%Y_o = \beta_0 + \beta_1\log(\mbox{CityAccess}_o) + \beta_2X_o + \phi_c + \epsilon_o
\end{eqnarray}
where
$Y_o$ is an outcome of interest regarding insurgent activities by violent Islamist organizations and $\beta_1$ is the coefficient of interest.\footnote{
%We claim that the IV results are not driven by a simplistic view that jihad tends to occur just in current populated locations.
%Although we control for the contemporary population for our meaningful thought experiment, contemporary population (the size of potential target) itself does not seem to be a key factor to explain jihadist violence.
As $X_o$ also includes contemporary populations, our specification effectively compares areas with similar contemporary population sizes but differing past prosperities, allowing us to isolate the persistent effects of historical Islamic civilization.
%\blue{Table \ref{tab_ols_jihad_city_today} reports positive correlations between jihadist events and proximity to contemporary cities of various sizes.}
Nevertheless, our results remain robust and virtually unchanged whether or not controlling for contemporary populations.
%\ST{No need to add the following nightlight sentence?}
In a later section of robustness checks, we also demonstrate that the main results are robust to controlling for nightlight luminosity, which captures local economic development.
%In this table, we observe weak correlations between jihads and contemporary cities of various sizes.
%We use the same dependent variables and a set of geographical controls in each panel of Table \ref{tab_iv_jihad_pop}.
%In column (4) through (9), we examine the correlations, restricting the countries covering the Sahara.
}
We report standard errors adjusting for spatial auto-correlation with distance cutoff at 100 km.\footnote{
With this distance cutoff, the standard error is equivalent to be clustered by 3 × 3 grid cell squares (the own grid cell at center and surrounding eight grid cells).
Section \ref{sec_alt_mechanisms} reports that the standard errors exhibit similar magnitudes across a range of higher distance cutoffs.
}
\par
In our baseline estimations, $Y_o$ takes the following three variables:
(A) a dummy which takes 1 if cell $o$ has at least one violent event by a jihadist organization in ACLED from 2010--2019; (B) log (number of violent events by jihadist organizations from 2010--2019 in cell $o$); (C) log (distance from the nearest point of violent event by a jihadist organization between 2010 and 2019 from the centroid of cell $o$).

\subsection{Logic and Validity of the Instrument}
%\subsection{\red{The Logic of the Instrumental Variable and Identification Assumptions}}
%We use ancient water access as an instrumental variable, as it can in theory predict past prosperity and decline of the landlocked cities as explained above.
We first argue that ancient water sources and their shrinkage can in theory explain \textit{both} the past prosperity and the subsequent decline of pre-colonial cities. %and their subsequent decline.  %past prosperity and decline of pre-colonial cities.
We then outline the identification assumptions underlying our IV strategy and present empirical support.
%We then present identification assumptions and their empirical support.

%\subsubsection*{Ancient water sources as origins of core pre-colonial cities and subsequent decline}
%\subsubsection*{Ancient water sources as the origins of pre-colonial cities and their subsequent decline}
%\subsubsection*{Ancient water sources and their shrinkage as origins of core pre-colonial cities and their decline}
%\subsubsection*{Ancient water sources and the rise and decline of pre-colonial cities}
\subsubsection*{Ancient water sources as drivers of pre-colonial city prosperity and decline}
%Ancient water sources as the foundations of pre-colonial city prosperity and decline
%Ancient water sources as drivers of pre-colonial city prosperity and decline
%Ancient water sources as the foundations of pre-colonial city prosperity and decline
%Ancient Water Sources as Drivers of the Rise and Decline of Pre-Colonial Cities
%\blue{We argue that water access in ancient periods predicts the locations of core pre-colonial cities along the trans-Saharan caravan routes, which thrived under Islamic states until the 1800s but declined thereafter.}
The logic begins with the role of ancient lakes in shaping early human settlements.
%The transportation cost of inland trade across the Sahara is arguably high, \blue{given that camels were the primary mode of transport}.
%Before the invention of modern trading technologies, the main transport mode was camel.
%The transportation cost in inland trade over the Sahara by camel is arguably high.
In the presence of high inland transportation costs across the Sahara, locations with a greater capacity to sustain human life were attractive for settlement. %\blue{and city formation}.
%Locations close to lakes and rivers are thus attractive.
Proximity to lakes and rivers provided direct benefits, such as drinking water and support for agriculture, as well as indirect benefits through their support of animal life.
Livestock depended critically on water sources, while fishing enhanced food security.
Moreover, terrestrial animals congregated near water, lowering hunting costs for humans.
This mechanism aligns with \citet{Bosker2021}, who emphasizes the availability of reliable water sources as the fundamental ``city seed'' under high transportation costs.
\citet{DBABW2011} provide further empirical support.\footnote{
By compiling records of refuges, sightings, fossils, and rock art sites, they show that the estimated spatial distribution of faunal species (such as fish, molluscs, and savannah mammals) overlaps significantly with ancient water sources.
Records of barbed bones, used exclusively for hunting large water-dependent animals, further reveal a strong correlation between human settlements and ancient lakes.
See the Supporting Information of \citet{DBABW2011} for technical details.
See also \citet{Dunne_etal2012} and \citet{Sereno_etal2008} for additional evidence of human settlements and animal use in the Sahara in the African Humid Period.
}

%Subsequent evolution of cities, thrived in particular under historical Islamic states, would then be based on the common factors shared by initial human settlements directly affected by the initial water sources, as long as transportation costs remained high prior to European colonization.
The subsequent evolution of cities that thrived under historical Islamic states was therefore shaped in part by factors common to initial human settlements---notably the hydrological and topographical continuities that made water accessible through lakes, rivers, and oases.
Despite the shrinkage of water sources, this spatial dependence persisted throughout the pre-colonial era; as long as camels served as the primary mode of transport and overland trade costs remained high, these locations retained their comparative advantage.
%This spatial dependence likely persisted prior to colonization---even as many water sources gradually shrank---so long as transportation costs remained high and camels served as the primary mode of transport.
%Subsequent evolution of cities and historical states would then be based on the initial human settlements directly affected by the initial water sources, as long as transportation costs had been kept high before European colonization.

%\blue{Post colonization, the loss of comparative environmental advantage due to a long-run decline in water resources, combined with the rise of modern trading technologies concentrated in coastal areas, helps explain the eventual decline of once-prominent inland cities.}

Post-colonization, the economic geography of West Africa underwent a profound transformation.
The comparative advantages that had sustained inland cities were eroded due to the disappearance of most water resources, combined with the rise of modern trading technologies that favored coastal areas, reorienting economic activity away from overland trade.
Together, these forces can explain the decline of once-prosperous cities, reducing them from the core under pre-colonial Islamic states to the periphery of the modern states.

\subsubsection*{Identification assumptions}
Beyond the predetermined nature of water access in ancient periods, we need the following identifying assumptions for causal inference.
The exclusion restriction requires that ancient water access influences contemporary outcomes only through its influence on economic activities in historical states prior to colonization.
The independence assumption posits that, after controlling for cell-level geographical characteristics, unobservable determinants of contemporary Islamist insurgencies are uncorrelated with ancient water access.
There is no direct way to test these assumptions, but the following two arguments support them.

First, most ancient lakes have now disappeared due to exogenous long-term climate and environmental changes.
%\ST{Redundant and inaccurate?}\blue{We presume that the direct effect of lakes is present as long as lakes exist. Our logic is that ancient lakes directly affected human settlements and economic activities when humans and animals depended heavily on these water sources. Subsequent evolution of historical states would then be based on the initial human settlements affected by ancient water sources.}
Paleoclimatic and paleohydrological studies indicate that most ancient lakes had desiccated long before the rise of pre-colonial Islamic civilizations and the colonial era, thereby strengthening the conceptual validity of the instrument's exclusion restriction.
%Paleoclimatic and paleohydrological studies indicate that
The African Humid Period ended around 5,000--6,000 years before present (e.g., \citealt{deMenocal_etal2000}).
Rather than disappearing abruptly, ancient lakes gradually contracted after this transition, as permanent high-level lacustrine systems gave way to more fragmented and shallow water bodies (\citealt{LHGBKQ2011}).
With the exception of megalakes, most notably Lake Chad, most ancient lakes had disappeared as permanent surface water bodies by about 2,500--2,700 years before present due to sustained aridification (\citealt{DB2006, KVLEC2008, LHGBKQ2011}).
Therefore, ancient lakes are not likely to have a direct influence on contemporary Islamist insurgencies.\footnote{Nevertheless, to further rule out any potential hydrological advantages as an alternative explanation, Section \ref{sec_alt_mechanisms} shows that controlling for contemporary groundwater availability does not explain our results.}
%In order to indirectly test the exclusion restriction, we run the OLS regressions on a set of geographical variables and a proxy of historical economic devlopement, controlling for the baseline geographial controls and country fixed effects.

Second, we check the correlation between ancient water access and a set of pre-determined characteristics, including geographical conditions and pre-colonial variables of culture and institutions, after controlling for the baseline geographical controls and contemporary country fixed effects.
Geographical variables include ecological diversity, temperature, precipitation, caloric suitability, and pastoralism suitability.
Pre-colonial variables include jurisdictional hierarchy of local community, polygamy as a marital composition, reliance on irrigation, degree of class stratification, and rules of political succession of local headman, drawing from the \textit{Ethnographic Atlas}.
Table \ref{tab_exclusion_geo_institution_pop}  reports that the accessibility of ancient water sources is mostly uncorrelated with these pre-determined characteristics.
In particular, we find no systematic relationship with key pre-colonial institutional features, suggesting that access to ancient lakes does not capture deep-rooted institutional differences.
Importantly, reliance on irrigation, highlighted by \cite{ABH2023} as an important coordination device in response to the loss of water sources, is also orthogonal to ancient water access, reinforcing that it does not capture irrigation-based institutional adaptation.\footnote{
We see significant correlations with precipitation and caloric suitability.
However, most variations of these variables are concentrated in coastal countries and the high correlations pick these effects.
We confirm the robustness of our main results when additionally controlling for average precipitation in section \ref{sec_alt_mechanisms}.
}

\subsection{Main Empirical Results}\label{sec_emp_IV_results}
%We begin by presenting the first-stage results, followed by the second-stage results.

\subsubsection*{The first stage}
%\ST{(Words from HELDRING 2019, empirical strategy) Aside from the exclusion restriction being met, the first-stage correlation between the instruments and state presence needs to be sufficiently strong. If this correlation is weak, the second-stage results become difficult to interpret (Staiger and Stock, 1997). I report first-stage regressions for all IV estimations and find that partial F-statistics of the excluded instrument are typically high enough to conclude that the first stages are sufficiently strong.}

%\ST{Report F-stats in word as well}\red{Table \ref{tab_first_stage_pop} reports the corresponding first-stage regression results, which are equivalent to those in Table \ref{tab_water_historical_city} except for additionally controlling for contemporary populations.}\footnote{Indeed, the IV has the highest power for predicting the set of cities captured in $\mbox{CityDecline}_o$ among possible combinations of pre-colonial cities.
%To show the validity of the IV, we conduct placebo tests, in which we run the same specification with proximity to a currently-populated inland city and a contemporary small coastal city as a dependent variable respectively.
%Table \ref{tab_placebo_first_stage_pop} reports that proximity to an ancient lake does not predict any of these cities better.
%Insignificant results with other combinations of cities are also available upon request.
%}

Columns (1)-(3) of Table \ref{tab_first_stage_pop} report the estimation results of the first stage regression (\ref{eq_city_origin}).
%\blue{Given the exogenous nature of the ancient lakes, the OLS regressions plausibly identify the causal estimates of proximity to the ancient lakes on the formation of core pre-colonial trading points.}
According to column (3), our preferred specification, a 1\% increase in proximity to an ancient lake increases proximity to a pre-colonial trade point by 0.13\% at the 1\% level of statistical significance.
The F-statistics for this first-stage specification is 50.81.
%F-statistics from alternative specifications reported in subsequent robustness checks are of similar magnitude, generally around 50.
The F-statistics from alternative specifications across robustness checks in a later section remain consistently around 30--50. %, which is considered to be high enough to conclude that the first stage correlation is sufficiently strong.
%Therefore, these results indicate a strong first-stage relationship between the instrument and the city decline is reasonably sufficiently strong to interpret the second-stage results.
These results indicate a strong first-stage relationship between the instrument and city decline, supporting the reliability of the second-stage estimates.

Columns (4)-(6) provide supplementary evidence.
Distance to ancient lakes also significantly predicts proximity to historical inland trade routes, which constituted the backbone of trans-Saharan commerce prior to colonization.
This result is consistent with the idea that ancient water sources shaped not only the locations of pre-colonial cities but also the spatial structure of inland trade networks that sustained them. Together, these findings reinforce the interpretation that ancient water access captures fundamental determinants of pre-colonial economic activity, rather than serving merely as a source of exogenous variation that narrowly predicts a single historical outcome.\footnote{
As Figure \ref{map_geography_history_jihad} shows, we also have information on ancient rivers.
%Recall from Figure \ref{map_geography_history_jihad} that some pre-colonial cities that are far from ancient lakes are located very close to the ancient river lines.
However, we observe the overall weak result of the ancient river effects.
The results are available upon request.
%The results show that proximity to an ancient lake has the highest power to predict proximity to a core trade city.
%This would be because overall there is a wide range of areas where ancient rivers were covering but there are no pre-colonial cities overall.
%These results indicate that proximity to an ancient lake has the highest power to predict proximity to a core trade city.
}

\input{tables/tab_first_stage_pop.tex}
The following three observations further support the validity of the instrument.
%Further analysis demonstrates that this instrument is actually meaningful.
%First, The IV has the highest power to predict the declined cities (rather than other sets of historical cities).
First, the IV has the highest power for predicting the set of declined cities captured in $CityDecline_o$ among possible combinations of pre-colonial cities.
Figure \ref{fig_first_stage} reports the same coefficient estimates for the first-stage regressions with alternative combinations of pre-colonial cities.
The first row represents our main specification, same as the estimate reported in Table \ref{tab_first_stage_pop}.
The third row of the figure reports that the coefficient size for the proximity to any pre-colonial city is 0.07, which is almost half of that for the proximity to a declined city.
%The coefficient size for the proximity to a declined city is almost twice as large as that for the proximity to any pre-colonial city.
Other rows also confirm that the ancient water access cannot well predict other sets of cities, including non-declined cities or coastal cities.
%To show the validity of the IV, we conduct placebo tests, in which we run the same specification with proximity to a currently-populated inland city and a contemporary small coastal city as a dependent variable respectively.
%Table \ref{tab_placebo_first_stage_pop} reports that proximity to an ancient lake does not predict any of these cities better.
%Insignificant results with other combinations of cities are also available upon request.

Second, ancient water sources do not predict contemporary city formation.
Table \ref{tab_ols_city_today_water_rev} presents the relationship between contemporary economic development and past and present water sources and contemporary economic development. % with the same other controls of specification (\ref{eq_city_origin}).
In columns (1)-(2), we use the log distance to the nearest city with a contemporary population exceeding 50,000. Proximity to an ancient lake has a small and statistically insignificant effect, indicating a lack of persistent influence of initial geography on contemporary city formation.
In columns (3)-(4), we proxy local economic activity using nighttime light intensity in 2015 from the Visible Infrared Imaging Radiometer Suite (VIIRS). %, which provides high-resolution measures.
We find a negative effect of proximity to ancient lakes, while proximity to contemporary water sources has a positive and statistically significant effect.
Together, these results highlight the importance of accounting for changes in first-nature geography over time, illustrated by the shrinking of ancient water sources and the rising economic relevance of the Niger River.
%\ST{Do we need this Panel (B)? Confusing?} In panel (B), restricting the grid cells in the countries covering the Sahara (i.e., Chad, Niger, Mali and Mauritania), we use the log of one plus distance (km) to the nearest city with contemporary population over 10,000 as an dependent variable, given that these landlocked countries have much smaller population densities than other West African countries many of which face coasts.
%\MK{REMOVE CURRENT PANEL (B) of Table \ref{tab_ols_city_today_water} (Sahara, 10000). New (B) = Nightlights. Also remove the headlines of entire West Africa in the table.}

Third, proximity to contemporary water sources has no explanatory power for proximity to a core pre-colonial city.
As Table \ref{tab_first_stage_pop} illustrates, the coefficients on a contemporary water source are statistically insignificant in most specifications and their sizes are also significantly smaller than those on proximity to an ancient water source.
These results imply that locations of the core trade cities reflect the initial geography from ancient periods rather than contemporary water sources.

\begin{figure}[t]
\begin{center}
\includegraphics[width=9cm]{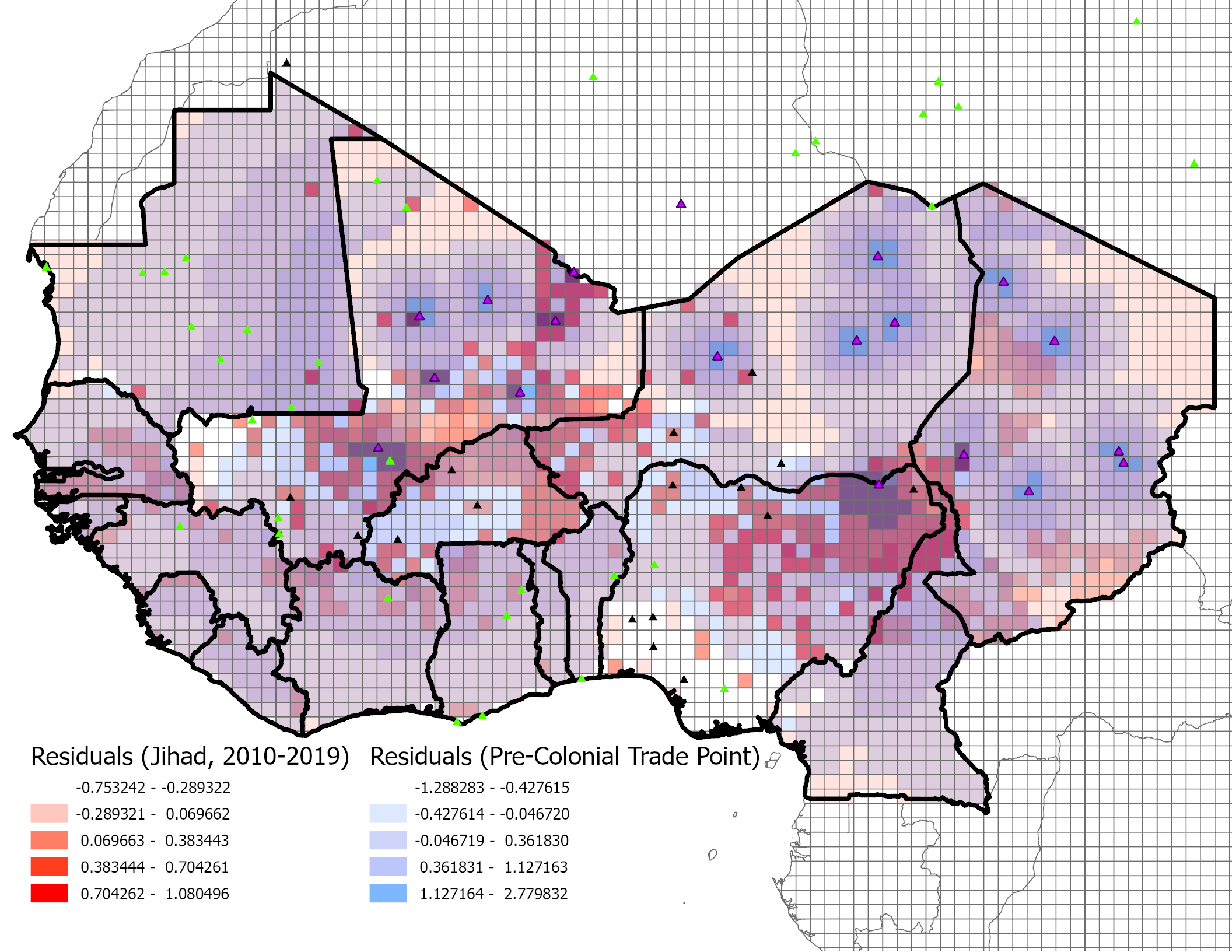} %元々14.5cm
\caption{Overlay of Residuals for Jihadist Violence and Pre-Colonial Trade Points}
\label{map_residuals}
{\parbox[t]{\textwidth}{
{\scriptsize\begin{singlespace}
\textit{Notes}:
This figure overlays two residuals: the red shading represents residuals from the regression of the jihad (2010--2019) dummy on the full set of controls; the blue represents (negative) residuals from the regression of log distance to pre-colonial inland trade points with less than 100,000 population today on the full controls.
The controls include landlocked dummy, malaria suitability, caloric suitability in post 1500, elevation, ruggedness, and country fixed effects.
The purple triangles indicate pre-colonial inland trade points with less than 100,000 population today, the light green ones indicate pre-colonial coastal trade points with less than 100,000 population today, and the black ones indicate the other pre-colonial trade points.
The color of cells where high residuals overlap turns purple.\end{singlespace}}}}
\end{center}
\end{figure}
\subsubsection*{The second stage}
We present the correlation between city decline and contemporary jihad, the reduced-form relationship between our instrument and the outcome, and then the main IV estimates.
%We first present the correlation between the city decline and the outcome of interest as well as the reduced-form relationship between the instrument and the outcome of interest. We then report the main instrumental variable estimates.

\textbf{The correlation.} Figure \ref{map_residuals} illustrates a strong correlation between pre-colonial cities and contemporary jihad.
This map overlays two sets of residuals: the red shading represents residuals from regressing the jihad dummy on the full set of controls, while the blue represents the (negative) residuals from regressing the log distance to pre-colonial core inland trade points on the same controls.
Cells where high residuals overlap appear in purple.
Notably, we observe clusters dark purple cells around several pre-colonial trade centers. %indicating the significant relationship between these colonial cities and contemporary jihad.%, after controlling for contemporary populations.

\textbf{The reduced-form results.}
Table \ref{tab_water_jihad_pop_rev} reports the regression results of contemporary jihadist incidents directly on the proximity to ancient water lakes.
%According to column (1) of all the panels,
A 1\% increase in proximity to ancient lakes from a grid cell increases the probability of experiencing a jihadist event in the cell by 0.04\%, the number of jihadist events in the cell by 0.16\%, and the proximity to a jihadist event from the cell by 0.15\% during 2001--2019 (all $p<0.01$).
This table also indicates that these effects are concentrated in 2010--2019, as we describe details later.
These results demonstrate that ancient water access—a key determinant of historical prosperity and subsequent decline—robustly predicts contemporary jihadist activity, reinforcing the relevance of the instrument and corroborating the causal interpretation of the following IV estimates.
%\ST{Takeaway draft 2} These results provide reassuring evidence that our instrument has strong predictive power for the final outcome, corroborating the causal chain established in the IV analysis.

%\input{tables/tab_iv_jihad_pop.tex}
\textbf{The main IV results.} Table \ref{tab_iv_jihad_pop} reports the IV estimates of the effects of the declined cities on contemporary jihad (2001--2019).
According to column (1) of all the panels, a 1\% increase in proximity to a declined city increases the probability of experiencing a jihadist event in the cell by 0.3\%, the number of jihadist events in the cell by 1.2\%, and the proximity to a jihadist event from the cell by 1.2\% during 2001--2019 (all $p<0.01$).
%All of these estimates are statistically significant at the 1\% level in our preferred specification.

These effects are mostly concentrated in 2010--2019, during which the intensity of jihadist events have significantly risen.
The mean of log intensity of jihadist events is 0.244 in 2010--2019 and 0.011 in 2001--2009.
From column (2) of all the panels, we observe statistically insignificant effects of the declined cities during 2001--2010 and their coefficient sizes are also significantly smaller.
In column (3) of all the panels, point estimates of the effects of the declined cities during 2010--2019 and their level of statistical significance are similar to those in column (1).
% 1\% increase in proximity to a pre-colonial core city from a grid cell increases the average proximity to a jihadist event from the cell by 1.3\%, the probability of experiencing a jihadist event in the cell by 3.3\%, and the average number of jihadist events in the cell by 1.2\% during 2010-2019.
%All of these estimates are statistically significant at the 5\% level in our preferred specification.
These results indicate that the jihadist events during 2010--2019 are closely linked to the historical Islamic places, the core trade cities back to more than 200 years ago. Section \ref{sec_mechanism} discusses the mechanism behind these results.

\input{tables/tab_iv_jihad_pop.tex}
We also use the same specification but with the explanatory variable being proximity to the trade route network up to 1800CE constructed by \citet{MNP2018}. %, calculating the distance to the nearest trade route from each grid cell.
%Similarly with the main interest variable (i.e., proximity to the inland historical trade cities),
%We pick the inland trade route network and calculate the distance to the nearest trade route from each grid cell.
Columns (4)-(6) in Table \ref{tab_iv_jihad_pop} show the results for this measure.
We find the qualitatively same results as in columns (1)-(3).
Importantly, the coefficient size of proximity to a trade route network is smaller than proximity to a core trading city, implying that proximity to a core trading \textit{city} matters more than proximity to a trade \textit{route}.

\section{Persistence of Jihadist Ideology as a Colonial Legacy}\label{sec_mechanism}
%This section explores the mechanism behind the above empirical results. %specific to the West African experiences. The global-scale discussion follows in the next section.
We argue that the persistence of jihadist ideology as a legacy of European colonization serves as the primary mechanism underlying the above results, supported in four ways.
%\blue{First, we emphasize that jihad has been cyclic, with distinct spatial distributions over the centuries.}
%First, we emphasize that jihad is not merely a contemporary phenomenon, but cyclic over the centuries.
%We also show that the spatial distribution of jihadist activity is inconsistent over time, despite this historical continuity.
%Second, we then argue that the empirical results are explained by the mechanism that jihadist ideology is persistent as a legacy of European colonization, a claim supported by both historical anecdotes and individual-level survey data.
%\blue{Second, we provide qualitative historical evidence on responses to colonization forces---shaped by military asymmetry and a prevalent religious practice---contributing to the ideological persistence. Quantitative evidence on historical weapon access further support the story.}
First, we provide quantitative and qualitative evidence on historical weapon access and military asymmetry, strategic responses to colonial invasion, and a prevalent religious practice, to substantiate the proposed mechanism of ideological persistence. %contributing to the ideological persistence.
Second, we rationalize the observed patterns using a simple multi-period model of conflict. %between a colonizer and an Islamic state.
%\blue{Third, MOSQUEs..summarized by one line. \textbf{Use up to 11 or 12 lines here. Maybe 11 is difficult, so 12? Actually 11 may be difficult, so shall we use at most 12 lines in total?}}
Third, we offer consistent quantitative evidence using individual-level survey data on extreme religious ideologies.
%and that how historical conflicts ended matter for explaining contemporary conflicts.
%how conflicts end matter for future conflicts.
Fourth, we highlight heterogeneity across contemporary jihadist groups, documenting differences in their reliance on localized recruitment strategies, which further reinforce the proposed mechanism.

\subsection{Strategic Adaptation to Colonial Rule through Religious Practice}

\subsubsection*{Military asymmetry during colonial invasion and the persistence of jihadist ideology}
The interpretation aligning with the set of empirical findings concerns the balance of military power between Islamic states and European military forces during colonial invasion.
\par
Islamic state forces with reasonable access to coastal areas could directly obtain arms and ammunition from European traders (e.g., \citealt{Kea1971, Law1976, Thornton1999}), influencing how they confronted European invasion.
Samori’s (Wassoulou) state forces exemplify this pattern.
%Benefiting from access to weapons from European traders,
They imported more modern weapons from Sierra Leone, and adopted a hard confrontation strategy against European forces, including guerilla tactics (\citealt{Legassick1966, Crowder1971}).
Such Islamic state forces with relatively good access to modern weapons engaged in more intense fighting with European colonizers.
As a result, European military forces decisively defeated such state forces.
Consequently, these states ceased to exist, and the seeds of jihadist ideology within them might also have diminished.
\par
By contrast, Islamic state forces with limited access to weapons did not engage as fiercely with European forces due to stark asymmetries in military capacity.
Two prominent examples are the Tukulor Empire and the Sokoto Caliphate, both of which ultimately adopted submission as primary strategy in the face of overwhelming military asymmetries.
For example, \citet{Crowder1971} notes, ``the reasons for the ineffectiveness of Tukulor military resistance are fairly obvious,'' further describing them as ``hopelessly outgunned'' by the French military, which resulted in no military resistance for most of the conquest period.
Similarly, in the Sokoto Caliphate, limited access to European weapons and the resulting inferiority in weapon quality ultimately left it with ``no tactics, no personal gallantry and no resistance'' against European conquest (\citealt{Crowder1971}, p. 294).
Under such circumstances, these state forces resorted to strategies of alliance, acquiescence, or submission to European powers.
While these states also ceased to exist, the seeds of jihadist ideology within them were {\it not} diminished, fueling future jihadist conflicts.
\par
At the same time, even when examining within-state variations, areas characterized by relatively better access to weapons witnessed direct confrontation. %, further consistent with the same underlying mechanism.
In the Tukulor Empire, the well-known gun-slave cycle operated not solely at the level of the state but at the level of individual warriors, generating substantial variation in access to weapons across territories (\citealt{Roberts1987}).
In its capital, Segu, where access to firearms was relatively better in the mid-19th century, state forces engaged in armed confrontation with French troops during the 1850s, as recorded in \cite{Brecke1999}. %, our source for historical conflict data.
In the Sokoto Caliphate, although submission became the dominant state-level outcome, military capacity varied across emirates (\citealt{Smaldone1977}).
Certain emirates, notably Sokoto and Kano, possessed relatively greater concentrations of firearms, and episodes of armed confrontation with British forces are documented in these areas, albeit for short durations (\citealt{Brecke1999}).
\par
These patterns indicate that both across and within states, military confrontation was more likely where access to weapons was comparatively better, whereas limited access was associated with the absence of sustained resistance, such that the seeds of jihadist ideology were not diminished and instead persisted beyond the colonial conquest.
\par
However, while the presence of jihadist ideology might be a necessary condition for contemporary jihadist activities, it is not sufficient on its own.
This persistence cannot solely explain why significant effects on contemporary jihad were observed only during 2010–2019 and not in the earlier period of 2001–2009.
The persistence of jihadist ideology, coupled with a contemporary shock that strengthens insurgent forces, such as the inflow of fighters and weapons from Libya following its security collapse (\citealt{Shaw2013}), could trigger a sudden surge in contemporary jihad.
Indeed, it has been well-documented that the influx of weapons from Libya since 2011 has contributed to a widespread violent conflict in West Africa (e.g., \citealt{Marsh2017}; \citealt{MN2019}).

\input{tables/tab_gun_slave.tex}
\subsubsection*{Empirical support for the military asymmetry}
%Table \ref{tab_gun_slave} provides empirical support for the military-asymmetry channel during colonial invasion by linking our geography-based variation to historical access to weapons and related logistics. The key point is that locations farther from ancient lakes—which are also farther from declined inland trade cities—systematically exhibit greater access to military-relevant resources prior to and during colonization.

Consistently, Table \ref{tab_gun_slave} provides direct empirical support for the military-asymmetry channel by linking proximity to ancient lakes—and, by construction, proximity to declined cities—to historical access to weapons prior to colonization.

Columns (1)-(2) examine gun access in 1757–1806.
The gun access measure is constructed as a quantity-weighted measure that weights gun imports by the inverse of the distance from each grid cell to coastal trading locations involved in the Anglo–West African gun trade, drawing on \cite{Inikori1977}, which reports the destinations and quantities of guns carried on 111 trading voyages from England to West Africa over this period.
For each destination, we aggregate the total number of guns shipped and assign geographic coordinates to representative trading locations when destinations are reported at a broad regional level.
Additional details on port assignments and data construction are provided in Appendix \ref{app_data_precolonial}.

These columns show that the core geography significantly predicts gun access: a 1\% increase in distance from an ancient lake and from a declined city leads to a 0.02\% ($p<0.01$) and 0.27\% ($p<0.05$) increase in gun access, respectively.
Importantly, these estimates are conditional on distance to the coast, which is strongly correlated with gun access, reflecting the fact that firearms were initially imported through coastal ports and subsequently transported inland.
These results imply that areas farther from the historical inland cores had systematically better access to firearms prior to colonial encounters.
%This indicates that areas farther from the ancient inland cores were better positioned to obtain firearms before and during early colonial encounters.

\if0
\ST{DROP because railway is NOT pre-colonial!}
\blue{Columns (3)-(4) focus on distance to colonial railways, which were critical for moving weapons, troops, and traded goods. The negative and significant coefficients indicate that areas farther from ancient lakes and declined inland cities were closer to colonial rail infrastructure.
This pattern is consistent with railways connecting coastal entry points to export-oriented regions, rather than serving the historically inland Islamic cores.}
%\blue{This pattern is consistent with railways being concentrated in coastal or outward-facing regions rather than in the historically inland cores.}
%\blue{This pattern reflects the fact that railways were built primarily to connect coastal entry points to export-oriented regions, rather than to serve the historically inland Islamic cores.}
\fi

Columns (3)-(6) examine slave exports in the 1700s and 1800s, a key channel through which weapons were obtained in exchange with European traders, as noted in Section
\ref{sec_history}.\footnote{ %\ref{sec_history_19th}.
%\ST{Pre-colonial (rather than colonial)?}
%\blue{The European colonizers built ports and engaged in colonial trade along the coast.The number of Atlantic slave exports proxy for the intensity of colonial trading activities.}
The data come from \citet{Nunn2008}.
Each grid cell, the unit of analysis, is assigned to its corresponding ethnic homeland, so that variation in the dependent variable operates at the ethnic homeland level.
As the data contain missing values for uninhabited areas, the number of observations is reduced accordingly.
}
Across both centuries, distance from ancient lakes and from declined cities significantly predicts slave exports.
Specifically, a 1\% increase in distance from an ancient lake and from a declined city leads to a 0.27\% and 4.17\% increase in slave exports in 1700s (both $p<0.01$), and with a 0.19\% ($p<0.01$) and 2.88\% ($p<0.05$) increase in 1800s, respectively.
This finding reinforces the interpretation that areas farther from the historical cores had greater access to European trade networks, and by extension, to weapons.

%\ST{TAKEAWAY DRAFT MAIN, combining DRAFT1 \& DRAFT2 below}
Taken together, these results confirm that the ``past-core'' areas faced a severe disadvantage in acquiring modern arms.
This lack of access to guns and the economic means to acquire them (slave trade) substantiates the historical narrative that state forces in these areas were ``hopelessly outgunned.''
This overwhelming military asymmetry vis-\`a-vis European forces forced them into strategies of outward submission rather than armed confrontation, thereby allowing the seeds of jihadist ideology to persist.

\begin{figure}[t]
\begin{center}
\includegraphics[width=9cm]{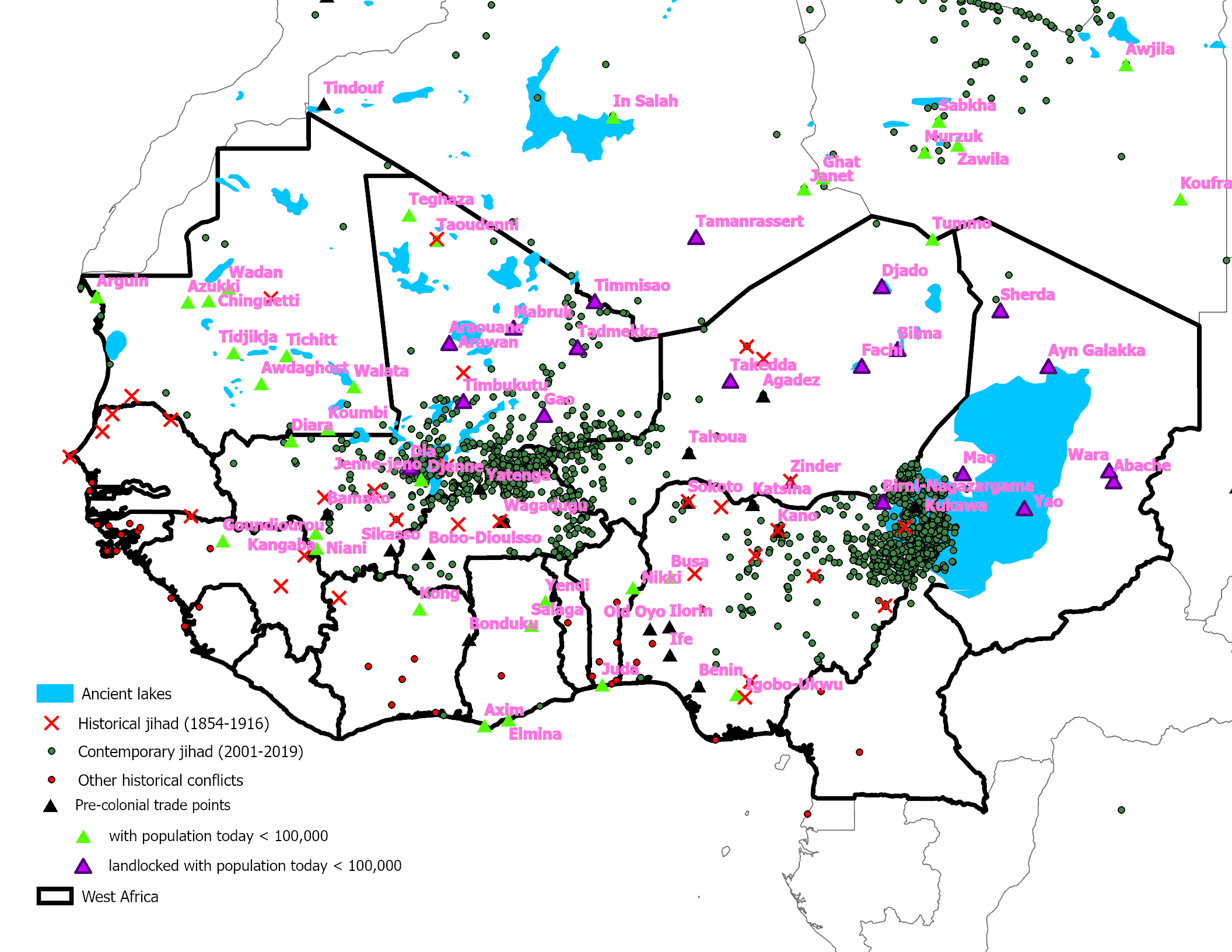}
\caption{Historical and Contemporary Jihad}
\label{map_cycles_jihad}
{\parbox[t]{\textwidth}{
{\scriptsize\begin{singlespace}
\textit{Notes}:
This figure shows the locations of historical (red crosses) and contemporary jihad (green points), along with pre-colonial cities, ancient lakes and other historical conflicts.
%Historical conflicts between 1854 and 1916 are based on \citet{Brecke1999}.
Data on historical conflicts between 1854 and 1916 are drawn from \citet{Brecke1999} and supplemented with additional geocoding based on multiple sources, as detailed in Section \ref{sec_data}.
Historical jihad refers to conflicts against European forces involving Islamic states or Muslim ethnic groups, whereas other historical conflicts include all the other conflicts against European forces.
The purple triangles indicate pre-colonial inland trade points with less than 100,000 population today, the yellow-green triangles indicate pre-colonial coastal trade points with less than 100,000 population today, and the black triangles indicate all the other pre-colonial trade points.\end{singlespace}
}}}
\end{center}
\end{figure}
\subsubsection*{The cycle of jihad with distinct spatial distributions}
The historical records on colonial-era conflicts offer supporting evidence for this asymmetry. %\blue{Jihad has been cyclic over the centuries, but with distinct spatial distributions across time.}
As Figure \ref{map_geography_history_jihad} shows, contemporary Islamist violence is concentrated in areas formerly governed by historical Islamic states, yet with substantial spatial variation across different states.\footnote{
\href{https://www.culturesofwestafrica.com/maps/}{Cultures of West Africa} creates maps of historical states before colonization as well as modern countries after independence by using multiple sources of references.
%\footnote{The references and maps are available in the \href{https://web.archive.org/web/20190603215326/https://www.culturesofwestafrica.com/wp-content/uploads/2018/09/HistoryWestAfrica.pdf}{website}.}
We digitize these maps from pre-colonial periods to the colonial era (Figure \ref{map_historical states}).
Appendix \ref{app_empires} describes how to identify Islamic states in detail.
}
A key distinction lies in the extent to which these states avoided direct armed confrontation against European forces. Contemporary jihadist violence is concentrated in the former territories of the Tukulor Empire and the Sokoto Caliphate—both of which ultimately offered little effective resistance against European conquest—while very little contemporary jihadist violence is observed in the former territory of the Samori (Wassoulou) Empire, which adopted a hard confrontation strategy against European forces (\citealt{Crowder1971}; \citealt{Legassick1966}).\footnote{
Appendix \ref{app_historical_conflict}
summarizes the strategies adopted by Islamic states in response to European colonial forces, drawing on multiple historical sources.}
%Figure \ref{fig: islamic_state_strategies} displays the spatial distribution of contemporary jihadist violence over historical Islamic states in 1860.}
This pattern stands in contrast to the widely recognized tendency for violence to recur in the same locations %through intergenerational transmission of grievances and distrust
%(e.g., \citealt{BR2014, DFO2019, RTZ2013a, VV2012}).
(e.g., \citealt{BR2014, DFO2019, VV2012}).

To further investigate this,
we use data on conflicts involving historical Islamic states and European powers from \citet{Brecke1999}, which records violent events with at least 32 deaths worldwide between 1400 and 2000,\footnote{
The raw data are available on this \href{https://brecke.inta.gatech.edu/research/conflict/}{website}.
For each event, it reports the actors involved, the start and end years, and the region of onset.
Actors are defined as political entities exercising effective sovereignty over territories (e.g., states, kingdoms, or sub-national groups).
Table \ref{tab_brecke} lists all recorded colonial conflicts involving historical states in West Africa.
In total, the database documents 42 such conflict events, of which 15 involve Islamic states.
However, several limitations of this dataset should be noted.
As is common for long-run historical conflict compilations, the dataset may not record all relevant events---a concern particularly salient for Africa, with comparatively limited historical documentation (\citealt{DFO2019}). %raises the possibility of underreporting (\citealt{DFO2019}).
Geographic information is also often coarse, with conflict locations reported at a broad regional level. %that may introduce geolocation error.
Individual conflicts may additionally be missing: for example, the military confrontation between Umar Tall, founder of the Tukulor Empire, and French forces at Médine in 1857 is not recorded (\citealt{Roberts1987}).
These limitations suggest the dataset should be interpreted as a partial rather than exhaustive account of historical conflicts.
} %, especially in regions with sparse archival coverage such as West Africa.}
%We focus on conflicts in West Africa in which the actors consist of historical states and European countries.
supplemented with additional geocoding based on multiple sources.\footnote{Specifically, we rely on \citet{FK2017}, which comprehensively geolocate conflict events between 1700 and 1900 in Africa.
For conflicts after 1901, we manually geocode the regions listed in \citet{Brecke1999} using web-based sources such as Wikipedia and Google Maps.}
Figure \ref{map_cycles_jihad} maps the locations of historical jihads and shows the distinct spatial distribution of jihads over time.
Table \ref{tab_colonial_jihad_lake} shows that neither ancient water access nor proximity to past-core cities significantly predicts the location of historical jihads.
The persistent influence of declined cities is therefore specific to contemporary jihadist violence, consistent with the idea that ideological seeds survived where outward submission, rather than armed confrontation, was the dominant colonial-era strategy.

\if0
\blue{However, several limitations of the (\citealt{Brecke1999}) dataset should be noted.
First, as is common for long-run historical conflict compilations, the dataset may not record all relevant events.
This concern is particularly salient for Africa, where the historical literature is comparatively limited, raising the possibility of underreporting of conflicts (\citealt{DFO2019}).
In addition, the geographic information provided in the dataset is often coarse, with conflict locations reported at a relatively broad regional level, which may introduce geolocation error.}

\blue{Second, individual conflicts may be missing from the dataset.
For example, the military confrontation between Umar Tall, founder of the Tukulor Empire, and French forces at Médine in 1857 is not recorded (\citealt{Roberts1987}).
These limitations suggest that the dataset should be interpreted as providing a partial, rather than exhaustive, account of historical conflicts, especially in regions with sparse archival coverage such as West Africa.}
\fi

\subsubsection*{The religious practice of \textit{taqiyya}}
The persistence of ideology can be interpreted as being sustained through the religious practice of \textit{taqiyya}, which allowed Muslims to outwardly adapt to situations they could not change while internally preparing to reassert Islamic purity.
The outward conformity coupled with inward resistance, reinforced through educational institutions, can be transmitted across generations until the time is ripe.\footnote{
Originating in the early Middle Ages, taqiyya was practiced among Shi'ite and other minority sects and transmitted across generations as a survival strategy under Sunni domination (e.g., \citealt{BSKC2012}). %, Daftary2007, Friedman2009}).
Moreover, analogous practices of dissimulation can be found beyond the Islamic context: among crypto-Jews in early modern Spain (\citealt{Yovel2009}), the 'hidden Christians' in Tokugawa Japan (\citealt{Turnbull1998}), and African diasporic religions under slavery (e.g., \citealt{MGG2005}).
%, which concealed African deities under Catholic forms (\citealt{MGG2005, Johnson2015}).
}
Importantly, \quotes{taqiyya is of fundamental importance in Islam and practically every Islamic sect agrees to it and practices it} (\citealt{Mukaram2004, Ibrahim2010}).
Furthermore, taqiyya was not confined to the colonial period; it has also been observed among contemporary jihadists.
To illustrate this continuity, Appendix \ref{app_qualitative_taqiyya} describes how taqiyya has been practiced both in the colonial period and in the strategies of present-day jihadist groups such as Al Qaeda, the Islamic State, and Boko Haram.

\subsection{The Dynamic Model of Conflict}
%\ST{Connect x axis to geography}
To rationalize the mechanism,
we use a simple dynamic model of conflict, %between an European colonizer and an Islamic state
built on \citet{BS2020}.
In this section, we outline the model, present simulation results, and map these to geography and history in our context.
Appendix \ref{app_model} provides details.

%We simulate the model to replicate the nature of historical and contemporary jihad that we observed.
%The model is built on \citet{BS2020}.

%\subsubsection*{The setup}
\textbf{The setup.}
There are two players, a colonizer ($F$) and an Islamic state ($M$), competing over a divisible territory.
There are two periods ($t=1,2$), with no discounting of future payoffs.
The total amount of the territory is normalized to 1, and the allocation at the beginning of period $t$ is given by $\omega_{Ft}$ and $\omega_{Mt}$ for players $F$ and $M$, respectively. %, where $\omega_{Ft} + \omega_{Mt} = 1$.
The relative military strength of player $i \in \{F, M\}$ is denoted by $\lambda_i$, where $\lambda_F+\lambda_M=1$.
Players incur a cost of conflict ($\phi_{i}$, $i \in \{F, M\}$) if they engage in fighting.
We make two important and realistic assumptions:
(i) player $F$ has a smaller amount of initial territory at the beginning of the game ($\omega_{F1} < \omega_{M1}$);
(ii) player $F$ is militarily stronger ($\lambda_F > \lambda_M$).
%The analysis will demonstrate that the degree of this asymmetric power relations matters to explain both the historical and contemporary jihads.
%The cost of conflict for player $i$ is denoted by $\phi_i$.
%We assume that these costs are equal, $\phi_F = \phi_M = \phi$.

In each period, the bargaining game has two stages.
In Stage 1, player $i \in \{F, M\}$ can either ``challenge'' by making a claim $\sigma_{it}$ where $\omega_{it}<\sigma_{it}\le 1$, or make no claim.
If neither player challenges, both players retain control of their respective shares ($\omega_{it}$).
The cost of making a challenge for player $i$ is $c_i$, which is private information and independently drawn from distribution $F(c)$ over $[\underline{c}, \bar{c}]$.
Stage 2 is reached if exactly one player makes a claim and the other does not.
The other player then decides whether to concede to the claim.
If both players make claims or if neither claims, Stage 2 is not reached and the game ends.

The winner of a conflict will take the full territory, %(i.e., $\sigma_{it}=1$ if $i$ wins)
and conflicts occur in the following two cases.
The first case is where only one player makes a claim in Stage 1 and the other player does not concede to in Stage 2.
Suppose $F$ is the only player who makes a claim.
Then, $F$'s winning probability is $\lambda_F+\theta$ while $M$'s winning probability becomes $1-(\lambda_F+\theta)=\lambda_M-\theta$, where $\theta$ is a parameter governing the first-mover advantage.
We assume $0 \le \lambda_i-\theta$ and $\lambda_i+\theta \le 1$ for all $i$ to ensure that the winning probabilities are well-defined.
The second case is where both players make claims in Stage 1.
In this case, the winning probability of player $i$ is $\lambda_i$, as both players are simultaneously challenging.

Player $i$'s instantaneous utility from controlling a share $\sigma_{it}$ in period $t$ is given by $u_i(\sigma_{it})$, where $u_i$ is an increasing, strictly concave, and differentiable function on $[0,1]$.
%Without loss of generality, we normalize the function such that $u_i(1)=1$ and $u_i(0)=0$.

%\subsubsection*{The two-period game}
The dynamic game proceeds as follows.
If conflict occurs in $t=1$, then the game ends and both players continue to receive utilities from the obtained (or lost) territory in $t=2$.
%If a territorial transfer from one player to the other due to concession occurred in $t=1$, the resulting allocation becomes the new status quo territory at the beginning of $t=2$ ($\omega_{F2}, \omega_{M2}$).
If a territorial transfer due to concession occurrs in $t=1$, the resulting allocation becomes the new status quo territory at the beginning of $t=2$ ($\omega_{F2}, \omega_{M2}$).
If neither player challenges in $t=1$, the status quo territory remains unchanged in $t=2$ ($\omega_{i2}=\omega_{i1}$).

%\subsubsection{Optimal strategies and the equilibrium}
\textbf{Optimal strategies and the equilibrium.}
We solve the game by backward induction, given the set of parameters. %, detailed in Appendix \ref{app_model}.
First, we solve for optimal strategies in $t=2$ as in the one-shot game,
given the new status quo territory ($\omega_{F2}, \omega_{M2}$) realized after $t=1$ (in the scenario where conflict has not occurred in $t=1$).
This game can be expressed as one in which both players simultaneously decide whether to challenge with the optimal amount of claim or not challenge, labelling the optimal challenge Hawk ($H$) and no challenge Dove ($D$).
In this game, player $i$'s strategy is defined as $s_i: [\underline{c}, \bar{c}] \to \{H,D\}$.
We can then obtain equilibrium strategies in $t=2$ for $i\in\{F,M\}$, based on the optimal cutoff $\hat{x}_{i2}(\omega_{i2})$ such that $s_i(c_i) = H$ if $c_i \le \hat{x}_{i2}(\omega_{i2})$.  %, depending on the newly realized territory in $t=2$.
Next, we solve for optimal strategies in $t=1$, taking into account the expected payoffs in $t=2$ under a potential new status quo $(\omega_{F2}, \omega_{M2})$. %, before the challenge cost $c_i$ is realized.
Solving this problem yields the equilibrium cutoff strategies in $t=1$, $\hat{x}_{i1}$ for $i\in\{F,M\}$.

\begin{figure}[t]
\begin{center}
\includegraphics[width=7.5cm]{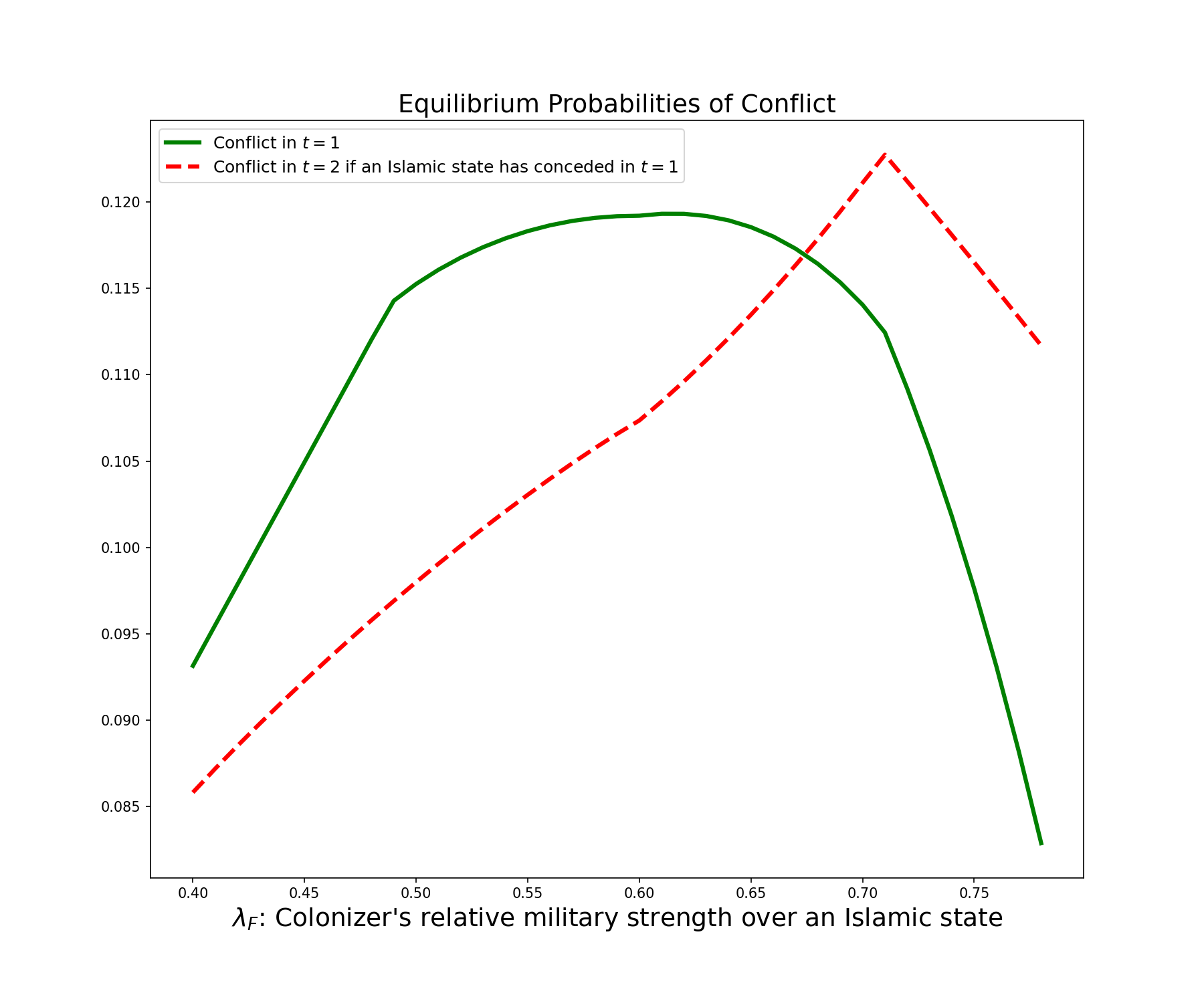}
\includegraphics[width=7.5cm]{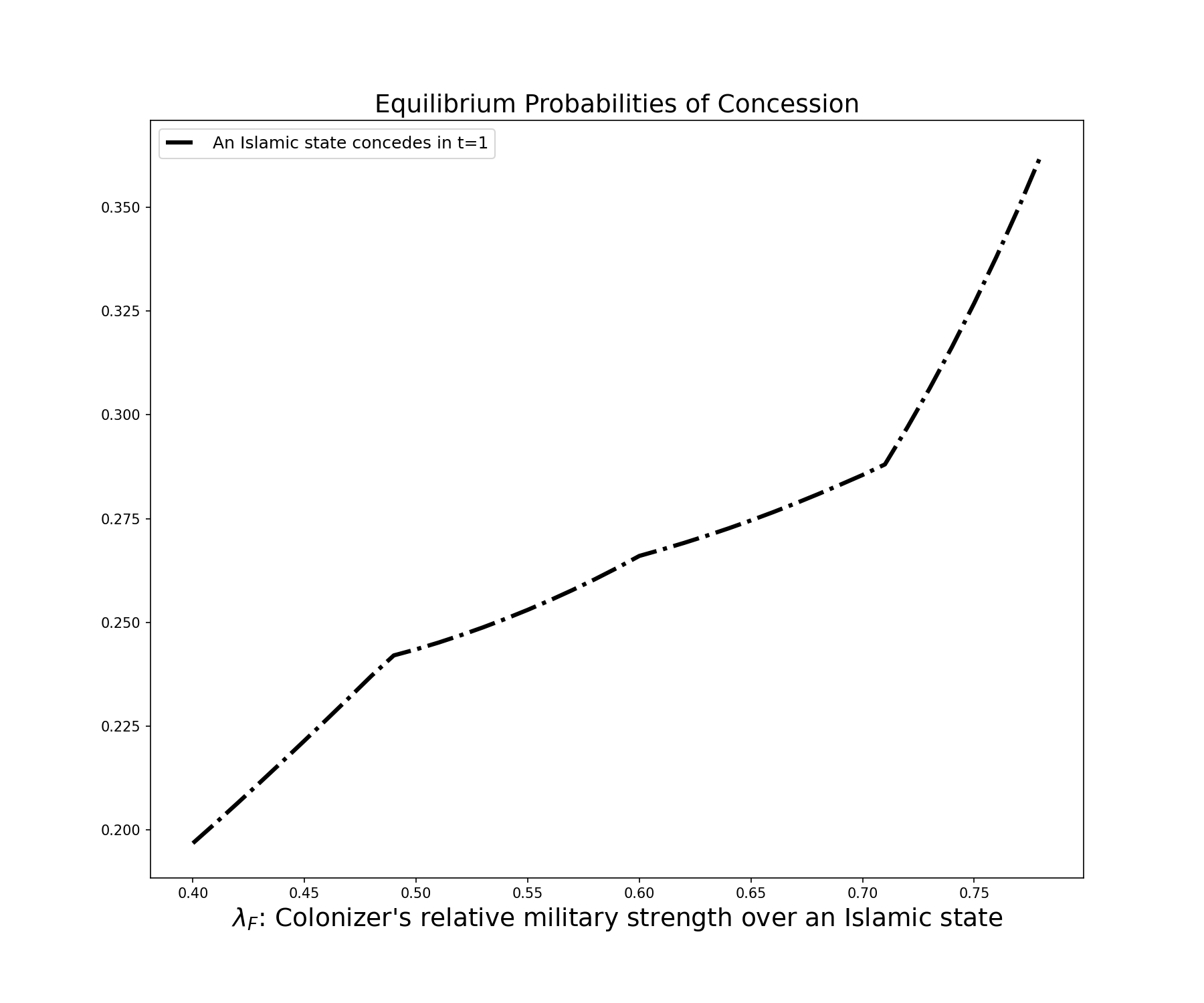}
\caption{Simulated Equilibrium Probabilities of Conflict and Concession}
\label{fig_model_simulation}
{\parbox[t]{\textwidth}{
{\scriptsize\begin{singlespace}
\textit{Notes}:
This figure shows the simulated equilibrium probabilities of conflict and concession from the presented dynamic model of conflict.
The horizontal axis in both panels represents the colonizer's relative military strength over an Islamic state ($\lambda_F$).
Note that the maximum value this parameter can take is 0.8 because $\theta=0.2$ (the first mover advantage), which also adds to the conflict winning probability.
\end{singlespace}}}}
\end{center}
\end{figure}

\subsubsection*{Simulating equilibrium probabilities of conflict and concession}
%\textbf{Simulating equilibrium probabilities of conflict and concession.}
Figure \ref{fig_model_simulation} plots the equilibrium probabilities of conflict and concession by varying $\lambda_F$ and setting $\theta=0.2$, $\omega_F=0.3$, $\phi_i = 0.2$, %$v_i=1.5$, $g_i=0.5$,
$u_i(\sigma_{it})=\sigma_{it}^{0.6}$, and $c_i$ uniformly distributed on [0, 1] for all $i$.
In equilibrium,
the probability of conflict in $t=1$ is $F(\hat{x}_{F1})\cdot F(\hat{x}_{M1})$,
the probability that player $F$ challenges and player $M$ concedes in $t=1$ is $F(\hat{x}_{F1})\cdot(1-F(\hat{x}_{M1}))$,
and the probability of conflict in $t=2$ conditional on $F$ having challenged and $M$ having conceded in $t=1$ is $F(\hat{x}_{F2}(\hat{\sigma}_{F1}))\cdot F(\hat{x}_{M2}(1-\hat{\sigma}_{F1}))$, where $\hat{\sigma}_{F1}$ represents $F$'s optimal claim in $t=1$.

%\subsubsection*{Simulating equilibrium probabilities of conflict and concession}
%The equilibrium probability of conflict in $t=1$ is $F(\hat{x}_{F1})\cdot F(\hat{x}_{M1})$,
%the equilibrium probability that player $F$ challenges and player $M$ concedes in $t=1$ is $F(\hat{x}_{F1})\cdot(1-F(\hat{x}_{M1}))$, and the equilibrium probability of conflict in $t=2$ conditional on that $F$ had challenged and $M$ had conceded in $t=1$ is $F(\hat{x}_{F2}(1-\eta_{M1}))\cdot F(\hat{x}_{M2}(\eta_{M1}))$.

The left panel illustrates that relative military strength has a non-monotonic effect on the conflict probability in $t=1$:
while it rises as the military strength of the ``rising power'' ($F$) increases from below to moderately above that of the ``status quo power'' ($M$), it drops sharply once $F$ becomes significantly stronger and the asymmetry becomes sufficiently large.
%As the military strength of the ``rising power'' ($F$) increase from below to moderately above that of the ``status quo power'' ($M$), the conflict probability in $t=1$ rises.
%However, once $F$ becomes significantly stronger and the asymmetry becomes sufficiently large, the conflict probability in $t=1$ drops sharply.
The right panel shows that the probability that $F$ challenges and $M$ concedes increases monotonically with the rising power's military strength.

The left panel also demonstrates that, after $M$ has conceded to $F$ in $t=1$, the conflict probability in $t=2$ is substantially higher under high military asymmetry than under moderate asymmetry.
Notably, this outcome arises even without a positive shock to $M$'s military capacity.
Incorporating such a shock further reinforces this relationship.

\subsubsection*{Mapping the model to geography and history}
%\textbf{Mapping the model to geography and history.}
In the context of jihad in West Africa, player $F$ corresponds to a colonizer and $M$ corresponds to an Islamic state, with the degree of military asymmetry captured by the colonizer's relative military strength, $\lambda_F$.
Conflict in $t=1$ reflects historical jihad against the colonizer, while conflict in $t=2$ reflects contemporary jihad broadly understood as resistance to Westernization.
In the historical context of West Africa, the military asymmetry was shaped in large part by geography: as documented in Table \ref{tab_gun_slave}, proximity to declined cities is a strong predictor of limited historical weapon access, with past-core areas systematically disadvantaged in acquiring firearms.
Empirically, proximity to declined cities---where the degree of historical military asymmetry is systematically higher---significantly predicts contemporary jihad (Table \ref{tab_iv_jihad_pop}) but not historical jihad (Table \ref{tab_colonial_jihad_lake}).
%(And qualitatively, areas with more moderate military asymmetry have more intense historical jihad.)
%\blue{Empirically, we observe more historical jihad and less contemporary jihad in regions with moderate military asymmetry at the time of colonial invasion, and conversely less historical jihad and more contemporary jihad in regions with higher military asymmetry, particularly where Islamic states did not intensely resist colonial forces.}
The simulation results in Figure \ref{fig_model_simulation} therefore illustrate that such empirical patterns can be rationalized within a simple two-period model of bargaining and conflict. %, where the degree of military asymmetry varied systematically with proximity to declined cities.

\subsection{Extreme Religious Ideologies from Individual-Level Surveys}
%change%
We use Afrobarometer surveys to support the ideological persistence.
%The Afrobarometer comprises nationally representative, individual-level surveys conducted across several African countries, with geo-coded information available in each enumeration area (EA).
%We use two waves (rounds 6 and 7), implemented between 2014 and 2018, which include relevant variables for this study. Appendix \ref{app_data} provides more details.
%\par
Although there are no direct questions on jihadist ideology or violent extremism, we use variables that capture extreme religious ideologies. % broadly related to jihadism.
This approach aligns with the widely held perspective that jihadist ideology seeks to preserve the purity of Islam against the secularization and Westernization of culture and institutions, as exemplified by movements such as Salafi jihadism (\citealt{Sounaye2017}).
These include views on excluding other religions, governance by religious law (particularly the Shariah law), and restrictions on female education,
which are the only ones we found in the Afrobarometer that closely relate to \textit{extreme} religious ideologies.
%Notably, apart from the three variables we utilize, we did not find other Afrobarometer variables that closely relate to jihadism, despite including many variables from religious practices.
%\blue{Notably, these three variables are the only ones we found in the Afrobarometer that closely relate to \textit{extreme} religious ideologies.}% jihadism, despite the presence of many variables on religious practices.}

%\input{tables/tab_mechanism_ideology.tex}
We estimate the following IV regression:
\begin{eqnarray}\label{eq_afrobarometer}
Y_{rei} = \beta_0 + \beta_1 CityDecline_e + \beta_2X_{rei} + \beta_3X_e + \phi_c + \phi_r + \epsilon_{rei}
\end{eqnarray}
where the unit of analysis is individual $i$ who resides in enumeration area $e$ and participated in the survey at round $r$.
$Y_{rei}$ is one of the three outcome variables introduced above.
$\mbox{CityDecline}_e$ is the log of one plus distance to a declined city from enumeration area $e$ and we instrument it by the ancient water access from $e$.
$X_i$ is a vector of individual-level controls and $X_e$ is a vector of enumeration area-level geographical controls.\footnote{
Individual-level controls include age, age squared, female dummy, nine categorical indicators of education, and four categorical indicators of living condition.
Enumeration area-level geographical controls include the logarithm of distance (km) to the nearest water sources today, landlocked dummy, average malaria suitability, average caloric suitability in post 1500, and average elevation. As this dataset provides us with point data about each enumeration, we create a buffer with 50km radii around the locations of enumeration points when we calculate average malaria suitability, average caloric suitability in post 1500, and average elevation.
}
We additionally control for the round fixed effects $\phi_r$ when an outcome variable is available in the both rounds.
We report standard errors clustered at the region level in this specification.
To capture variations within the Muslim population, rather than across different religious groups, we restrict the sample for analysis to Muslim respondents.
%We restrict the sample to Muslim respondents in the analysis to capture the difference within Muslim not across different religions.
%To isolate variations within the Muslim population, rather than across different, we limi

\input{tables/tab_mechanism_ideology.tex}
Table \ref{tab_mechanism_ideology} reports the IV regression results.\footnote{See Appendix \ref{app_Afrobarometer} for detailed sources and definitions of the variables that we use in this analysis.
The Afrobarometer data is available in most African countries.
Chad is an exceptional country where we observe jihadist events but the data is not available.
}
The dependent variables are ordered and standardized.
The empirical results show that as a respondent's EA is closer to a declined city, he or she is more likely to dislike people of a different religion as neighbors (column 1, $p<0.01$), agree with governing a country by religious law rather than secular law (column 3, $p<0.05$), and disagrees with girls and boys having equal opportunities to education (column 5, $p<0.01$).
%All of these three effects are statistically significant at the 1\% level.
%These effects are statistically significant at the 1\% level (columns 1 and 5) and at the 5\% level (column 3).
%Hence, these results, coupled with the qualitative evidence above, support that the persistence of jihadist ideology is the primary mechanism behind the main results.
Hence, this individual-level survey evidence also supports that the persistence of jihadist ideology is the primary mechanism behind the main results.
%\label{sec_history_colonial}
%\input{tables/tab_mechanism_ideology.tex}

\subsection{Organizational Heterogeneity and Localized Recruitment}\label{sec_disc_heteg}
%We showed that contemporary jihadist violence is concentrated around declined historical cities and argued that persistent jihadist ideology in these areas provides a plausible underlying mechanism.
We have established that contemporary jihadist violence is concentrated around declined historical cities, arguing that persistent jihadist ideology in these locales serves as the underlying mechanism.
If this persistence operates through place-based transmission, then organizations that rely more on localized recruitment would exhibit a stronger spatial association with these declined cities.
%A natural implication is that jihadist organizations operating in these areas rely on localized recruitment.
To assess this implication, we examine heterogeneity across contemporary jihadist groups, focusing on differences in their recruitment strategies.
%\blue{As shown above, we observe persistent jihadist ideology near declined cities. But does this imply that recruitment is actively taking place in these areas, thereby leading to increased jihadist violence? To investigate this, we examine heterogeneity across contemporary jihadist organizations over time and provide supporting evidence for localized recruitment.}

Specifically, we focus on three primary jihadist factions active in West Africa: Al Qaeda affiliates, the Islamic State (IS) affiliates, and Boko Haram. The first two factions operate within a broader global framework dominated by the rivalry between Al Qaeda and the Islamic State, particularly following their split in 2014 (\citealt{Hamming2017}; \citealt{Novenario2016}; \citealt{Zelin2014}). Although rival groups rarely engage in full-scale military conflict against one another, they compete for supremacy over the jihadist movement.
We classify specific West African groups and their affiliations in Table \ref{tab_jihadist_groups}.
%\footnote{Note that the relationship between Boko Haram and Al Qaeda is complicated and time-variant. See \citet{Cummings2017} and \citet{Zenn2020} for details. Note also that the Islamic State in West Africa was established in 2015 after splitting from Boko Haram (\citealt{Bohm2020}).}
%Most jihadist groups in the world are networked.
%The two cores are Al Qaeda and the Islamic State, which are rivalries after their split in 2014 due to a difference in their ideologies (\citealt{Hamming2017}; \citealt{Novenario2016}; \citealt{Zelin2014}).\footnote{
%Note that the relationship between Boko Haram and Al Qaeda is complicated and time-variant. See \citet{Cummings2017} and \citet{Zenn2020} for details. Note also that the Islamic State in West Africa was established in 2015 after splitting from Boko Haram (\citealt{Bohm2020}).}
%Although it is rare that rival jihadist groups militarily fight, these groups compete for supremacy of the global jihadist movement.
%Particularly, we focus on the three largest factions that are active in West Africa---Al Qaeda, the Islamic State (IS), and Boko Haram. %---and assign each jihadist group to each of these factions or none of them based on multiple information sources.
%%Recall that the left map of Figure \ref{app_fig_groups_AQ_IS_WA} shows violent events involving each of major jihadist groups in West Africa.
%Table \ref{tab_jihadist_groups} categorizes the jihadist groups affiliated with Al Qaeda and the Islamic State.

Motivated by the organizational evolution, we further divide the relevant decade into 2010--2015 and 2016--2019.
%These periods correspond to initial and later stages for the three large factions we focus on.
%The initial stage period of Al Qaeda-affiliated groups corresponds to 2001-2009.
The former corresponds to the initial period of IS-linked groups, given that the split of the Islamic extremist groups into Al Qaeda and Islamic States as two competing factions occurred in February 2014.
In the later period, violent events by IS-linked groups became more prevalent compared to the earlier period until 2015.
%\par
%Figure \ref{map_residuals_heteg} maps the overlay of residuals in the same way as Figure \ref{map_residuals} but separately for the IS-linked and Al-Qaeda-linked groups over 2010-2019. While we observe similar areas of dark purple cells in these maps, locations of red cells (i.e., the spatial spillovers beyond the declined cities) look highly heterogeneous both across organizations and across time.
\par
\begin{figure}[t]
\begin{center}
\includegraphics[width=10cm]{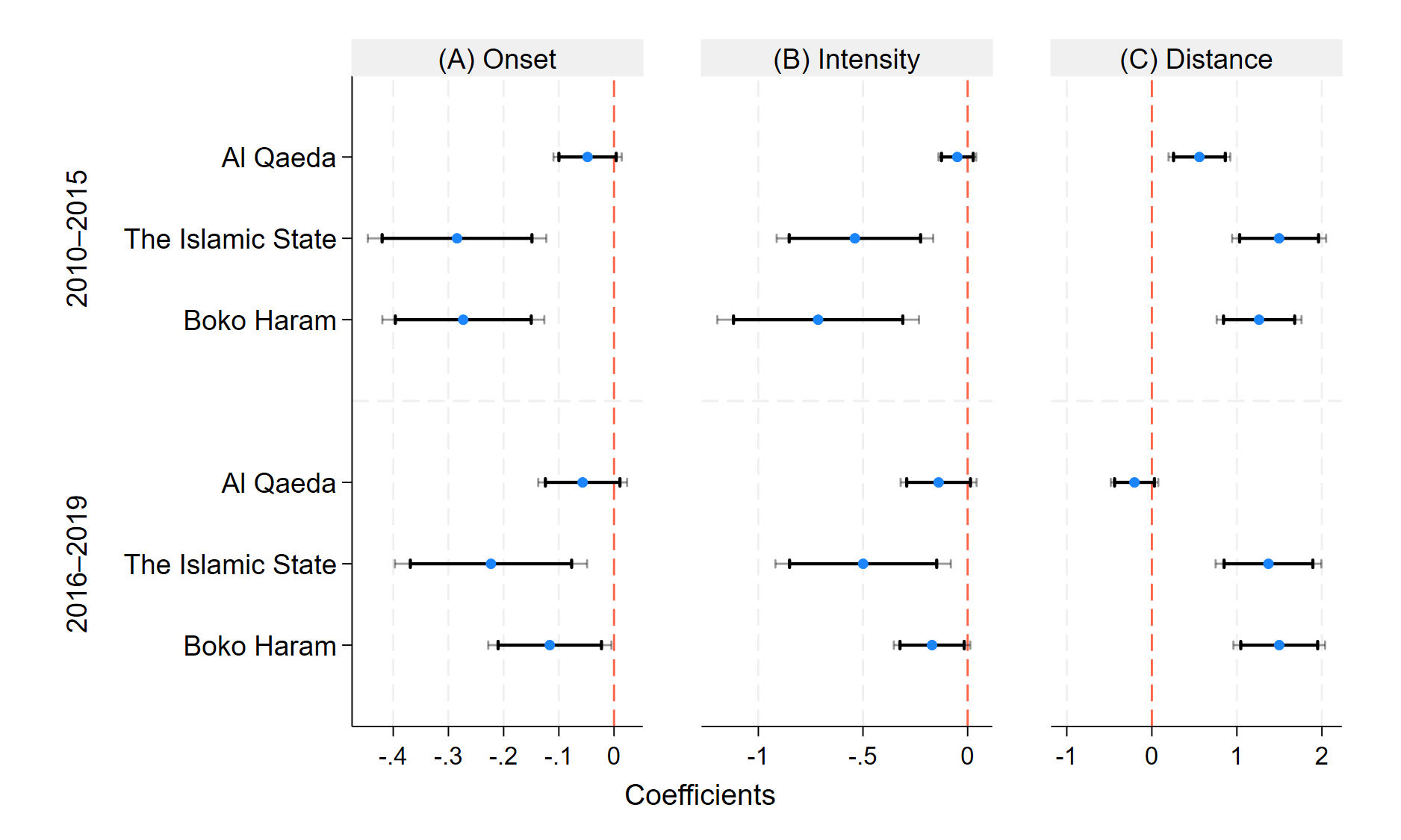}
\caption{Heterogeneous Persistent Effects across Contemporary Jihadist Factions}
\label{fig_heterogeneity}
{\parbox[t]{\textwidth}{
{\scriptsize\begin{singlespace}
\textit{Notes}:
All regressions are estimated using an IV of logarithm of one plus distance (km) to the nearest ancient lake. The unit of observation is a grid cell (about 55km × 55km).
The dependent variables are (A) a dummy which takes 1 if jihad involving each faction occurred, (B) logarithm of one plus the number of such jihadist events, and (C) logarithm of one plus distance (km) to the nearest such jihadist event.
The interest variable is the logarithm of one plus distance (km) to the nearest pre-colonial landlocked trade point whose contemporary population is less than 100,000, where landlocked is defined as being over 1,000 km from the nearest coast point.
%We control for landlocked dummy, average malaria suitability, average caloric suitability in post 1500, average elevation, ruggedness, and logarithm of one plus population in 2010 in all the specifications.
We use standard errors adjusting for spatial auto-correlation with the 100km cutoff.
The darker black lines represent 90\% confidence intervals, whereas the lighter black lines represent 95\% confidence intervals.\end{singlespace}}}}
\end{center}
\end{figure}
\par
Figure \ref{fig_heterogeneity} reports results of the main empirical IV specification for Al Qaeda-linked groups, IS-linked groups, and Boko Haram separately.
The signs of the three coefficients (for onset, intensity, and distance) for all three factions during 2010--2015 are consistent with the main results (Table \ref{tab_iv_jihad_pop}).
However, the explanatory power of declined cities is highly heterogeneous across factions: while the effects for IS-linked groups and Boko Haram are statistically significant and stable across both periods (2010--2015 and 2016--2019), the effects for Al Qaeda-linked groups are largely statistically insignificant in both periods, with notably smaller point estimates in magnitude.\footnote{
For IS-linked groups and Boko Haram, the effects are somewhat stronger during the initial emergence period (2010--2015), suggesting potential spatial diffusion to wider areas in the subsequent period.
}
\textbf{Supportive evidence of localized recruitment.}
%Historical and locational dependence and persistence are highly heterogeneous across different factions.
%Why do we observe the heterogeneity across different factions in the persistent influence of declined landlocked cities?
The above results indicate that Al Qaeda-affiliated groups depend less on the locations of declined cities compared to IS-affiliated ones or Boko Haram. %, as evidenced by the comparison of results between factions during 2016-2019.
While uncovering the precise mechanism behind this heterogeneity is beyond the scope of this paper, one plausible interpretation follows from differences in organizational structures and operation strategies.
If Al Qaeda-linked groups maintain a mobile core membership, particularly during their evolutionary stages, while other groups rely more heavily on local recruitment, and as local recruitment might be facilitated in areas with stronger persistence of jihadist ideology, then this across-faction heterogeneity is readily explainable.
%\blue{If Al Qaeda-linked groups have a core set of members to move around especially in their evolutional stages and the other groups depend more on local recruitment at any given time, and if local recruitment is facilitated in areas with a stronger persistence of jihadist ideology, then this across-faction heterogeneity is reasonably explainable.}
Consistent with this interpretation, it is documented that, in contrast to Al Qaeda’s lesser reliance on local constituencies, %and limited ability to recruit local populations,
Boko Haram and IS have adopted more localized recruitment strategies, drawing on existing community networks to expand their support base (\citealt{Bloom2017}; \citealt{BESA2017}).
\par
Finally, we provide quantitative evidence highlighting the different recruitment strategies by Al Qaeda-affiliated groups, IS-affiliated groups, and Boko Haram.
Simply assuming that jihadist groups attack both Muslim and non-Muslim %(mostly Christian in West Africa)
populations and that operating members of jihadist groups are Muslim, we define two concepts of the ``insurgent's (market) access'' measures. Insurgent's ``target access'' (ITA) in district $o$ is approximately defined as
$ITA_o \approx \sum_d \frac{\mbox{population}_d}{\mbox{distance}_{od}}$
and insurgent's ``labor market access'' (ILMA) in district $o$ is approximately defined as:
$ILMA_o \approx \sum_d \frac{\mbox{Muslim population}_d}{\mbox{distance}_{od}}$
where populations are measured using the WorldPop datasets and the World Religion Database (WRD), both of which are introduced in Appendix \ref{app_data_contemporary}.
Note that the unit of analysis for this exercise is a second-level subnational administrative division (not a grid cell), given the measurability of Muslim populations.
The right map of Figure \ref{app_fig_groups_AQ_IS_WA} shows these units, contemporary Muslim population shares, and violent events by Al Qaeda- and IS-linked groups in 2016-19.
\par
Table \ref{tab_jihad_ma} reports the regression results showing the correlation between violent events by each faction and ILMA, the key measure of accessibility to the potential pool of insurgent labor.
The coefficient on log (ILMA) is strikingly negative only for Al Qaeda-linked groups but positive for IS-linked groups and Boko Haram.
This contrast is consistent with the previous heterogeneity results in Figure \ref{fig_heterogeneity}, as well as with qualitative evidence on the differential reliance on localized recruitment strategies.

%\subsection{Interaction with Contemporary Shocks and Spatial Diffusion of Violent Events}\label{sec_disc_shocks}
%{\color{red}Under construction. This section will be added later iff we still think we need it after completing the other parts first.}
%%%%%%%%%%%%%%%%%%%%%%%%%%%%%%%%%%%%%%%%%%%%%%%%%%%%%%%%%%%%%%%%%%%%%%%%%%%%%%%%%%%%%%%%%%%%
\section{Discussions}\label{sec_disc}

\subsection{Alternative Mechanisms and Robustness Checks}\label{sec_alt_mechanisms}
We rule out prominent alternative mechanisms and conduct extensive robustness checks.

\textbf{Water resources.}
Are our results driven by contemporary water resources?
While our main specification already controls for modern access to lakes and rivers, one might still be concerned that the current availability of other water resources could be influencing our findings.
Motivated by prior research linking groundwater availability to local conflict (\citealt{CJL2025}) and precipitation variability to violence (e.g., \citealt{DM2023,MN2025}), %we examine whether these factors account for our results. Specifically,
we additionally control for shallow groundwater availability (0–50 m depth) and average precipitations for the period 2001--2017 with the detailed explanation in Appendix \ref{app_data_geographical}.
In Figure \ref{fig_plus_controls}, our main estimates remain virtually unchanged in magnitude and statistical significance, which confirms that contemporary groundwater access or precipitation patterns do not drive our results.

\textbf{The spread of Islam.}
A natural alternative interpretation is that our results simply reflect the historical spread of Islam rather than the decline of pre-colonial cities or colonial legacies. %In particular,
Trade centers may have facilitated the diffusion of Islam, implying that locations historically connected to trade networks could remain more religious today and thus exhibit more Islamist violence, even in the absence of subsequent decline or colonial resistance.

%We assess this possibility by explicitly controlling for the historical spread of Islam.
To address this concern, we collect data on mosques constructed prior to 1920 from \citet{Pradines2022} and use proximity to historical mosques as a proxy for early Islamic penetration, with details explained in Appendix \ref{app_data_precolonial}.
Figure \ref{fig_plus_controls} shows that after controlling for distance to historical mosques constructed before 1860, our main results remain robust: the coefficients of interest are nearly identical in magnitude and significance to those in the benchmark specification. This indicates that early Islamic diffusion alone cannot account for the persistent relationship between declined cities and contemporary jihadist violence.

%Although we do find a moderately positive correlation between proximity to historical core trade points and proximity to historical mosques (\MK{Table A.X}), this correlation is far from perfect, as mosques have been more widely distributed across peripheral and rural areas over time.
We find an only moderately positive correlation between proximity to historical core trade points and proximity to historical mosques (Table \ref{tab_historical_mosque}), as mosques have been more widely distributed across peripheral and rural areas over time.
This pattern aligns with existing evidence that the spread of Islam in Africa was shaped by multiple forces, including the slave trade (\citealt{PUAS2025}). Overall, these findings suggest that the historical spread of Islam itself %, while related to trade networks,
does not constitute the dominant mechanism driving our main results.
%\begin{itemize}
%\item Referee's comment: More fundamentally, I interpret this to be rather a story of access to trade centres. Trade centers caused the spread of Islam which means that ancient trade centres might still be associated with higher islamist ideology and hence higher islamist violence. The story does not need the decline, nor the colonial resistance part.
%\item To do so, we collected information on historical mosques constructed prior to 1900 from \citet{Pradines2022}, presuming that the presence of historical mosques is a good proxy of historical spread of Islam.
%\item Figure \ref{fig_plus_controls} shows that after controlling for distance to historical mosques, all the persistence effects are robust and the coefficient sizes are almost the same as the benchmark specification.
%\item This indicates that the spread of Islam cannot simply explain the main results.
%\item Indeed, not surprisingly, we have a moderately positive correlation between the proximity to historical mosques and historical core trade points.
%\item However, the correlation is not very strong as historical mosques are more widespread. This is reasonable given that we have mosques and the spread of Islam in peripheral or rural areas
%\item \citet{PUAS2025} also use the same data on historical mosques to understand the determinants of the spread of Islam in Africa, attributing to slave trade
%\end{itemize}

\textbf{Colonial activities and state capacity.} %\ST{``Colonial activities''? ``Colonial legacy'' sounds confusing because our primary mechanism also points to colonial legacy..}
%We next investigate the roles of colonial legacy and state capacity. \red{Better starting sentence?}
We next investigate the roles of colonial activities and state capacity, as highlighted in previous studies, to address concerns that these historical and institutional factors may be driving our results.
%\textbf{Colonial activities.}
We first consider whether features inherited from the later colonial period could account for our findings.
One concern is proximity to present-day national borders, as \citet{MP2016} document that colonial border design often partitioned ethnic groups in ways that shaped post-independence conflict.
Another possibility is the spatial distribution of colonial infrastructure investments—railways, roads, and related projects—highlighted in previous studies (e.g., \citealt{Huillery2009, Ricart-Huguet2021, JKM2017}), which link these investments to long-term economic and political outcomes.
A further channel emphasized in the literature is the pattern of European settlement, shown by \citet{AJR2001}, %and \citet{Huillery2011}
to have persistent effects on institutions and development.
%Christian missionary activities are well-known entities which fostered education and built health facilities in colonial Africa (e.g., \citealt{Nunn2014}; \citealt{CR2020}).
We incorporate measures for each of these dimensions into our baseline specification, yet in all cases, Figure \ref{fig_plus_controls} shows that our main results remain stable in magnitude and statistical significance.
%\ST{WORDS FROM A DIFFERENT PREVIOUS SECTION: We use three variables that capture colonial activities. In panel (A), the dependent variable is proximity to a colonial railway which is one of the direct measure of colonial investment. In panel (B), the dependent variable is proximity to a Christian missionary activity. Christian missionary activities are well-known entities which fostered education and built health facilities in colonial Africa (e.g., \citealt{Nunn2014}; \citealt{CR2020}).}
%\textbf{State capacity.}
We next examine the role of ``state capacity,'' proxied by distance to the capital city.
This measure is motivated by evidence that distance to the seat of government is negatively associated with the provision of public goods and administrative reach (e.g., \citealt{MP2014}).
Once included in our regressions, this variable does not alter our main coefficient.

\textbf{Christian-Muslim difference.}
The main results may be capturing differences between Christian and Muslim populations.
In West Africa, the spatial distribution of the Christian population largely reflects missionary activities during the colonial period.
They influenced not only religious affiliation, but also broader outcomes (e.g., \citealt{CR2020}; \citealt{Nunn2014}). %; \citealt{Okoye2021}).
%They influenced not only religious affiliation, but also human capital (\citealt{Nunn2014}), culture and trust (\citealt{Okoye2021}), and political engagement (\citealt{HLS2025}).
Figure \ref{fig_plus_controls} shows that our results are unchanged when we additionally control for proximity to historical mosques and missionary activities, which capture the relative dominance of Islam and Christianity across space, suggesting that the main results are not driven by differences in the religious composition of the population.
Nevertheless, there are several areas relatively close to the coast where we observe jihadist violence in regions with a mixed presence of Christian and Muslim populations (e.g., Nigeria and Burkina Faso).
To address this concern, Figure \ref{fig_sahara_pop_less_005m} and Table \ref{tab_iv_jihad_nonjihad_sahara_pop} panel (A) restrict the sample to only countries covering the Sahara (Mauritania; Mali; Niger; Chad), where the Muslim population share is nearly 100\% (Figure \ref{app_fig_groups_AQ_IS_WA}),
and confirm the robustness of the main results.
This exercise also shows that the main results are not solely driven by the large ancient Lake Chad or by the concentration of jihad events associated with Boko Haram around northeastern Nigeria.
%Last but not least,
Moreover, all four Sahara-covering countries were former French colonies, which rules out differences in colonial policies and institutions, particularly those between British and French rule (e.g., \citealt{AFJS2019}), as an alternative explanation.

%\red{We also excluded the prominent Chad Lake Effects}
%Since the significant persistent effects on contemporary jihad concentrate in 2010-2019, we report results in 2010-2019, 2010-2015, and 2016-2019 in the first three robustness checks that change geographical coverage of the sample or the treatment variable.

%\subsubsection*{Timbuktu effect or Chad lake effect\red{Better general wording?}}
%Bring the robustness check using small cities $<50,000$
%Prominent site effects
%Beyond regional outliers: The Chad lake and Timbuktu effects
%Beyond prominent sites: The Chad lake and Timbuktu effects
%Beyond landmark places: The Chad lake and Timbuktu effects

\textbf{General factors driving conflicts.} %Another concern is that
The main results may not reflect the persistent effect of declined cities, but rather other unobserved structural factors that drive conflicts around these locations.
If this were the case, %dominant mechanism
we would expect similar IV estimates for non-jihadist conflict events as well.
%To address this concern, as a placebo test, Table \ref{tab_iv_nonjihad_sahara_pop} reports the results of the same main specification except for changing the dependent variable to violent events involving non-state and non-jihadist organizations.
To address this concern, panel (B) of Table \ref{tab_iv_jihad_nonjihad_sahara_pop} presents results of a placebo test in which we re-estimate our main specification replacing the dependent variable with violent events involving non-state, non-jihadist organizations.%Table \ref{tab_iv_nonjihad_sahara_pop} results the results.
\footnote{We again restrict the sample to countries covering the Sahara to focus on Muslim-dominated areas and to exclude coastal areas where clearly different types of conflicts occur (e.g., conflict in the Niger Delta).}
%That is, the sample grid cells correspond to those in Table \ref{tab_iv_jihad_sahara_pop}.
%This test works as a placebo test.
For all three outcomes, the coefficients are statistically insignificant and substantially smaller in magnitude than the corresponding estimates for jihadist events reported in panel (A). %of the same table. %\ref{tab_iv_jihad_sahara_pop}. % Figure \ref{fig_plus_IVs_add}.

Note that this contrast does not imply that localized and contemporary factors are unimportant for jihad.
Proximity to contemporary water sources, which matters for conflict (\citealt{DM2023}), similarly affects both jihadist and non-jihadist events (Table \ref{tab_iv_jihad_nonjihad_sahara_pop}), droughts impact both event types through disruptions between transhumant pastoralists and sedentary agriculturalists (\citealt{MN2025}), and domestic contexts such as political marginalization and grievances also matter (\citealt{DR2013}). Quantifying the relative importance of these factors remains an important agenda for future research.

\if0
Note that this contrast between jihadist and non-jihadist events does not mean that typical localized and contemporary factors underlying insurgency are not important for jihad.
For example, as \citet{DM2023} also argue in detail, proximity to contemporary water sources has similar effects on jihadist events (column 3 of Table \ref{tab_iv_jihad_sahara_pop}) and non-jihadist events (column 3 of Table \ref{tab_iv_nonjihad_sahara_pop}). % during 2010-2019.
%\citet{Dowd2015} and
\citet{DR2013} point out that not only the global jihadist ideology but also domestic contexts, such as political marginalization and grievances, are important factors driving jihadist movements in West Africa.
%\blue{\citet{BB2019}, through anecdotes and interviews taken in multiple cities in Mali, argue that local land-use conflicts lead pastoral groups to support or join jihadist groups.}
\citet{MN2025} show that droughts in Africa, through the mechanism that they disrupt the arrangement between transhumant pastoralists and sedentary agriculturalists, impact both jihadist and non-jihadist events similarly.
%Combining our findings with these research, it is apparent that both jihad-specific and general factors underlying insurgency matter to explain contemporary jihad.
Quantifying relative importance of these factors and understanding local dynamics of jihad are important future research agendas.
\fi

\textbf{Other robustness checks.}
Appendix \ref{app_results} illustrates that the main results are also robust to alternative measures of the ancient water access, alternative measures of declined cities, alternative conflict event data, and alternative adjustments for spatial auto-correlations.

\subsection{Global Perspective: Jihad in ``Past-Core-and-Present-Periphery''}\label{sec_disc_global}
%\ST{The first three paragraphs could be shortened}
%Why do jihadist activities emerge in some places but not in other places {\it within} the Islamic world?
%%While the previous sections have focused on a local mechanism specific to West Africa, this section broadens the perspective to explore global patterns across the Islamic world.
%While the previous sections have focused on a local mechanism specific to West Africa,
Finally, we broaden the scope to examine whether the pattern of ``past-core-and-present-periphery,'' and the associated colonial legacies, extends globally.
%In other words, contemporary jihads tend to emerge in areas that experienced reversals of fortune over the centuries.
\if0
\blue{Many contemporary hotspots of jihadist violence are located in regions that were historically central to overland trade and Islamic civilization but have since declined relative to modern geopolitical centers.
%We observe that many contemporary hotspots of jihadist violence are located in regions that were historically central to overland trade and Islamic civilization but have since declined relative to modern geopolitical centers.
Crucially, consistent with our proposed mechanism in West Africa, the transition of these regions from core to periphery was frequently cemented by colonial interventions that reshaped local power dynamics.}
\fi
%This pattern advances the broader argument of reversal of fortune in \citet{AJR2001, AJR2002}, by linking historical reversals to the eruption of ideological violence.
This pattern advances the broader argument of reversal of fortune in \citet{AJR2001, AJR2002}, by linking historical reversals to the eruption of violent religious revivalism.
%It is not surprising that jihadist activity is more prevalent in relatively underdeveloped areas within the Islamic world, corresponding to categories (ii) and (iv).
%What is less obvious, However, is why jihads are especially concentrated in the ``past-core-and-present-periphery'' locations among underdeveloped Islam areas.

As mapped in Figure \ref{map_global} and detailed in Appendix \ref{app_qualitative_global}, contemporary jihadist violence is also concentrated in regions—most notably Syria, Iraq, Afghanistan, and Pakistan—that were once prosperous hubs of overland trade and Islamic civilization but were subsequently marginalized through colonial disruption.
In each case, colonial interventions restructured local power hierarchies in ways that eroded the political and religious authority of historically dominant groups, creating the conditions for ideological persistence and eventual resurgence. While establishing a causal relationship at the global scale lies beyond the scope of this paper, the qualitative evidence across these cases is broadly consistent with the mechanism identified in West Africa.

At the same time, detailed pathways translating long-run decline into violence are likely context-specific. Understanding Islamist violence therefore requires combining a global perspective on structural reversals with close attention to historical trajectories. %through which cultural revival emerges.
Further research is warranted to systematically examine the relationship between past prosperity, decline, and various forms of cultural revival globally, including non-Islamic cases. %, both jihadist and non-jihadist,
\section{Conclusion}\label{sec_conclusion}
This paper explores the origins and persistence of jihadist insurgencies in West Africa through the lens of historical geography, colonial disruption, and religious ideology. We demonstrate that contemporary Islamist violence is disproportionately concentrated in areas that were once central to Islamic political and economic life but have since become peripheral.
These regions illustrate the enduring influence of initial geographic advantage, even after its material importance has diminished.

A key underlying mechanism is the transmission of jihadist ideology as a colonial legacy, which we argue is shaped by both historical conflict and religious practice.
In particular, the religious practice, concealment of one’s beliefs in the face of threat, offers a powerful explanation for how ideological continuity could survive colonial suppression and later reemerge in culturally resonant ways.
We support this mechanism from multiple angles:
%This mechanism is supported by multiple strands of evidence:
historical weapon access, a dynamic model, %\blue{mosque names,}
individual-level survey data, and the localized recruitment by Islamist groups.

The spatial pattern uncovered in West Africa resonates with a broader global phenomenon of violent ideological revival in formerly prosperous but later marginalized regions.
Together, these findings provide a unified framework for understanding the spatial persistence and ideological continuity of Islamist violence, with implications for both historical political economy and contemporary security policy.
At the same time, the spatial diffusion of jihadist violence beyond historically significant locations reflects complex contemporary dynamics that our framework does not fully capture.
Understanding how historical legacies interact with contemporary factors that are driving the spatial propagation of conflict represents a promising direction for future research. %, through recruitment networks, arms flows, and state capacity,
\end{onehalfspace}
%\end{doublespace}
%\end{spacing}
%%%%%%%%%%%%%%%%%REFERENCE%%%%%%%%%%%%%%%%%%%%%%%%%%%%%%%%%%%%%%%%%%%%%%%%%%%%%%%%%%%%%%%%%%%%%%%%%%%%%%%%%%%%%%%%%%%%%%
%\clearpage
\begin{small}
\begin{spacing}{0.95}
\bibliographystyle{ecta}
\bibliography{ref_IslamHistory}
%\nocite{*}  % this code is used only when you want to cite paper which is not cited in the main article.
\end{spacing}
\end{small}

%%%%%%%%%%%%%%%%%FIGUREs%%%%%%%%%%%%%%%%%%%%%%%%%%%%%%%%%%%%%%%%%%%%%%%%%%%%%%
% Built in the main texts

%%%%%%%%%%%%%%%%%TABLEs%%%%%%%%%%%%%%%%%%%%%%%%%%%%%%%%%%%%%%%%%%%%%%%%%%%%%%
\clearpage
%\section*{Tables}

%%% Historical city origin
%%\input{tables/tab_water_historical_city.tex}   % in App

%%% Changing natural geography
%%\input{tables/tab_ols_city_today_water.tex}               % in App
%%\input{tables/tab_ols_city_today_colonial_activity.tex}  % in App
%%\input{tables/tab_ols_colonial_activity_water.tex}        % Moved to the mechanism section

%%% Main iv
%\input{tables/tab_exclusion_geo_institution_pop.tex}
%\input{tables/tab_first_stage_pop.tex}
%%\input{tables/tab_placebo_first_stage.tex}  % in App
%\input{tables/tab_iv_jihad_pop.tex}
%%\input{tables/tab_ols_jihad_city_today.tex} % in App
%%\input{tables/tab_iv_nonjihad_sahara.tex}   % in App
%%\input{tables/tab_iv_lnnightlight.tex}      % in App

%%% Mechanism
%\input{tables/tab_cycles_lake.tex}
%%\input{tables/tab_cycles_iv_conflcit_trade.tex}       %% Moved to App
%%\input{tables/tab_cycles_iv_ratio_conflcit_trade.tex}    % Moved to App
%\input{tables/tab_cycles_ols_ratio_jihad_history_today_pop.tex}
%\input{tables/tab_ols_colonial_activity_water.tex}        % Moved in!
%\input{tables/tab_mechanism_ideology.tex}

%%%%%%%%%%%%%%%%%%%%%%%%%%%%%%
% Appendix
%%%%%%%%%%%%%%%%%%%%%%%%%%%%%%%
\clearpage
\appendix
\pagenumbering{arabic}

%\begin{spacing}{1.2}
\begin{onehalfspace}
%\section*{SUPPLEMENTAL APPENDIX}
%\section*{Jihad over Centuries}
\section*{Supplement to ``Jihad over Centuries''}
%\section*{{\large Masahiro Kubo \& Shunsuke Tsuda}}
\begin{center}
\author{Masahiro Kubo, \textit{CERDI}
\\
Shunsuke Tsuda, \textit{University of Essex}}
\end{center}
%%%%%%%%%%%%  Preliminary   %%%%%%%%%%%%%%%%%%%%%
%\section*{Contents}
%\startcontents[sections]
%\printcontents[sections]{l}{1}{\setcounter{tocdepth}{2}}

%\clearpage
\setcounter{figure}{0}
\setcounter{table}{0}
\setcounter{equation}{0}
\renewcommand{\thefigure}{\Alph{section}.\arabic{figure}}
\renewcommand{\thetable}{\Alph{section}.\arabic{table}}
\renewcommand{\theequation}{\Alph{section}.\arabic{equation}}
\section{Additional Figures and Tables}\label{app_fig_tab}

\begin{figure}[htbp]
\begin{center}
\includegraphics[width=7.5cm]{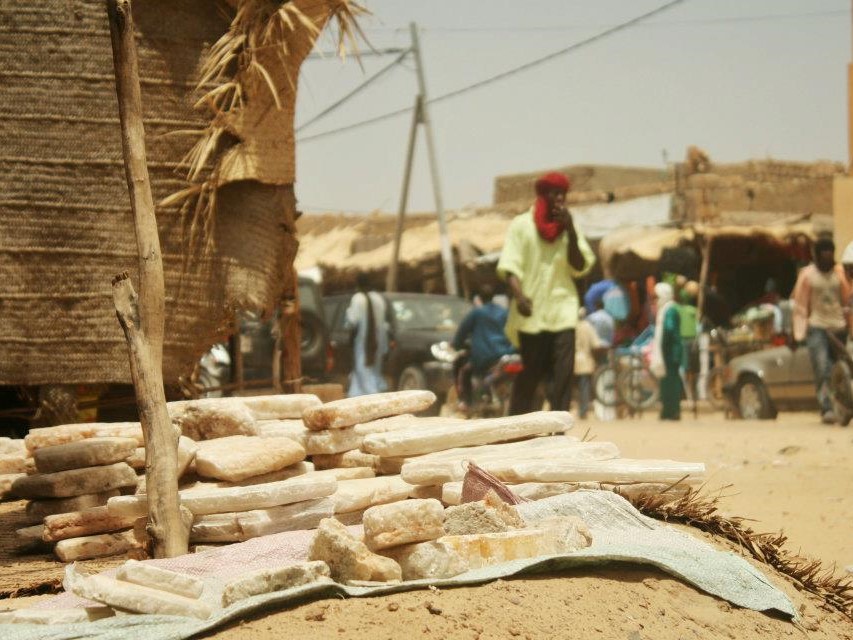}
\includegraphics[width=7.5cm]{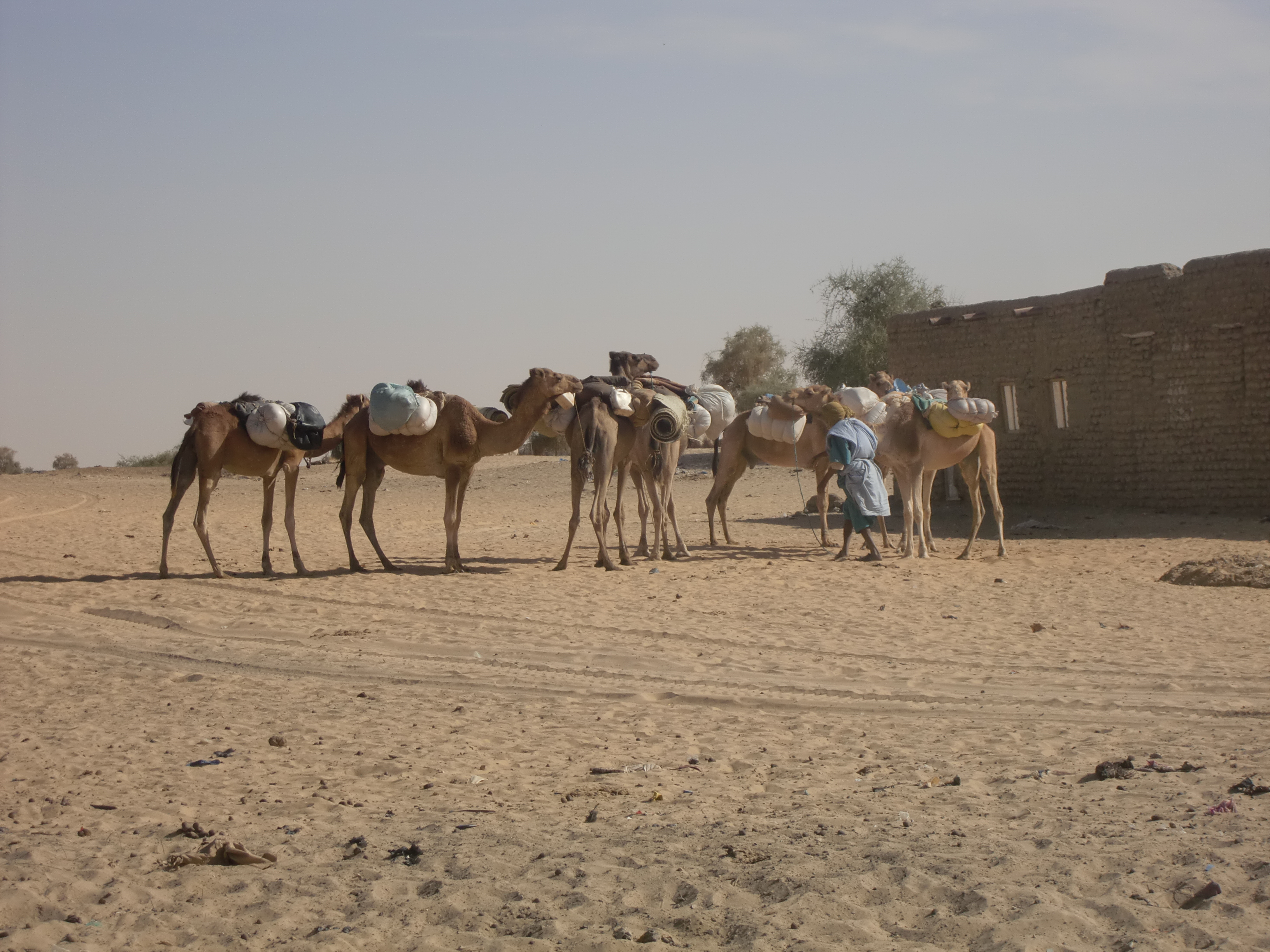}\\
\includegraphics[width=7.5cm]{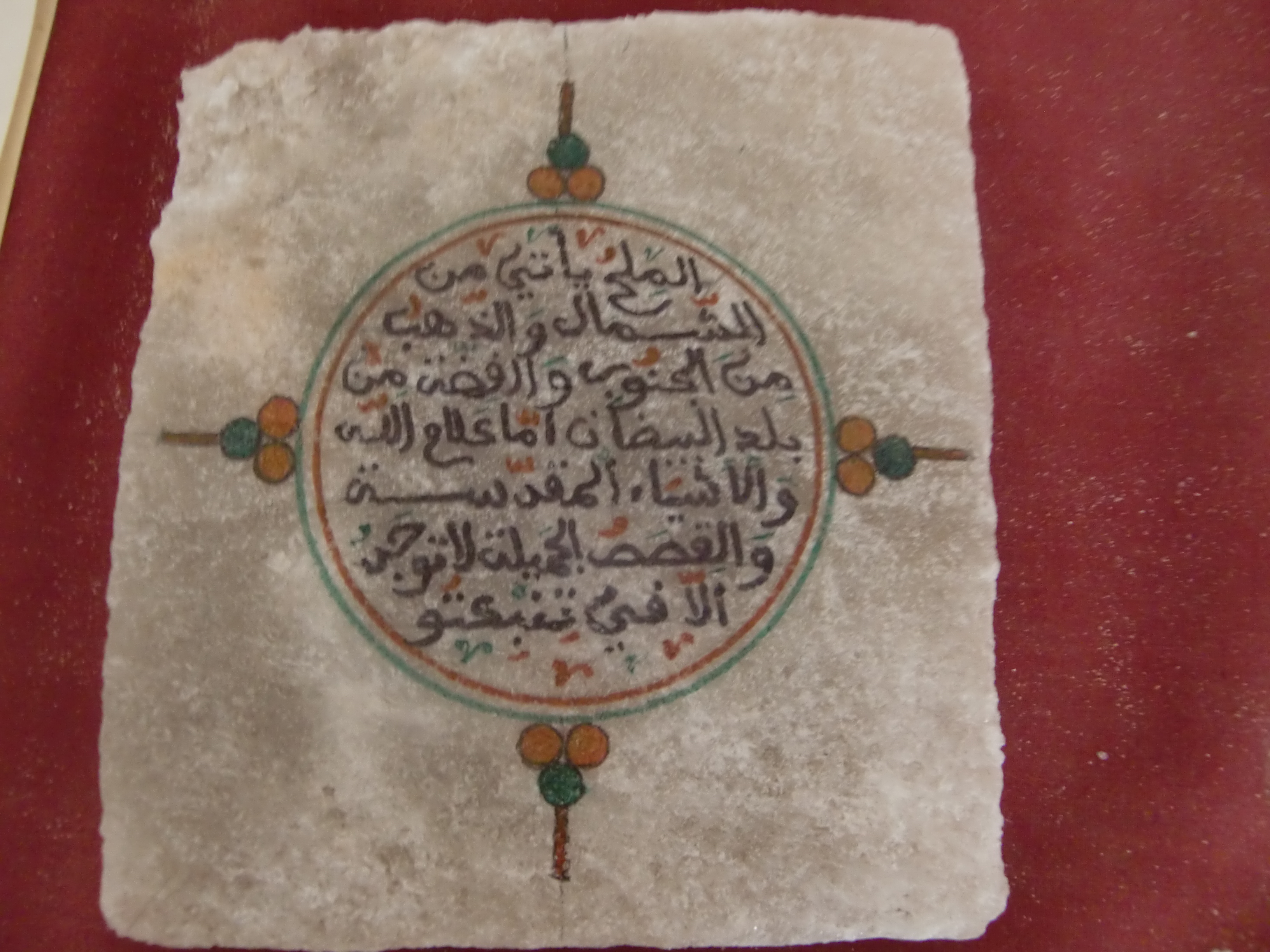}
\includegraphics[width=7.5cm]{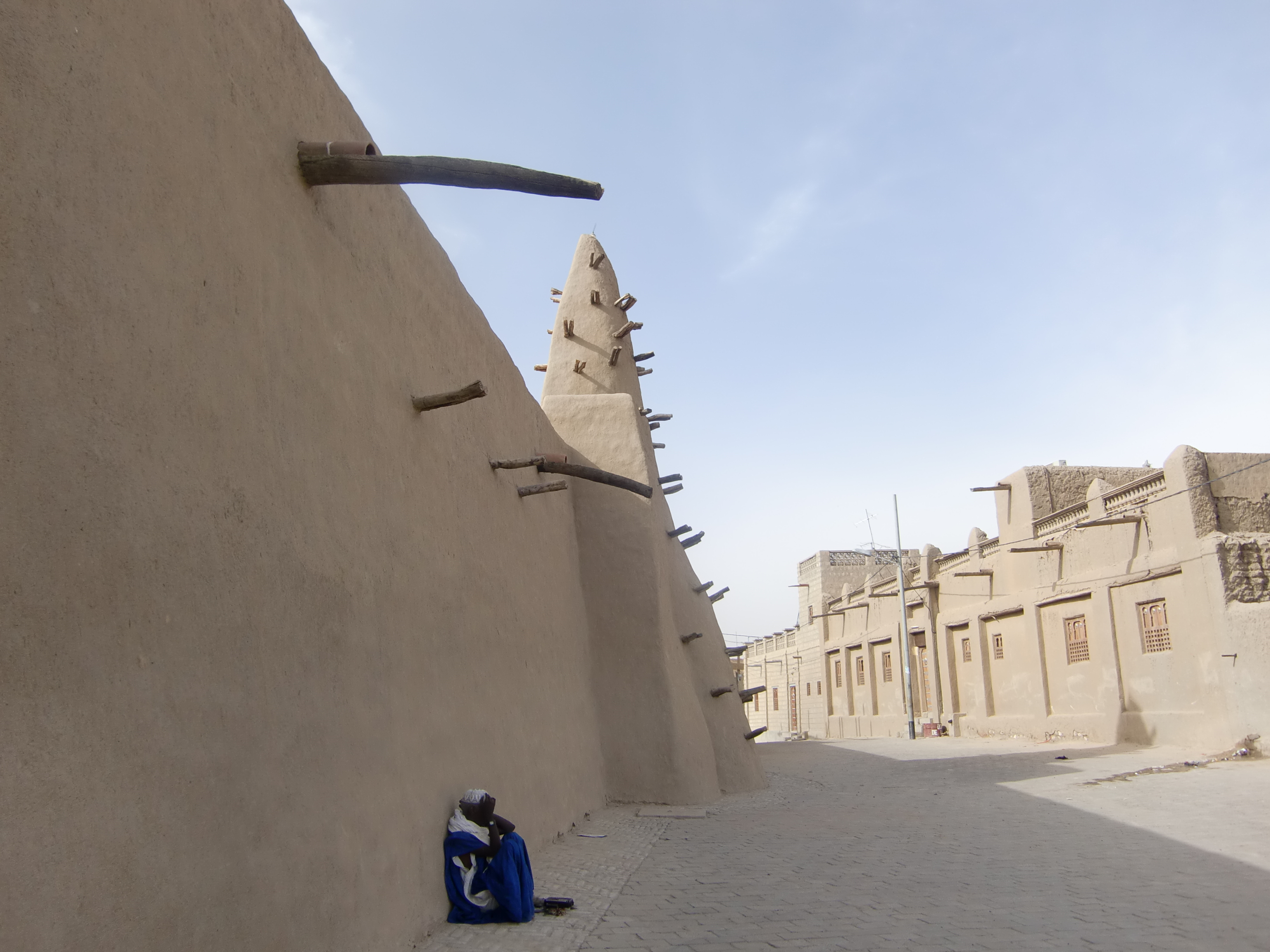}
\caption{Timbuktu}
\label{app_fig_Timbuktu}
{\parbox[t]{\textwidth}{
{\scriptsize\begin{singlespace}
\textit{Notes}:
Salt (top-left), which has been used for various purposes (bottom-left), is still transported by caravan (top-right) even today.
The bottom-right picture shows the Djinguereber mosque, which was broken by a jihadist group Ansar Dine in 2012.
The top-left picture was taken by Daiki Kobayashi and the other three ones were taken by Shunsuke Tsuda in 2010, before the surge of Islamist insurgencies in Mali.\end{singlespace}}}}
\end{center}
\end{figure}

\if0
\begin{figure}[htbp]
\begin{center}
\includegraphics[width=8cm]{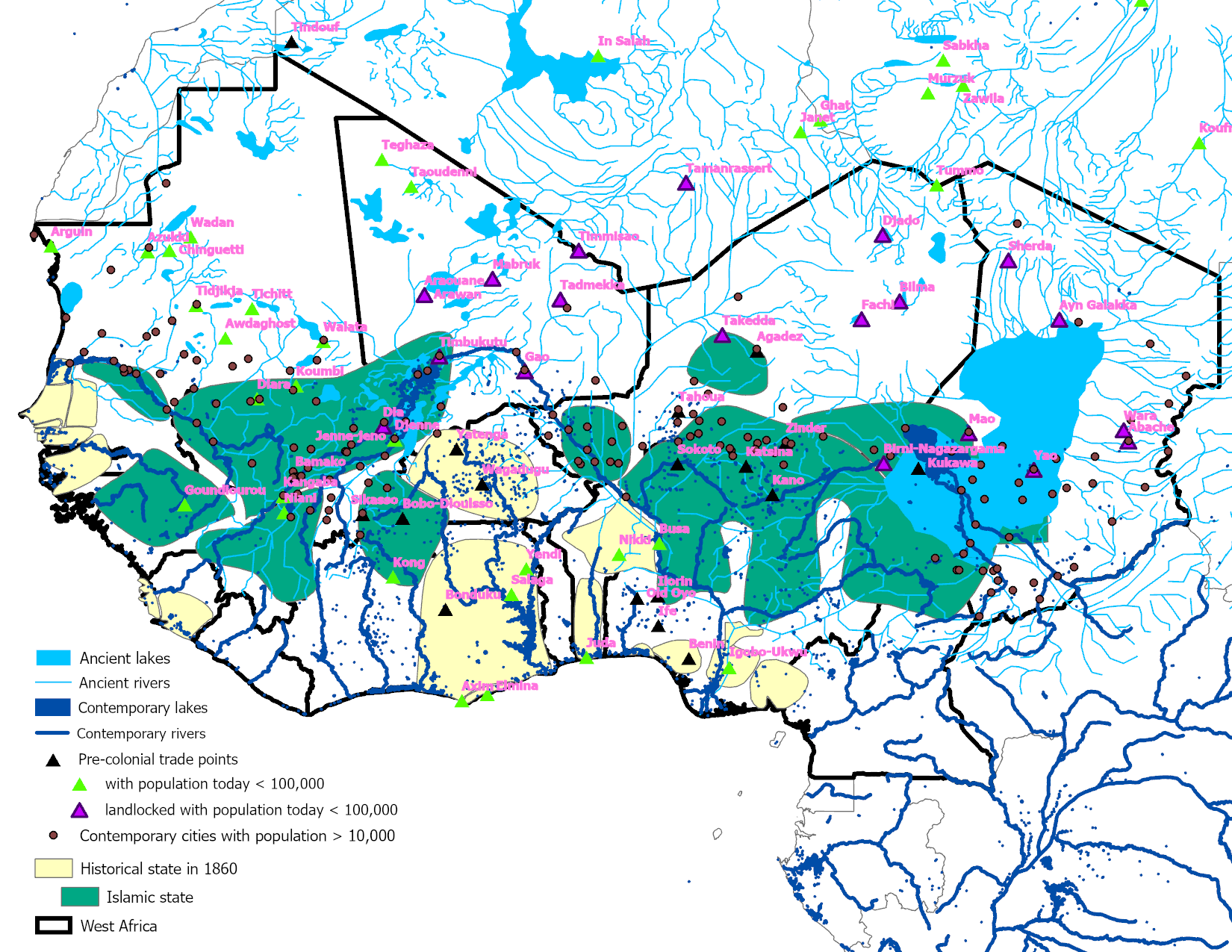}
\caption{Contemporary Cities with Populations $>$ 10,000 in Countries covering the Sahara %\ST{Typo? Over 100,000? No, over 10,000 and below 50,000. Complex.. Also combine this with Figure 1 in Intro?}
}\label{app_map_sahara_water_sources_cities_past_present}
{\parbox[t]{170mm}{{\footnotesize{\it Notes}:
\ST{Data sources?; NO NEED TO HAVE THIS AS WE END UP DROPPING CURRENT PANEL (B) of TABLE A.1 \& TABLE A.3?}
}}}
\end{center}
\end{figure}
\fi

\begin{figure}[htbp]
\begin{center}
\includegraphics[width=9cm]{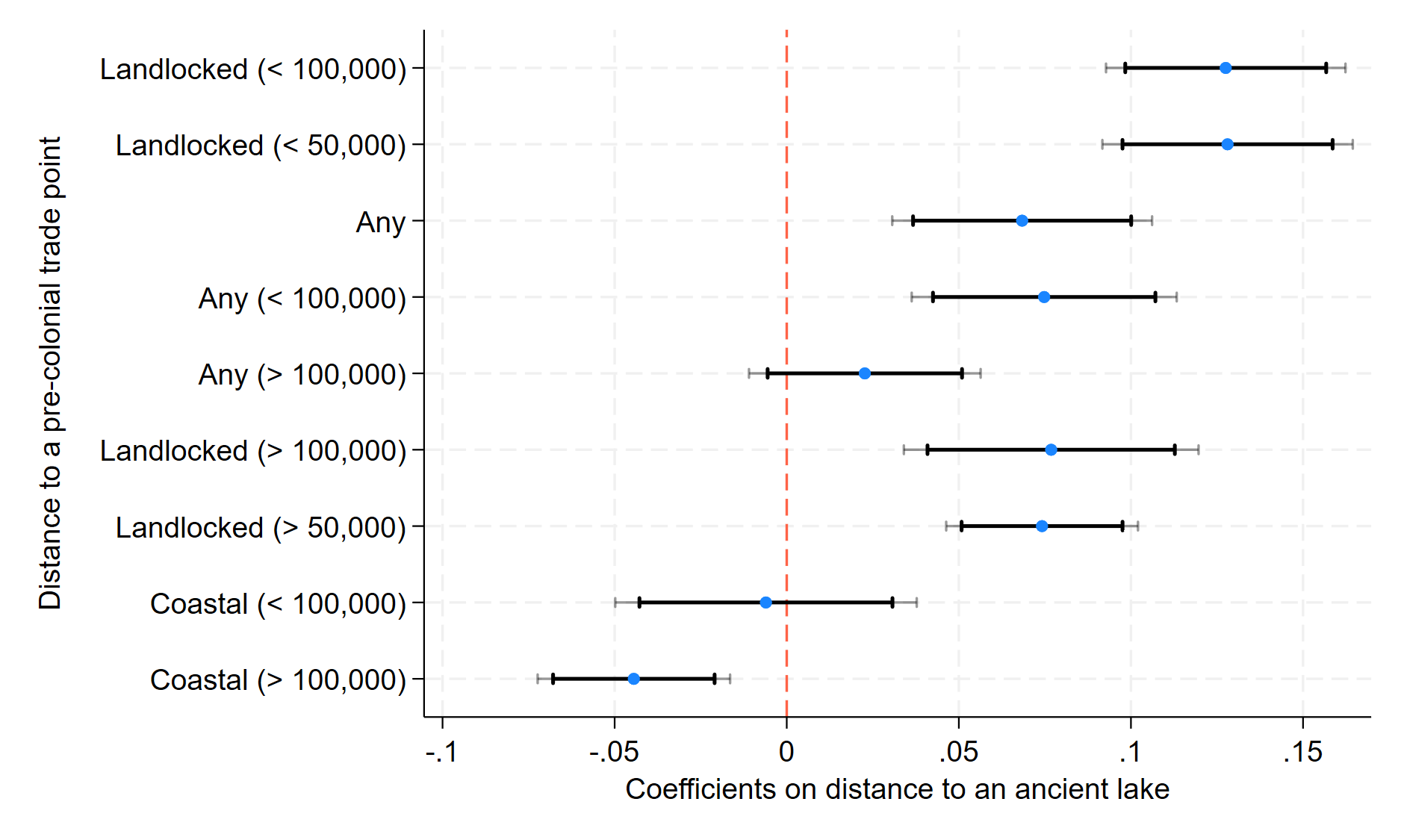}\\
\caption{Placebo First Stage---Ancient Water Sources and Historical Cities}
\label{fig_first_stage}
{\parbox[t]{\textwidth}{
{\scriptsize\begin{singlespace}
\textit{Notes}:
All regressions are estimated using OLS.
%The unit of observation is a grid cell (about 55km × 55km).
All distance variables indicate the log distance (km) to the nearest object.
The dependent variables are represented in the y-axis.
%(A) the logarithm of one plus distance (km) to the nearest pre-colonial landlocked trade point whose contemporary population is more than 100,000,
%(B) the logarithm of one plus distance (km) to the nearest pre-colonial coastal (non-landlocked) trade point whose contemporary population is less than 100,000,
%(C) the logarithm of one plus distance (km) to the nearest pre-colonial coastal (non-landlocked) trade route up to 1800.
Landlocked is defined as being over 1,000 km from the nearest coast point. We control for landlocked dummy, average malaria suitability, average caloric suitability in post 1500, average elevation, ruggedness, and logarithm of one plus population in 2010 in all the specifications.
We use standard errors adjusting for spatial auto-correlation with the 100km cutoff.
The darker (lighter) black lines represent 90\% (95\%) confidence intervals.
\end{singlespace}}}}
\end{center}
\end{figure}

\begin{figure}[htbp]
\begin{center}
\includegraphics[width=9cm]{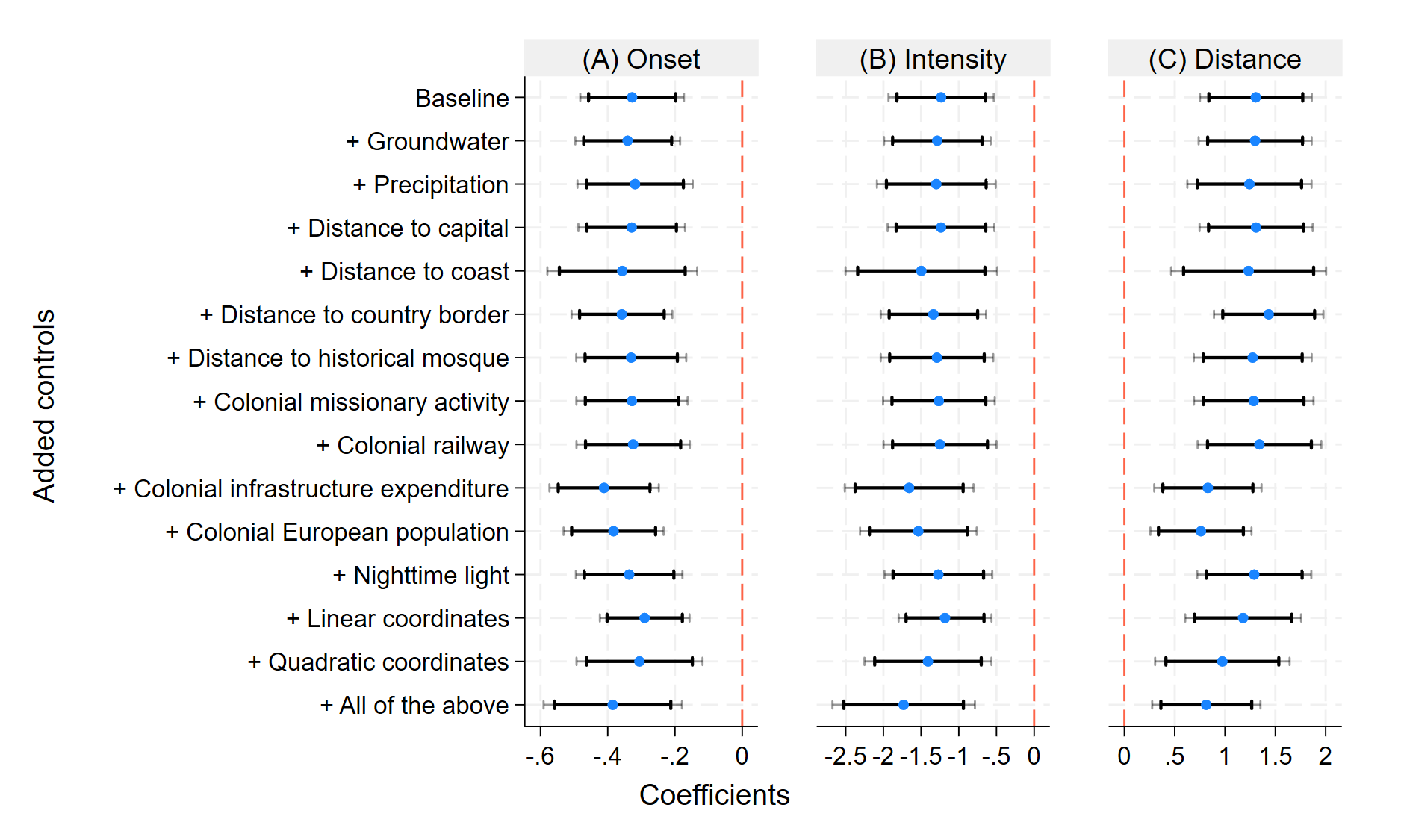}\\
\caption{Robustness Checks with Additional Controls}
\label{fig_plus_controls}
{\parbox[t]{\textwidth}{
{\scriptsize\begin{singlespace}
\textit{Notes}:
%\ST{ADD Historical mosque, Nighttime light, Colonial railway}
All regressions are estimated using an IV of log distance (km) to the nearest ancient lake.
%The unit of observation is a grid cell (about 55km × 55km).
%The dependent variables are (A) jihadist event onset indicator, (B) logarithm of the number of jihadist events, and (C) log distance (km) to the nearest jihadist event.
The interest variable is the log distance (km) to the nearest pre-colonial landlocked trade point whose contemporary population is less than 100,000, where landlocked is defined as being over 1,000 km from the nearest coast point.
%As a baseline, we control for landlocked dummy, average malaria suitability, average caloric suitability in post 1500, average elevation, ruggedness, and population in 2010 in all the specifications.
In addition to the baseline controls, % (landlocked dummy, average malaria suitability, average caloric suitability in post 1500, average elevation, ruggedness, and population in 2010),
we control for groundwater availability, average precipitation today, distance to the capital, distance to the coast, distance to the country border, distance to the nearest historical mosque established before 1860, distance to the Christian missionary acitivity, distance to the colonial railway, infrastructure expenditures, and European population in the colonial period, the logarith of nighttime luminocity in 2009, and latitude and longitude in both linear and quadratic forms, each entered separately as control variables. %(see Appendix \ref{app_data} for variable definitions).
Finally, we include all these variables jointly in the specification.
We use standard errors adjusting for spatial auto-correlation with the 100km cutoff.
The darker (lighter) black lines represent 90\% (95\%) confidence intervals.
\end{singlespace}}}}
\end{center}
\end{figure}

\begin{figure}[htbp]
\begin{center}
\includegraphics[width=10cm]{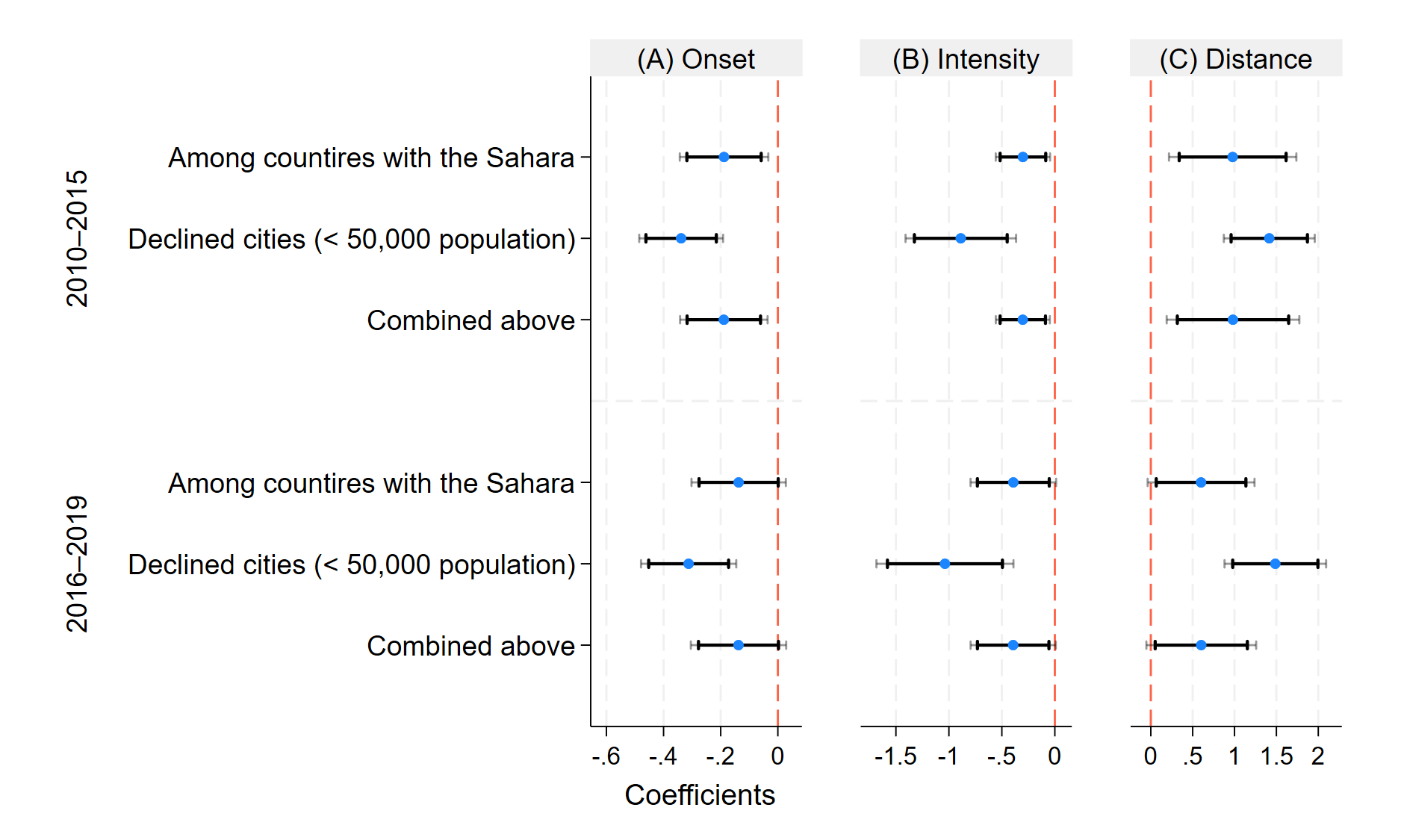}\\
\caption{Robustness Checks with Alternative Geographical Specifications}
\label{fig_sahara_pop_less_005m}
{\parbox[t]{\textwidth}{
{\scriptsize\begin{singlespace}
\textit{Notes}:
%\ST{(I) (A) Onset (B) Intensity (C) Distance (NO NEED TO WRITE ``Log''); (II) Same dot colors; (III) Why wide gap between years and row names? If we cannot reduce this gap, then better to use only one line rather than two lines; (IV) By the way, this fig does not look very good..? To defend against Christian-Muslim difference mechanism, do NOT rely SOLELY on this figure. One of multiple arguments. Others include, country FEs already controlled, missionary activity controlled (bring this again?), else?}
All regressions are estimated using an IV of log distance (km) to the nearest ancient lake.
%The unit of observation is a grid cell (about 55km × 55km).
%The dependent variables are (A) jihadist event onset indicator, (B) logarithm of the number of jihadist events, and (C) log distance (km) to the nearest jihadist event.
In each panel and period, the top row reports results for grid cells located in countries that cover the Sahara (Mauritania; Mali; Niger; Chad).
The middle row reports results using an interest variable defined as the log distance (km) to the nearest pre-colonial landlocked trade point whose contemporary population is below 50,000.
The bottom row reports results that combine both restrictions.
%The interest variable is the logarithm of one plus distance (km) to the nearest pre-colonial landlocked trade point whose contemporary population is less than 100,000, where landlocked is defined as being over 1,000 km from the nearest coast point.
%We control for landlocked dummy, average malaria suitability, average caloric suitability in post 1500, average elevation, ruggedness, and logarithm of one plus population in 2010 in all the specifications.
We use standard errors adjusting for spatial auto-correlation with the 100km cutoff.
The darker (lighter) black lines represent 90\% (95\%) confidence intervals.
%\ST{By the way, this fig does not look very good..? To defend against Christian-Muslim difference mechanism, do NOT rely SOLELY on this figure. One of multiple arguments. Others include, country FEs already controlled, missionary activity controlled (bring this again?), else?}
\end{singlespace}}}}
\end{center}
\end{figure}

\begin{figure}[htbp]
\begin{center}
\includegraphics[width=10cm]{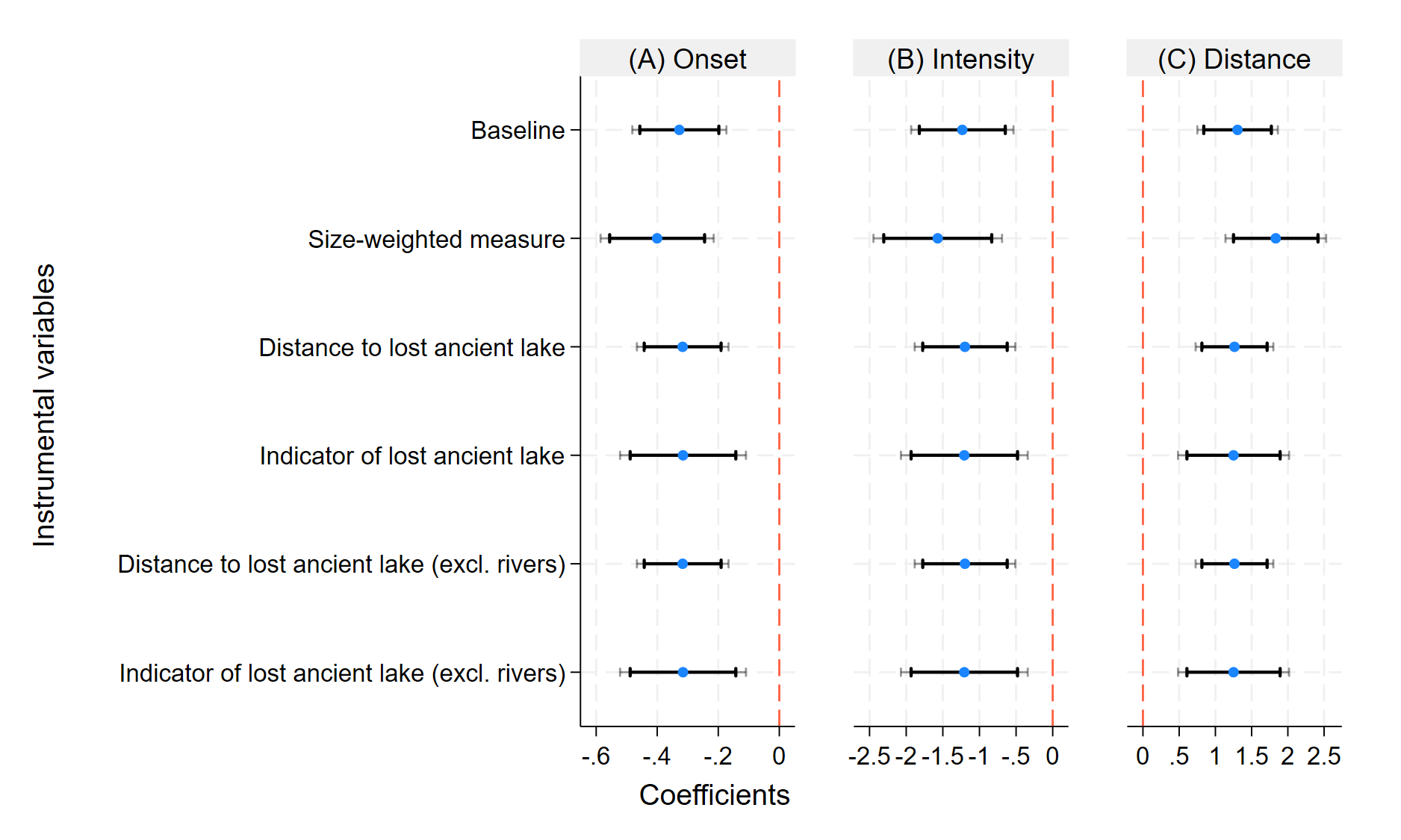}\\
\caption{Robustness Checks with Alternative IV Definitions}
\label{fig_plus_IVs_add}
{\parbox[t]{\textwidth}{
{\scriptsize\begin{singlespace}
\textit{Notes}:
%\ST{(A) Onset (B) Intensity (C) Distance (NO NEED TO WRITE ``Log''); excluding --> excl.; Size-weighted measure}
All regressions are estimated using alternative IVs.
In addition to the log distance (km) to the nearest ancient lake and a size-weighted measure of ancient water access, we employ two alternative definitions of lost ancient lakes and use both distance-based and indicator-based instruments.
The first definition identifies areas where ancient lakes existed but no contemporary lakes remain; the corresponding instruments are the distance to the nearest such lost ancient lake and a dummy that takes a value of 1 if a grid cell lies in such an area.
The second definition further restricts these areas to those where neither contemporary lakes nor rivers are present; again, we use the distance to the nearest such lost ancient lake and a dummy that takes 1 if a grid cell lies in such an area.
These are labeled as "lost ancient lake (excl. rivers)" in the figure.
%The unit of observation is a grid cell (about 55km × 55km).
%The dependent variables are (A) jihadist event onset indicator, (B) logarithm of the number of jihadist events, and (C) log distance (km) to the nearest jihadist event.
We use standard errors adjusting for spatial auto-correlation with the 100km cutoff.
The darker (lighter) black lines represent 90\% (95\%) confidence intervals.
\end{singlespace}}}}
\end{center}
\end{figure}

\if0
\begin{figure}[htbp]
\begin{center}
\includegraphics[width=12cm]{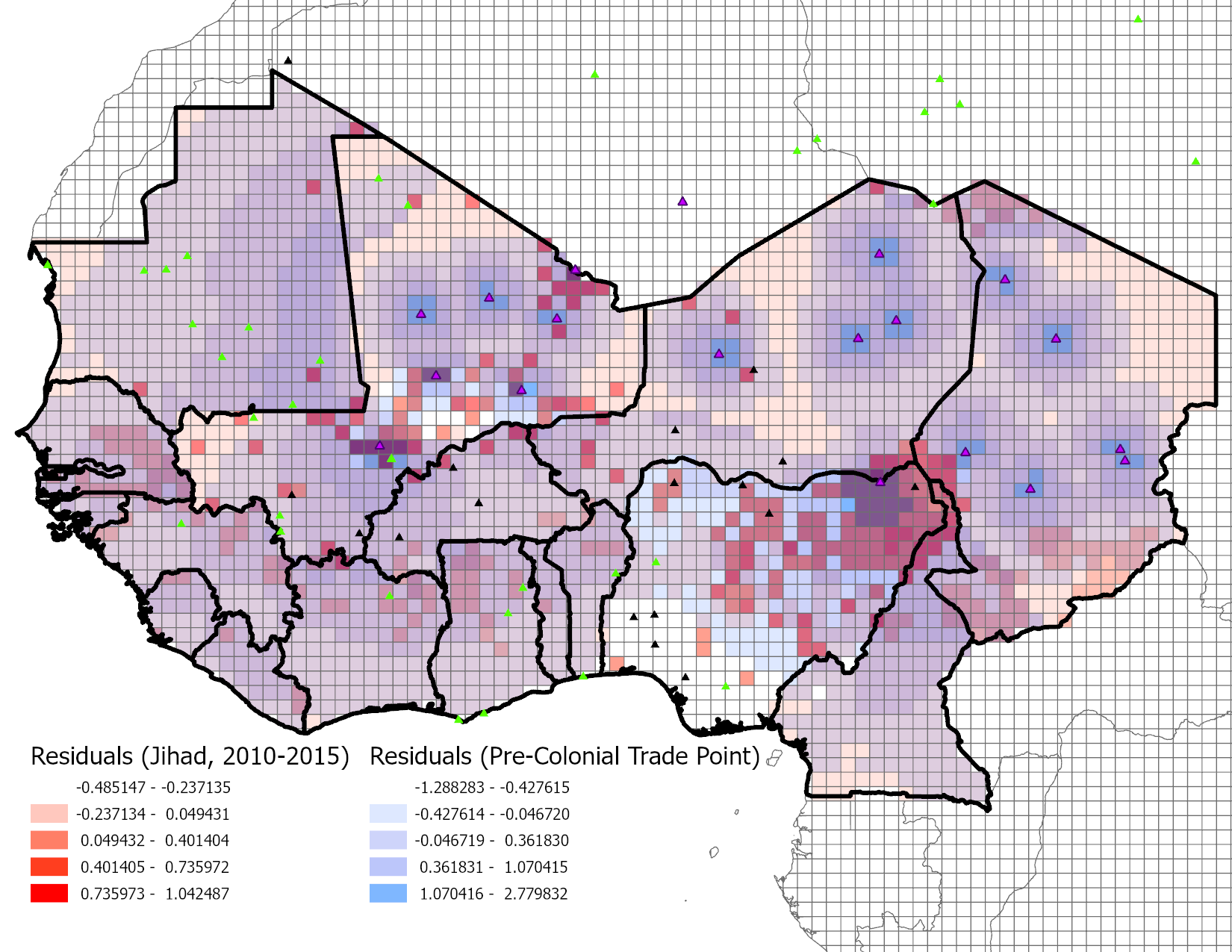}\\
\includegraphics[width=12cm]{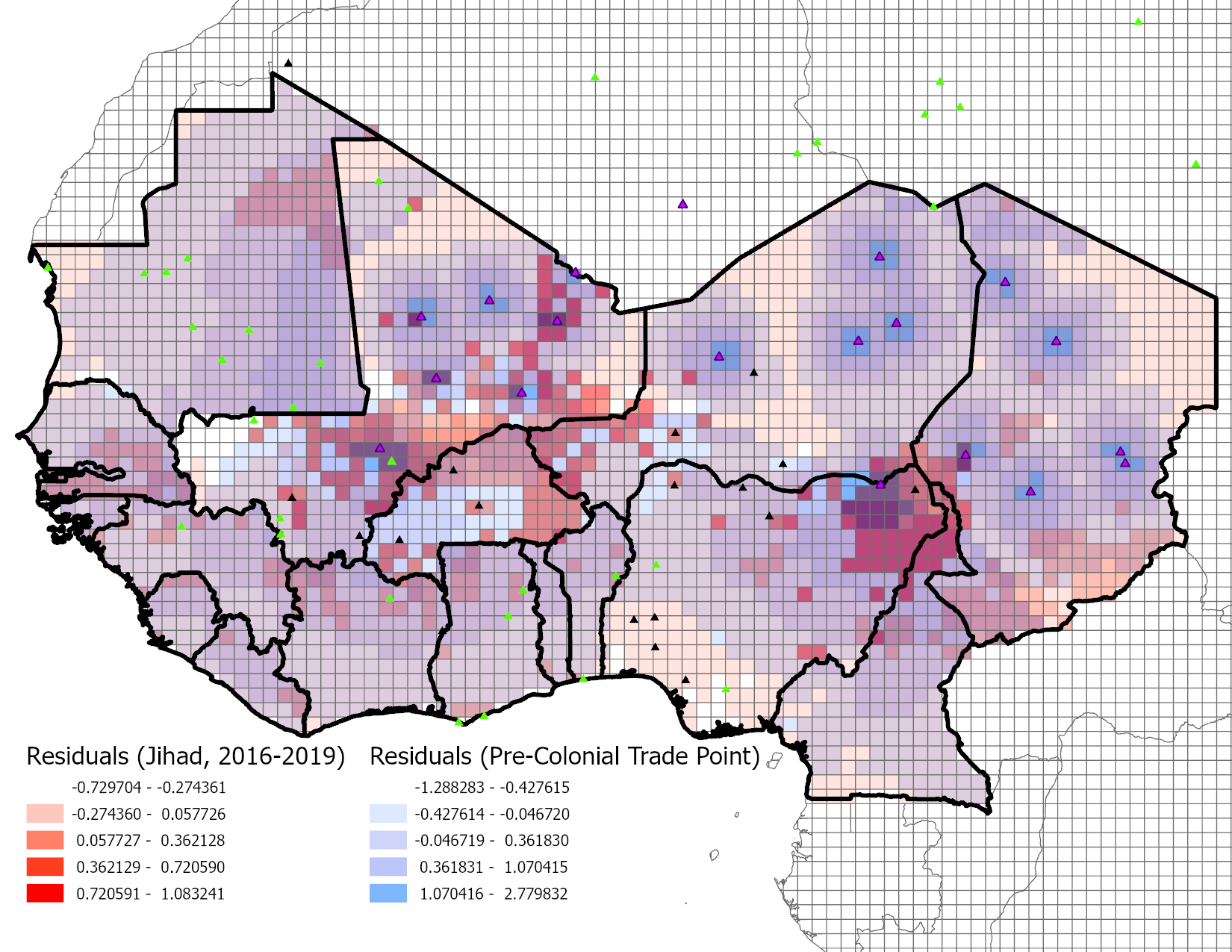}
\caption{Overlay of Residuals over Time Controlling for Population}
\label{map_residuals_over_time}
{\parbox[t]{170mm}{{\footnotesize{\it Notes}:
Both maps show the overlay of the two residuals---the red scheme represents residuals from the regression of a dummy variable of jihad (2010-2015 in the top map; 2016-2019 in the bottom map) on log population (in 2015) with full controls; the blue scheme indicates negative residuals from the regression of log distance to a pre-colonial inland trade point with less than 100,000 population today on log population (in 2015) with full controls.
Full controls include landlocked dummy, average malaria suitability, average caloric suitability in post 1500, average elevation and country fixed effects.
The purple triangles indicate pre-colonial inland trade points with less than 100,000 population today, the yellow green triangles indicate pre-colonial coastal trade points with less than 100,000 population today, and the black triangles indicate the other pre-colonial trade points.
The color of cells where high residuals overlap turns purple (a mix of red and blue).}}}
\end{center}
\end{figure}
\fi

\clearpage
%%% 4.1.
%%\input{tables/tab_water_historical_city.tex}
\input{tables/tab_exclusion_geo_institution_pop.tex}

%%% 4.2.
\input{tables/tab_ols_city_today_water_rev.tex}
%%\input{tables/tab_ols_city_today_colonial_activity.tex}

%%% 4.3.
%%\input{tables/tab_ols_jihad_city_today.tex}
%\input{tables/tab_first_stage_pop.tex}    % Moved to main
%%\input{tables/tab_placebo_first_stage_pop.tex}
\input{tables/tab_water_jihad_pop_rev.tex}  % BY THE WAY, SHOULD WE SHOW THIS "REDUCED FORM"?
%%\input{tables/tab_iv_lnnightlight.tex}

%%% Mechanism
%%\input{tables/tab_cycles_iv_conflcit_trade.tex}        % MERGED WITH THE OTHER TABLE
%%\input{tables/tab_cycles_iv_ratio_conflcit_trade.tex}  % DROPPED because this is mechanical
%\input{tables/tab_cycle_jihad_pop.tex}     % DROPPED because this is mechanical
\input{tables/tab_colonial_jihad_lake.tex}

%%% Alternative Mechanisms
\input{tables/tab_historical_mosque.tex}
\input{tables/tab_iv_jihad_nonjihad_sahara_pop.tex}

%%\input{tables/tab_first_stage_pop_less_005m_pop.tex}
%%\input{tables/tab_iv_jihad_pop_less_005m_pop.tex}

%%\input{tables/tab_first_stage_sahara_pop_less_005m_pop.tex}
%%\input{tables/tab_first_stage_sahara_pop_less_005m_pop_iv2.tex}
%%\input{tables/tab_iv_jihad_sahara_pop_less_005m_pop.tex}
%%\input{tables/tab_iv_jihad_sahara_pop_less_005m_pop_iv2.tex}

%%\input{tables/tab_water_historical_city_WA2.tex}
%%\input{tables/tab_iv_jihad_pop_iv2.tex}

%%% Other robustness
%\input{tables/tab_iv_jihad_ucdp_pop.tex}
\input{tables/tab_jihad_ucdp_pop_rev.tex}
\input{tables_manual/tab_spatial_se_manual.tex}

%\section{Additional Robustness Checks}\label{app_robust}
%{\color{red} Under construction.}

%%%%%%%%%%%%%%%%%%%%%%%%%%%%%%%%%%%%%%%%%%%%%%%%%%%%%%%%%%%%%%%%%%%%%%%%%%%%%%%%%%%%%%%%%%%%%%
\clearpage
\setcounter{figure}{0}
\setcounter{table}{0}
\setcounter{equation}{0}
\renewcommand{\thefigure}{\Alph{section}.\arabic{figure}}
\renewcommand{\thetable}{\Alph{section}.\arabic{table}}
\renewcommand{\theequation}{\Alph{section}.\arabic{equation}}
\section{Construction of Historical States}\label{app_empires}
\begin{figure}[htbp]
\begin{center}
\includegraphics[width=5cm]{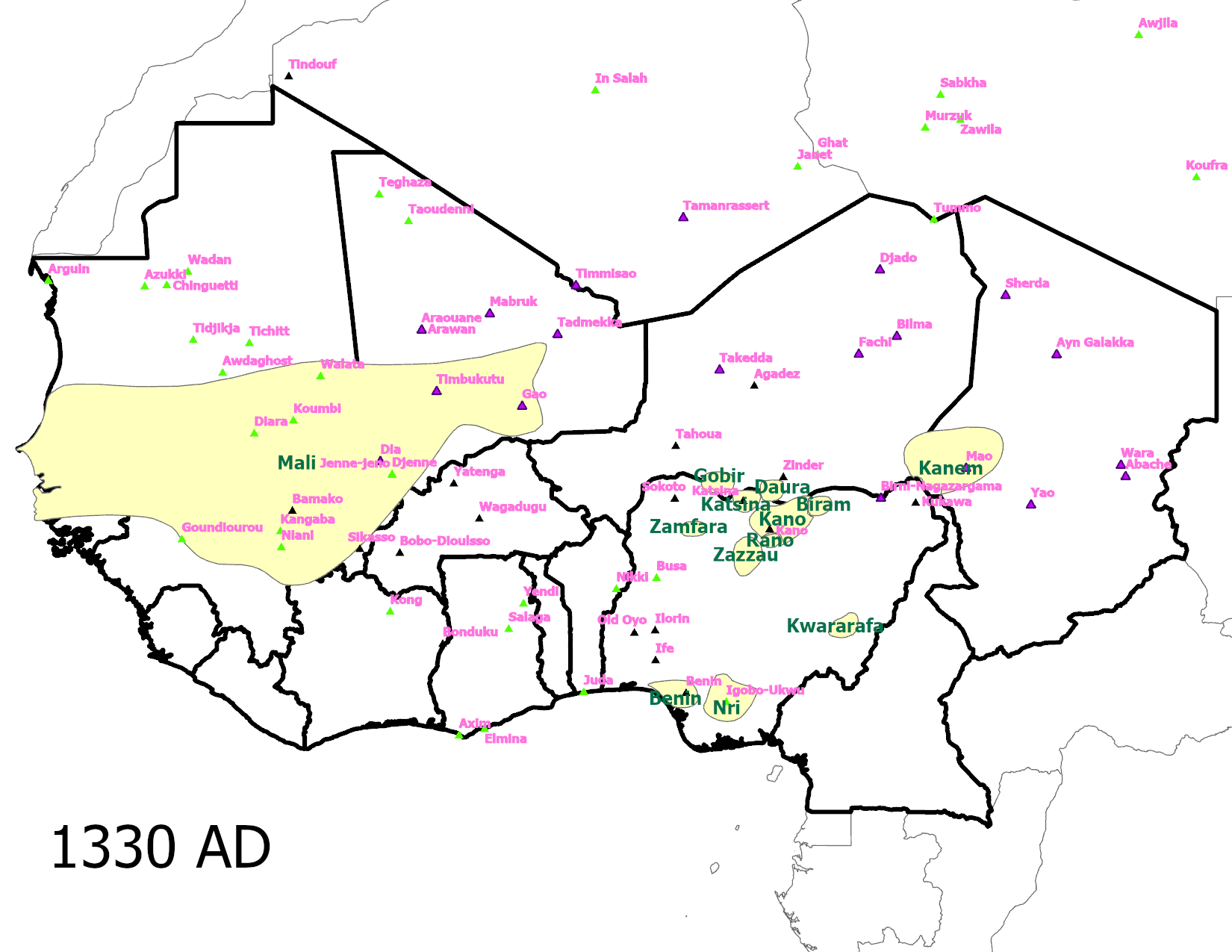}
\includegraphics[width=5cm]{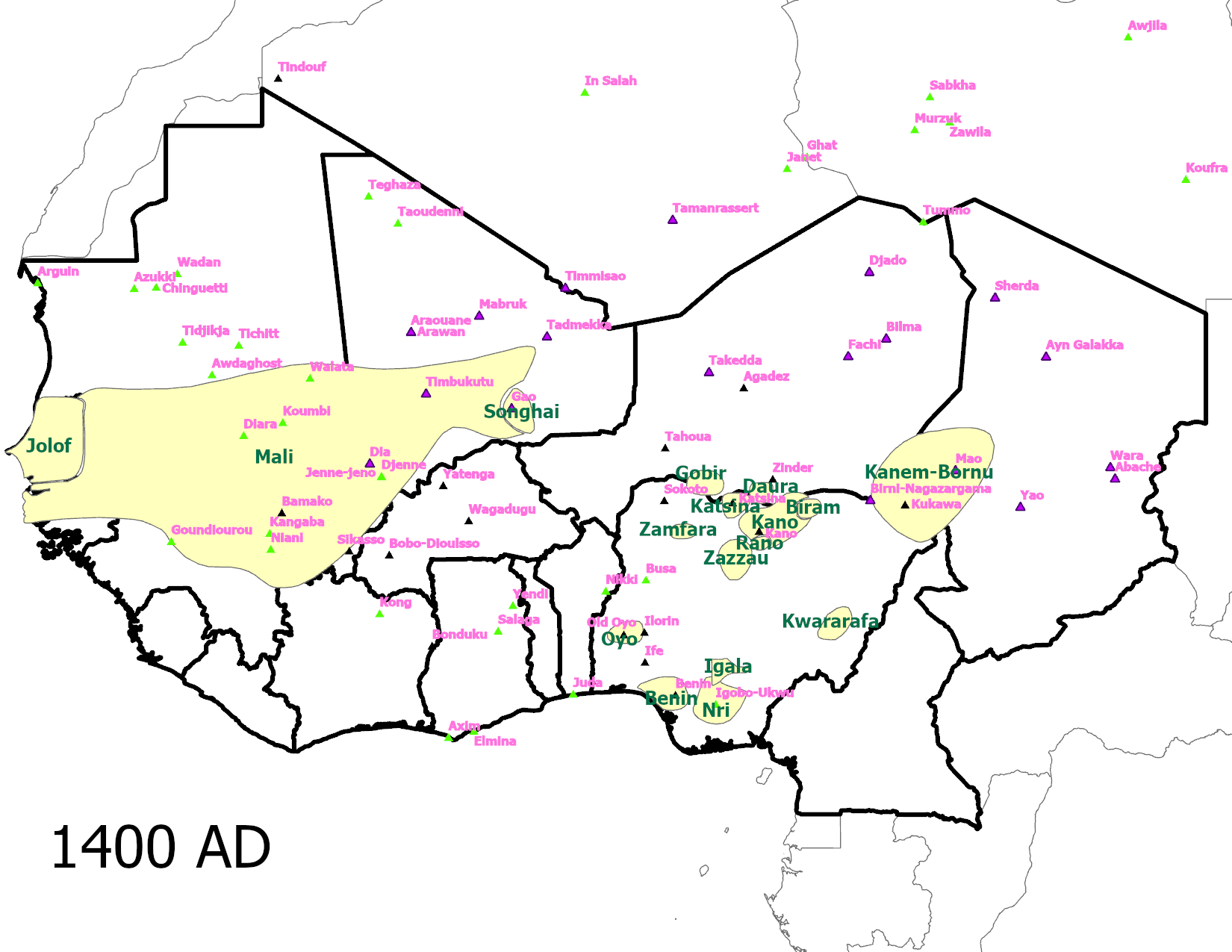}
\includegraphics[width=5cm]{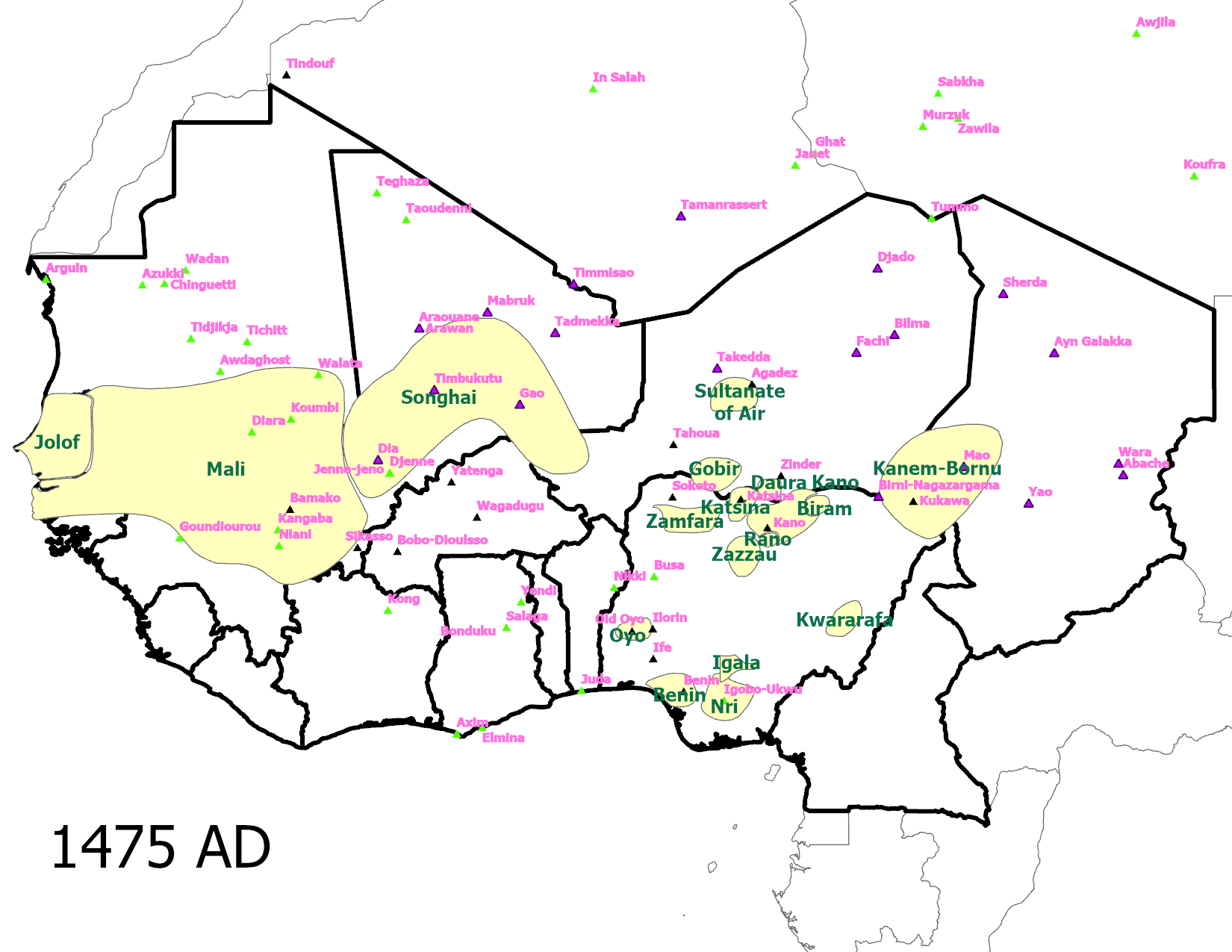}\\
\includegraphics[width=5cm]{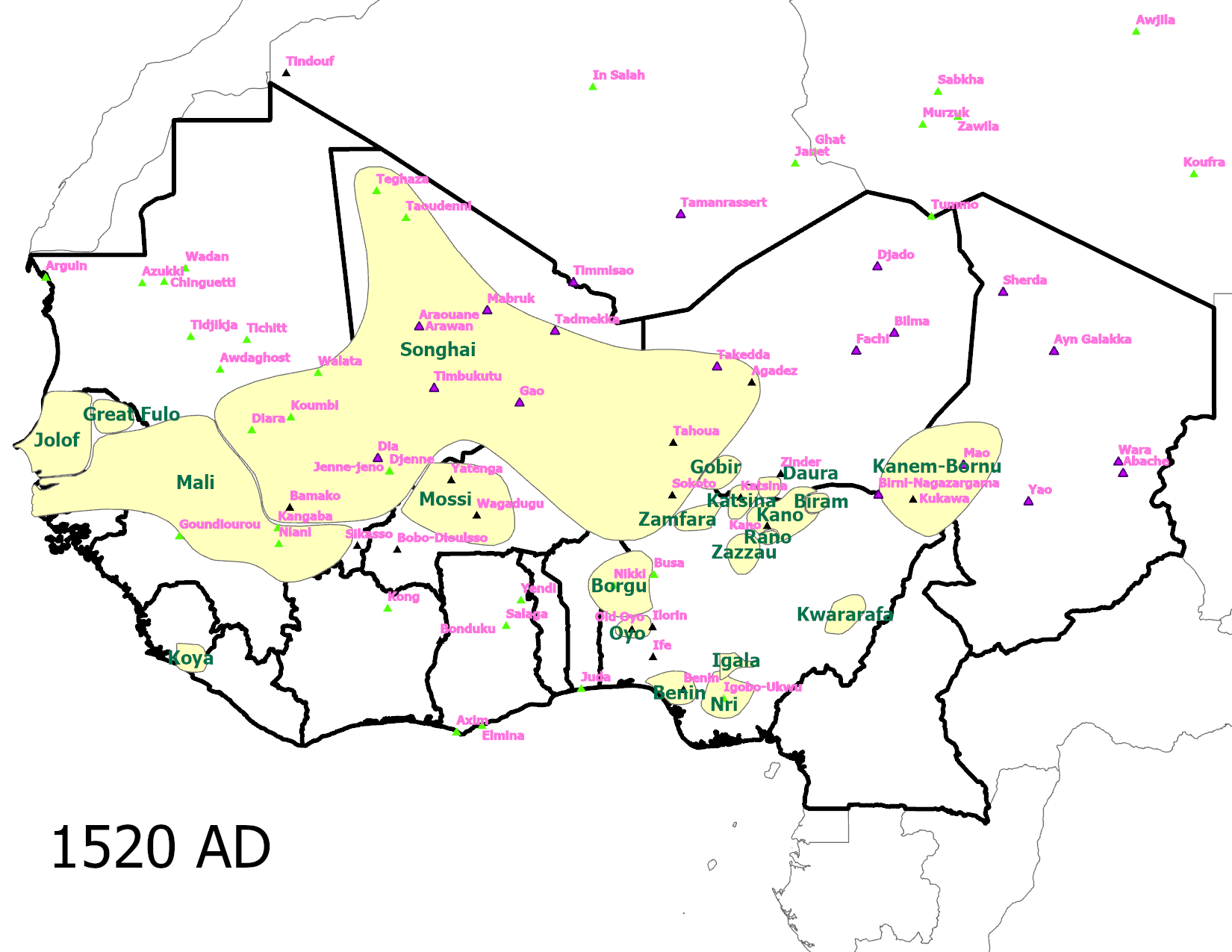}
\includegraphics[width=5cm]{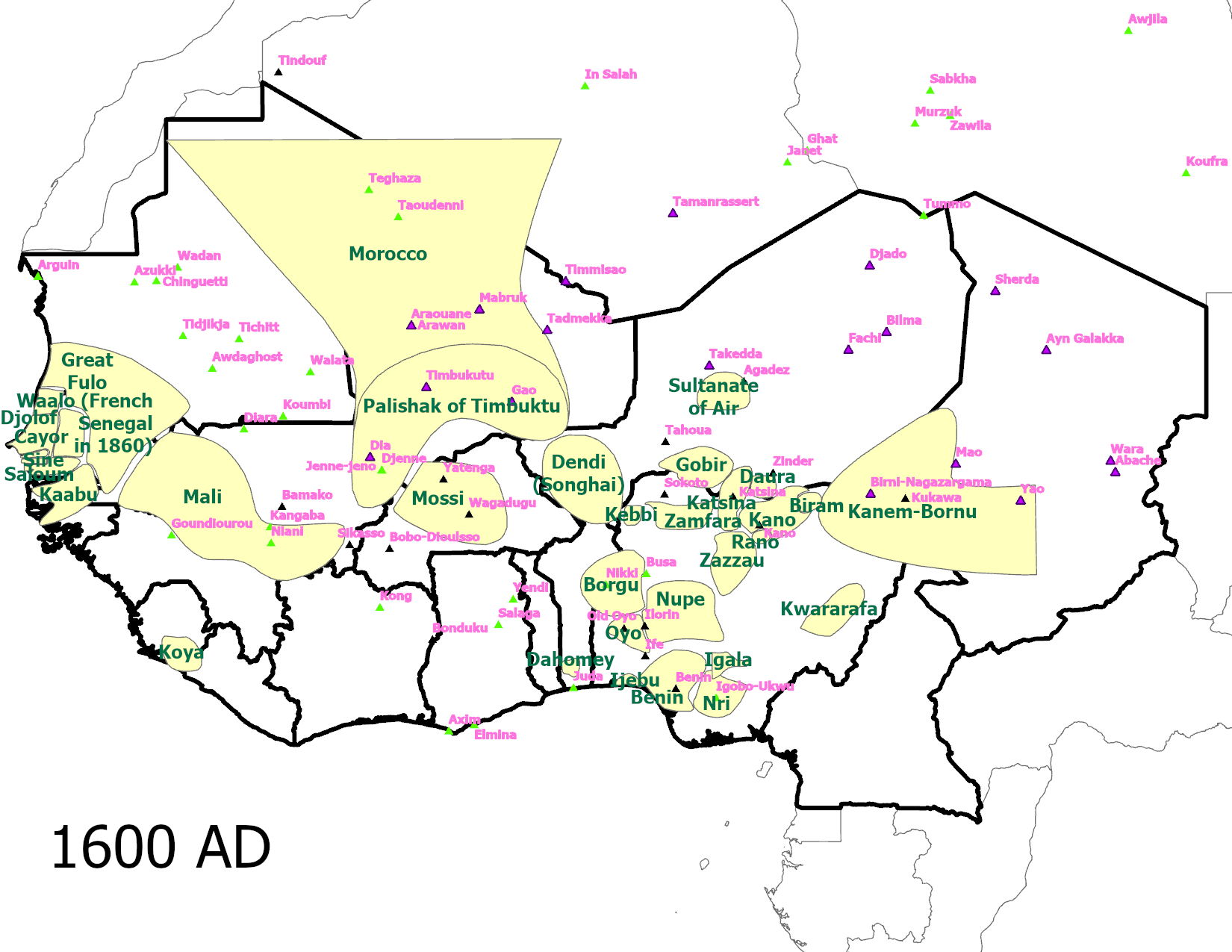}
\includegraphics[width=5cm]{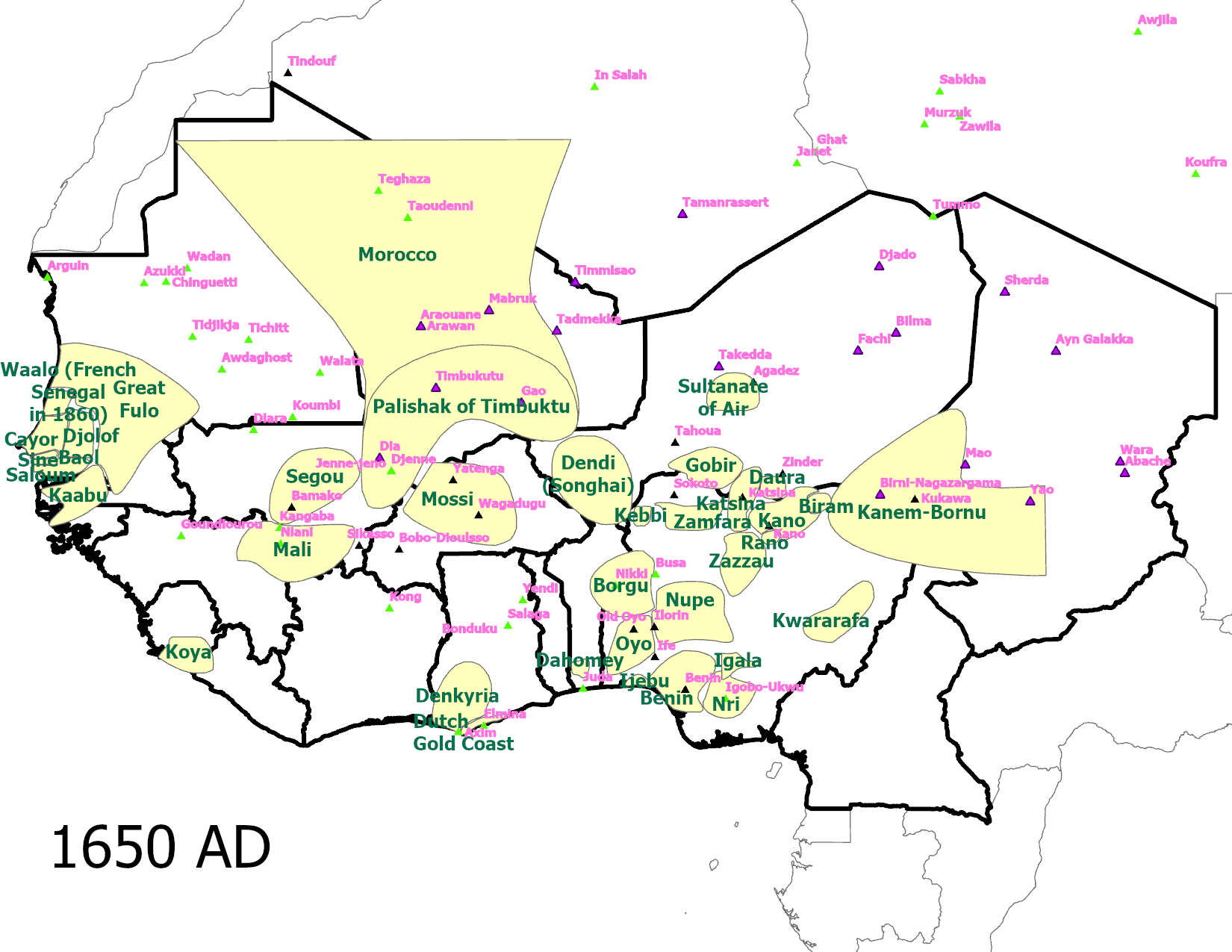}\\
\includegraphics[width=5cm]{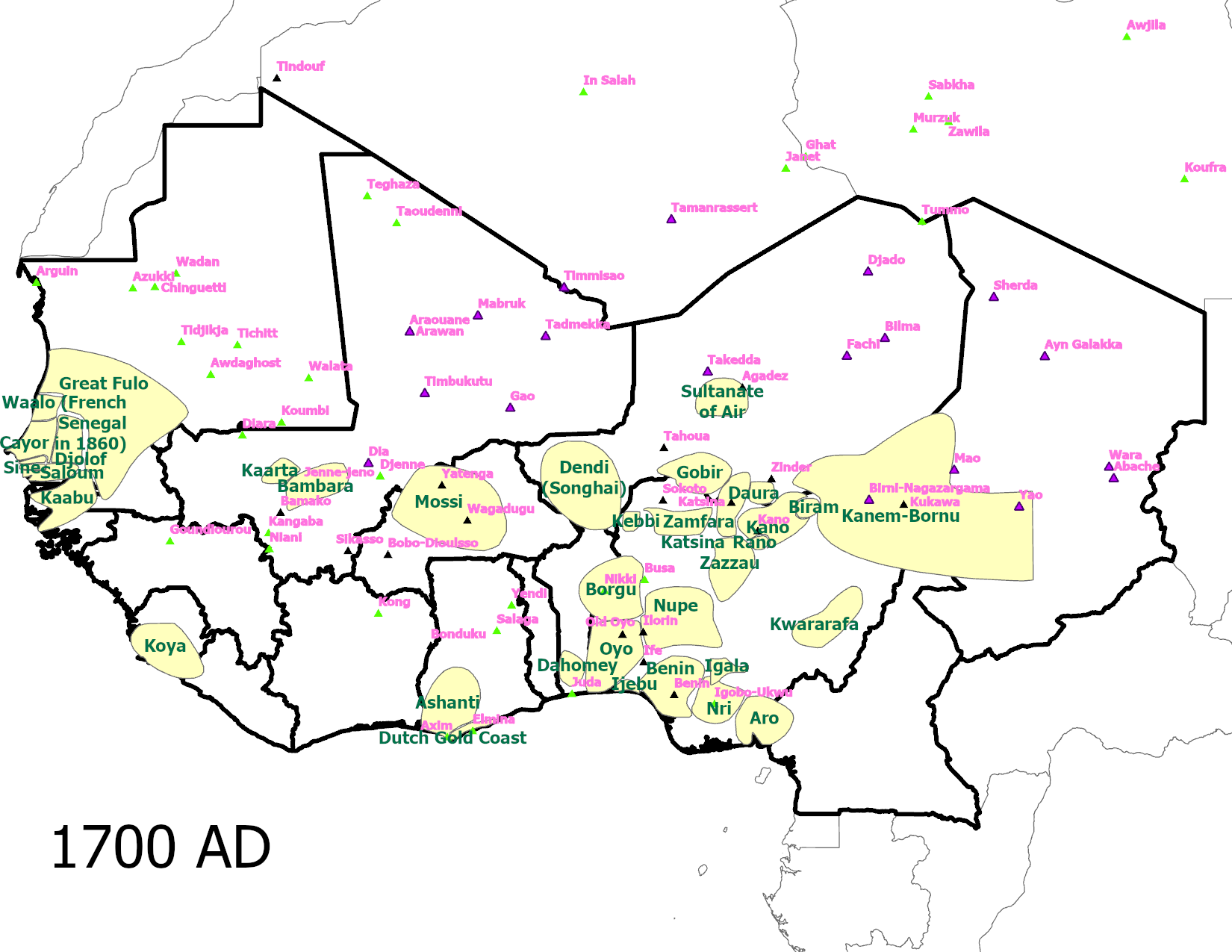}
\includegraphics[width=5cm]{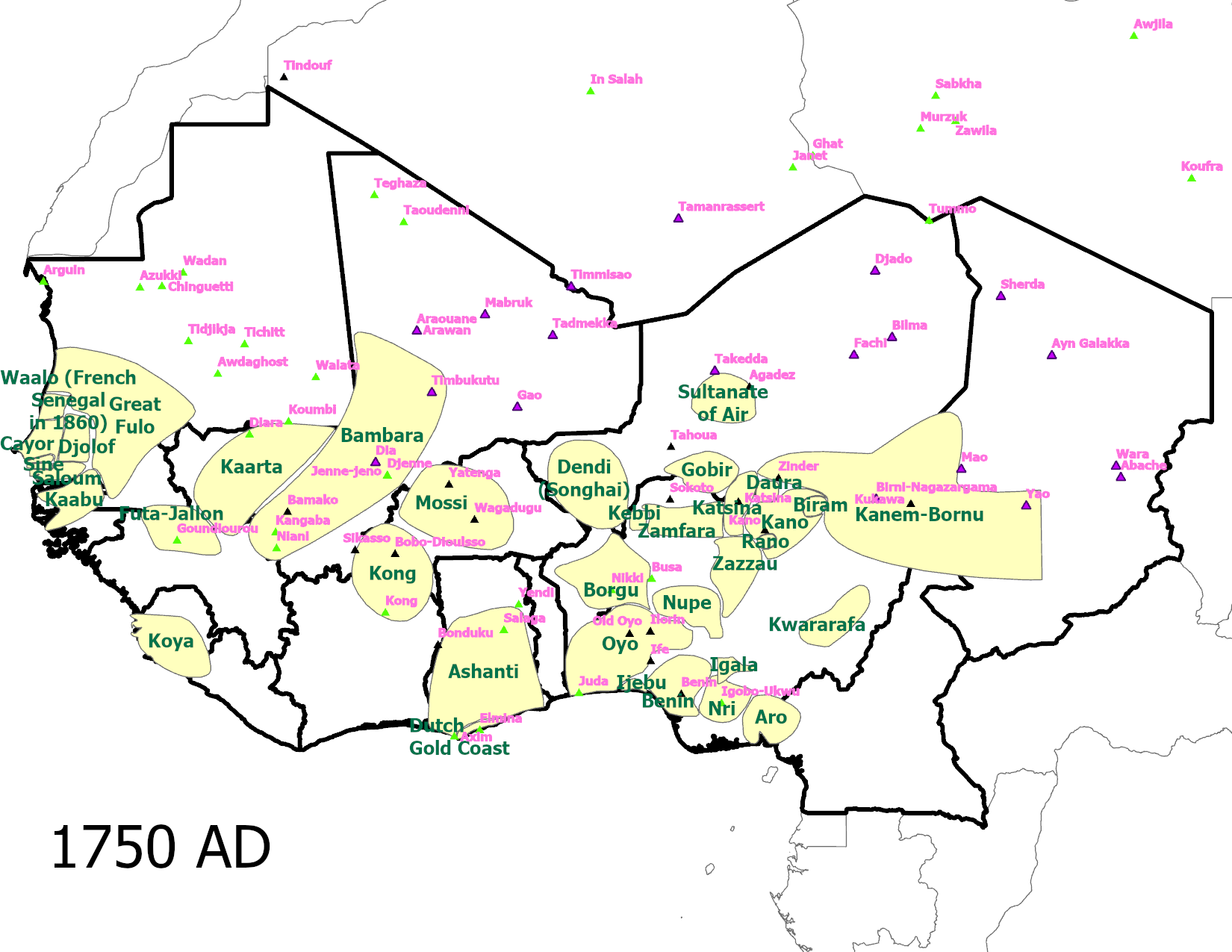}
\includegraphics[width=5cm]{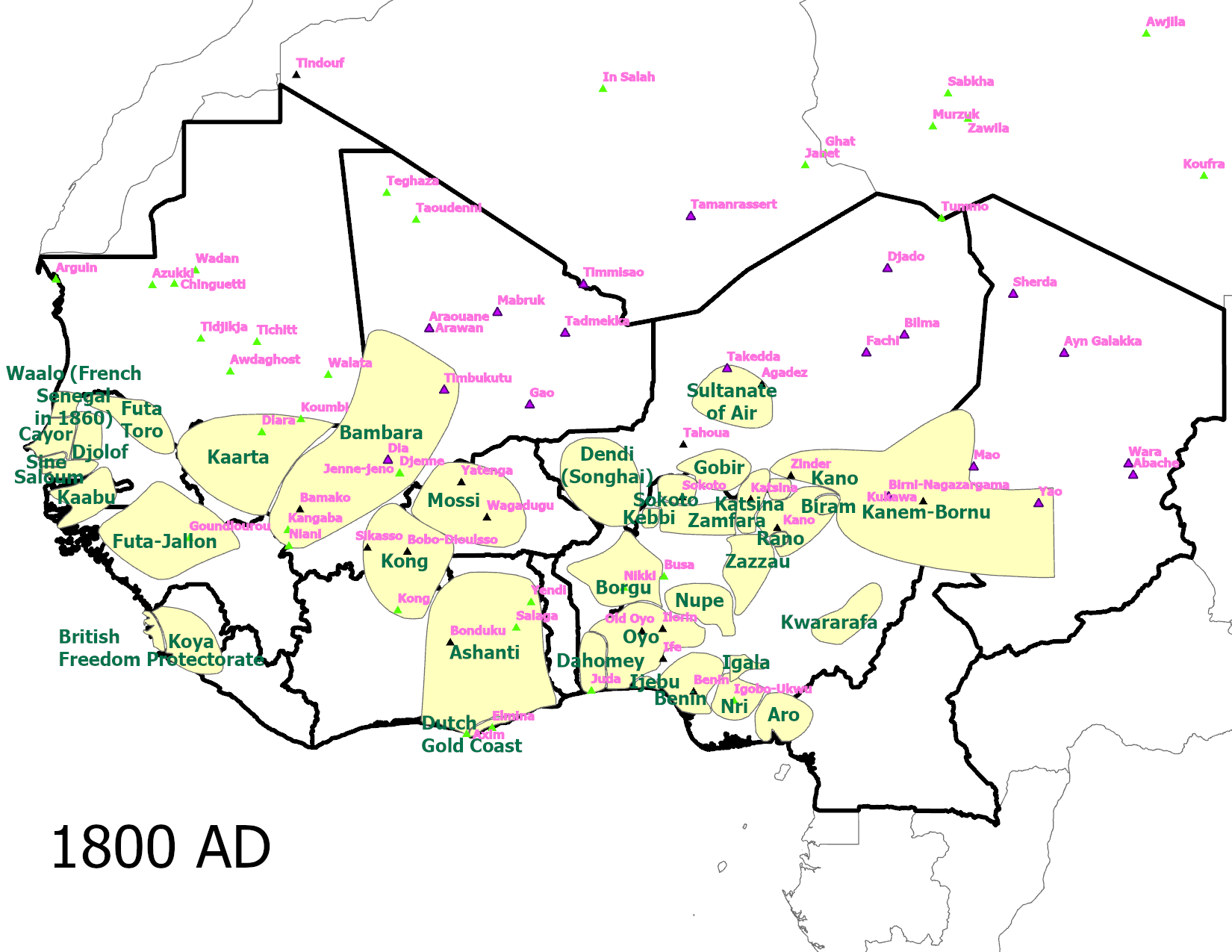}\\
\includegraphics[width=5cm]{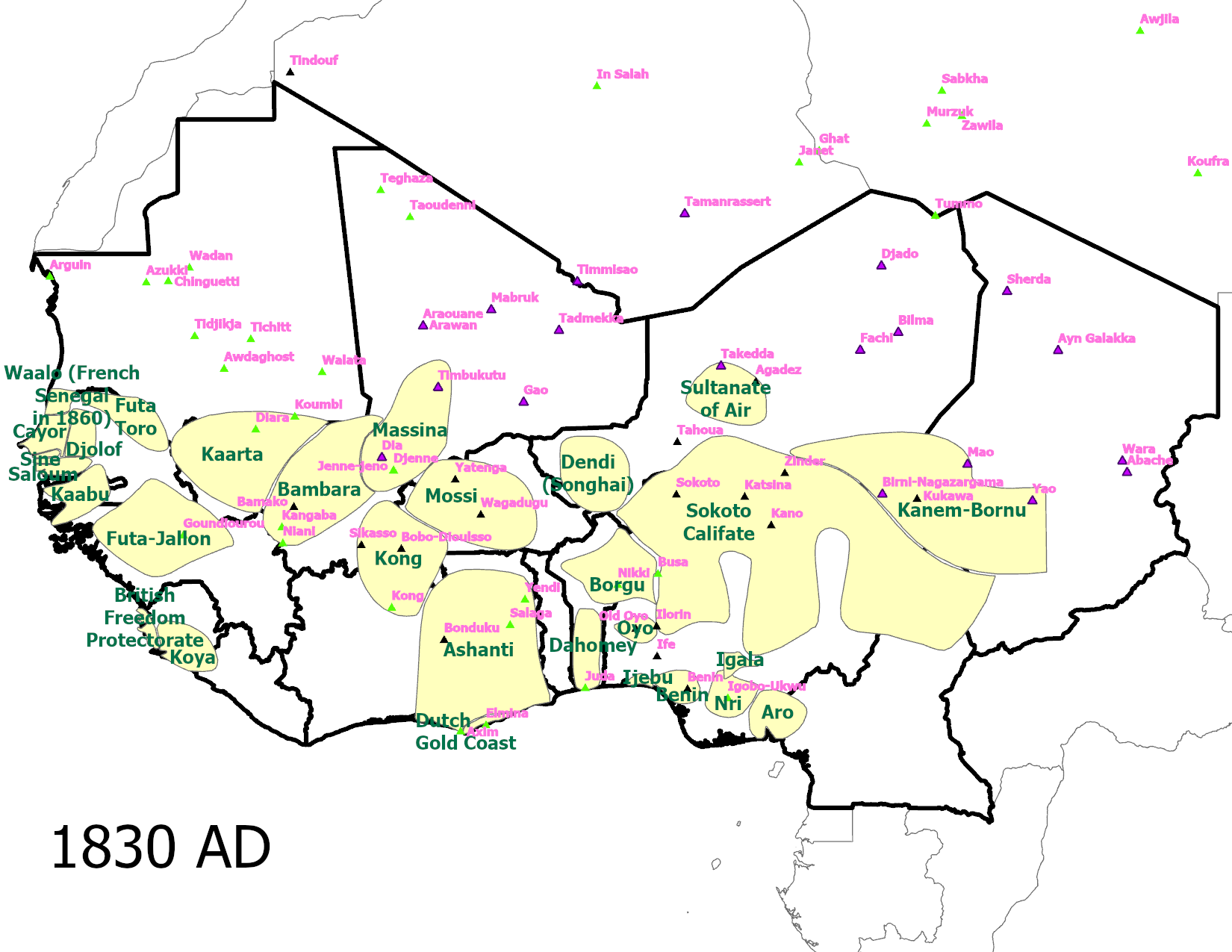}
\includegraphics[width=5cm]{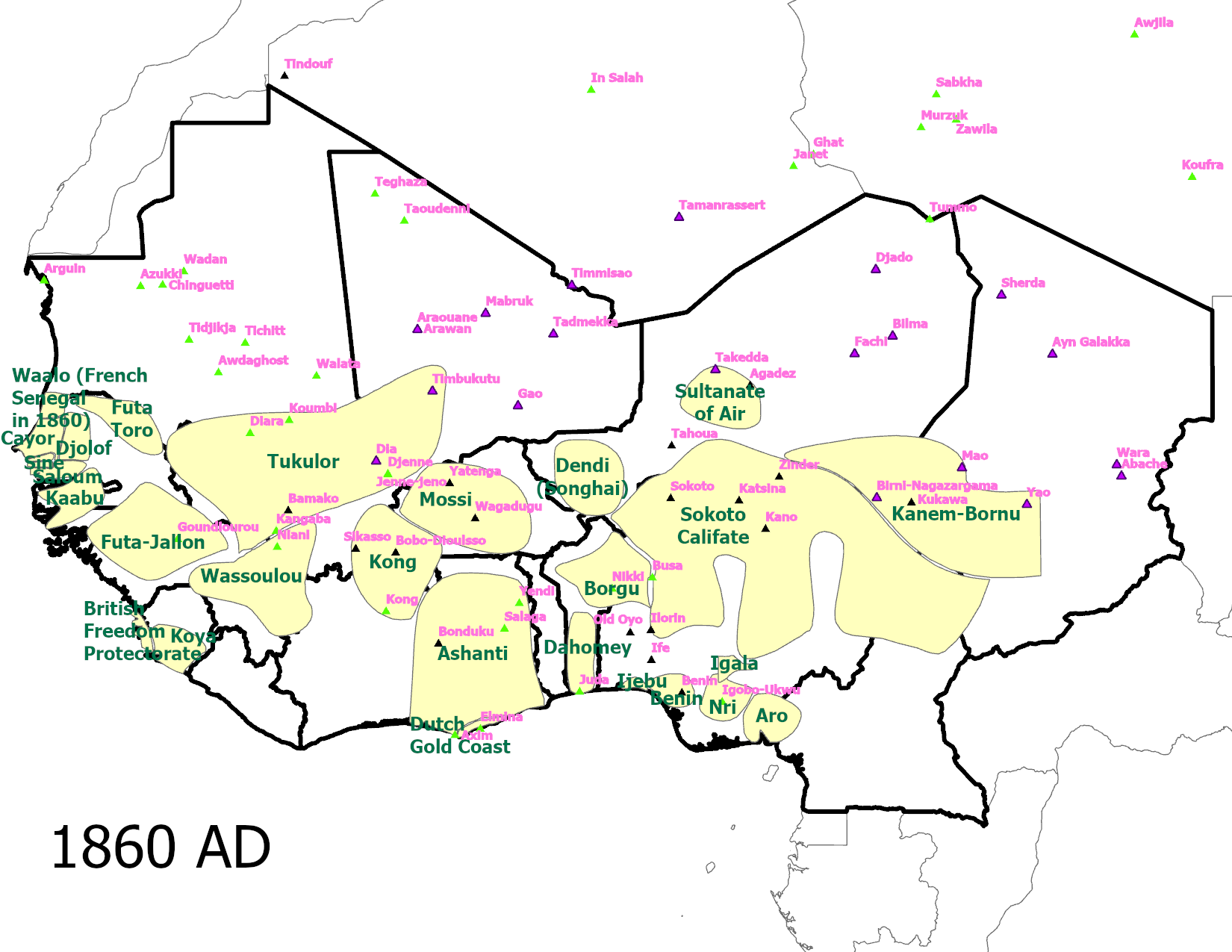}
\includegraphics[width=5cm]{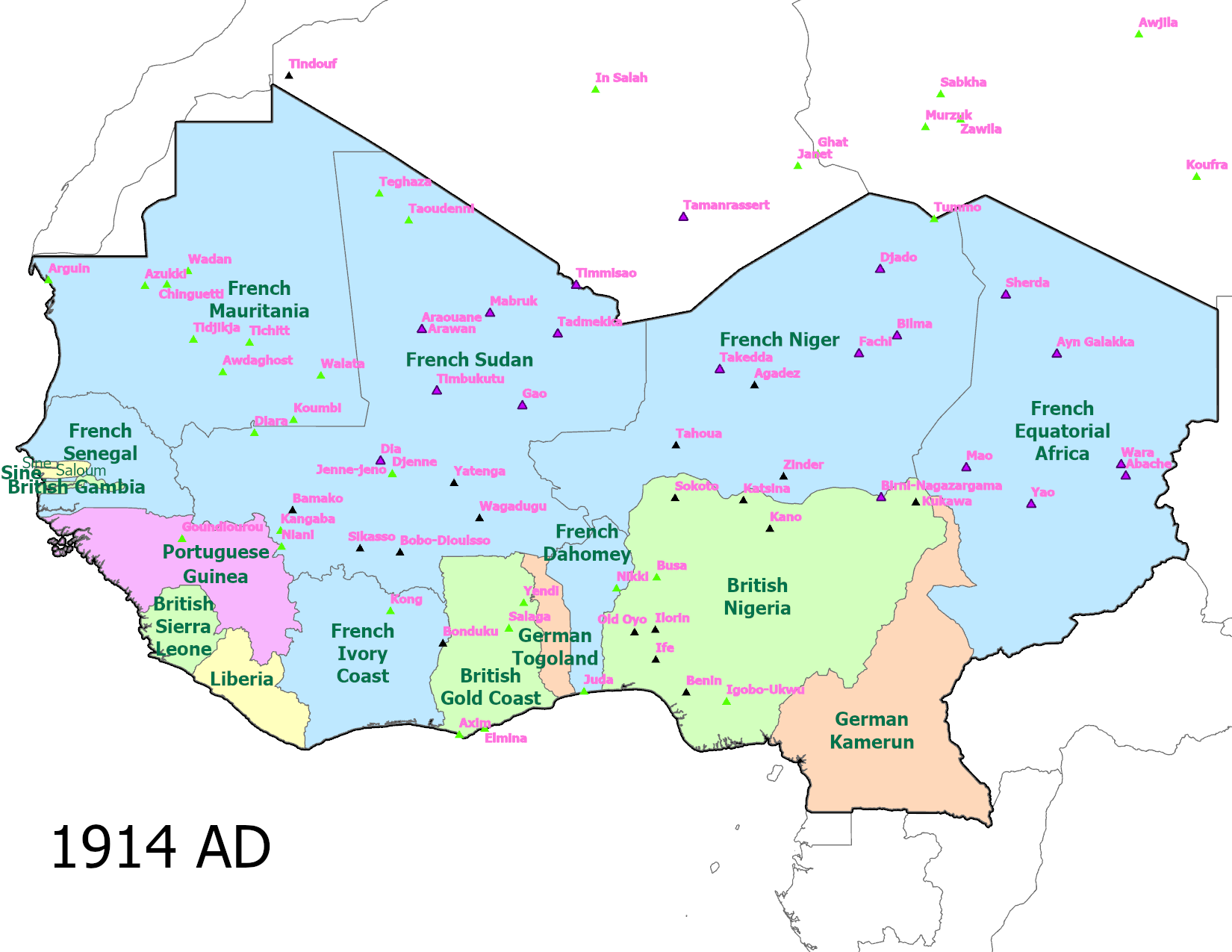}
\caption{Historical States over the Centuries}
\label{map_historical states}
{\parbox[t]{\textwidth}{
{\scriptsize\begin{singlespace}
\textit{Notes}:
These maps show the evolution of historical states from 1330 AD to 1914 AD.
Yellow areas indicate historical states before 1860 AD.
In 1914 AD, the light blue regions indicate the French territory, the light green regions indicate British territory, the light purple region indicates Portuguese territory, the light orange regions indicate German territory, and the yellow regions indicate independent states.
We digitized the maps from \href{https://www.culturesofwestafrica.com/maps/}{Cultures of West Africa}.
The purple triangles indicate pre-colonial inland trade points with less than 100,000 population today, the yellow green triangles indicate pre-colonial coastal trade points with less than 100,000 population today, and the black triangles indicate the other pre-colonial trade points.\end{singlespace}}}}
\end{center}
\end{figure}

%{\color{red} Under construction.}
\href{https://www.culturesofwestafrica.com/maps/}{Cultures of West Africa} provides maps that show spatial locations of historical states before colonization as well as modern countries after independence by using multiple sources of references.
They provide maps about state landscapes from 0 AD to 1980 AD.
We digitized maps for historical states from 1330 AD to 1914 AD just after colonial conquest, using Arc GIS.
Figure \ref{map_historical states} %and Figure \ref{map_historical states_contd}
shows our digitized maps.
\par
To identify which states consisted predominantly of Muslims, we rely on \citetApp{Kasule1998} (p.58) and \citetApp{RNF2004} (p.74-75) that show the extent of Islam circa 1800 AD and locations of states.
However, the Mossi and Kong states extended out of Islamic extent.
Hence, we rely on additional resources to judge if they were Islamic states.
According to \citetApp{Azarya1980} (p.425), since the Kong state was ruled by Muslim, we judge it as the Islamic state.
On the other hand, according to \citetApp{Skinner1958} (p. 1102), since the Mossi state was pagan until European conquests,  it was not the Islamic state.

\section{Additional Data Sources and Variables}\label{app_data}
\subsection{Pre-Colonial Variables}\label{app_data_precolonial}
The pre-colonial variables we use are defined as follows.
%\textbf{Slave Exports.} Log of slave exports per homeland area ($km^2$) is controlled. Source: \citetApp{Nunn2008}.\\
%\textbf{European Explorers.} Indicator for whether or not European explorers passed. Source: \citetApp{NW2011}.\\

\textbf{Cities in 1400.} Indicator for whether a city with a population larger than 200,000 in 1400 was in a given area. Source: \citetApp{Chandler1987}.
%\textbf{Pre-Colonial Conflict.} Indicator for whether a major pre-colonial ethnic conflict occurred in a given area. Source: \citetApp{BR2014}.\\

\textbf{Population in 800.} We use the History Database of the Global Environment (HYDE, version 3.1) constructed by \citetApp{KBJ2010}.
%The resolution of grid cells is 5' $\times$ 5' (corresponding to 9.3 $\times$ 9.3 km at the equator).
%{\color{red} We will no longer use it.}\\

\textbf{Historical Islamic states.}  The indicator which takes a value of one if a grid cell locates in historical Islamic states (given by Appendix \ref{app_empires}), otherwise takes a value of 0. If a grid cell strands in both Islamic states and Non-Islamic states, we assign the state information with larger size in a grid cell.

\textbf{Jurisdictional hierarchy.} We use ``Jurisdictional Hierarchy beyond the Local Community'' (v33) in \textit{Ethnographic Atlas}.
This is an ordered variable which indicates 1. ``No political authority beyond local community  (e.g., autonomous bands and villages),'' 2. ``One level  (e.g., petty chiefdoms),'' 3.``Two levels (e.g., larger chiefdoms),'' 4.``Three levels  (e.g., states),'' and 5.``Four levels (e.g., large states).''

\textbf{Polygamy.} We use ``Marital composition: monogamy and polygamy'' (v9) in \textit{Ethnographic Atlas}.
This is a categorical variable which indicates 1. ``Monogamous,'' 2. ``Polygynous, with polygyny occasional or limited,'' 3.``Polygynous, with polygyny common and preferentially sororal, and co-wives not reported to occupy separate quarters,'' 4.``Polygynous, with polygyny common and preferentially sororal, and co-wives typically occupying separate quarters,'' 5.``Polygynous, with polygyny general and not reported to be preferentially sororal, and co-wives typically occupying separate quarters,'' 6. ``Polygynous, with polygyny general and not reported to be preferentially sororal, and co-wives not reported to occupy separate quarters,'' and 7.``Polyandrous.''
For our analyses, we use an indicator variable which takes a value of 0 if the categorical variable takes a value of either 0 or 7, otherwise takes a value of 1.

\textbf{Irrigation.} We use ``Agriculture: intensity'' (v28) in \textit{Ethnographic Atlas}.
This is an ordered variables which indicates 1. ``Complete absence of agriculture,'' 2. ``Casual agriculture, i.e., the slight or sporadic cultivation of food or other plants incidental to a primary dependence upon other subsistence practices,'' 3.``Extensive or shifting cultivation, as where new fields are cleared annually, cultivated for a year or two, and then allowed to revert to forest or brush for a long fallow period,'' 4.``Horticulture, i.e., semi-intensive agriculture limited mainly to vegetable gardens or groves of fruit trees rather than the cultivation of field crops,'' 5.``Intensive agriculture on permanent fields, utilizing fertilization by compost or animal manure, crop rotation, or other techniques so that fallowing is either unnecessary or is confined to relatively short periods,'' and 6.``Intensive cultivation where it is largely dependent upon irrigation.''
For our analyses, we use an indicator variable which takes a value of 1 if the ordered variable takes a value of 6, otherwise takes a value of 0.

\textbf{Class stratification.} We use ``Class differentiation: primary'' (v66) in \textit{Ethnographic Atlas}.
This is a categorical variable which indicates 1. ``Absence of significant class distinctions among freemen (slavery is treated in EA070), ignoring variations in individual repute achieved through skill, valor, piety, or wisdom,'' 2. ``Wealth distinctions, based on the possession or distribution of property, present and socially important but not crystallized into distinct and hereditary social classes,'' 3.``Elite stratification, in which an elite class derives its superior status from, and perpetuates it through, control over scarce resources, particularly land, and is thereby differentiated from a property-less proletariat or serf class,'' 4.``Dual stratification into a hereditary aristocracy and a lower class of ordinary commoners or freemen, where traditionally ascribed noble status is at least as decisive as control over scarce resources,'' and 5.``Complex stratification into social classes correlated in large measure with extensive differentiation of occupational statuses.''
For our analyses, we use an ordered variable which takes a value of 0 if the categorical variable takes a value of 1, takes a value of 1 if the categorical variable takes a value of either 2 or 3, and takes a value of 2 otherwise.

\textbf{Local headman.} We use ``Political succession'' (v72) in \textit{Ethnographic Atlas}.
This is a categorical variable which indicates 1. ``Patrilineal heir,'' 2. ``Matrilineal heir,'' 3.``Nonhereditary succession through appointment by some higher authority,'' 4.``Nonhereditary succession on the basis primarily of seniority or age,'' 5.``Nonhereditary succession through influence, e.g., of wealth or social status,'' 6.``Nonhereditary succession through election or some other mode of formal consensus,'' 7.``Nonhereditary succession through informal consensus,'' and 8.``Absence of any office resembling that of a local headman.''
For our analyses, we use an indicator variable which takes a value of 1 if the categorical variable takes a value of either 6 or 7, otherwise takes a value of 0.

\textbf{Atlantic slave exports.} The logarithm of one plus the number of the Atlantic slave trade exports at ethnic homeland in the 1700s and 1800s.
Source: \citetApp{Nunn2008}.

\textbf{Gun access.} To capture pre-colonial gun access, we construct a quantity-weighted measure of gun access, $\sum_w \frac{GunImports_{w}}{Distance_{ow}}$, where $Distance_{ow}$ is the distance from grid cell $o$ to the nearest coastal trading location $w$.
The underlying data on gun imports are drawn from \cite{Inikori1977}, specifically \textit{Appendix II, An analysis of 111 trading voyages made from England to West Africa between 1757 and 1806, showing names of the vessels, year of voyage, destination in West Africa and the quantity of guns carried.}
For each destination, we aggregate the total number of guns shipped across all recorded voyages between 1757 and 1806, and use this cumulative quantity to weight gun access.

The destination list in \cite{Inikori1977} occasionally refers to broad regional categories rather than specific ports (e.g., Senegambia, Gambia, Windward Coast, Sierra Leone, Gold Coast, Benin, and Cameroon).
In such cases, we assign a representative coastal trading location that was central to Anglo–African commerce during the late eighteenth century.
Specifically, we map Senegambia to Saint-Louis, Gambia to Bintang, the Windward Coast to Assinie-Mafia, Sierra Leone to Sherbro Island, the Gold Coast to Cape Coast Castle, Benin to Ouidah, and Cameroon to Bimbia.
Latitude and longitude coordinates for these locations are obtained from Wikipedia and GeoNames.

\textbf{Historical mosques.} We use the logarithm of one plus distance (km) from the centroid of each grid cell to the nearest historical mosque established before 1860 and before 1900.
The underlying data are drawn from \cite{Pradines2022}, specifically \textit{Annex 2: Inventory and Atlas of Historical Mosques in Sub-Saharan Africa Listed by Contemporary States}.
This annex provides maps with regional labels as well as a consolidated list of place names associated with historical mosques.
We rely exclusively on the list of places rather than the regional names displayed on the maps, as the latter do not necessarily correspond to locations that actually contain historical mosques (for example, Accra in Ghana appears as a regional label despite the absence of a documented historical mosque).

On pages 3–4, \cite{Pradines2022} defines "historical mosques" as African mosques constructed before 1920.
However, among the place names that appear in the annexed list, several mosques are explicitly described in the main text as having been built after 1920 (for example, Doumga Ouro Alfa, discussed on page 109).
To address this inconsistency and to focus on pre-colonial mosques, we focus on the mosques before 1860 and before 1900 and cross-checked mosque construction dates using available web-based sources, including Wikipedia, \href{https://www.archnet.org/}{Archnet}, and \href{https://www.islamicarchitecturalheritage.com/}{the Islamic Architectural Heritage Database}.

Each place name in the annexed list is interpreted as indicating the presence of at least one historical mosque, rather than a precise count or geolocated inventory of individual structures.
For instance, for N'Djamena, the main text (p. 144) notes that \quotes{according to a nineteenth-century Arabic manuscript, the mosques of N'Djamena, the new capital of the region, amounted to 38.}
Nevertheless, the annexed list records only \quotes{N'Djamena} without any further subdivision or enumeration of mosque locations within the city.
Given this aggregation and the lack of consistent information on the number and exact locations of mosques within each place, we do not construct a count of historical mosques at the grid-cell level.
Instead, we use the distance to the nearest historical mosque as our primary measure.

\subsection{Colonial Variables}\label{app_data_colonial}
The colonial variables we use are defined as follows.
%\textbf{Split by National Border.} Indicator for whether an ethnic homeland being split by national border or not. Source: Intersecting \citetApp{Murdock1959} map with the Digital Chart of the World (DCW) shapefile.\\

\textbf{Distance to a mission station.} The logarithm of one plus distance (km) to the nearest mission station from a centroid of each grid cell.
Source: \citetApp{Nunn2010}.

\textbf{Distance to a colonial railway.} The logarithm of one plus distance (km) to the nearest colonial railway from a centroid of each grid cell.
Source: \citetApp{NW2011}.

\textbf{Infrastructure expenditures.} The logarithm of total expenditures (in 1910–1940)  on construction, transportation, sewage and electricity adjusted to 1910 French francs at colonial district level. Source: \citetApp{Ricart-Huguet2021}.\footnote{We thank Joan Ricart-Huguet for sharing the data of historical administrative boundaries in West Africa.}

\textbf{European population.} The logarithm of European population at colonial district level in the 1920s-1930s. Source: \citetApp{Ricart-Huguet2021}.

\subsection{Geographical Variables}\label{app_data_geographical}
The geographical variables we use are defined as follows.

\textbf{Distance to water sources today.} The minimum distance to either river or lakes from a unit of analysis. The source of river centerlines and lakes come from \href{https://www.naturalearthdata.com}{Natural Earth}, \quotes{Rivers + lake centerlines} version4.1.0. and from \href{https://www.hydrosheds.org/products/hydrolakes}{HydroLAKES}, respectively.
%\href{https://hub.arcgis.com/datasets/0abb136c398942e080f736c8eb09f5c4_0/about}{HydroLAKES}.

\textbf{Elevation.} Mean elevation within a given area in kilometers. Source: Four Tiles: ``GT30W020N40,'' ``GT30E020N40,'' ``GT30W020S10,'' and ``GT30E020S10'' from  \href{https://earthexplorer.usgs.gov/}{GTOPO30}.

\textbf{Agricultural suitability.} Mean land quality for agriculture within a given area. Source: \citetApp{Michalopoulos2012}.

\textbf{Pastoralism suitability.} The average suitability within the unit of analysis, calculated based on nomadic pastoralism suitability from \citetApp{BS2010}.

\textbf{Caloric suitability.} The average caloric suitability (1000 Cal) within the unit of analysis. Source: \citetApp{GO2015}, \citetApp{GO2016} and \citetApp{GOS2017}.

\textbf{Ecological diversity.} Ecological diversity constructed by \citetApp{Fenske2014}.

\textbf{Temperature.} The average temperature within the unit of analysis for the period 2001-2017, calculated based on Terrestrial Air Temperature: 1900-2017 Gridded Monthly Time Series (V 5.01) from \citetApp{MW2018}.

\textbf{Precipitation.} The average precipitation within the unit of analysis for the period 2001-2017, calculated based on Terrestrial Precipitation: 1900-2017 Gridded Monthly Time Series (V 5.01) from \citetApp{MW2018}.

\textbf{Groundwater availability.}  The share of each cell with groundwater depth ranging from 0 to 50 meters. Source: \citetApp{MBDT2012}.

\textbf{Malaria suitability.} Mean malaria suitability index within a given area. Source: \citetApp{SKMSMS2004}

\textbf{Distance to the coast.} The logarithm of one plus distance (km) to the nearest coastal point from a centroid of each grid cell. Source: \href{https://www.naturalearthdata.com}{Natural Earth}. ``Coastline'' version 4.1.0.

\textbf{Ruggedness.} Index of terrain ruggedness as constructed by \citetApp{NP2012} for cells at 30 arc-second resolution. The variable used in the analysis is the average value of the index within the unit of analysis.

\textbf{Coordinates.} The coordinates---latitude and longitude---of each grid cell centroid are defined according to the World Geodetic System 1984 (WGS84).

%{\bf Ethnic Homeland Area.} We control for log land area ($km^2$) of each ethnic homeland. Source: \citetApp{Murdock1959}\\
%\textbf{Migratory Distance From Addis Ababa.}  The geodesic distance from Addis Ababa to the centroid of a given area in kilometers. I use the coordinate (9N, 38E) in \citetApp{RDRRFC2005} as the location of Addis Ababa.\\

\subsection{Contemporary Variables}\label{app_data_contemporary}
The contemporary variables we use are defined as follows.

\textbf{Contemporary cities.}
We have two sets of contemporary cities.
As the first contemporary city data, we use \href{https://ghsl.jrc.ec.europa.eu/ucdb2018Overview.php}{Urban Centre Database UCDB R2019A}.
This database identifies the cities with over 50,000 population in 2015 all over the world and provides their geolocations.
%\ST{CUT THE FOLLOWING?}\blue{In addition, we supplement these data with cities with populations above 10,000 in countries covering the Sahara (i.e., Chad, Niger, Mali, and Mauritania), by using information from Wikipedia (Figure \ref{app_map_sahara_water_sources_cities_past_present}).
%Population figures from Wikipedia are primarily based on available national censuses conducted between 2005 and 2013.}

\textbf{Nighttime lights.} We rely on two data sources. %\ST{DMSP NOT used anymore?}
One of data sources comes from the Visible Infrared Imaging Radiometer Suite (VIIRS), which covers from 2012 to 2020.
The other comes from the Defense Meteorological Satellite Program’(DMSP) s Operational Linescan System, which covers from 1992 to 2013.
\footnote{
The detailed explanations and discussions about the night light luminosity data can be found in \citetApp{CN2015}, \citetApp{EZGHT2021}, \citetApp{GOB2020}, and \citetApp{GOBL2021}
}

\textbf{Distance to the capital.}  The logarithm of one plus distance (km) from each grid cell centroid to the capital of its country. Source: \citetApp{UN2018}

\textbf{Distance to the country border.} The logarithm of one plus distance (km) from each grid cell centroid to the nearest country border. Source: Source: \href{https://www.naturalearthdata.com}{Natural Earth}. ``Admin 0 – Countries'' version 4.1.0.

\textbf{Ethnologue.}
\href{https://worldgeodatasets.com/language/}{World Language Mapping System} (WLMS) Database maps the location of ethnic groups' homelands.
It maps the traditional homelands which correspond to the ones covered by the 15th edition of \citetApp{Ethnologue2005}.
However, the WLMS does not map in the following: populations away from their homelands (e.g., in cities, refugee populations, etc.), immigrant languages, ethnic groups of unknown location, widespread ethnicities (i.e., groups whose boundaries are essentially identical to a countries boundary) and extinct languages.
We match between Ethnologue and WRD based on the unique Ethnologue identifier for each ethnic group within a country.\footnote{
There are fifteen groups which cannot be matched to WRD.
For the ethnic groups, we utilize the percent of each religion from \href{https://joshuaproject.net/}{Joshua Project}.}

\textbf{Muslims in 2005.}
\href{https://worldreligiondatabase.org/}{World Religion Database} (WRD) provides us with fractions of Muslims at ethnic group level within a country in 2005.

\textbf{Population.} \href{https://www.worldpop.org/}{WorldPop datasets} provide approximately 100m$\times$100m cell-level estimated population density (\citealtApp{Tatem2017}).
%To my knowledge, this dataset is the most reliable publicly-available information on population density in the world.
See \citetApp{SGLT2015} and \citetApp{Lloyd2019} for the technical detail for constructing this dataset. From this raw data, we construct approximately 1km$\times$1km cell-level estimated population density in the both countries.

\textbf{Alternative conflict events data.} Additional data on conflict comes from Uppsala Conflict Data Program Georeferenced Event Dataset Version 21.1 (UCDP GED) (\citealtApp{CS2013}; \citealtApp{PTO2020}; \citealtApp{Pettersson_etal2021}; \citealtApp{SM2013}). The UCDP GED codes geo-locations of events, times of events, and names of conflict actors which engage in each event, covering the period between 1989 and 2020.
We follow a similar strategy as what we did with the ACLED to pick jihadist organizations.

\subsection{Afrobarometer}\label{app_Afrobarometer}
We use respondents in West African countries available in rounds 6 and 7.
The West African countries include Benin, Burkina Faso, Cabo Verde, Cameroon, Côte d'Ivoire, Gambia (round 7 only), Ghana, Guinea, Liberia, Mali, Niger, Nigeria, Senegal, Sierra Leone and Togo.
Round 6 was surveyed between 2014 and 2015.
Round 7 was surveyed between 2016 and 2018.
The variables we use are defined as follows.

\textbf{Age.}
A respondent's age. Survey questions: Q1 (rounds 6 and 7).

\textbf{Female.}
An indicator for female respondents.
Survey questions: Q101 (rounds 6 and 7).

\textbf{Education.}
The ten categories of educational attainment.
They are classified as ``no formal schooling,'' ``Informal schooling only,'' ``Some primary schooling,'' ``Primary school completed,'' ``Some secondary school/high school,'' ``Secondary school completed/high school,'' ``Post-secondary qualifications, not univ,'' ``Some university,'' ``University completed,'' or ``Post-graduate.''
Survey questions: Q97 (rounds 6 and 7).

\textbf{Living conditions.}
The five categories of present living conditions.
They are classified as ``Very Bad,'' ``Fairly bad,'' ``Neither good nor bad,'' ``Fairly good,'' or ``Very good.''
Survey questions: Q4B (rounds 6 and 7).

\textbf{Religion.}
A respondent's religion was asked in Q98A (round 6) and Q98 (round 7).
They are condensed as ``Christian,'' ``Muslim,'' or ``Other'' in the variable ``RELIG\_COND'' of rounds 6 and 7.
We use the variable to restrict the sample to Muslim respondents.

\textbf{Neighbors from different religion.}
A respondent was asked whether she would like having people of a different religion as neighbors, dislike it, or not care.
A respondent chose one of the following answers: 1.``Strongly dislike,'' 2.``Somewhat dislike,'' 3.``Would not care,'' 4.``Somewhat like,'' or 5.``Strongly like.'' We use the variable which takes the values of 1 through 5 as a dependent variable.
Survey questions: Q89A (round 6) and Q87A (round 7).
In Table \ref{tab_mechanism_ideology}, we re-scaled the variable to the following for the clearer interpretation of the results: 1.``Strongly like,'' 2.``Somewhat like,'' 3.``Would not care,'' 4.``Somewhat dislike,'' and 5.``Strongly like.''

\textbf{Governed by religious law.}
A respondent was asked which of the following statements is closest to her view:
``Our country should be governed primarily by religious law'' (Statement 1)
or ``Our country should be governed only by civil law'' (Statement 2).
A respondent chose one of the following answers: 1.``Agree very strongly with statement 2,'' 2.``Agree with statement 2,'' 3.``Agree with neither,'' 4.``Agree with statement 1,'' or 5.``Agree very strongly with statement 1.''
We use the variable which takes the values of 1 through 5 as a dependent variable.
Survey question: Q65 (round 7).

\textbf{Equal opportunities to education.}
A respondent was asked whether she disagrees or agrees with the following statement:
``In our country today, girls and boys have equal opportunities to get an education.''
A respondent chose one of the following answers:  1.``Strongly disagree,'' 2.``Disagree,'' 3. ``Neither agree nor disagree,'' 4.``Agree,'' or 5.``Strongly agree.''
We use the variable which takes the values of 1 through 5 as a dependent variable.
Survey question: Q77A (round 7).
In Table \ref{tab_mechanism_ideology}, we re-scaled the variable to the following for the clearer interpretation of the results: 1.``Strongly agree,'' 2.``Agree,'' 3.``Neither agree nor disagree,'' 4.``Disagree,'' and 5.``Strongly disagree.''

%%%%%%%%%%%%%%%%%%%%%%%%%%%%%%%%%%%%%%%%%%%%%%%%%%%%%%%%%%%%%%%%%%%%%%%%%%%%%%%%%%%%%%%%%%%%%%
%\clearpage
\setcounter{figure}{0}
\setcounter{table}{0}
\setcounter{equation}{0}
\renewcommand{\thefigure}{\Alph{section}.\arabic{figure}}
\renewcommand{\thetable}{\Alph{section}.\arabic{table}}
\renewcommand{\theequation}{\Alph{section}.\arabic{equation}}
\section{Historical Conflicts}\label{app_historical_conflict}
\subsection{Conflict Catalogues}\label{app_brecke}
In this section, we construct conflict confrontation sample between European actors and historical states from conflict catalogue.
We match conflict actors in the database with historical states in 1860 from \href{https://www.culturesofwestafrica.com/maps/}{Cultures of West Africa} by using online resources.\footnote{
We mainly depend on wikipedia, and \href{https://joshuaproject.net/}{Joshua Project}.}
%For the main analysis, we use matched conflicts based on direct match or alternate name match.
We follow the procedures below:

1. We match a historical state with a conflict actor by its name (direct match).

2. If a state name cannot be found in conflict catalogue, we match by its alternate names or spellings (alternate name match). Matched conflicts must occur following the establishment years.

%\begin{itemize}
%\item[1.] We match a historical state with a conflict actor by its name (direct match).
%\item[2.] If a state name cannot be found in conflict catalogue, we match by its alternate names or spellings (alternate name match). Matched conflicts must occur following the establishment years.
%\end{itemize}

However, \citetApp{Brecke1999} includes many conflicts that indicate only larger ethnic groups (e.g., Fulani) or larger areas (e.g., Sierra Leone) that may have been related to historical states.
As additional sources, we make use of information about locations of conflict from \citetApp{FK2017} and \citetApp{BDL2019}.
\citetApp{FK2017} provides the information about conflicts between 1700 and 1900.
For conflicts after 1901, we use digitized information by ourselves, using web sources (e.g., wikipedia and google maps).
%Following \citetApp{FK2017}, historical jihad is defined as conflicts involving African entities and European colonial powers in the Islamic zone mapped by \citetApp{BB1918}.
Table \ref{tab_brecke} lists all the conflicts related to historical states in West Africa from \citetApp{Brecke1999}.\footnote{
We exclude the following conflicts from the list since they are not identified in alternative sources: \quotes{France-England (Benin), 1792}, \quotes{Nikki-France (Borgou, Benin), 1916} and \quotes{Adja-France (Mono, Benin), 1918-19}.}
\par
\input{tables/table_brecke.tex}

%%%%%%%%%%%%%%%%%%%%%%%%%%%%%%%%%%%%%%%%%%%%%%%%%%%%%%%%%%%%%%%%%%%%%%%%%%%%%%%%%%%%%%%%%%%%%%
%\clearpage
%\setcounter{figure}{0}
%\setcounter{table}{0}
%\setcounter{equation}{0}
%\renewcommand{\thefigure}{\Alph{section}.\arabic{figure}}
%\renewcommand{\thetable}{\Alph{section}.\arabic{table}}
%\renewcommand{\theequation}{\Alph{section}.\arabic{equation}}
\subsection{Strategies against Colonization}\label{app_strategies}
% Why there are diffs in strategies: millitant strategy and availability to weapons
%\textbf{Dendi (Songhai).}  \\
\if0
\begin{figure}[htbp]
\begin{center}
%\includegraphics[width=15cm]{maps/map_Islam_groups_Muslim_WA_2010_2015.png}
\includegraphics[width=9cm]{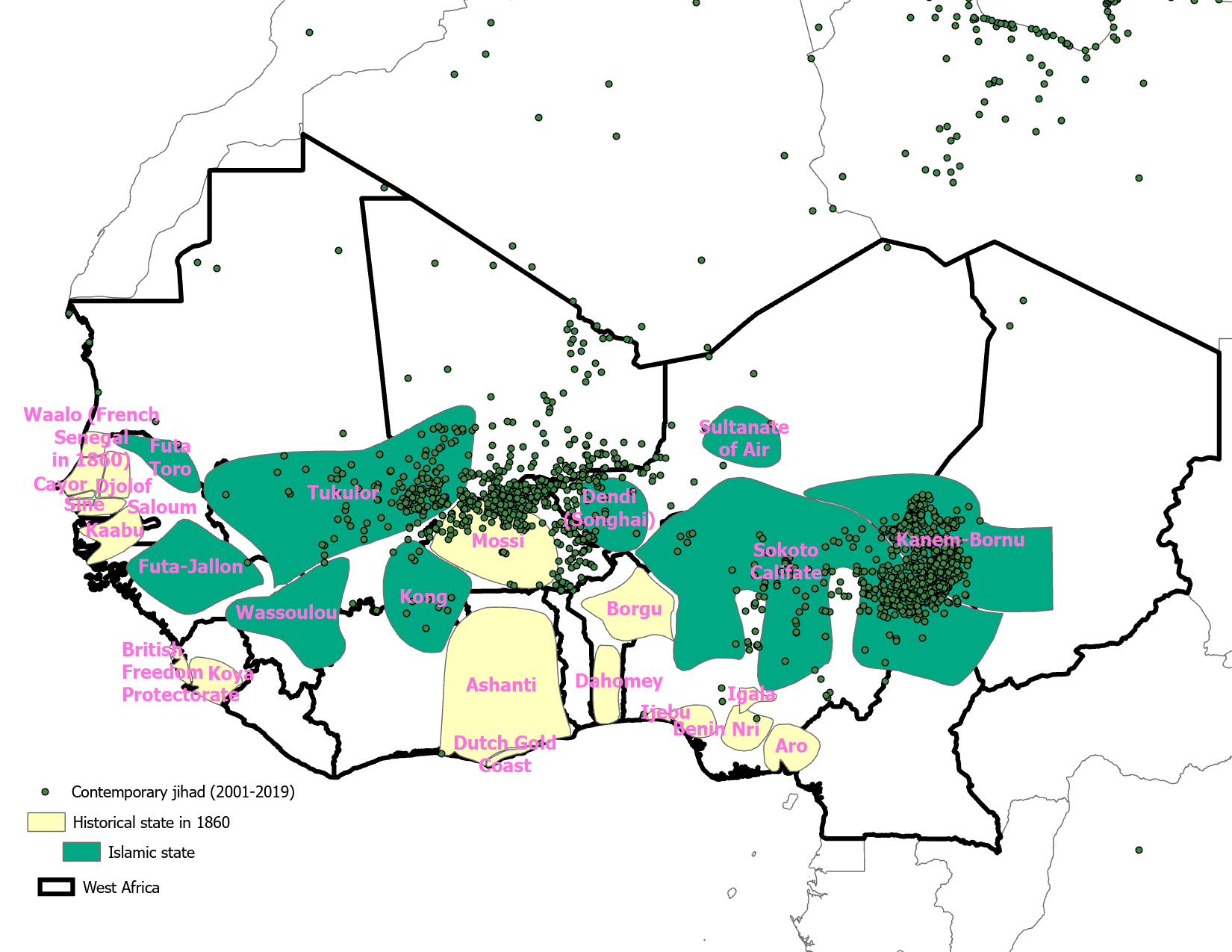}
\caption{Islamic States in 1860 and Contemporary Jihad}
%\ST{Cut this? Doesn't look beautiful? Also Figure 1 already provides mostly similar info except for the Islamic state names}
{\scriptsize(Source: ACLED and  \href{https://www.culturesofwestafrica.com/maps/}{Cultures of West Africa})}
\label{fig: islamic_state_strategies}
\end{center}
\end{figure}
\fi

\textbf{Futa Toro.} Elites often pursued negotiation and accommodation with the French, while many local fighters joined the jihad movement of Umar Tal, the leader of Tukulor empire, and resisted French forces, reflecting a mixed strategy of diplomatic accommodation and intermittent armed resistance (\citealtApp{Barry1998}).
%The leader of Tukulor empire, Ahmadu, recruited {\it talib\'{e}s} (i.e., students of religion who formed the backbone of his father's army) from Futa Toro (\citealtApp{Boahen1985} p.120).

\textbf{Futa Jallon.} Futa Jallon resistance to French expansion relied on the use of diplomacy rather than military measures (\citealtApp{Mcgowan1981}, p.246).
%\textbf{Kanem-Bornu.}

\textbf{Kong Empire.} Kong empire was destroyed in 1895 by Samori who accused of soliciting French protection (\citealtApp{Azarya1980} p.434).
%\textbf{Sultanate of Air.}
%Ikime, Obaro (1977). Fall of Nigeria. Heinemann

\textbf{Sokoto Califate.} The Sokoto Caliphate was conquered by British colonial forces in 1903.
Although it was military conquest (\citealtApp{Boahen1985} p.137), Sokoto Caliphate had limited access to the European weapons, which ended up in ``no tactics, no personal gallantry and no resistance'' against the conquest (\citealt{Crowder1971}, p.294).
The territory was divided between British, French, and German powers.

\textbf{Tukulor Empire.} The leader Ahmadu, who succeeded the founder of the empire, chose strategies of alliance and militant confrontation and relied more on the alliance than confrontation (\citealtApp{Boahen1985} p.119).  Besides the French, he was forced to fight on other two fronts: against his brothers who contested his authority and rebellions of his subjects. To deal with these two, he needed arms and ammunition as well as financial resources through trade, both of which necessitated friendly relations with the French (\citealtApp{Boahen1985} p.119-120).

\textbf{Wassoulou (Mandingo/Mandinka) Empire.} The ruler, Samori Ture, chose the strategy of confrontation (\citealtApp{Boahen1985} p.123).
\setcounter{figure}{0}
\setcounter{table}{0}
\setcounter{equation}{0}
\renewcommand{\thefigure}{\Alph{section}.\arabic{figure}}
\renewcommand{\thetable}{\Alph{section}.\arabic{table}}
\renewcommand{\theequation}{\Alph{section}.\arabic{equation}}
\section{Model Appendix}\label{app_model}

\subsubsection*{The model setup}
There are two players, a colonization force ($F$) and an Islamic state ($M$), who are competing over a divisible territory.
There are two periods ($t=1,2$), with no discounting of future payoffs.
The total amount of the territory is normalized to 1, and the allocation at the beginning of period $t$ is given by $\omega_{Ft}$ and $\omega_{Mt}$ for players $F$ and $M$, respectively, where $\omega_{Ft} + \omega_{Mt} = 1$.
We assume that player $F$, a colonizer, has a smaller amount of initial territory at the beginning of the game, thereby $\omega_{F1} < \omega_{M1}$.
The relative military strength of player $i \in \{F, M\}$ is denoted by $\lambda_i$ where $\lambda_F+\lambda_M=1$.
We assume that player $F$ is militarily stronger, thereby $\lambda_F > \lambda_M$.
The analysis will demonstrate that the degree of this asymmetric power relations matters to explain both the historical and contemporary jihads.
The cost of conflict for player $i$ is denoted by $\phi_i$.
%We assume that these costs are equal, $\phi_F = \phi_M = \phi$.

In each period, the bargaining game has two stages.
In Stage 1, player $i \in \{F, M\}$ can either ``challenge'' by making a claim $\sigma_{it}$ where $\omega_{it}<\sigma_{it}\le 1$, or make no claim.
If neither player challenges, both players retain control of their respective shares ($\omega_{it}$).
The cost of making a challenge for player $i$ is $c_i$, which is private information and independently drawn from distribution $F(c)$ over $[\underline{c}, \bar{c}]$.
Stage 2 is reached if exactly one player makes a claim and the other does not.
The other player then decides whether to concede to the claim.
If both players make claims or if neither claims, Stage 2 is not reached and the game ends.

The winner of a conflict will take the full territory, %(i.e., $\sigma_{it}=1$ if $i$ wins)
and conflicts occur in the following two cases.
The first case is where only one player makes a claim in Stage 1 and the other player does not concede to in Stage 2.
Suppose $F$ is the only player who makes a claim.
Then, $F$'s winning probability is $\lambda_F+\theta$ while $M$'s winning probability becomes $1-(\lambda_F+\theta)=\lambda_M-\theta$, where $\theta$ is a parameter governing the first-mover advantage.
We assume $0 \le \lambda_i-\theta$ and $\lambda_i+\theta \le 1$ for all $i$ to ensure that the winning probabilities are well-defined.
The second case is where both players make claims in Stage 1.
In this case, the winning probability of player $i$ is $\lambda_i$, as both players are simultaneously challenging.

Player $i$'s instantaneous utility from controlling a share $\sigma_{it}$ in period $t$ is given by $u_i(\sigma_{it})$, where $u_i$ is an increasing, strictly concave, and differentiable function on $[0,1]$.
Without loss of generality, we normalize the function such that $u_i(1)=1$ and $u_i(0)=0$.

%\subsubsection*{The two-period game}
The dynamic game proceeds as follows.
If conflict has occurred in $t=1$, then the game ends and both players continue to get utilities from the obtained (or lost) territory in $t=2$.
If territorial transfer from one player to the other due to concession has occurred in $t=1$, then the territory after the transfer becomes the new status quo territory at the beginning of $t=2$ ($\omega_{F2}, \omega_{M2}$).
If neither player challenges in $t=1$, then the status quo territory remains the same in $t=2$ ($\omega_{i2}=\omega_{i1}$).
We solve the dynamic game by backward induction.

\subsubsection*{Optimal strategies in $t=2$ as in the one-shot game}
Given the new status quo territory ($\omega_{F2}, \omega_{M2}$) realized after $t=1$ (in the scenario where conflict has not occurred in $t=1$), we solve for optimal strategies in $t=2$ as in the one-shot game.

We begin with deriving the optimal amount of the claim.
Suppose $j$ makes a claim $\sigma_{j2}$ in Stage 1 and $i$ is the second mover.
Then, player $i$ concedes if:
\[ u_i(1-\sigma_{j2}) \ge (\lambda_j + \theta) \cdot u_i(0) + (1-(\lambda_j + \theta)) \cdot u_i(1) - \phi_i = \lambda_i - \theta - \phi_i \]
When $\sigma_{j2}=1$, this condition becomes $\phi_i \ge \lambda_i - \theta$.
If this condition is violated, player $j$'s optimal challenge is $\sigma_{j2} = 1-\eta_{i2}$ such that $u_i(\eta_{i2}) = \lambda_i - \theta - \phi_i$.
To sum, the optimal challenge by $j$ in $t=2$ is $\hat{\sigma}_{j2}\equiv 1-\eta_{i2}$ where:
\[ \eta_{i2} = \begin{cases} u^{-1}[\lambda_i - \theta - \phi_i] & \text{if } \phi_i < \lambda_i-\theta \\ 0 & \text{if } \phi_i \ge \lambda_i-\theta \end{cases} \]
Then, in any perfect Bayesian equilibrium, each player $i$ either make the optimal claim $\hat{\sigma}_{i2}\equiv 1-\eta_{j2}$ or no challenge.
This game can thus be expressed as a game in which both players simultaneously decide whether to challenge with the optimal amount of claim or not challenge.
Labelling the optimal challenge Hawk ($H$) and no challenge Dove ($D$), the payoff matrix for the row player, $i$, becomes as follows:
\begin{center}
\begin{tabular}{|c|c|c|}
\hline
 & H & D \\
\hline
\textbf{$H$} & $\lambda_i - \phi_i - c_i$ & $u_i(1-\eta_{j2}) - c_i$ \\
\hline
\textbf{$D$} & $u_i(\eta_{i2})$ & $u_i(\omega_{i2})$ \\
\hline
\end{tabular}
\end{center}

From this payoff matrix, the difference in player $i$'s net gains from choosing $H$ over $D$ between when $j$ chooses $H$ and when $j$ chooses $D$ is as follows, denoting player $i$'s instantaneous payoff in period $t$ when $i$ chooses $A_i$ and $j$ chooses $A_j$ where $A_i,A_j\in\{H,D\}$ by $\pi_{it}(A_i,A_j)$:
\begin{align*}\Omega_i(\omega_{i2}) &\equiv [\pi_{i2}(H,H) - \pi_{i2}(H,D)] - [\pi_{i2}(D,H) - \pi_{i2}(D,D)] \\
&= \lambda_i - \phi_i - u(\eta_{i2}) - u(1-\eta_{j2}) + u(\omega_{i2})
\end{align*}
Using this function, we can define that actions are strategic complements for player $i$ if $\Omega_i(\omega_{i2})>0$ and strategic substitutes for player $i$ if $\Omega_i(\omega_{i2})<0$.

Player $i$'s strategy is $s_i: [\underline{c}, \bar{c}] \to \{H,D\}$.
Suppose player $i$ thinks player $j$ will choose H with probability $p_j$.
The expected payoffs from choosing $H$ and $D$ in the one-shot game, and their difference are:
\begin{align*}
E[\pi_{i2}(H)]
&= p_j\pi_{i2}(H,H)+(1-p_j)\pi_{i2}(H,D)
=  p_j(\lambda_i - \phi_i) + (1-p_j)u_i(1-\eta_{j2})-c_i \\
E[\pi_{i2}(D)] &= p_j\pi_{i2}(H,H)+(1-p_j)\pi_{i2}(H,D)
= p_j u(\eta_{i2}) + (1-p_j)u(\omega_{i2}) \\
E[\pi_{i2}(H)] - E[\pi_{i2}(D)] &= \underbrace{p_j[\lambda_i - \phi_i - u_i(\eta_{i2})] + (1-p_j)[u_i(1-\eta_{j2}) - u_i(\omega_{i2})]}_{\equiv x_{i2}} -c_i
\end{align*}
Therefore, $s_i(c_i) = H$ if $c_i \le x_{i2}$, so that the probability of playing $H$ is $p_i = F(x_{i2})$.
Hence, the best response function is given by:
\[ x_{i2}=\Gamma_i(x_{j2}) \equiv F(x_{j2})[\lambda_i - \phi_i - u_i(\eta_{i2})] + (1-F(x_{j2}))[u(1-\eta_{j2}) - u(\omega_{i2})] \]
In equilibrium, we have $\hat{x}_{i2} = \Gamma_i(\hat{x}_{j2})$ for $i,j\in\{F,M\}$.
\citet{BS2020} confirm the existence and uniqueness of the Bayesian Nash equilibrium.
Solving the system by assuming that $c_i$ is uniformly distributed on [0, 1], we get:
\[ \hat{x}_{i2}(\omega_{i2}) = \frac{u_i(1-\eta_{j2}) - u_i(\omega_{i2}) + \Omega_i(\omega_{i2})[u_j(1-\eta_{i2}) - u_j(\omega_{j2})]}{1 - \Omega_i(\omega_{i2})\Omega_j(\omega_{j2})} \]

\subsubsection*{Optimal strategies in t = 1 under the shadow of the future}
Denote the optimal challenge and equilibrium cutoff in $t=1$ for player $i$ by $1-\eta_{j1}$ and $\hat{x}_{i1}$.
We first derive $(\eta_{F1},\eta_{M1})$ and then obtain $(\hat{x}_{F1},\hat{x}_{M1})$.

When player $i$ makes its decision in $t=1$, it takes into account the ``shadow of the future.''
In particular, player $i$ takes into account the expected payoff in $t=2$, depending on a potential new status quo $(\omega_{F2}, \omega_{M2})$, before the challenge cost $c_i$ is realized:
Under the assumption that $c_i$ uniformly distributed over [0,1], this expected payoff becomes:
%\begin{align*}
%E\pi_{i2}(\omega_{i2}) &= F(\hat{x}_{i2}) F(\hat{x}_{j2}) [\lambda_i u_i(1) + (1-\lambda_i)u_i(0) - \phi_i - E\{c_i: c_i \le \hat{x}_{i2}\}] \\
%&+ F(\hat{x}_{i2})(1-F(\hat{x}_{j2}))[u_i(1-\eta_{j2}) - E\{c_i: c_i \le \hat{x}_{i2}\}] \\
%&+ (1-F(\hat{x}_{i2}))F(\hat{x}_{j2})u_i(\eta_{i2}) \\
%&+ (1-F(\hat{x}_{i2}))(1-F(\hat{x}_{j2}))u_i(\omega_{i2})
%\end{align*}
{\footnotesize
\begin{align*}
E\pi_{i2}(\omega_{i2})
&= F(\hat{x}_{i2}) F(\hat{x}_{j2}) [\lambda_i - \phi_i - E\{c_i: c_i \le \hat{x}_{i2}\}]
+ F(\hat{x}_{i2})(1-F(\hat{x}_{j2}))[u_i(1-\eta_{j2}) - E\{c_i: c_i \le \hat{x}_{i2}\}]    \\
&+ (1-F(\hat{x}_{i2}))F(\hat{x}_{j2})u_i(\eta_{i2})
+ (1-F(\hat{x}_{i2}))(1-F(\hat{x}_{j2}))u_i(\omega_{i2})  \\
&= \hat{x}_{i2} \hat{x}_{j2} [\lambda_i - \phi_i - \frac{\hat{x}_{i2}}{2}]
+ \hat{x}_{i2}(1-\hat{x}_{j2})[u_i(1-\eta_{j2}) - \frac{\hat{x}_{i2}}{2}]    \\
&+ (1-\hat{x}_{i2})\hat{x}_{j2}u_i(\eta_{i2})
+ (1-\hat{x}_{i2})(1-\hat{x}_{j2})u_i(\omega_{i2})
\end{align*}
}
where $\hat{x}_{i2} = \hat{x}_{i2}(\omega_{i2})$ and $\hat{x}_{j2} = \hat{x}_{j2}(\omega_{j2})$.
That is, the new territory allocation in $t=2$, which depends on players' actions in $t=1$, can determine the equilibrium cutoffs in $t=2$, from which the expected payoff is calculated.
Then, the value functions for player $i$ are as follows:
\begin{align*}
V_i(H,H) &= \pi_{i1}(H,H) + \lambda_i u_i(1) + (1-\lambda_i)u_i(0)
= 2\lambda_i - \phi_i - c_i \\
V_i(H,D) &= \pi_{i1}(H,D) + E\pi_{i2}(\hat{w}_{i2})
= u_i(1-\eta_{j1}) - c_i + E\pi_{i2}(1-\eta_{j1}) \\
V_i(D,H) &= \pi_{i1}(D,H) + E\pi_{i2}(\hat{w}_{i2})
= u_i(\eta_{i1}) + E\pi_{i2}(\eta_{i1}) \\
V_i(D,D) &= \pi_{i1}(D,D) + E\pi_{i2}(\omega_{i1})
= u_i(\omega_{i1}) + E\pi_{i2}(\omega_{i1})
\end{align*}

To derive $\eta_{M1}$, suppose player $M$ is the second mover.
If $M$ concedes to $F$, $M$'s value function is $V_M(D,H)$.
If $M$ does not concede to $F$, $M$'s value function is:
\[ V_M(\text{not } D,H) = 2[1-(\lambda_F+\theta)] - \phi_M = 2(\lambda_M-\theta) - \phi_M \]
Hence, $M$ concedes to $F$ if $V_M(D,H) \ge V_M(\text{not }D,H)$.
For $\sigma_{F1}=1$, $M$ concedes if $\pi_{M2}(\omega_{M2}=0) \ge V_M(\text{not }D,H)$. If this holds, $\eta_{M1}=0$.
If not, the optimal challenge by $F$, $\hat{\sigma}_{F1} \equiv 1-\eta_{M1}$, satisfies:
\[ u_M(\eta_{M1}) + E\pi_{M2}(\eta_{M1}) = V_M(\text{not } D,H) \]
%\[ \Pi_{M_2}(\hat{\lambda}_F(\bar{\eta}_M), \hat{\lambda}_M(\bar{\eta}_M), \bar{\eta}_M) \]
We can derive the similar conditions for $\eta_{F1}$ as well.
We can numerically solve for $(\eta_{F1}, \eta_{M1})$.

%We can numerically solve for $\bar{\eta}_M$ (and similarly for $\bar{\eta}_F$) using the following iterative procedure:
%\begin{description}
%    \item[Step 1:] Given $\bar{\eta}_M$, solve for $\hat{\lambda}_F(1-\bar{\eta}_M)$ and $\hat{\lambda}_M(\bar{\eta}_M)$.
%    \item[Step 2:] Substitute $\hat{\lambda}_F(\bar{\eta}_M)$ and $\hat{\lambda}_M(\bar{\eta}_M)$ into the LHS of the equation above.
%    \item[Step 3:] Stop if $|LHS - RHS| < \epsilon$. Otherwise, return to Step 1.
%\end{description}
Next, given the obtained optimal challenges $(\eta_{F1}, \eta_{M1})$, we obtain the equilibrium cutoff strategies in $t=1$.
Suppose player $i$ thinks $j$ will choose $H$ in $t=1$ with probability $p_j$.
The expected values that $i$ gains from choosing $H$ and $D$, and their difference are:
\begin{align*}
E[V_i(H)] &= -c_i + p_j[2\lambda_i - \phi_i] + (1-p_M)[u_i(1-\eta_{j1}) + E\pi_{i2}(1-\eta_{j1})] \\
E[V_i(D)] &= p_j[u_i(\eta_{i1}) + \pi_{i2}(\eta_{i1})] + (1-p_j)[u_i(\omega_{i1})+E\pi_{i2}(\omega_{i1})] \\
E[V_i(H)] - E[V_i(D)] &= -c_i + p_j[2\lambda_i - \phi_i - u_i(\eta_{i1}) - E\pi_{i2}(\eta_{i1})] \\
&+ (1-p_j)[u_i(1-\eta_{j1}) + \pi_{i2}(1-\eta_{j1}) - u_i(\omega_{i1}) - E\pi_{i2}(\omega_{i1})]
\end{align*}
Hence, the best response function for player $i$ in $t=1$ is given by:
{\small
\[ x_{i1}=\tilde{\Gamma}_i(x_{j1}) \equiv F(x_{j1})[2\lambda_i - \phi_i - u_i(\eta_{i1}) - E\pi_{i2}(\eta_{i1})] + (1-F(x_{j1}))[u_i(1-\eta_{j1}) + \pi_{i2}(1-\eta_{j1}) - u_i(\omega_{i1}) - E\pi_{i2}(\omega_{i1})] \]}
In equilibrium, whose existence we numerically verify under reasonable ranges of parameter values, we have $\hat{x}_{i1} = \tilde{\Gamma}_i(\hat{x}_{j1})$ for $i,i\in\{F,M\}$.
Solving this, we get:
{\footnotesize
\[ \hat{x}_{i1} = \frac{u_i(1-\eta_{j1}) + E\pi_{i2}(1-\eta_{j1}) - u_i(\omega_{i1}) - E\pi_{i2}(\omega_{i1}) + \tilde{\Omega}_{i}(\eta_{i1})[u_j(1-\eta_{i1}) + E\pi_{j2}(1-\eta_{i1}) - u_j(\omega_{j1}) - E\pi_{j2}(\omega_{j1})]}{1 - \tilde{\Omega}_{i}(\eta_{i1})\tilde{\Omega}_{j}(\eta_{j1})} \]
}
where
\[ \tilde{\Omega}_{i}(\eta_{i1}) \equiv 2\lambda_i -\phi_i - u_i(\eta_{i1}) - E\pi_{i2}(\eta_{i1}) - u_i(1-\eta_{j1}) - E\pi_{i2}(1-\eta_{j1}) + u_i(\omega_{i1}) + E\pi_{i2}(\omega_{i1})  \]
\setcounter{figure}{0}
\setcounter{table}{0}
\setcounter{equation}{0}
\renewcommand{\thefigure}{\Alph{section}.\arabic{figure}}
\renewcommand{\thetable}{\Alph{section}.\arabic{table}}
\renewcommand{\theequation}{\Alph{section}.\arabic{equation}}
\section{Appendix for the Organizational Heterogeneity}\label{app_heteg_groups}
\begin{figure}[htbp]
\begin{center}
\includegraphics[width=7.5cm]{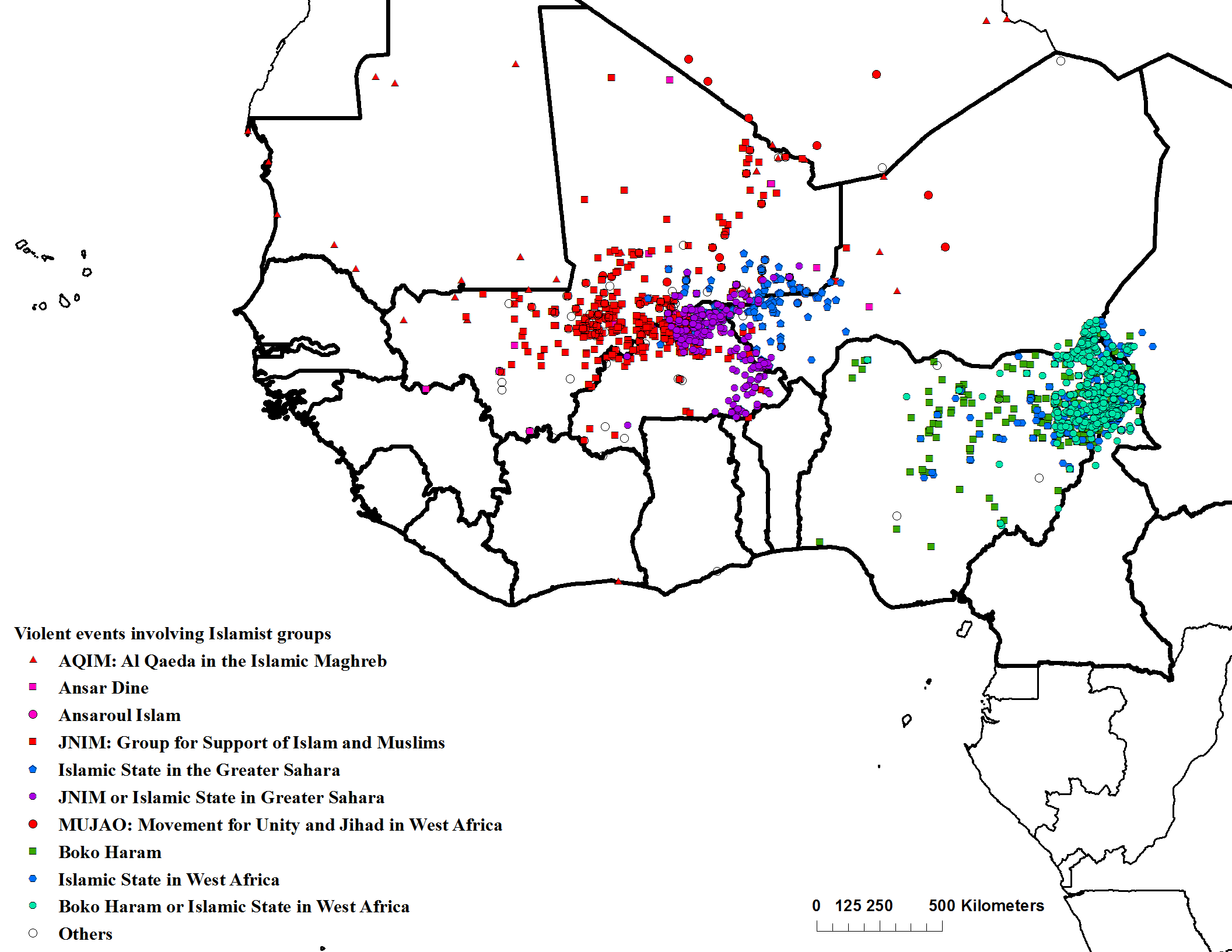}
\includegraphics[width=7.5cm]{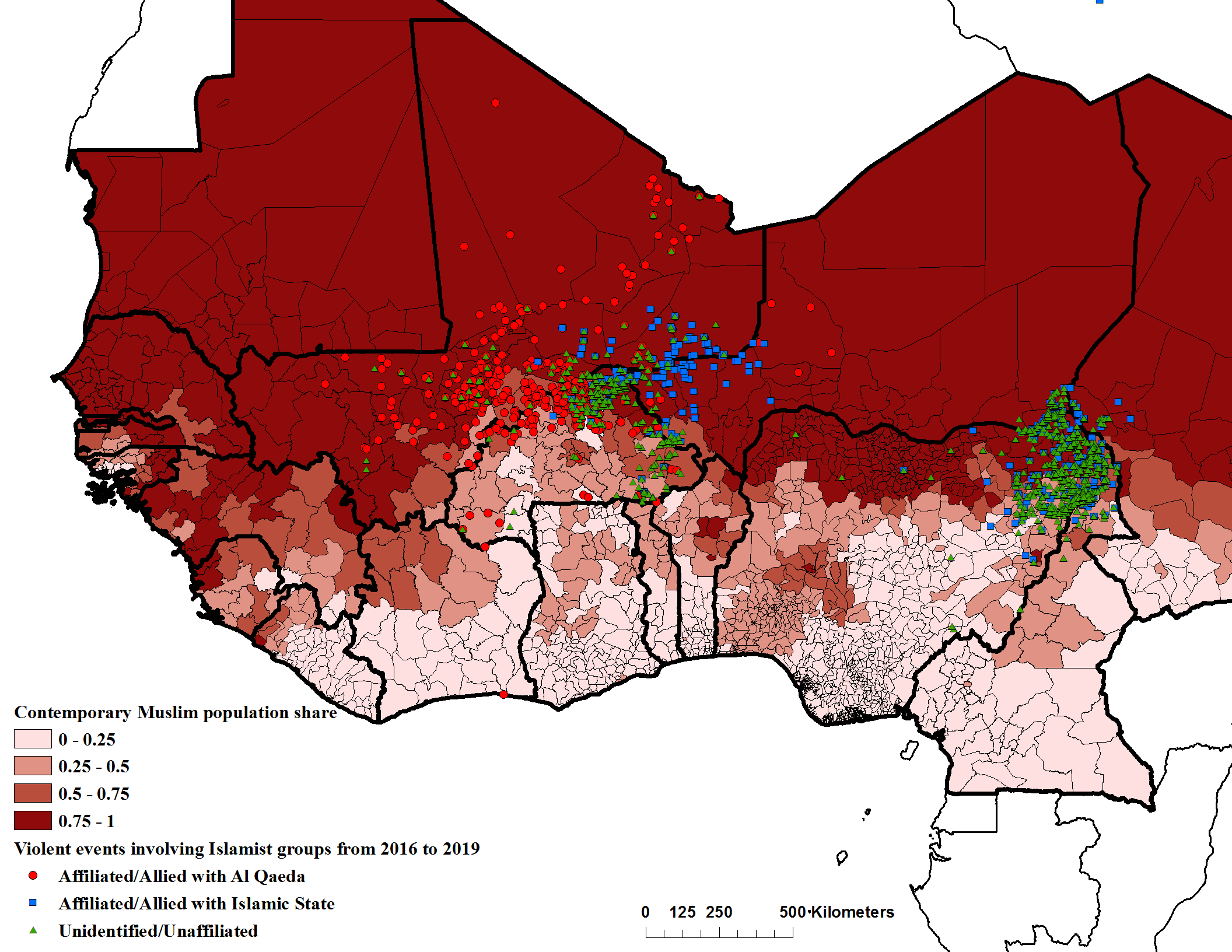}
\caption{Contemporary Muslim Population and Jihad in West Africa}
\label{app_fig_groups_AQ_IS_WA}
{\parbox[t]{\textwidth}{
{\scriptsize\begin{singlespace}
\textit{Notes}:
The left map shows violence and conflict events by all identified jihadist/Islamist groups in West Africa from 2001 to 2019.
The right map shows contemporary Muslim population and violence and conflict events by groups affiliated with Al Qaeda and the Islamic State from 2016 to 2019 (because the split occurred in 2015).
Sources: ACLED and the World Religion Database.\end{singlespace}}}}
\end{center}
\end{figure}

Recall that the left map of Figure \ref{app_fig_groups_AQ_IS_WA} shows violence events by jihadist groups in West Africa.
Table \ref{tab_jihadist_groups} lists groups affiliated with Al Qaeda and the Islamic State, the two largest factions under global competition.
Drawing directly from \href{https://cisac.fsi.stanford.edu/mappingmilitants}{Mapping Militants Project (MMP)}, we list stated ideologies and goals of major jihadist organizations below.

{\bf Al Qaeda.}
``Al Qaeda aims to rid the Muslim world of Western influence, to destroy Israel, and to create an Islamic caliphate stretching from Spain to Indonesia that imposes strict Sunni interpretation of Shariah law.''

{\bf The Islamic State.}
``The Islamic State’s ideology is rooted in Salafism---a fundamentalist movement within Sunni Islam---and Jihadism---a modern interpretation of the Islamic concept of struggle, often used in the context of defensive warfare...Salafis believe the most pure, virtuous form of Islam was practiced by the early generation of Muslims (known as Salaf) who lived around the lifetime of the prophet Muhammed...Since its inception, the Islamic State has sought to establish an Islamic caliphate based on its Salafi philosophy and fundamentalist interpretation of Shariah law.''

{\bf AQIM: Al Qaeda in the Islamic Maghreb.}
``The group’s main focus was the overthrow of the Algerian government and establishment of an Islamic caliphate in the Maghreb that would enforce Shariah law...expanded this goal in the early 2000s to include the overthrow of the governments of Mauritania, Morocco, Tunisia, and Mali, and the reclamation of lost Islamic lands in southern Spain.''

{\bf Ansar Dine.}
``Ansar Dine was a Salafi-jihadist group that aimed to establish Shariah law across Mali and targeted western civilians, especially peacekeepers in Mali.
Ansar Dine’s ideology closely mirrored that of AQIM, which came to view Ansar Dine as its southern arm in Mali. Unlike the MNLA, Ansar Dine did not seek independence for northern Mali but rather a country unified under Islam.''

{\bf Ansaroul Islam.}
``Ansaroul Islam’s main goal is allegedly to reconquer and rebuild Djeelgodji, an ancient Fulani empire that disappeared after French colonization in the late 19th century...Ansaroul Islam interacts closely with AQ front groups and affiliates in North Africa; Ansaroul Islam activity has purportedly created a front allowing AQ to achieve its primary aim—inspiring Muslims globally to attack enemies of Islam—in Burkina Faso.''

{\bf JNIM: Group for Support of Islam and Muslims.}
``The group’s goals and ideological basis are closely aligned with those of AQIM, and it seeks to build up a Salafi-Islamist state while restoring the caliphate.
The merger of various AQ-affiliates into the JNIM was consistent with AQ’s new operational focus on “unity” as a means to fully and effectively implement Shariah law in areas where the jihadists previously had not possessed complete control.''

{\bf MUJAO: Movement for Unity and Jihad in West Africa.}
``MUJAO’s stated goal was to engage in and encourage the spread of jihad in West Africa, as well as establish Shariah law in the region...MUJAO’s ideology and goals closely mirrored those of AQ and AQIM, the group it broke off from.''

{\bf Al Mulathamun Battalion.}
``Despite its split from AQIM, the AMB claimed to remain loyal to the ideology and command of Al Qaeda Central.
The militant group aimed to spread jihad through all of the Sahara and impose Shariah law in North Africa.''

{\bf Islamic State in the Greater Sahara.}
``The ISGS draws much of its strategic direction and ideological goals from the IS. As an affiliate of the Islamic State, the ISGS has pledged loyalty to the IS’s goal of restoring the Islamic caliphate.''

{\bf Boko Haram.}
``Boko Haram, which translates roughly to “Western education is forbidden,” is a Sunni Islamist militant organization that opposes Western education and influence in Nigeria.
Its founder Mohammad Yusuf...originally followed and preached the Izala doctrine, which advocates the establishment of a Muslim society that follows the lessons of its pious ancestors.
After his initial radicalization in 2002, Yusuf’s ideology evolved and radicalized into a philosophy that rejected all Western and secular aspects of Nigerian society.
Boko Haram originally advocated a doctrine of withdrawal from society but did not aim to overthrow the Nigerian government.
Yusuf’s death and increased conflict with the Nigerian government in 2009 sparked the political opposition and violent campaign that Boko Haram became known for.
Under the leadership of Abubakar Shekau and Abu Musab al-Barnawi, the group sought to establish an Islamic caliphate to replace the Nigerian government.''\footnote{Note that the relationship between Boko Haram and Al Qaeda is complicated and time-variant.
See \citetApp{Cummings2017} and \citetApp{Zenn2020} for details.
Note also that the Islamic State in West Africa was established in 2015 after splitting from Boko Haram (\citealtApp{Bohm2020}).}

\begin{table}[htbp]
\caption{Jihadist Groups in West Africa Affiliated with Al Qaeda and the Islamic State}
\label{tab_jihadist_groups}
\begin{center}
\begin{adjustbox}{max width=\textwidth,max totalheight=\textheight} 
\begin{tabular}{cc}
\hline\hline
{\bf Al Qaeda-affiliated groups}              & {\bf IS-affiliated groups}                \\\hline
AQIM: Al Qaeda in the Islamic Maghreb                & Islamic State in West Africa        \\
Ansar Dine                                           & Islamic State in the Greater Sahara \\
Ansaroul Islam                                       &                                     \\
JNIM: Group for Support of Islam and Muslims       &                                     \\
MUJAO: Movement for Unity and Jihad in West Africa &                                     \\
Katiba Macina                                        &                                     \\
Al Mourabitoune Battalion                            &                                     \\
GMA: Mourabitounes Group of Azawad                   &                                     \\
Ansaru                                               &                                     \\
Katiba Salaheddine                                   &                                     \\
MIA: Islamic Movement of Azawad                      &
\\
\hline
\end{tabular}
\end{adjustbox}
\end{center}
\end{table}

\input{tables/tab_reg_jihad_ma_panel2.tex}

%%%%%%%%%%%%%%%%%%%%%%%%%%%%%%%%%%%%%%%%%%%%%%%%%%%%%%%%%%%%%%%%%%%%%%%%%%%%%%%%%%%%%%%%%%%%%%
%\clearpage
\setcounter{figure}{0}
\setcounter{table}{0}
\setcounter{equation}{0}
\renewcommand{\thefigure}{\Alph{section}.\arabic{figure}}
\renewcommand{\thetable}{\Alph{section}.\arabic{table}}
\renewcommand{\theequation}{\Alph{section}.\arabic{equation}}
\section{Other Robustness Checks}\label{app_results}
%\subsection{Robustness Checks}
We provide various robustness checks using alternative data sources and inferences.

\textbf{Alternative measures of the ancient water access.}
%\MK{Show correlations between different IVs?}
We further assess the robustness of our results using alternative IV measures, as illustrated in Figure \ref{fig_plus_IVs_add}.
In addition to the size-weighted measure of ancient water access defined in Section \ref{sec_emp_IV_strategy}, we consider four alternative instruments based on the presence of lost ancient lakes, all of which exclude areas covered by contemporary lakes.
Specifically, we use: (i) the distance to the nearest lost ancient lake, without excluding contemporary rivers; (ii) a dummy equal to one if the area of lost ancient lakes within a grid cell is positive, again without excluding contemporary rivers; (iii) the distance to the nearest lost ancient lake after additionally excluding areas within a 1 km buffer of contemporary river lines; and (iv) a dummy equal to one if the area of lost ancient lakes within a grid cell is positive under the same river-exclusion criterion.
Across all alternative IV specifications, the estimated effects are very similar to the main results in terms of onset, intensity, and distance outcomes, both in magnitude and statistical significance.

\textbf{Alternative measures of declined cities.}
Figure \ref{fig_sahara_pop_less_005m} confirms the robustness of the main results under a stricter definition of the ``decline'', restricting to historical core cities with present-day populations below 50,000.\footnote{
The results are also robust to a range of alternative population thresholds between 50,000 and 100,000 and to alternative definitions of landlocked areas using different distance thresholds (benchmarked at 1000 km), with results available upon request.
}
We further apply the same restriction within Muslim-dominated areas, thereby excluding not only the ``Chad Lake effect'' in Nigeria but also the ``Timbuktu effect'' in Mali.
The results remain qualitatively unchanged.
Note that the estimated effects are not simply driven by contemporary underdevelopment by controlling for contemporary population levels.
Additionally, while Table \ref{tab_ols_city_today_water_rev} shows that proximity to ancient lakes has a small and statistically insignificant effect on contemporary city formation, Figure \ref{fig_plus_controls} also demonstrates that the results are robust when controlling for nightlight luminosity.
This additional test further strengthens our argument, ruling out contemporaneous urbanization as a confounding factor.

%\blue{We examine if the estimated effects are driven by currently-populated locations by focusing on the declined historical cities whose present-day populations are below 50,000.
%Although we confirmed proximity to ancient lakes has a small and statistically insignificant effect on contemporary city formation in Table \ref{tab_ols_city_today_water} and, Figure \ref{fig_plus_controls} demonstrates that the results are robust when controlling for nightlight luminosity, this additional test further strengthens our argument, ruling out contemporaneous urbanization as a confounding factor.
%Figure \ref{fig_sahara_pop_less_005m} confirms the robustness of the main results under this stricter definition of the declined historical cities.
%We further apply the same restriction within the sphere of Muslim-dominated areas (Mauritania; Mali; Niger; Chad), thereby excluding not only the ``Chad Lake effect'' in Nigeria but also the ``Timbuktu effect'' in Mali. The results remain qualitatively unchanged.}

\textbf{Uppsala Conflict Data Program Georeferenced Event Dataset (UCDP-GED).}
We use the UCDP-GED, an alternative conflict event dataset, to check robustness of the main results where we used the ACLED.
There are two key differences between these two datasets.
First, the UCDP-GED contains conflict and violent events that caused at least 1 fatality with the pair of actors (including the one-sided violence, in which case ``civilians'' is another side of actors) involved in the conflict that caused at least 25 fatalities in at least one calendar year.
Given that the ACLED contains events regardless of the number of fatalities, we check the robustness of our results with relatively severe events.
%\red{Second, the UCDP-GED contains conflict events in the entire world from 1989-2020 (though we pick events from 2001-2019). That is the reason why we used this dataset in our previous discussion of the global scale.[ST: Move these to the next subsection. I'll do it.]}
%Table \ref{tab_iv_jihad_ucdp_pop} confirms the robustness of the main results.
Table \ref{tab_jihad_ucdp_pop_rev} confirms the robustness of the main results.
Surprisingly, restricting to the relatively severe events by construction of this data, we also find statistically significant effects of the historical trade points on contemporary jihad (1\% for the distance and 5\% for the other two measures) during 2001-2009, unlike the main results from the ACLED.
Note also that the coefficient size for the distance is similar between 2001-2009 and 2010-2019 and that the coefficient sizes for the onset and intensity in 2001-2009 are smaller than those in 2010-2019.

\textbf{Adjusting for spatial auto-correlations.}
To address spatial autocorrelation, we additionally control for latitude and longitude and adjust the standard errors  (\citealtApp{CK2025}).
Figure \ref{fig_plus_controls} shows that our estimates are robust even when controlling for latitude and longitude in both linear and quadratic forms.
Moreover, Table \ref{tab_spatial_se_manual} reports standard errors allowing for spatial correlation with higher distance cutoffs (200km, 300km, 400km, 500km, and 1000km) for the main IV estimation results.
The size of standard error is non-monotonic in the size of distance cutoff.
The main results in terms of statistical significance are robust to different cutoffs.
At the largest standard errors, the coefficients from the three main outcomes (distance, onset, and intensity) are statistically significant at 5\% levels.

%%%%%%%%%%%%%%%%%%%%%%%%%%%%%%%%%%%%%%%%%%%%%%%%%%%%%%%%%%%%%%%%%%%%%%%%%%%%%%%%%%%%%%%%%%%%%%
%\clearpage
\setcounter{figure}{0}
\setcounter{table}{0}
\setcounter{equation}{0}
\renewcommand{\thefigure}{\Alph{section}.\arabic{figure}}
\renewcommand{\thetable}{\Alph{section}.\arabic{table}}
\renewcommand{\theequation}{\Alph{section}.\arabic{equation}}
\section{Appendix for Qualitative Evidence and Case Studies}\label{app_qualitative}

\subsection{The Religious Practice of \textit{Taqiyya} in Jihadist Movements}\label{app_qualitative_taqiyya}

This section describes how taqiyya has been practiced both in the colonial period and in the strategies of present-day jihadist groups.

%\textbf{Historical jihad.}
\subsubsection*{Historical jihad}
Islamic radicalisation in Northern Nigeria can be traced back to the onset of colonial rule (\citealtApp{PerouseDeMontclos2014}).
Under British domination, Muslim actors engaged in strategic accommodation through taqiyya---religiously sanctioned dissimulation.
After the fall of the Sokoto Caliphate in 1903, many Muslim elites outwardly accepted colonial rule while covertly preserving Islamic authority and identity.
As \citetApp{Last2008} notes, \quotes{full closure was not possible, but a degree of closure was feasible, whether through dissimulation or social distance.}
Leaders such as the Vizier of Sokoto chose to remain under colonial rule and participated in reconfiguring a new form of \textit{jama'a}---a collective Islamic life---under Christian overlordship, thus keeping alive the idea of \textit{Dar al-Islam} as a mental and spiritual space.

This practice was not merely individual but systematic.
\citetApp{PerouseDeMontclos2014} emphasizes that during much of the early colonial period, \quotes{taqiyya ('dissimulation') became a main pillar of resistance against colonial rule.} \citetApp{Naniya1993}, in her study of the Kano Emirate, observes that the `ulama', while publicly complying with the new order, ``resorted to clandestine activities in opposition to the new society being created by the British.''
In this sense, taqiyya functioned as a long-term method of ideological survival and quiet resistance beneath the surface of colonial governance.
\citetApp{Hiskett1994} explains, ``Taqiyya enables the Muslim outwardly to accept a situation he is powerless to change, while inwardly waiting for the tide to turn. It condones dissimulation. It allows Muslims to cooperate with an infidel authority when there is no alternative, while reserving the moral right to restore Islam to its proper position of dominance when the time is ripe'' (p.115).

Beyond Northern Nigeria, similar dynamics appeared in the Western Sudan.
In Timbuktu, local marabouts and \textit{Alfa} mediated between Islamic authority and French colonial power.
As \citetApp{Marty1920} remarked, with Timbuktu's submission ``it can be said that they are now almost on our side. It should be remembered that holy war was not an absolute obligation. Many documents prove this, and the Alfas of Timbuktu are well acquainted with those documents.''
This observation highlights the ways in which local Muslim leaders employed dissimulation and pragmatic accommodation to preserve religious authority while outwardly cooperating with colonial administrators.

%\textbf{Contemporary jihad.}
\subsubsection*{Contemporary jihad}
The practice of taqiyya is also well-documented in the context of contemporary jihadist movements, including both Al Qaeda (\citealtApp{SB2004, Campbell2005}) and the Islamic State (\citealtApp{Bunzel2019}).
\citetApp{Campbell2005} describes jihadists practicing taqiyya as ``super sleepers'' committed to becoming embedded in target societies, and ``permitted'' to drink alcohol, live together, pray together, eat during Ramadan, dress in Western style and socialise with women, to avoid suspicion or detection.
The use of taqiyya has also been repeatedly highlighted in European court cases against French jihadists (\citealtApp{HT2017, Hecker2018, Hecker2021, Louarn2013}).
For instance, \citetApp{Hecker2018} documents \quotes{the prosecutor noted that one of the defendants claimed to want to return to a normal life, but in reality he continued to visit jihadist websites and attempt to recruit sympathizers.}
Similar references to taqiyya also appear in Dutch Salafist court cases (\citealtApp{Koning2020}).

This tactic has also shaped the operations of Boko Haram, active not only in Nigeria but also in neighboring states such as Niger and Chad.
Following intensified military pressure after 2009, Boko Haram leaders adopted a strategy of dispersion and withdrawal.
\citetApp{Antimbom2016} interprets this as a conscious invocation of the Sultan of Sokoto's tactic in 1903, noting: \quotes{Boko Haram's leaders settled to a strategy of scattering (taqiyya), like the Sultan of Sokoto in 1903, who called his followers to disband to avoid being defeated by the British.}
In this way, taqiyya functions not merely as a theological doctrine but as a transhistorical strategic logic, linking colonial-era Islamic non-resistance to contemporary jihadist insurgency.

\subsection{The Global Perspective}\label{app_qualitative_global}
To provide an illustrative global perspective, we draw on three data sources.
First, we use global-scale information about historical overland trade routes from \citetApp{MNP2018}.
Second, we use worldwide population estimates over the centuries from the History Database of the Global Environment (HYDE).
Third, contemporary jihadist events (2001–2019) come from the Uppsala Conflict Data Program Georeferenced Event Dataset (UCDP GED), which contains conflict events for the entire world.
Figure H.1 maps this information, overlaying historical trade routes and past population centers with the Muslim share in 2005 from the World Religion Database at the ethnic homeland level (\citealtApp{Ethnologue2005}) and contemporary jihadist violent events.
We observe concentrations of jihadist events in Syria, Iraq, Afghanistan, and Pakistan. While these countries are today often regarded as peripheral, they were historically core hubs of global overland trade networks. The following case studies examine each in turn, drawing on qualitative historical evidence to assess whether the mechanisms identified in West Africa—colonial disruption of prosperous Islamic cores and the subsequent persistence of jihadist ideology—find parallels in these global cases.

\begin{figure}[t]
\begin{center}
\includegraphics[width=7.5cm]{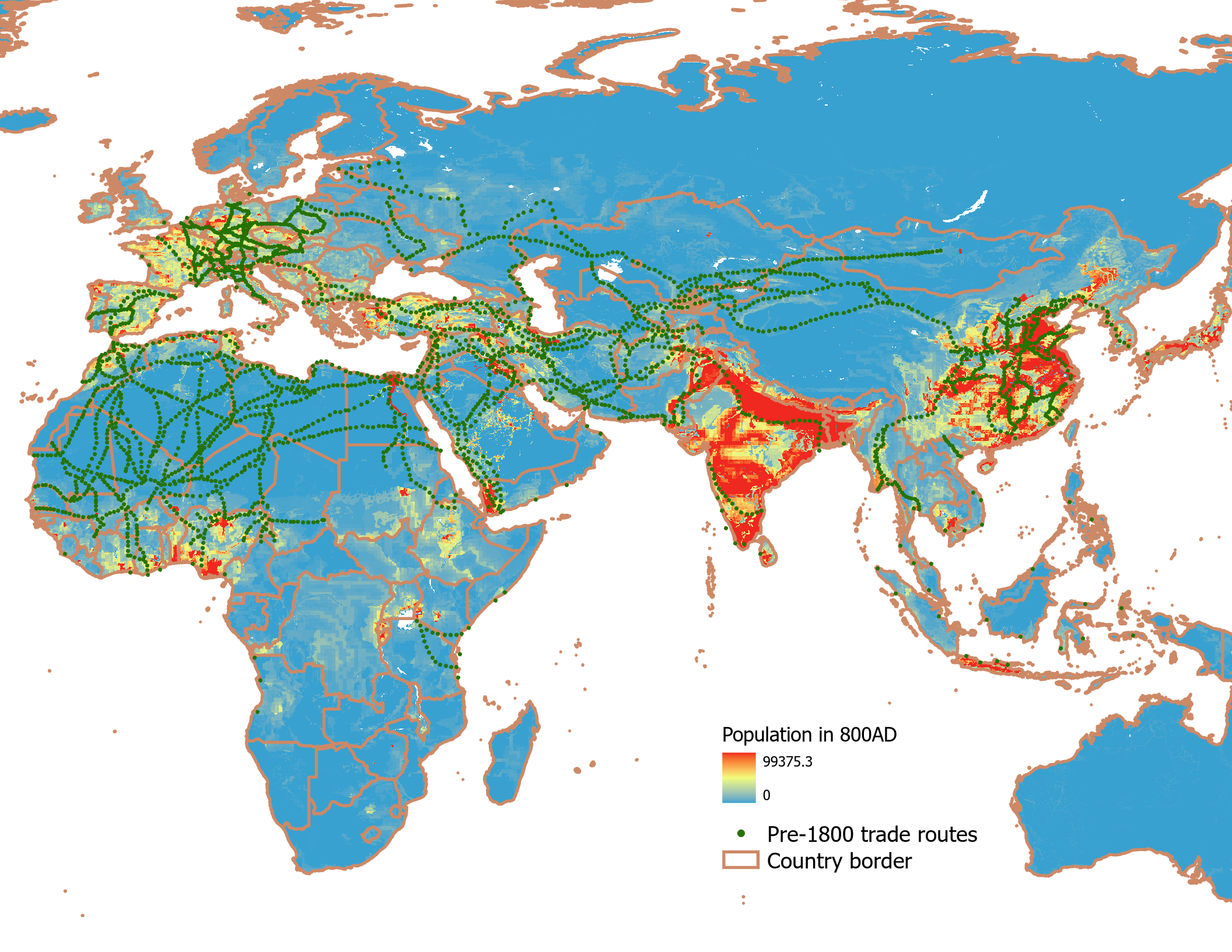}
\includegraphics[width=7.5cm]{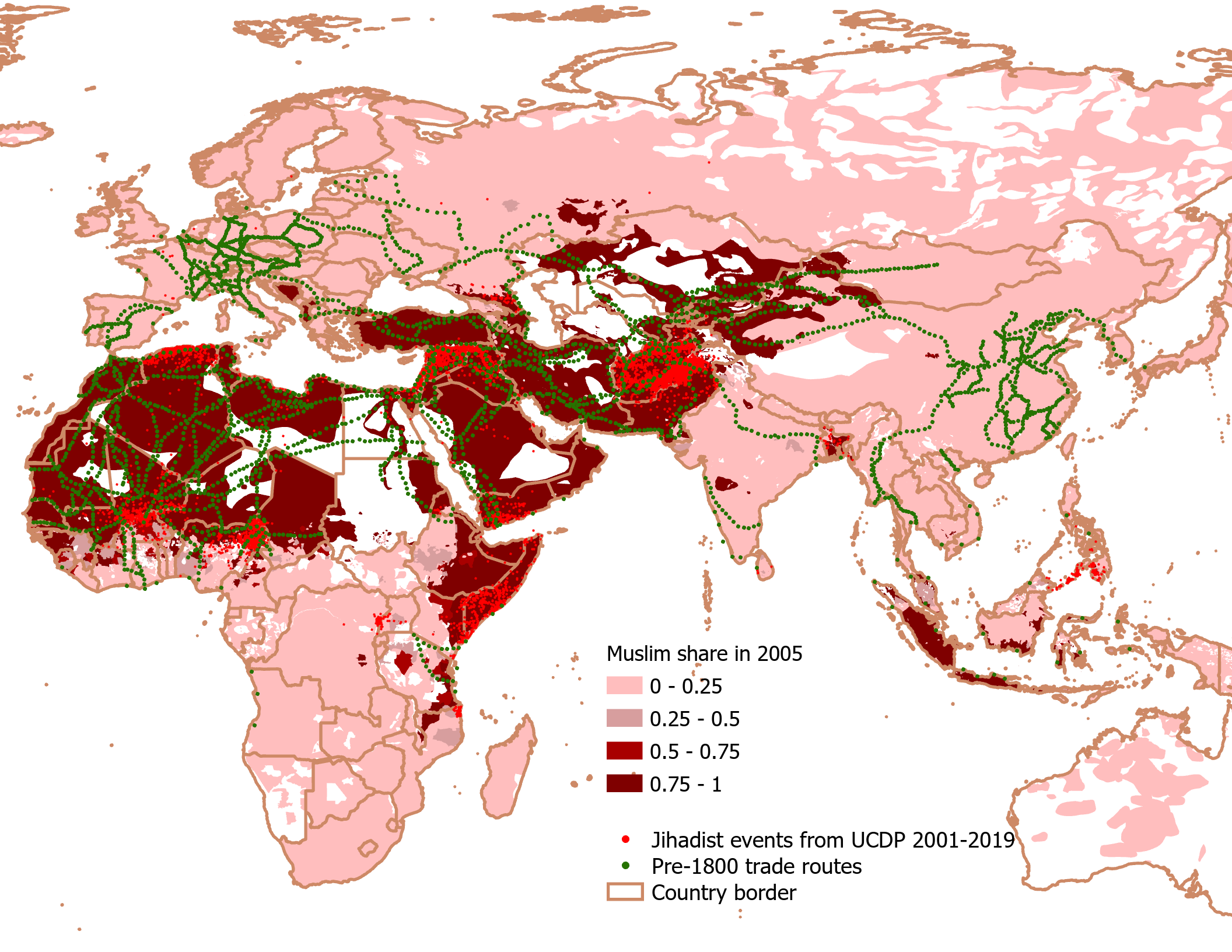}\\
% 元々14.5cmだったのでAppに持ってくる場合そうしよう
\caption{Historical Populations, Overland Trade, and Contemporary Islam and Jihad}
\label{map_global}
{\parbox[t]{\textwidth}{
{\scriptsize\begin{singlespace}
\textit{Notes}:
Both panels show the overland trade routes up to 1800AD: the left panel adds population in 800AD (\href{https://landuse.sites.uu.nl/datasets/}{HYDE 3.1}), while the right panel overlays the Muslim share in 2005 (\href{https://worldreligiondatabase.org/}{World Religion Database}) at ethnic homeland (\citealt{Ethnologue2005}) together with contemporary jihadist violent events  (UCDP GED).
%See section \ref{sec_data} for detailed data sources.
\end{singlespace}}}}
\end{center}
\end{figure}

%\textbf{The Levant: French Mandate and the Syrian Reversal.} %\MK{Benchmark: Alternative drafts 2 \& 4}
\subsubsection*{The Levant: French mandate and the Syrian reversal}
Cities such as Damascus, Aleppo, Raqqa, and Deir ez-Zor constituted the historical centers of Sunni political, religious, and economic power---the country’s ``past core.''
During the Great Syrian Revolt (1925–27), these core areas, particularly Damascus, experienced intense repression by the French mandatory authorities (\citealtApp{Provence2005}).
The bombardment of Damascus and the prolonged counterinsurgency campaign not only caused extensive physical destruction but also produced profound psychological effects.
As Khoury notes, \quotes{the physical and psychological exhaustion produced by nearly two years of full-scale rebellion led to a general demoralization of the Syrian masses} (\citealtApp{Khoury2014}).
This episode marked a decisive rupture in the political centrality of Syria's traditional urban core.

At the same time, the French Mandate fundamentally restructured the relationship between religion and political authority. During the interwar period, the influence of Sunni religious leaders was progressively eroded as Islamic institutions were brought under closer administrative control of the state (\citealtApp{Khoury2014}).
Through the regulation of waqf property, religious courts, and clerical appointments, Islam was not abolished but increasingly incorporated into a framework of state management.
This transformation laid the foundations of what later scholars describe as a system of state-controlled Islam, a legacy that survived independence and was further consolidated under subsequent regimes.

In parallel, French military and administrative policies empowered peripheral social groups, most notably the Alawites, who were disproportionately recruited into the Troupes Spéciales du Levant—the colonial force that later evolved into the Syrian army (\citealtApp{Khoury2014}; \citealtApp{Dam2011}).
This policy produced a long-term inversion of Syria’s socio-political hierarchy: the historically dominant Sunni urban elite increasingly found themselves governed by a military establishment rooted in formerly marginal regions (\citealtApp{Perlmutter1969}).

Under the Alawi-dominated Ba'thist regime, this inversion translated into a persistent pattern of Sunni political exclusion and grievance.
As \citetApp{Dam2011} emphasizes, Sunni opposition did not disappear but remained a recurrent and latent feature of Syrian politics, periodically surfacing under conditions of crisis.
Over time, the persistent and deep-seated grievances help to fuel the resurgence of Sunni jihadist movements, which framed their violence not as a reaction to secularism but as an assault on an illegitimate, state-managed Islam and a sectarian political order (\citealtApp{Naji2006, Lister2015}).

While there is no direct empirical evidence, it is notable that the organization's principal strongholds emerged not in the heavily repressed urban cores of the Mandate period, but in eastern and peripheral regions such as Raqqa and Deir ez-Zor, with Raqqa serving as the Islamic State’s de facto capital and Deir ez-Zor forming a key zone of territorial expansion, areas that historically lay outside the main theaters of colonial counterinsurgency (\citealtApp{Provence2005, Lister2015}).
This spatial contrast between the heavily bombed urban core---most notably Damascus---and later jihadist strongholds in peripheral regions parallels the pattern observed in West Africa, where regions that did not experience decisive military defeat and political exhaustion later became the epicenters of contemporary Islamist insurgency.
Together, this contrast suggests that the long-term weakening of the urban Sunni core---through wartime devastation, political demoralization, and the erosion of religious authority---combined with uneven state penetration in past-core-and-present-peripheral regions, may have shaped the terrain on which jihadist mobilization later became possible.

%\blue{While there is no direct empirical evidence, it is notable that the organization's principal strongholds emerged not in the heavily repressed urban cores of the Mandate period, but in eastern and peripheral regions that historically lay outside the main theaters of colonial counterinsurgency (\citealtApp{Provence2005, Lister2015}). This pattern suggests that the long-term weakening of the urban Sunni core---through wartime devastation, political demoralization, and the erosion of religious authority---combined with uneven state penetration in peripheral regions to shape the terrain on which jihadist mobilization later became possible.}

%\textbf{Mesopotamia: British Legacy in Iraq.}
\subsubsection*{Mesopotamia: British legacy in Iraq}
Neighboring Iraq had similar experience over the centuries.
In the 8th century, the rise of the Abbasid dynasty transformed Baghdad into a premier center of trade and Islamic culture.
\citetApp{BBV2013} point out that ``In 800, only four decades after its founding, Baghdad had become a metropolis of more than 300,000 inhabitants... [and] the center of economic and political power in the Islam world.''
Later in 1920s, the British Mandate constructed a state that forced together diverse ethnic and sectarian groups—Sunni, Shia, and Kurd—under a centralized administration in Baghdad.
The British relied heavily on the Sunni minority to staff the administration and military, institutionalizing a sectarian imbalance to govern the Shia majority (\citealtApp{Dodge2003}; \citealtApp{Fieldhouse2006}). This colonial reliance on a specific sect created a fragile state structure where political exclusion became a central feature of governance.

Following the eventual collapse of this order in the post-2003 era, the former Sunni ``core'', now marginalized and stripped of power (a ``present periphery'' in political terms), became the primary breeding ground for insurgency and jihadist organizations like Al-Qaeda in Iraq and later the Islamic State.
This dynamic was particularly visible in northern Iraq, notably in Mosul, a historically Sunni ``core'' that later became the most important urban base of the Islamic State.
The rise of the Islamic State in Mosul was facilitated by the participation of former Iraqi military and intelligence officers such as Haji Bakr, many of whom were based in the city and had been marginalized following the collapse of the Baathist state (\citealtApp{JSSBJRW2016, Haddad2020}).
The Iraqi case illustrates how colonial rule reshaped political hierarchies and how the later collapse of this order transformed former Sunni political cores into key sites of insurgency.

\if0
\STmemo{\citet{Dodge2003} memo \\
- British in the 1920s \\
- The U.S. occupation of Iraq in 2003 \\
- ``Britain's failed attempt, during the 1920s and 1930s, to build a liberal state in Iraq forms the historical backdrop against which the removal of Saddam Hussein in 2003 and its aftermath should be understood.'' \\
- Between 1914 and 1932, the British government created the modern state of Iraq.
\\
\ST{Should also mention about Mosul? Historical core, and recent base for ISIS; Same logic as Raqqa in Syria}
}
\fi

%\textbf{Afghanistan and Pakistan.} %\MK{Benchmark: Alternative draft 2 \& 4}
\subsubsection*{Afghanistan and Pakistan}
Afghanistan and Pakistan offer a parallel case of colonial distortion.
Historically, the Pashtun areas served as a gateway for trade, and conquest into India.
However, the British Empire, seeking to secure its Indian frontier against Russia during the ``Great Game'' established the Durand Line in 1893.
This artificial border bisected Pashtun tribal lands, transforming a historical center of connectivity into a marginalized borderland, later institutionalized as the Federally Administered Tribal Areas (FATA) (\citealtApp{Hopkins2008}).
Although Afghanistan gained formal independence after successive wars against the British army, British policy after 1842 focused on isolating Afghanistan from imperial economic and political circuits rather than incorporating it into the colonial order (\citealtApp{Hopkins2008, Dalrymple2013}).
The British ``Forward Policy'' further reinforced this isolation by leaving frontier regions intentionally underdeveloped and autonomously governed in order to function as a buffer zone. Periods from the late nineteenth century through the interwar and early Cold War years up to 1978 were largely characterized by political stability rather than continuous jihadist mobilization while Islamic revival movement occasionally occurred (\citealtApp{Hopkins2008}).

However, the Soviet invasion in 1979 triggered Afghan jihad, highlighting the importance of external shocks in reactivating latent ideological resources.
As documented in existing scholarship, this mobilization was embedded in transnational flows of fighters, finance, and weapons shaped by Cold War geopolitics rather than arising solely from domestic conditions (\citealtApp{Rashid2002, Abbas2014}).
This pattern aligns closely with our broader argument that large-scale jihadist mobilization emerges through the interaction between persistent ideological repertoires and major external shocks including invasion into land, foreign fighter inflows or arms transfers.

Pakistan followed a distinct yet complementary trajectory.
Large parts of present-day Pakistan were directly incorporated into British India and thus experienced prolonged colonial rule, missionary activity, and administrative restructuring.
After the failure of the 1857 rebellion, many Muslim religious leaders concluded that sustained rebellion against British rule was no longer viable and increasingly emphasized religious education, moral reform, and communal life rather than direct political confrontation (\citealtApp{Metcalf1982}).

During and after the Soviet–Afghan war, militant networks based not only in Pakistan's frontier zones but also in colonized interior regions supplied fighters, training, and ideological resources to jihad in Afghanistan.
As existing scholarship emphasizes, Afghan jihad relied heavily on recruitment, sanctuary, and mobilization on the Pakistani side of the border (\citealtApp{Kepel2002, Rashid2002}).
Taken together, these patterns suggest not that jihadist ideology persisted through continuous mobilization under colonial rule, but that religious institutions and ideological resources were not eradicated and could later be reactivated with external shocks.

\end{onehalfspace}

%%%%%%APPENDIX_REFERENCE%%%%%%%%%%%%%%%%%%%%%%%%%%%%%%%%%%%%%%%%%%%%%%%%%%%%%%%%%%%%%%%%%%%%%%%%%%%%%%%%%%%%%%%%%%%%%%
%\clearpage
\begin{small}
\begin{spacing}{0.95}
%\bibliographystyleApp{aer}
\bibliographystyleApp{ecta}
\bibliographyApp{Ref_IslamHistory}
%%\nocite{*}
\end{spacing}
\end{small}
\end{document}

%% file: tables/tab_first_stage_pop.tex
\begin{table}[htbp]\centering \caption{First Stage---Ancient Water Sources and Historical Landlocked Cities \label{tab_first_stage_pop}} \begin{adjustbox}{max width=\textwidth,max totalheight=\textheight}\begin{threeparttable}\def\sym#1{\ifmmode^{#1}\else\(^{#1}\)\fi}\begin{tabular}{l*{7}{c}}\hline\hline & \multicolumn{6}{c}{Log distance to}\\
                &\multicolumn{3}{c}{Inland trade point (< 100,000)}      &\multicolumn{3}{c}{Inland trade route}                  \\\cmidrule(lr){2-4}\cmidrule(lr){5-7}
                &\multicolumn{1}{c}{(1)}         &\multicolumn{1}{c}{(2)}         &\multicolumn{1}{c}{(3)}         &\multicolumn{1}{c}{(4)}         &\multicolumn{1}{c}{(5)}         &\multicolumn{1}{c}{(6)}         \\
\hline
Log (Distance to an ancient lake)&    0.108\sym{***}&    0.108\sym{***}&    0.128\sym{***}&   0.0898\sym{**} &   0.0879\sym{**} &    0.170\sym{***}\\
                & (0.0161)         & (0.0162)         & (0.0177)         & (0.0352)         & (0.0354)         & (0.0286)         \\
Log (Distance to a lake/river today)&  -0.0162         &  -0.0140         &  0.00787         &  -0.0304         &  -0.0291         &  0.00791         \\
                & (0.0222)         & (0.0222)         & (0.0190)         & (0.0412)         & (0.0414)         & (0.0294)         \\
\hline
F-stat          &    44.52         &    44.08         &    50.81         &     6.41         &     6.08         &    34.84         \\
Observations    &     2616         &     2616         &     2616         &     2616         &     2616         &     2616         \\
Colonizer FE & No & Yes & No & No & Yes & No \\ Country FE & No & No & Yes & No & No & Yes \\ Geographic Controls & Yes & Yes & Yes & Yes & Yes & Yes\\ \hline\end{tabular}\begin{tablenotes}[flushleft]\footnotesize \item {\it Notes}: All regressions are estimated using OLS. The unit of observation is a grid cell (about 55km × 55km). All Log(Distance) variables indicate the logarithm of one plus distance (km) to the nearest object. The dependent variables are the logarithm of one plus distance (km) to the nearest pre-colonial landlocked trade point whose contemporary population is less than 100,000 (columns 1-3) and the logarithm of one plus distance (km) to the nearest pre-colonial landlocked trade route up to 1800 (columns 4-6). Landlocked is defined as being over 1,000 km from the nearest coast point. We control for landlocked dummy, average malaria suitability, average caloric suitability in post 1500, average elevation, ruggedness, and logarithm of one plus population in 2010 in all the specifications. We report standard errors adjusting for spatial auto-correlation with distance cutoff at 100km in parentheses. \sym{*} \(p<0.1\), \sym{**} \(p<0.05\), \sym{***} \(p<0.01\). \end{tablenotes}\end{threeparttable}\end{adjustbox}\end{table} 

%% file: tables/tab_iv_jihad_pop.tex
\begin{table}[htbp]\centering         \def\sym#1{\ifmmode^{#1}\else\(^{#1}\)\fi}         \caption{IV Estimates of Persistent Effects on Jihad \label{tab_iv_jihad_pop}}         \begin{adjustbox}{max width=\textwidth,max totalheight=\textheight}         \begin{threeparttable}         \footnotesize         \begin{tabular}{l*{7}{c}}         \hline\hline         {\bf (A)} & \multicolumn{6}{c}{{\bf Onset}}\\
                &\multicolumn{1}{c}{(1)}&\multicolumn{1}{c}{(2)}&\multicolumn{1}{c}{(3)}&\multicolumn{1}{c}{(4)}&\multicolumn{1}{c}{(5)}&\multicolumn{1}{c}{(6)}\\
                &\multicolumn{1}{c}{All}&\multicolumn{1}{c}{2001-09}&\multicolumn{1}{c}{2010-19}&\multicolumn{1}{c}{All}&\multicolumn{1}{c}{2001-09}&\multicolumn{1}{c}{2010-19}\\
\hline
Log (Distance to a landlocked trade point ($<$ 100,000))&   -0.325\sym{***}& -0.00422         &   -0.327\sym{***}&                  &                  &                  \\
                & (0.0786)         & (0.0135)         & (0.0787)         &                  &                  &                  \\
Log (Distance to a landlocked trade route up to 1800)&                  &                  &                  &   -0.244\sym{***}& -0.00317         &   -0.246\sym{***}\\
                &                  &                  &                  & (0.0550)         &(0.00996)         & (0.0548)         \\
\hline
Observations    &     2616         &     2616         &     2616         &     2616         &     2616         &     2616         \\
Mean (Dep. Var.)&    0.133         &    0.011         &    0.129         &    0.133         &    0.011         &    0.129         \\
SD (Dep. Var.)  &    0.339         &    0.106         &    0.335         &    0.339         &    0.106         &    0.335         \\
\hline         {\bf (B)} & \multicolumn{6}{c}{{\bf Intensity}}\\
                &\multicolumn{1}{c}{(1)}&\multicolumn{1}{c}{(2)}&\multicolumn{1}{c}{(3)}&\multicolumn{1}{c}{(4)}&\multicolumn{1}{c}{(5)}&\multicolumn{1}{c}{(6)}\\
                &\multicolumn{1}{c}{All}&\multicolumn{1}{c}{2001-09}&\multicolumn{1}{c}{2010-19}&\multicolumn{1}{c}{All}&\multicolumn{1}{c}{2001-09}&\multicolumn{1}{c}{2010-19}\\
\hline
Log (Distance to a landlocked trade point ($<$ 100,000))&   -1.233\sym{***}&  -0.0236         &   -1.234\sym{***}&                  &                  &                  \\
                &  (0.357)         & (0.0218)         &  (0.357)         &                  &                  &                  \\
Log (Distance to a landlocked trade route up to 1800)&                  &                  &                  &   -0.926\sym{***}&  -0.0177         &   -0.926\sym{***}\\
                &                  &                  &                  &  (0.235)         & (0.0158)         &  (0.235)         \\
\hline
Observations    &     2616         &     2616         &     2616         &     2616         &     2616         &     2616         \\
Mean (Dep. Var.)&    0.249         &    0.011         &    0.244         &    0.249         &    0.011         &    0.244         \\
SD (Dep. Var.)  &    0.775         &    0.109         &    0.771         &    0.775         &    0.109         &    0.771         \\
\hline         {\bf (C)} & \multicolumn{6}{c}{{\bf Distance}}\\
                &\multicolumn{1}{c}{(1)}&\multicolumn{1}{c}{(2)}&\multicolumn{1}{c}{(3)}&\multicolumn{1}{c}{(4)}&\multicolumn{1}{c}{(5)}&\multicolumn{1}{c}{(6)}\\
                &\multicolumn{1}{c}{All}&\multicolumn{1}{c}{2001-09}&\multicolumn{1}{c}{2010-19}&\multicolumn{1}{c}{All}&\multicolumn{1}{c}{2001-09}&\multicolumn{1}{c}{2010-19}\\
\hline
Log (Distance to a landlocked trade point ($<$ 100,000))&    1.164\sym{***}&    0.310         &    1.306\sym{***}&                  &                  &                  \\
                &  (0.274)         &  (0.211)         &  (0.283)         &                  &                  &                  \\
Log (Distance to a landlocked trade route up to 1800)&                  &                  &                  &    0.874\sym{***}&    0.233\sym{*}  &    0.980\sym{***}\\
                &                  &                  &                  &  (0.167)         &  (0.139)         &  (0.173)         \\
\hline
Observations    &     2616         &     2616         &     2616         &     2616         &     2616         &     2616         \\
Mean (Dep. Var.)&    4.816         &    5.698         &    4.932         &    4.816         &    5.698         &    4.932         \\
SD (Dep. Var.)  &    1.101         &    0.743         &    1.157         &    1.101         &    0.743         &    1.157         \\
\hline     \\ Country FE & Yes & Yes & Yes & Yes & Yes & Yes \\ Geographic Controls & Yes & Yes & Yes & Yes & Yes & Yes         \\\hline         \end{tabular}         \begin{tablenotes}[flushleft]         \footnotesize         \item {\it Notes}: All regressions are estimated using IV with logarithm of one plus distance (km) to the nearest ancient lake as an instrument. The unit of observation is a grid cell (about 55km × 55km). The dependent variables are (A) dummy variables which take a value of 1 if jihad occurred during a period given in each column, otherwise take a value of 0, (B) logarithm of one plus the number of jihad events during a given period in each column, and (C) logarithm of one plus distance (km) to the nearest jihad during a period given in each column. All Log(Distance) variables indicate the logarithm of one plus distance (km) to the nearest object. Landlocked is defined as being over 1,000 km from the nearest coast point. The interest variables are the logarithm of one plus distance (km) to the nearest pre-colonial landlocked trade point whose contemporary population is less than 100,000 in columns (1)-(3), and the logarithm of one plus distance (km) to the nearest pre-colonial landlocked trade route up to 1800 in columns (4)-(6). We control for the logarithm of distance (km) to the nearest water sources today, landlocked dummy, average malaria suitability, average caloric suitability in post 1500, average elevation, and ruggedness in all the specifications. We report standard errors adjusting for spatial auto-correlation with distance cutoff at 100km in parentheses. \sym{*} \(p<0.1\), \sym{**} \(p<0.05\), \sym{***} \(p<0.01\).         \end{tablenotes}\end{threeparttable}\end{adjustbox}\end{table}         

%% file: tables/tab_gun_slave.tex
\begin{table}[htbp]\centering         \caption{Weapon Access in the Pre-Colonial Period \label{tab_gun_slave}}     \begin{adjustbox}{max width=\textwidth,max totalheight=\textheight}         \begin{threeparttable}         \def\sym#1{\ifmmode^{#1}\else\(^{#1}\)\fi}         \begin{tabular}{l*{7}{c}}         \hline\hline
                &\multicolumn{2}{c}{\textbf{\shortstack[c]{Gun access \\ in 1757-1806}}}&\multicolumn{2}{c}{\textbf{\shortstack[c]{Slave exports \\ in 1700s}}}&\multicolumn{2}{c}{\textbf{\shortstack[c]{Slave exports \\ in 1800s}}}\\\cmidrule(lr){2-3}\cmidrule(lr){4-5}\cmidrule(lr){6-7}
                &\multicolumn{1}{c}{(1)}&\multicolumn{1}{c}{(2)}&\multicolumn{1}{c}{(3)}&\multicolumn{1}{c}{(4)}&\multicolumn{1}{c}{(5)}&\multicolumn{1}{c}{(6)}\\
                &\multicolumn{1}{c}{OLS}&\multicolumn{1}{c}{IV}&\multicolumn{1}{c}{OLS}&\multicolumn{1}{c}{IV}&\multicolumn{1}{c}{OLS}&\multicolumn{1}{c}{IV}\\
\hline
Log (Distance to an ancient lake)&   0.0168\sym{***}&                  &    0.270\sym{***}&                  &    0.186\sym{***}&                  \\
                &(0.00621)         &                  & (0.0712)         &                  & (0.0714)         &                  \\
Log (Distance to a landlocked trade point ($<$ 100,000))&                  &    0.269\sym{**} &                  &    4.169\sym{***}&                  &    2.884\sym{**} \\
                &                  &  (0.121)         &                  &  (1.581)         &                  &  (1.415)         \\
Log (Distance to the coast)&   -0.241\sym{***}&   -0.155\sym{**} &   -0.481\sym{*}  &    0.858         &   -0.886\sym{***}&   0.0401         \\
                & (0.0383)         & (0.0660)         &  (0.288)         &  (0.659)         &  (0.234)         &  (0.571)         \\
\hline
R$^2$           &    0.740         &    0.642         &    0.302         &    0.087         &    0.322         &    0.176         \\
Adj-R$^2$       &    0.739         &    0.641         &    0.299         &    0.084         &    0.320         &    0.173         \\
Observations    &     2616         &     2616         &     2489         &     2489         &     2489         &     2489         \\
Mean (Dep. Var.)&    3.747         &    3.747         &    2.319         &    2.319         &    2.720         &    2.720         \\
SD (Dep. Var.)  &    0.465         &    0.465         &    4.407         &    4.407         &    3.827         &    3.827         \\
Colonizer
FE
&
Yes
&
Yes
&
Yes
&
Yes
&
Yes
&
Yes
\\
Geographic
Controls
&
Yes
&
Yes
&
Yes
&
Yes
&
Yes
&
Yes
\\\hline
\end{tabular}
\begin{tablenotes}[flushleft]
\footnotesize
\item
{\it
Notes}:
Columns (1), (3), and (5) report OLS estimates, while columns (2), (4), and (6) report IV estimates using the logarithm of one plus distance (km) to the nearest ancient lake as an instrument.
The
unit
of
observation
is
a
grid
cell
(about
55km
×
55km).
All
Log(Distance)
variables
indicate
the
logarithm
of
one
plus
distance
(km)
to
the
nearest
object.
Landlocked
is
defined
as
being over 1,000 km
from
the
nearest
coast
point.
The dependent variables are the logarithm of a quantity-weighted measure of gun access in 1757–1806 (columns 1–2), the logarithm of one plus the number of Atlantic slave trade exports in the 1700s (columns 3–4), and in the 1800s (columns 5–6).
We
control
for
the
logarithm
of
distance
(km)
to
the
nearest
water
sources
today,
landlocked
dummy,
average
malaria
suitability,
average
caloric
suitability
in
post
1500,
average
elevation,
and
ruggedness
in
all
the
specifications.
We
report
standard
errors
adjusting
for
spatial
auto-correlation
with
distance
cutoff
at
100km
in
parentheses.
\sym{*} \(p<0.1\), \sym{**} \(p<0.05\), \sym{***} \(p<0.01\).
\end{tablenotes}\end{threeparttable}\end{adjustbox}\end{table}

%% file: tables/tab_mechanism_ideology.tex
\begin{table}[htbp]\centering         \caption{IV Estimates of Persistent Effects on Religious Ideology of Muslims \label{tab_mechanism_ideology}}         \begin{adjustbox}{max width=\textwidth,max totalheight=\textheight}\begin{threeparttable}\def\sym#1{\ifmmode^{#1}\else\(^{#1}\)\fi}\begin{tabular}{l*{7}{c}}\hline\hline
                &\multicolumn{2}{c}{\shortstack{{\bf Neighbors from} \\ {\bf different religion:} \\ 1 (strongly like) - \\ 5 (strongly dislike)}}&\multicolumn{2}{c}{\shortstack{{\bf Governed by} \\ {\bf religious law:} \\ 1 (strongly disagree) - \\5 (stronlgy agree)}}&\multicolumn{2}{c}{\shortstack{{\bf Equal opportunities} \\ {\bf to education:} \\ 1 (strongly agree) - \\5 (stronlgy disagree)}}\\\cmidrule(lr){2-3}\cmidrule(lr){4-5}\cmidrule(lr){6-7}
                &\multicolumn{1}{c}{(1)}         &\multicolumn{1}{c}{(2)}         &\multicolumn{1}{c}{(3)}         &\multicolumn{1}{c}{(4)}         &\multicolumn{1}{c}{(5)}         &\multicolumn{1}{c}{(6)}         \\
\hline
Log (Distance to a landlocked trade point ($<$ 100,000))&   -0.477\sym{***}&                  &   -0.445\sym{**} &                  &   -0.321\sym{***}&                  \\
                &  (0.132)         &                  &  (0.202)         &                  & (0.0940)         &                  \\
Log (Distance to a landlocked trade route in 1800)&                  &   -0.398\sym{***}&                  &   -0.342\sym{**} &                  &   -0.245\sym{***}\\
                &                  &  (0.114)         &                  &  (0.139)         &                  & (0.0759)         \\
\hline
Observations    &    17427         &    17427         &     9166         &     9166         &     9252         &     9252         \\
Mean (Dep. Var.)&    2.322         &    2.322         &    2.611         &    2.611         &    1.637         &    1.637         \\
SD (Dep. Var.)  &    1.416         &    1.416         &    1.702         &    1.702         &    0.960         &    0.960         \\
F-stat          &   38.196         &   54.627         &   44.525         &   69.565         &   45.756         &   72.583         \\
No. of Clusters &      191         &      191         &      172         &      172         &      172         &      172         \\
Round FE        &      Yes         &      Yes         &       No         &       No         &       No         &       No         \\
Country
FE
&
Yes
&
Yes
&
Yes
&
Yes
&
Yes
&
Yes
\\
Geographic
Controls
&
Yes
&
Yes
&
Yes
&
Yes
&
Yes
&
Yes
\\
Individual
Controls
&
Yes
&
Yes
&
Yes
&
Yes
&
Yes
&
Yes\\
\hline\end{tabular}\begin{tablenotes}[flushleft]\footnotesize
\item
{\it
Notes}:
All
regressions
are
estimated
using
IV
with
logarithm
of
one
plus
distance
(km)
to
the
nearest
ancient
lake
as
an
instrument.
The
unit
of
observation
is
a
respondent
who
is
Muslim
in
countries
of
West
Africa
surveyed
in
Afrobarometer.
In
columns
1-2,
the
dependent
variable
is
the
ordered
variable
which
indicates
how
much
a
respondent
would
dislike
having
people
of
a
different
religion
as
neighbors.
This
variable
is
available
in
round
6
(surveyed
between
2014
and
2015)
and
7
(surveyed
between
2016
and
2018).
In
columns
3-4,
the
dependent
variable
is
the
ordered
variable
which
indicates
how
much
a
respondent
agrees
with
governance
by
religious
law
rather
than
civil
law.
This
variable
is
available
in
round
7.
In
columns
5-6,
the
dependent
variable
is
the
ordered
variable
which
indicates
how
much
a
respondent
disagrees
with
girls
and
boys
having
equal
opportunities
to
get
an
education.
This
variable
is
available
in
round
7.
The
interest
variables
are
the
logarithm
of
one
plus
distance
(km)
to
the
nearest
pre-colonial
landlocked
trade
point
whose
contemporary
population
is
less
than
100,000
and
the
logarithm
of
one
plus
distance
(km)
to
the
nearest
pre-colonial
landlocked
trade
route
up
to
1800.
Landlocked
is
defined
as
being over 1,000 km
from
the
nearest
coast
point.
Geographic
controls
include
the
logarithm
of
distance
(km)
to
the
nearest
water
sources
today,
landlocked
dummy,
average
malaria
suitability,
average
caloric
suitability
in
post
1500,
and
average
elevation.
Individual
controls
include
age,
age
squared,
female
dummy,
nine
categorical
indicators
of
education,
and
four
categorical
indicators
of
living
condition.
If
a
dependent
variable
is
available
from
multiple
rounds
of
Afrobarometer,
we
additionally
control
for
round
fixed
effects.
Standard errors clustered at the region levels in parentheses.
\sym{*} \(p<0.1\), \sym{**} \(p<0.05\), \sym{***} \(p<0.01\).
\end{tablenotes}\end{threeparttable}\end{adjustbox}\end{table}

%% file: tables/tab_exclusion_geo_institution_pop.tex
\begin{table}[htbp]\centering \def\sym#1{\ifmmode^{#1}\else\(^{#1}\)\fi} \caption{Correlations---Access to Ancient Lakes and Pre-Determined Characteristics \label{tab_exclusion_geo_institution_pop}} \begin{adjustbox}{max width=\textwidth,max totalheight=\textheight} \begin{threeparttable} \footnotesize \begin{tabular}{l*{6}{c}} \hline\hline \multicolumn{5}{l}{{\bf (A) Geography}}\\
                &\multicolumn{1}{c}{(1)}&\multicolumn{1}{c}{(2)}&\multicolumn{1}{c}{(3)}&\multicolumn{1}{c}{(4)}&\multicolumn{1}{c}{(5)}\\
                &\multicolumn{1}{c}{\shortstack[c]{Ecological \\ diversity}}&\multicolumn{1}{c}{Temperature}&\multicolumn{1}{c}{Precipitation}&\multicolumn{1}{c}{\shortstack[c]{Caloric \\ suitability}}&\multicolumn{1}{c}{\shortstack[c]{Pastoralism \\ suitability}}\\
\hline
Log (Distance to an ancient lake)&  0.00144         &  -0.0149         &    3.356\sym{***}&  0.00522\sym{***}& -0.00426         \\
                &(0.00526)         & (0.0411)         &  (0.654)         &(0.00127)         &(0.00642)         \\
\hline
R$^2$           &    0.011         &    0.606         &    0.875         &    0.994         &    0.519         \\
Adj-R$^2$       &    0.008         &    0.605         &    0.874         &    0.994         &    0.518         \\
Observations    &     2571         &     2615         &     2615         &     2616         &     2616         \\
Mean (Dep. Var.)&    0.416         &   28.065         &   54.530         &    0.532         &    0.375         \\
SD (Dep. Var.)  &    0.422         &    2.007         &   52.202         &    0.533         &    0.194         \\
\hline \multicolumn{5}{l}{{\bf (B) Pre-colonial culture and institution}}\\
                &\multicolumn{1}{c}{(1)}&\multicolumn{1}{c}{(2)}&\multicolumn{1}{c}{(3)}&\multicolumn{1}{c}{(4)}&\multicolumn{1}{c}{(5)}\\
                &\multicolumn{1}{c}{\shortstack[c]{Jurisdictional \\ hierarchy}}&\multicolumn{1}{c}{Polygamy}&\multicolumn{1}{c}{Irrigation}&\multicolumn{1}{c}{\shortstack[c]{Class \\ stratification}}&\multicolumn{1}{c}{\shortstack[c]{Local \\ headman}}\\
\hline
Log (Distance to an ancient lake)&  -0.0279         &  -0.0190         & 0.000665         &  -0.0119         &  -0.0104         \\
                & (0.0357)         & (0.0125)         & (0.0118)         & (0.0213)         &(0.00789)         \\
\hline
R$^2$           &    0.356         &    0.336         &    0.563         &    0.189         &    0.616         \\
Adj-R$^2$       &    0.353         &    0.333         &    0.561         &    0.185         &    0.614         \\
Observations    &     1749         &     1869         &     1880         &     1700         &     1345         \\
Mean (Dep. Var.)&    2.443         &    0.836         &    0.272         &    1.332         &    0.182         \\
SD (Dep. Var.)  &    0.888         &    0.371         &    0.445         &    0.705         &    0.386         \\
\hline \\ Country FE & Yes & Yes & Yes & Yes & Yes \\ Geographic Controls & Yes & Yes & Yes & Yes & Yes \\\hline \end{tabular} \begin{tablenotes}[flushleft] \footnotesize \item {\it Notes}: All regressions are estimated using OLS. The unit of observation is a grid cell (about 55km × 55km). In panel (A), the dependent varibles are (1) ecological diversity (2) average temperature (3) average precipitation (4) average caloric suitability (5) average pastoralism suitability. In panel (B), the dependent varibles from the {\it Ethnographic Atlas} are (1) jurisdictional hierarchy (v33) (2) polygamy (v9) (3) irrigation (v28) (4) class stratification (v66) (5) local headman (v72). We control for the logarithm of distance (km) to the nearest water sources today, landlocked dummy, average malaria suitability, average caloric suitability in post 1500, average elevation, ruggedness, and logarithm of one plus population in 2010 in all the specifications. We report standard errors adjusting for spatial auto-correlation with distance cutoff at 100km in parentheses. \sym{*} \(p<0.1\), \sym{**} \(p<0.05\), \sym{***} \(p<0.01\). \end{tablenotes}\end{threeparttable}\end{adjustbox}\end{table}

%% file: tables/tab_ols_city_today_water_rev.tex
\begin{table}[htbp]\centering         \def\sym#1{\ifmmode^{#1}\else\(^{#1}\)\fi}         \caption{Water Sources and Contemporary Economic Activity \label{tab_ols_city_today_water_rev}}     \begin{adjustbox}{max width=\textwidth,max totalheight=\textheight}         \begin{threeparttable}         \def\sym#1{\ifmmode^{#1}\else\(^{#1}\)\fi}         \begin{tabular}{l*{5}{c}}         \hline\hline
                &\multicolumn{2}{c}{\textbf{Log (Distance to a city ($>$ 50,000))}}&\multicolumn{2}{c}{\textbf{Nightlight luminosity in 2015}}\\\cmidrule(lr){2-3}\cmidrule(lr){4-5}
                &\multicolumn{1}{c}{(1)}         &\multicolumn{1}{c}{(2)}         &\multicolumn{1}{c}{(3)}         &\multicolumn{1}{c}{(4)}         \\
\hline
Log (Distance to an ancient lake)&  -0.0259         & -0.00935         &    0.252\sym{***}&    0.140\sym{***}\\
                & (0.0212)         & (0.0179)         & (0.0402)         & (0.0467)         \\
Log (Distance to a lake/river today)&    0.346\sym{***}&    0.315\sym{***}&   -0.522\sym{***}&   -0.439\sym{***}\\
                & (0.0254)         & (0.0243)         & (0.0704)         & (0.0643)         \\
\hline
R$^2$           &    0.660         &    0.730         &    0.432         &    0.531         \\
Adj-R$^2$       &    0.659         &    0.729         &    0.430         &    0.529         \\
Observations    &     2616         &     2616         &     2616         &     2616         \\
Mean (Dep. Var.)&    4.497         &    4.497         &    1.875         &    1.875         \\
SD (Dep. Var.)  &    1.167         &    1.167         &    2.636         &    2.636         \\
     Country FE & No & Yes & No & Yes \\ Geographic Controls & Yes & Yes & Yes & Yes         \\\hline         \end{tabular}         \begin{tablenotes}[flushleft]         \footnotesize     \item {\it Notes}: All regressions are estimated using OLS. The unit of observation is a grid cell (about 55km × 55km). All Log(Distance) variables indicate the logarithm of one plus distance (km) to the nearest object. The dependent variables are the logarithm of one plus distance (km) to the nearest city whose contemporary population over 50,000 in columns (1)-(2), and the logarithm of one plus total night light luminosity (VIISR) in 2015 in columns (3)-(4). We control for landlocked dummy, average malaria suitability, average caloric suitability in post 1500, average elevation, and ruggedness in all the specifications. We report standard errors adjusting for spatial auto-correlation with distance cutoff at 100km in parentheses. \sym{*} \(p<0.1\), \sym{**} \(p<0.05\), \sym{***} \(p<0.01\).         \end{tablenotes}\end{threeparttable}\end{adjustbox}\end{table}         

%% file: tables/tab_water_jihad_pop_rev.tex
\begin{table}[htbp]\centering         \def\sym#1{\ifmmode^{#1}\else\(^{#1}\)\fi}         \caption{Reduced Form---Ancient Water Sources and Jihad \label{tab_water_jihad_pop_rev}}     \begin{adjustbox}{max width=\textwidth,max totalheight=\textheight}         \begin{threeparttable}         \def\sym#1{\ifmmode^{#1}\else\(^{#1}\)\fi}         \begin{tabular}{l*{10}{c}}         \hline\hline
                &\multicolumn{3}{c}{\textbf{Onset}}                      &\multicolumn{3}{c}{\textbf{Intensity}}                  &\multicolumn{3}{c}{\textbf{Distance}}                   \\\cmidrule(lr){2-4}\cmidrule(lr){5-7}\cmidrule(lr){8-10}
                &\multicolumn{1}{c}{(1)}&\multicolumn{1}{c}{(2)}&\multicolumn{1}{c}{(3)}&\multicolumn{1}{c}{(4)}&\multicolumn{1}{c}{(5)}&\multicolumn{1}{c}{(6)}&\multicolumn{1}{c}{(7)}&\multicolumn{1}{c}{(8)}&\multicolumn{1}{c}{(9)}\\
                &\multicolumn{1}{c}{All}&\multicolumn{1}{c}{2001-09}&\multicolumn{1}{c}{2010-19}&\multicolumn{1}{c}{All}&\multicolumn{1}{c}{2001-09}&\multicolumn{1}{c}{2010-19}&\multicolumn{1}{c}{All}&\multicolumn{1}{c}{2001-09}&\multicolumn{1}{c}{2010-19}\\
\hline
Log (Distance to an ancient lake)&  -0.0415\sym{***}&-0.000539         &  -0.0418\sym{***}&   -0.157\sym{***}& -0.00301         &   -0.157\sym{***}&    0.148\sym{***}&   0.0396         &    0.167\sym{***}\\
                &(0.00971)         &(0.00173)         &(0.00969)         & (0.0448)         &(0.00282)         & (0.0448)         & (0.0312)         & (0.0270)         & (0.0303)         \\
Log (Distance to a lake/river today)&  -0.0430\sym{***}&  0.00341         &  -0.0465\sym{***}&   -0.101\sym{***}&  0.00334         &   -0.105\sym{***}&    0.148\sym{***}&   0.0231         &    0.140\sym{***}\\
                & (0.0122)         &(0.00318)         & (0.0120)         & (0.0333)         &(0.00342)         & (0.0331)         & (0.0395)         & (0.0319)         & (0.0397)         \\
\hline
R$^2$           &    0.328         &    0.027         &    0.338         &    0.328         &    0.028         &    0.331         &    0.560         &    0.378         &    0.611         \\
Adj-R$^2$       &    0.326         &    0.024         &    0.336         &    0.326         &    0.025         &    0.329         &    0.559         &    0.376         &    0.610         \\
Observations    &     2616         &     2616         &     2616         &     2616         &     2616         &     2616         &     2616         &     2616         &     2616         \\
Mean (Dep. Var.)&    0.133         &    0.011         &    0.129         &    0.249         &    0.011         &    0.244         &    4.816         &    5.698         &    4.932         \\
SD (Dep. Var.)  &    0.339         &    0.106         &    0.335         &    0.775         &    0.109         &    0.771         &    1.101         &    0.743         &    1.157         \\
         Country FE & Yes & Yes & Yes & Yes & Yes & Yes & Yes & Yes & Yes\\ Geographic Controls & Yes & Yes & Yes & Yes & Yes & Yes & Yes & Yes & Yes         \\\hline         \end{tabular}         \begin{tablenotes}[flushleft]         \footnotesize         \item {\it Notes}: All regressions are estimated using OLS. The unit of observation is a grid cell (about 55km × 55km). The dependent variables are dummy variables which take a value of 1 if jihad occurred during a period given in each column, otherwise take a value of 0 in columns (1)-(3), logarithm of one plus the number of jihad events during a given period in each column in columns (4)-(6), and logarithm of one plus distance (km) to the nearest jihad during a period given in each column in columns (7)-(9). All Log(Distance) variables indicate the logarithm of one plus distance (km) to the nearest object. We control for landlocked dummy, average malaria suitability, average caloric suitability in post 1500, average elevation, ruggedness, and logarithm of one plus population in 2010 in all the specifications. We report standard errors adjusting for spatial auto-correlation with distance cutoff at 100km in parentheses. \sym{*} \(p<0.1\), \sym{**} \(p<0.05\), \sym{***} \(p<0.01\).         \end{tablenotes}\end{threeparttable}\end{adjustbox}\end{table}         

%% file: tables/tab_colonial_jihad_lake.tex
\begin{table}[htbp]\centering         \caption{Ancient Water Sources, Historical Core Cities, and Historical Jihad \label{tab_colonial_jihad_lake}}     \begin{adjustbox}{max width=\textwidth,max totalheight=\textheight}         \begin{threeparttable}         \def\sym#1{\ifmmode^{#1}\else\(^{#1}\)\fi}         \begin{tabular}{l*{7}{c}}         \hline\hline
                &\multicolumn{2}{c}{\textbf{Onset}}   &\multicolumn{2}{c}{\textbf{Duration}}&\multicolumn{2}{c}{\textbf{Distance}}\\\cmidrule(lr){2-3}\cmidrule(lr){4-5}\cmidrule(lr){6-7}
                &\multicolumn{1}{c}{(1)}&\multicolumn{1}{c}{(2)}&\multicolumn{1}{c}{(3)}&\multicolumn{1}{c}{(4)}&\multicolumn{1}{c}{(5)}&\multicolumn{1}{c}{(6)}\\
                &\multicolumn{1}{c}{OLS}&\multicolumn{1}{c}{IV}&\multicolumn{1}{c}{OLS}&\multicolumn{1}{c}{IV}&\multicolumn{1}{c}{OLS}&\multicolumn{1}{c}{IV}\\
\hline
Log (Distance to an ancient lake)&0.0000728         &                  &  0.00371         &                  &   0.0263         &                  \\
                &(0.00151)         &                  &(0.00541)         &                  & (0.0224)         &                  \\
Log (Distance to a landlocked trade point ($<$ 100,000))&                  & 0.000685         &                  &   0.0349         &                  &    0.248         \\
                &                  & (0.0142)         &                  & (0.0523)         &                  &  (0.203)         \\
\hline
R$^2$           &    0.009         &    0.008         &    0.003         &   -0.001         &    0.214         &    0.242         \\
Adj-R$^2$       &    0.006         &    0.006         &    0.000         &   -0.004         &    0.212         &    0.240         \\
Observations    &     2616         &     2616         &     2616         &     2616         &     2616         &     2616         \\
Mean (Dep. Var.)&    0.013         &    0.013         &    0.036         &    0.036         &    5.612         &    5.612         \\
SD (Dep. Var.)  &    0.112         &    0.112         &    0.450         &    0.450         &    0.802         &    0.802         \\
Colonizer
FE
&
Yes
&
Yes
&
Yes
&
Yes
&
Yes
&
Yes
\\
Geographic
Controls
&
Yes
&
Yes
&
Yes
&
Yes
&
Yes
&
Yes
\\\hline
\end{tabular}
\begin{tablenotes}[flushleft]
\footnotesize
\item
{\it
Notes}:
Columns (1), (3), and (5) report OLS estimates, while columns (2), (4), and (6) report IV estimates using the logarithm of one plus distance (km) to the nearest ancient lake as an instrument.
The
unit
of
observation
is
a
grid
cell
(about
55km
×
55km).
The
dependent
variables
are
dummy
variables
which
take
a
value
of
1
if
any
historical
jihad,
otherwise
take
a
value
of
0
(Onset),
total
duration
years
of
historical
jihad
(Duration),
and
logarithm
of
one
plus
distance
(km)
to
the
nearest
historical
jihad
(Distance).
All
Log(Distance)
variables
indicate
the
logarithm
of
one
plus
distance
(km)
to
the
nearest
object.
Landlocked
is
defined
as
being over 1,000 km
from
the
nearest
coast
point.
We
control
for
the
logarithm
of
distance
(km)
to
the
nearest
water
sources
today,
landlocked
dummy,
average
malaria
suitability,
average
caloric
suitability
in
post
1500,
average
elevation,
and
ruggedness
in
all
the
specifications.
We
report
standard
errors
adjusting
for
spatial
auto-correlation
with
distance
cutoff
at
100km
in
parentheses.
\sym{*} \(p<0.1\), \sym{**} \(p<0.05\), \sym{***} \(p<0.01\).
\end{tablenotes}\end{threeparttable}\end{adjustbox}\end{table}

%% file: tables/tab_historical_mosque.tex
\begin{table}[htbp]\centering         \caption{Historical Trade Networks and Historical Mosques \label{tab_historical_mosque}}     \begin{adjustbox}{max width=\textwidth,max totalheight=\textheight}         \begin{threeparttable}         \def\sym#1{\ifmmode^{#1}\else\(^{#1}\)\fi}         \begin{tabular}{l*{9}{c}}         \hline\hline          & \multicolumn{8}{c}{Log distance to historical mosque before}\\
                &\multicolumn{4}{c}{1860}                                                   &\multicolumn{4}{c}{1900}                                                   \\\cmidrule(lr){2-5}\cmidrule(lr){6-9}
                &\multicolumn{1}{c}{(1)}         &\multicolumn{1}{c}{(2)}         &\multicolumn{1}{c}{(3)}         &\multicolumn{1}{c}{(4)}         &\multicolumn{1}{c}{(5)}         &\multicolumn{1}{c}{(6)}         &\multicolumn{1}{c}{(7)}         &\multicolumn{1}{c}{(8)}         \\
\hline
Log (Distance to a landlocked trade point ($<$ 100,000))&   -0.129         &    0.362\sym{**} &                  &                  &   -0.154         &    0.368\sym{**} &                  &                  \\
                &  (0.218)         &  (0.181)         &                  &                  &  (0.219)         &  (0.181)         &                  &                  \\
Log (Distance to a landlocked trade route up to 1800)&                  &                  &   -0.158         &    0.272\sym{**} &                  &                  &   -0.187         &    0.276\sym{**} \\
                &                  &                  &  (0.292)         &  (0.135)         &                  &                  &  (0.299)         &  (0.136)         \\
\hline
Observations    &     2616         &     2616         &     2616         &     2616         &     2616         &     2616         &     2616         &     2616         \\
Mean (Dep. Var.)&    5.196         &    5.196         &    5.196         &    5.196         &    5.181         &    5.181         &    5.181         &    5.181         \\
SD (Dep. Var.)  &    0.887         &    0.887         &    0.887         &    0.887         &    0.894         &    0.894         &    0.894         &    0.894         \\
Colonizer
FE
&
Yes
&
No
&
Yes
&
No
&
Yes
&
No
&
Yes
&
No
\\
Country
FE
&
No
&
Yes
&
No
&
Yes
&
No
&
Yes
&
No
&
Yes
\\
Geographic
Controls
&
Yes
&
Yes
&
Yes
&
Yes
&
Yes
&
Yes
&
Yes
&
Yes
\\\hline
\end{tabular}
\begin{tablenotes}[flushleft]
\footnotesize
\item
{\it
Notes}:
All
regressions
are
estimated
using
IV
with
logarithm
of
one
plus
distance
(km)
to
the
nearest
ancient
lake
as
an
instrument.
The
unit
of
observation
is
a
grid
cell
(about
55km
×
55km).
The dependent variables are the logarithm of one plus distance (km) to the nearest historical mosque built before 1860 (columns 1–4), and before 1900 (columns 5–8).
All
Log(Distance)
variables
indicate
the
logarithm
of
one
plus
distance
(km)
to
the
nearest
object.
Landlocked
is
defined
as
being over 1,000 km
from
the
nearest
coast
point.
We
control
for
the
logarithm
of
distance
(km)
to
the
nearest
water
sources
today,
landlocked
dummy,
average
malaria
suitability,
average
caloric
suitability
in
post
1500,
average
elevation,
and
ruggedness
in
all
the
specifications.
We
report
standard
errors
adjusting
for
spatial
auto-correlation
with
distance
cutoff
at
100km
in
parentheses.
\sym{*} \(p<0.1\), \sym{**} \(p<0.05\), \sym{***} \(p<0.01\).
\end{tablenotes}\end{threeparttable}\end{adjustbox}\end{table}

%% file: tables/tab_iv_jihad_nonjihad_sahara_pop.tex
\begin{table}[htbp]\centering         \def\sym#1{\ifmmode^{#1}\else\(^{#1}\)\fi}         \caption{Persistent Effects on Jihadist and Non-Jihadist Violence in Saharan Countries \label{tab_iv_jihad_nonjihad_sahara_pop}}         \begin{adjustbox}{max width=\textwidth,max totalheight=\textheight}         \begin{threeparttable}         \footnotesize         \begin{tabular}{l*{10}{c}}         \hline\hline         \multicolumn{6}{l}{\textbf{(A) Jihad}}\\
                &\multicolumn{3}{c}{\textbf{Onset}}                      &\multicolumn{3}{c}{\textbf{Intensity}}                  &\multicolumn{3}{c}{\textbf{Distance}}                   \\\cmidrule(lr){2-4}\cmidrule(lr){5-7}\cmidrule(lr){8-10}
                &\multicolumn{1}{c}{(1)}&\multicolumn{1}{c}{(2)}&\multicolumn{1}{c}{(3)}&\multicolumn{1}{c}{(4)}&\multicolumn{1}{c}{(5)}&\multicolumn{1}{c}{(6)}&\multicolumn{1}{c}{(7)}&\multicolumn{1}{c}{(8)}&\multicolumn{1}{c}{(9)}\\
                &\multicolumn{1}{c}{2010-19}&\multicolumn{1}{c}{2010-15}&\multicolumn{1}{c}{2016-19}&\multicolumn{1}{c}{2010-19}&\multicolumn{1}{c}{2010-15}&\multicolumn{1}{c}{2016-19}&\multicolumn{1}{c}{2010-19}&\multicolumn{1}{c}{2010-15}&\multicolumn{1}{c}{2016-19}\\
\hline
Log (Distance to a landlocked trade point ($<$ 100,000))&   -0.139\sym{*}  &   -0.188\sym{**} &   -0.137         &   -0.458\sym{**} &   -0.301\sym{**} &   -0.392\sym{*}  &    0.575\sym{*}  &    0.978\sym{**} &    0.600\sym{*}  \\
                & (0.0841)         & (0.0790)         & (0.0842)         &  (0.224)         &  (0.131)         &  (0.206)         &  (0.337)         &  (0.388)         &  (0.326)         \\
Log (Distance to a lake/river today)&  -0.0791\sym{***}&  -0.0643\sym{***}&  -0.0895\sym{***}&   -0.200\sym{***}&  -0.0888\sym{***}&   -0.184\sym{***}&    0.217\sym{***}&    0.128\sym{*}  &    0.277\sym{***}\\
                & (0.0176)         & (0.0194)         & (0.0171)         & (0.0482)         & (0.0284)         & (0.0436)         & (0.0632)         & (0.0722)         & (0.0573)         \\
\hline
Observations    &     1616         &     1616         &     1616         &     1616         &     1616         &     1616         &     1616         &     1616         &     1616         \\
Mean (Dep. Var.)&    0.110         &    0.051         &    0.095         &    0.186         &    0.070         &    0.154         &    5.038         &    5.270         &    5.333         \\
SD (Dep. Var.)  &    0.312         &    0.220         &    0.293         &    0.619         &    0.355         &    0.551         &    1.114         &    0.961         &    1.157         \\
\hline         \multicolumn{6}{l}{\textbf{(B) Non-Jihad}}\\
                &\multicolumn{3}{c}{\textbf{Onset}}                      &\multicolumn{3}{c}{\textbf{Intensity}}                  &\multicolumn{3}{c}{\textbf{Distance}}                   \\\cmidrule(lr){2-4}\cmidrule(lr){5-7}\cmidrule(lr){8-10}
                &\multicolumn{1}{c}{(1)}&\multicolumn{1}{c}{(2)}&\multicolumn{1}{c}{(3)}&\multicolumn{1}{c}{(4)}&\multicolumn{1}{c}{(5)}&\multicolumn{1}{c}{(6)}&\multicolumn{1}{c}{(7)}&\multicolumn{1}{c}{(8)}&\multicolumn{1}{c}{(9)}\\
                &\multicolumn{1}{c}{All}&\multicolumn{1}{c}{2001-09}&\multicolumn{1}{c}{2010-19}&\multicolumn{1}{c}{All}&\multicolumn{1}{c}{2001-09}&\multicolumn{1}{c}{2010-19}&\multicolumn{1}{c}{All}&\multicolumn{1}{c}{2001-09}&\multicolumn{1}{c}{2010-19}\\
\hline
Log (Distance to a landlocked trade point ($<$ 100,000))&   0.0185         &    0.132\sym{*}  &  -0.0732         &    0.137         &    0.201         &  -0.0391         &   -0.259         &   -0.867\sym{***}&  0.00606         \\
                & (0.0947)         & (0.0711)         & (0.0826)         &  (0.193)         &  (0.126)         &  (0.153)         &  (0.275)         &  (0.309)         &  (0.248)         \\
Log (Distance to a lake/river today)&  -0.0397\sym{*}  &   0.0168         &  -0.0613\sym{***}&  -0.0953\sym{**} &   0.0368         &   -0.133\sym{***}&    0.138\sym{***}&   0.0857         &    0.201\sym{***}\\
                & (0.0217)         & (0.0142)         & (0.0190)         & (0.0462)         & (0.0234)         & (0.0393)         & (0.0500)         & (0.0619)         & (0.0447)         \\
\hline
Observations    &     1616         &     1616         &     1616         &     1616         &     1616         &     1616         &     1616         &     1616         &     1616         \\
Mean (Dep. Var.)&    0.166         &    0.062         &    0.130         &    0.256         &    0.085         &    0.194         &    4.433         &    5.047         &    4.576         \\
SD (Dep. Var.)  &    0.372         &    0.242         &    0.336         &    0.676         &    0.383         &    0.590         &    0.930         &    0.934         &    0.892         \\
\hline
\\
Country
FE
&
Yes
&
Yes
&
Yes
&
Yes
&
Yes
&
Yes
&
Yes
&
Yes
&
Yes\\
Geographic
Controls
&
Yes
&
Yes
&
Yes
&
Yes
&
Yes
&
Yes
&
Yes
&
Yes
&
Yes
\\\hline
\end{tabular}
\begin{tablenotes}[flushleft]
\footnotesize
\item
{\it
Notes}:
All
regressions
are
estimated
using
IV
with
logarithm
of
one
plus
distance
(km)
to
the
nearest
ancient
lake
as
an
instrument.
The
unit
of
observation
is
a
grid
cell
(about
55km
×
55km). The observations restrict grid cells in Mauritania, Mali, Niger and Chad. 
In Panels (A) and (B), the dependent variables are defined analogously for jihadist and non-jihadist events, respectively. In columns (1)–(3), they are dummy variables equal to 1 if at least one event occurred during the period given in each column, and 0 otherwise. In columns (4)–(6), they are the logarithm of one plus the number of events during the corresponding period. In columns (7)–(9), they are the logarithm of one plus the distance (in km) to the nearest event during the corresponding period.
All
Log(Distance)
variables
indicate
the
logarithm
of
one
plus
distance
(km)
to
the
nearest
object.
The
interest
variable
is
the
logarithm
of
one
plus
distance
(km)
to
the
nearest
pre-colonial
landlocked
trade
point
whose
contemporary
population
is
less
than
100,000.
Landlocked
is
defined
as
being over 1,000 km
from
the
nearest
coast
point.
We
control
for
the
logarithm
of
distance
(km)
to
the
nearest
water
sources
today,
landlocked
dummy,
average
malaria
suitability,
average
caloric
suitability
in
post
1500,
average
elevation,
ruggedness,
and
logarithm
of
one
plus
population
in
2010
in
all
the
specifications.
We
report
standard
errors
adjusting
for
spatial
auto-correlation
with
distance
cutoff
at
100km
in
parentheses.
\sym{*} \(p<0.1\), \sym{**} \(p<0.05\), \sym{***} \(p<0.01\).
\end{tablenotes}\end{threeparttable}\end{adjustbox}\end{table}

%% file: tables/tab_jihad_ucdp_pop_rev.tex
\begin{table}[htbp]\centering         \caption{IV Estimates of Persistent Effects on Jihad (UCDP) \label{tab_jihad_ucdp_pop_rev}}     \begin{adjustbox}{max width=\textwidth,max totalheight=\textheight}         \begin{threeparttable}         \def\sym#1{\ifmmode^{#1}\else\(^{#1}\)\fi}         \begin{tabular}{l*{10}{c}}         \hline\hline
                &\multicolumn{3}{c}{\textbf{Onset}}                      &\multicolumn{3}{c}{\textbf{Intensity}}                  &\multicolumn{3}{c}{\textbf{Distance}}                   \\\cmidrule(lr){2-4}\cmidrule(lr){5-7}\cmidrule(lr){8-10}
                &\multicolumn{1}{c}{(1)}&\multicolumn{1}{c}{(2)}&\multicolumn{1}{c}{(3)}&\multicolumn{1}{c}{(4)}&\multicolumn{1}{c}{(5)}&\multicolumn{1}{c}{(6)}&\multicolumn{1}{c}{(7)}&\multicolumn{1}{c}{(8)}&\multicolumn{1}{c}{(9)}\\
                &\multicolumn{1}{c}{All}&\multicolumn{1}{c}{2001-09}&\multicolumn{1}{c}{2010-19}&\multicolumn{1}{c}{All}&\multicolumn{1}{c}{2001-09}&\multicolumn{1}{c}{2010-19}&\multicolumn{1}{c}{All}&\multicolumn{1}{c}{2001-09}&\multicolumn{1}{c}{2010-19}\\
\hline
Log (Distance to a landlocked trade point ($<$ 100,000))&   -0.363\sym{***}&  -0.0840\sym{**} &   -0.374\sym{***}&   -1.248\sym{***}&  -0.0827\sym{**} &   -1.250\sym{***}&    1.322\sym{***}&    1.112\sym{***}&    1.473\sym{***}\\
                & (0.0808)         & (0.0406)         & (0.0814)         &  (0.354)         & (0.0402)         &  (0.355)         &  (0.270)         &  (0.214)         &  (0.280)         \\
\hline
Observations    &     2616         &     2616         &     2616         &     2616         &     2616         &     2616         &     2616         &     2616         &     2616         \\
Mean (Dep. Var.)&    0.123         &    0.015         &    0.120         &    0.217         &    0.013         &    0.214         &    4.959         &    5.866         &    5.070         \\
SD (Dep. Var.)  &    0.328         &    0.123         &    0.325         &    0.712         &    0.112         &    0.708         &    1.099         &    0.779         &    1.169         \\
         Country FE & Yes & Yes & Yes & Yes & Yes & Yes & Yes & Yes & Yes\\ Geographic Controls & Yes & Yes & Yes & Yes & Yes & Yes & Yes & Yes & Yes         \\\hline         \end{tabular}         \begin{tablenotes}[flushleft]         \footnotesize         \item {\it Notes}: All regressions are estimated using IV with logarithm of one plus distance (km) to the nearest ancient lake as an instrument. The unit of observation is a grid cell (about 55km × 55km). In this table, the jihadistic events data comes from Uppsala Conflict Data Program (UCDP) in 2001-2019. The dependent variables are dummy variables which take a value of 1 if jihad occurred during a period given in each column, otherwise take a value of 0 in columns (1)-(3), logarithm of one plus the number of jihad events during a given period in each column in columns (4)-(6), and logarithm of one plus distance (km) to the nearest jihad during a period given in each column in columns (7)-(9). All Log(Distance) variables indicate the logarithm of one plus distance (km) to the nearest object. Landlocked is defined as being over 1,000 km from the nearest coast point. The interest variable is  the logarithm of one plus distance (km) to the nearest pre-colonial landlocked trade point whose contemporary population is less than 100,000. We control for the logarithm of distance (km) to the nearest water sources today, landlocked dummy, average malaria suitability, average caloric suitability in post 1500, average elevation, ruggedness, and logarithm of one plus population in 2010 in all the specifications. We report standard errors adjusting for spatial auto-correlation with distance cutoff at 100km in parentheses. \sym{*} \(p<0.1\), \sym{**} \(p<0.05\), \sym{***} \(p<0.01\).         \end{tablenotes}\end{threeparttable}\end{adjustbox}\end{table}         

%% file: tables_manual/tab_spatial_se_manual.tex
\begin{table}[htbp]\centering \def\sym#1{\ifmmode^{#1}\else\(^{#1}\)\fi} \caption{Robustness to Spatial-Autocorrelation \label{tab_spatial_se_manual}} \begin{adjustbox}{max width=\textwidth,max totalheight=\textheight} \begin{threeparttable} \footnotesize \begin{tabular}{l*{4}{c}} \hline\hline {\bf (A)} & \multicolumn{3}{c}{{\bf Onset}}\\
                &\multicolumn{1}{c}{(1)}&\multicolumn{1}{c}{(2)}&\multicolumn{1}{c}{(3)}\\
                &\multicolumn{1}{c}{2010-19}&\multicolumn{1}{c}{2010-15}&\multicolumn{1}{c}{2016-19}\\
\hline
Log (Distance to a landlocked trade point ($<$ 100,000))&   -0.327         &   -0.340         &   -0.314         \\
\hspace{20em} \textit{200km cutoff}                &  [0.125]\sym{***}&  [0.121]\sym{***}&  [0.142]\sym{**} \\
\hspace{20em} \textit{300km cutoff}                &  [0.134]\sym{**} &  [0.132]\sym{**} &  [0.155]\sym{**} \\
\hspace{20em} \textit{400km cutoff}                &  [0.122]\sym{***}&  [0.121]\sym{***}&  [0.140]\sym{**} \\
\hspace{20em} \textit{500km cutoff}                &  [0.102]\sym{***}&  [0.102]\sym{***}&  [0.116]\sym{***}\\
\hspace{19.4em} \textit{1000km cutoff}                &  [0.108]\sym{***}& [0.0927]\sym{***}&  [0.121]\sym{***}\\
\hline
Observations    &     2616         &     2616         &     2616         \\
Mean (Dep. Var.)&    0.129         &    0.071         &    0.100         \\
SD (Dep. Var.)  &    0.335         &    0.258         &    0.301         \\
\hline {\bf (B)} & \multicolumn{3}{c}{{\bf Intensity}}\\
                &\multicolumn{1}{c}{(1)}&\multicolumn{1}{c}{(2)}&\multicolumn{1}{c}{(3)}\\
                &\multicolumn{1}{c}{2010-19}&\multicolumn{1}{c}{2010-15}&\multicolumn{1}{c}{2016-19}\\
\hline
Log (Distance to a landlocked trade point ($<$ 100,000))&   -1.234         &   -0.892         &   -1.042         \\
\hspace{20em} \textit{200km cutoff}                &  [0.576]\sym{**} &  [0.425]\sym{**} &  [0.540]\sym{*}  \\
\hspace{20em} \textit{300km cutoff}                &  [0.613]\sym{**} &  [0.449]\sym{**} &  [0.570]\sym{*}  \\
\hspace{20em} \textit{400km cutoff}                &  [0.558]\sym{**} &  [0.419]\sym{**} &  [0.510]\sym{**} \\
\hspace{20em} \textit{500km cutoff}                &  [0.474]\sym{***}&  [0.364]\sym{**} &  [0.429]\sym{**} \\
\hspace{19.4em} \textit{1000km cutoff}                &  [0.474]\sym{***}&  [0.339]\sym{***}&  [0.425]\sym{**} \\
\hline
Observations    &     2616         &     2616         &     2616         \\
Mean (Dep. Var.)&    0.244         &    0.118         &    0.188         \\
SD (Dep. Var.)  &    0.771         &    0.520         &    0.668         \\
\hline {\bf (C)} & \multicolumn{3}{c}{{\bf Distance}}\\
                &\multicolumn{1}{c}{(1)}&\multicolumn{1}{c}{(2)}&\multicolumn{1}{c}{(3)}\\
                &\multicolumn{1}{c}{2010-19}&\multicolumn{1}{c}{2010-15}&\multicolumn{1}{c}{2016-19}\\
\hline
Log (Distance to a landlocked trade point ($<$ 100,000))&    1.306         &    1.421         &    1.493         \\
\hspace{20em} \textit{200km cutoff}                &  [0.466]\sym{***}&  [0.415]\sym{***}&  [0.515]\sym{***}\\
\hspace{20em} \textit{300km cutoff}                &  [0.525]\sym{**} &  [0.445]\sym{***}&  [0.595]\sym{**} \\
\hspace{20em} \textit{400km cutoff}                &  [0.520]\sym{**} &  [0.420]\sym{***}&  [0.598]\sym{**} \\
\hspace{20em} \textit{500km cutoff}                &  [0.498]\sym{***}&  [0.388]\sym{***}&  [0.573]\sym{***}\\
\hspace{19.4em} \textit{1000km cutoff}                &  [0.588]\sym{**} &  [0.427]\sym{***}&  [0.666]\sym{**} \\
\hline
Observations    &     2616         &     2616         &     2616         \\
Mean (Dep. Var.)&    4.932         &    5.211         &    5.264         \\
SD (Dep. Var.)  &    1.157         &    1.021         &    1.180         \\
\hline \\ Country FE & Yes & Yes & Yes \\ Geographic Controls & Yes & Yes & Yes \\\hline \end{tabular} \begin{tablenotes}[flushleft] \footnotesize \item {\it Notes}: All regressions are estimated using IV with logarithm of the Ancient Water Access as an instrument. The unit of observation is a grid cell (about 55km × 55km). The dependent variables are (A) dummy variables which take a value of 1 if jihad occurred during a period given in each column, otherwise take a value of 0, (B) logarithm of one plus the number of jihad events during a given period in each column, and (C) logarithm of one plus distance (km) to the nearest jihad during a period given in each column. All Log(Distance) variables indicate the logarithm of one plus distance (km) to the nearest object. Landlocked is defined as being over 1,000 km from the nearest coast point. The interest variables are the logarithm of one plus distance (km) to the nearest pre-colonial landlocked trade point whose contemporary population is less than 100,000 in columns (1)-(3). We control for landlocked dummy, average malaria suitability, average caloric suitability in post 1500, average elevation, ruggedness, and logarithm of one plus population in 2010 in all the specifications. We report standard errors adjusting for spatial auto-correlation with distance cutoffs in brackets. \sym{*} \(p<0.1\), \sym{**} \(p<0.05\), \sym{***} \(p<0.01\). \end{tablenotes}\end{threeparttable}\end{adjustbox}\end{table}

%% file: tables/table_brecke.tex
\begin{table}[htbp] \centering \begin{adjustbox}{max width=\textwidth,max totalheight=\textheight} \begin{threeparttable}\caption{Colonial Conflicts in West Africa involving Historical States}\begin{tabular}{ccccc}
\toprule
Historical State & Islamic & European Enemy &  Start Year &  End Year \\
\midrule
          Djolof &       N &       Portugal &        1697 &      1697 \\
         Ashanti &       N &        Britain &        1711 &      1712 \\
         Dahomey &       N &        Britain &        1727 &      1729 \\
         Ashanti &       N &        Denmark &        1742 &      1742 \\
         Ashanti &       N &        Denmark &        1743 &      1743 \\
         Ashanti &       N &        Britain &        1823 &      1826 \\
           Ijebu &       N &        Britain &        1851 &      1851 \\
         Tukulor &       Y &         France &        1854 &      1861 \\
          Saloum &       N &         France &        1856 &      1858 \\
            Koya &       N &        Britain &        1861 &      1861 \\
       Futa Toro &       Y &         France &        1862 &      1862 \\
         Ashanti &       N &        Britain &        1863 &      1864 \\
           Cayor &       N &         France &        1864 &      1864 \\
         Dahomey &       N &        Britain &        1864 &      1865 \\
         Ashanti &       N &        Britain &        1865 &      1865 \\
         Ashanti &       N &        Britain &        1868 &      1869 \\
           Cayor &       N &         France &        1869 &      1869 \\
         Ashanti &       N &        Britain &        1873 &      1874 \\
       Futa Toro &       Y &         France &        1875 &      1875 \\
         Dahomey &       N &        Britain &        1878 &      1878 \\
       Wassoulou &       Y &         France &        1881 &      1888 \\
       Wassoulou &       Y &         France &        1885 &      1886 \\
       Wassoulou &       Y &         France &        1888 &      1891 \\
         Dahomey &       N &         France &        1889 &      1890 \\
         Dahomey &       N &         France &        1892 &      1893 \\
         Ashanti &       N &        Britain &        1893 &      1894 \\
       Wassoulou &       Y &         France &        1894 &      1895 \\
         Ashanti &       N &        Britain &        1895 &      1896 \\
           Benin &       N &        Britain &        1897 &      1897 \\
       Wassoulou &       Y &         France &        1898 &      1898 \\
     Kanem-Bornu &       Y &         France &        1899 &      1901 \\
         Ashanti &       N &        Britain &        1900 &      1903 \\
             Aro &       N &        Britain &        1901 &      1901 \\
     Kanem-Bornu &       Y &        Britain &        1902 &      1902 \\
     Kanem-Bornu &       Y &        Britain &        1902 &      1902 \\
 Sokoto Califate &       Y &        Britain &        1903 &      1903 \\
 Sokoto Califate &       Y &        Britain &        1906 &      1906 \\
           Ijebu &       N &        Britain &        1912 &      1913 \\
           Benin &       N &         France &        1914 &      1914 \\
           Benin &       N &         France &        1915 &      1916 \\
       Wassoulou &       Y &         France &        1915 &      1915 \\
       Wassoulou &       Y &         France &        1916 &      1916 \\
\bottomrule
\end{tabular}
\label{tab_brecke}\begin{tablenotes} \begin{spacing}{0.5}\item {\scriptsize \selectfont {\it Note:} The names of the historical states come from Culture of West Africa.                                                        Y indicates a state is Islamic and N indicates it is not.                                                        The two conflicts against Britain in 1902 involving with Kanem-Bornu are not the same.}\end{spacing} \end{tablenotes}\end{threeparttable}\end{adjustbox}\end{table} 

%% file: tables/tab_reg_jihad_ma_panel2.tex
\begin{table}[htbp]\centering \begin{adjustbox}{max width=\textwidth,max totalheight=\textheight} \begin{threeparttable} \def\sym#1{\ifmmode^{#1}\else\(^{#1}\)\fi} \caption{Market Access and Jihadist Violence by Major Organizations \label{tab_jihad_ma}} \begin{tabular}{l*{7}{c}} \hline\hline & \multicolumn{6}{c}{log (Number of Jihadist Violence)}\\
                &\multicolumn{2}{c}{Al Qaeda}         &\multicolumn{2}{c}{Islamic State}    &\multicolumn{2}{c}{Boko Haram}       \\\cmidrule(lr){2-3}\cmidrule(lr){4-5}\cmidrule(lr){6-7}
                &\multicolumn{1}{c}{(1)}         &\multicolumn{1}{c}{(2)}         &\multicolumn{1}{c}{(3)}         &\multicolumn{1}{c}{(4)}         &\multicolumn{1}{c}{(5)}         &\multicolumn{1}{c}{(6)}         \\
\hline
log (ILMA)      & -0.00554\sym{**} & -0.00321         &   0.0256\sym{***}&   0.0171\sym{***}&   0.0454\sym{***}&   0.0286\sym{***}\\
                &(0.00225)         &(0.00213)         &(0.00404)         &(0.00369)         &(0.00561)         &(0.00497)         \\
log (ITA)      &  -0.0189\sym{***}&  0.00409         &  -0.0591\sym{***}&   -0.143\sym{***}&  -0.0831\sym{***}&   -0.249\sym{***}\\
                &(0.00734)         & (0.0102)         &(0.00826)         & (0.0181)         &(0.00888)         & (0.0224)         \\
log (Population)&                  &  -0.0109\sym{***}&                  &   0.0399\sym{***}&                  &   0.0789\sym{***}\\
                &                  &(0.00397)         &                  &(0.00559)         &                  &(0.00768)         \\
\hline
Country $\times$ Year FE&      Yes         &      Yes         &      Yes         &      Yes         &      Yes         &      Yes         \\
R$^2$           &    0.307         &    0.307         &    0.069         &    0.075         &    0.052         &    0.069         \\
Adjusted R$^2$  &    0.299         &    0.299         &    0.059         &    0.064         &    0.041         &    0.058         \\
Mean (Dep. Var.)&    0.020         &    0.020         &    0.018         &    0.018         &    0.024         &    0.024         \\
SD (Dep. Var.)  &    0.194         &    0.194         &    0.185         &    0.185         &    0.209         &    0.209         \\
Observations    &    15320         &    15320         &    15320         &    15320         &    15320         &    15320         \\
\hline \end{tabular} \begin{tablenotes}[flushleft] \footnotesize \item {\it Notes}: Robust standard errors in parentheses. The sample includes all districts in West Africa from 2010 to 2019. Other controls include district area size. ITA $\equiv$ Insurgent's Target Market Access. ILMA $\equiv$ Insurgent's Labor Market Access. \item \sym{*} \(p<0.1\), \sym{**} \(p<0.05\), \sym{***} \(p<0.01\). \end{tablenotes}\end{threeparttable}\end{adjustbox}\end{table}